\documentclass{amsart}

\usepackage{amssymb,amsfonts,latexsym,amsthm,amsmath,amscd}

\def\aind{{\underline{a} }}

\def\Bra{\Big\langle}
\def\bra{\big\langle}
\def\bch{\bar\chi}

\def\C{{\mathbb C}}
\def\cL{{\mathcal L}}

\def\coe{c_1}
\def\cob{c_2}

\def\cm{c_3}
\def\ccontr{c_4}

\def\Dom{{\rm Dom}}

\def\der{\delta}

\def\e{\epsilon}
\def\eps{\e}
\def\err{{\rm err}}

\def\Ez{E_\rho}

\def\Fo{{\mathcal F}}
\def\Fpairs{{\mathfrak F \mathfrak P}}

\def\H{{\mathcal H}}
\def\h{{\underline{w}}}

\def\I{D_{\frac{1}{10}}}

\def\Ket{\Big\rangle}
\def\ket{\big\rangle}
\def\knull{\kappa}
\def\ks{\knull_\sigma}

\def\lTnl{\lambda}

\def\N{{\mathbb N}}

\def\opp{{\underline{\mathcal P}}}

\def\Ppar{P_f^\parallel}
\def\piop{\Upsilon}
\def\Polyd{{\mathfrak D}}
\def\Polpar{(\e ,\delta,\lTnl)}
\def\Polydrho{\Polyd(\frac{\rho}{20},\frac{\rho}{20},\frac12)}

\def\R{{\mathbb R}}
\def\Ran{{\rm Ran}}

\def\ren{{\mathcal R}_\rho}
\def\renop{R_\rho}
\def\resc{S_\rho}
\def\rvar{x}

\def\spvar{{\underline{X}}}

\def\sbsrm{{\rm{\bf SR}}[\mu]}
\def\sbsr{{\bf SR}$[\mu]$ }

\def\tuk{{\underline{\tilde k}}}

\def\tT{\tilde T}

\def\z{z}

\def\ua{\underline{a}}

\def\uk{{\underline{k}}}

\def\unull{{\underline{0}}}

\def\vac{\Omega}

\def\vk{{ k}}

\def\vp{{ p}}

\def\cB{{\mathcal B}}

\def\Hspace{\underline{{\mathfrak W}}}

\def\Wspace{{\mathfrak W}}

\def\Tspace{{\mathfrak T}}

\def\1{{\bf 1}}

\def\eqnn{\begin{eqnarray*}}
\def\eeqnn{\end{eqnarray*}}
\def\eqn{\begin{eqnarray}}
\def\eeqn{\end{eqnarray}}

\theoremstyle{plain}
\newtheorem{theorem}{Theorem}[section]
\newtheorem{definition}[theorem]{Definition}
\newtheorem{proposition}[theorem]{Proposition}
\newtheorem{hypothesis}{Condition}[section]
\newtheorem{lemma}[theorem]{Lemma}

\newtheorem{remark}[theorem]{Remark}

\numberwithin{equation}{section}

\def\prf{\begin{proof}}
\def\endprf{\end{proof}}

\begin{document}
\bibliographystyle{plain}

\title[Renormalized Electron Mass in Non-Relativistic QED]
{The Renormalized Electron Mass in Non-Relativistic Quantum Electrodynamics}

\author[V. Bach]{Volker Bach}
\address{V. Bach,
FB Mathematik, Universit\"at Mainz, D-55099 Mainz,  Germany.}
\email{vbach@mathematik.uni-mainz.de}

\author[T. Chen]{Thomas Chen}
\address{T. Chen,
Department of Mathematics, Princeton University, Fine Hall, Washington Road, Princeton, NJ 08544, USA.}
\email{tc@math.princeton.edu}

\author[J. Fr\"ohlich]{J\"urg Fr\"ohlich}
\address{J. Fr\"ohlich,
Institut f\"ur Theoretische Physik, ETH H\"{o}nggerberg,  CH-8093 Z\"{u}rich, Switzerland
and IH\'ES, Bures-sur-Yvette, France.}
\email{juerg@itp.phys.ethz.ch}

\author[I.M. Sigal]{Israel Michael Sigal}
\address{I.M. Sigal, 
Department of Mathematics,
University of Toronto,
100 St. George Street,
Sidney Smith Hall,
Toronto, ON  M5S 3G3, CANADA
and
Department of Mathematics, University of Notre Dame, Hurley Hall, Notre Dame, IN 46556-4618, USA.}
\email{sigal@math.toronto.edu }


\begin{abstract}
This work addresses the problem of infrared mass renormalization
for a scalar electron in a translation-invariant model of non-relativistic QED.
We assume that the interaction of the electron with the quantized
electromagnetic field comprises a fixed ultraviolet regularization
and an infrared regularization parametrized by $\sigma>0$.
For the value $p=0$ of the conserved total momentum of electron and photon field,
bounds on the renormalized mass are established which are uniform in
$\sigma\rightarrow0$, and the existence of a ground state is proved.
For $|p|>0$ sufficiently small, bounds
on the renormalized mass are derived for any fixed $\sigma>0$.
A key ingredient of our proofs is the operator-theoretic renormalization
group using the isospectral smooth Feshbach map. It provides an
explicit, finite algorithm that determines the renormalized
electron mass at $p=0$ to any given precision. 
\end{abstract}

\maketitle
 
\tableofcontents

\subsubsection*{Conventions:}

We use units in which the velocity of light $c$, Planck's constant $\hbar$,
and the bare electron mass $m$
have the values $c=\hbar=m=1$. 
\\
The letters
$C$ or $c$ denote various
constants whose values may change from one estimate to another.
\\
$\cL(\H_1,\H_2)$ and $\cB(\H_1,\H_2)$ denote the linear, and
the bounded linear operators  $\H_1\rightarrow\H_2$ for Banach spaces $\H_1$, $\H_2$.
\\
$D_r(z)\subset\C$ is the closed disc of radius $r$ centered at $z$,
and $D_r\equiv D_r(0)$.
\\
$B_r(x)\subset\R^3$ is the closed ball of radius $r$ centered
at $x\in\R^3$, and $B_r\equiv B_r(0)$.
\\
$\langle v,v'\rangle_{\R^3}$ denotes the Euclidean
scalar product for vectors $v,v'\in\R^3$.
\\
$v^2\equiv\langle v,v\rangle_{\R^3}\equiv|v|^2$.

\newpage

\parskip = 8 pt


\section{Definition of the model}

We study the translation-invariant system of a  
non-relativistic, spin-0 electron in $\R^3$ that is minimally
coupled to the ultraviolet
regularized quantized electromagnetic field. This model 
of quantum electrodynamics (QED) with non-relativistic matter has
originally been proposed by Dirac and Jordan.

The Hilbert space of the (spinless) electron is given by
\eqn
        \H_{el}=L^2(\R^3,d^3x_{el})\;.
\eeqn
The field quanta of the quantized electromagnetic field are massless, relativistic
bosons, referred to as {\em photons}. The Hilbert space of one-photon states
is given by
\eqn
        {\mathfrak h}:= L^2(\R^3,d^3 k)\otimes\C^2 \;,
\eeqn
where $k\in\R^3$ is the momentum
of a photon. The factor $\C^2$ accommodates its transverse polarization
states in the Coulomb gauge.
One can choose a pair of
polarization vectors $\e(k,\lambda)\in \R^3$, $|\e(k,\lambda)|=1$, 
for  $\lambda\in\{+,-\}$, associated to every $k\in\R^3\setminus\{0\}$,  
such that the triple $(\e(k,+),\e(k,-), n_k:=\frac{k}{|k|})$
is an orthonormal basis in $\R^3$. The label $\lambda\in\{+,-\}$ indicates 
the polarization of the photon.

The bosonic Fock space describing the pure states of the
quantized electromagnetic field is defined by
\eqn
        \Fo&=&\bigoplus_{n\geq0} {\rm Sym}_n {\mathfrak h}^{\otimes n}\;,
\eeqn
where ${\rm Sym}_n$ is the orthogonal projection onto the subspace
of totally symmetric $n$-particle wave functions in the $n$-fold
tensor product of ${\mathfrak h}$. The zero photon sector is given by
${\rm Sym}_0{\mathfrak h}^{\otimes 0}:=\C\{\vac\}$, where
$\vac$ is the vacuum vector in $\Fo$. Vectors $\Psi\in\Fo$
are identified with sequences $(\psi_0,\psi_1, \dots)$, where $\psi_0\in\C$, and where
$\psi_n(k_1,\lambda_1,\dots,k_n,\lambda_n)$ are $n$-particle
wave functions that are totally symmetric with respect to
their $n$ arguments.

For convenience, we introduce the notation
\eqn
        K:=(k,\lambda) \; \;, \; \int dK:=\sum_{\lambda\in\{+,-\}}
        \int d^3k \;,
\eeqn
and
\eqn
        \mu K:=(\mu k,\lambda) \; \; {\rm for } \; \mu\in\R\;.
\eeqn
The inner product on $\Fo$
is then defined by
\eqn
        \langle\,\Psi\,,\,\Phi\,\rangle := \sum_{n=0}^\infty
        \int dK_1\cdots dK_n \, \overline{\psi_n(K_1,\dots,K_n)} \,
        \phi_n(K_1,\dots,K_n)
        \;.
\eeqn
Given $\lambda\in\{+,-\}$ and $f\in L^2(\R^3, d^3k)$, we
define an {\em annihilation operator} $a(f,\lambda)$ acting on
any $\Psi=(\psi_n)_{n=0}^\infty$ with only finitely many non-zero entries by
\eqn
        \;\;\;\;
        (a(f,\lambda)\Psi)_n(K_1,\dots,K_n):=
        \sqrt{n+1}\int d^3k \bar f(k) \psi_{n+1}(k,\lambda,K_1,\dots,K_n)
\eeqn
and
\eqn
        a(f,\lambda)\vac=0 \;.
\eeqn
This defines a closable operator $a(f,\lambda)$
on $\Fo$ whose closure is also denoted by $a(f,\lambda)$. The adjoint of
$a(f,\lambda)$ with respect to the scalar product on $\Fo$ is the
{\em creation operator} $a^*(f,\lambda)$. With $a(f,\lambda)$ being anti-linear,
and $a^*(f,\lambda)$ being linear in $f$, it is possible to write
\eqn
        a(f,\lambda)=\int d^3k \, \overline{f(k)} \, a(k,\lambda)
        \; \; , \; \;
        a^*(f,\lambda)=\int d^3 k \, f(k) \, a^*(k,\lambda) \;,
\eeqn
where $a(K)$ and $a^*(K)$ are unbounded, operator-valued distributions satisfying
the canonical commutation relations
\eqnn
        \big[a(K),a^*(K')\big] =\delta_{\lambda,\lambda'}\delta^{(3)}(k-k')
        \;  \; , \; \;
        \big[a^\sharp(K),a^\sharp(K')\big] = 0 \; .
\eeqnn
For brevity, $a^\sharp$ henceforth denotes either $a$ or $a^*$.

The energy and momentum of a single photon with wave vector $k$ is given by
$|k|$ and $k$, respectively (recalling that in our units, the speed of light and
Planck's constant have the value 1). A configuration of $n$ non-interacting
photons with wave vectors $k_1,\dots,k_n$ has energy $\sum_{j=1}^n|k_j|$,
and momentum $\sum_{j=1}^n k_j$.
We define the free field Hamiltonian $H_f$, and the free field momentum
operator $P_f$, by
\eqn
        (H_f\Psi)_n(K_1,\dots,K_n)&=&\Big(\sum_{j=1}^n|k_j|\Big)\psi_n(K_1,\dots,K_n)\;,
        \nonumber\\
        (P_f\Psi)_n(K_1,\dots,K_n)&=&\Big(\sum_{j=1}^n k_j\Big)\psi_n(K_1,\dots,K_n)\;,
\eeqn
and $(H_f\Psi)_0=0$, $(P_f\Psi)_0=0$.
Expressed in terms of creation and annihilation operators,
\eqn
        H_f&=& \int dK \,
        a^*(K)\,|k|\,a_\lambda(K) \;,
        \nonumber\\
        P_f &=&\int dK \,
        a^*(K) \, k \, a (K) \;,
\eeqn
which are defined as weak integrals.

The states of an electron coupled to the quantized electromagnetic field are
elements of the tensor product Hilbert space
$$
        \H=\H_{el}\otimes \Fo \;.
$$
The model studied in this paper is defined by the Hamiltonian
\eqn
        H_{QED}=\frac{1}{2}\big(i\nabla_{x_{el}}\otimes\1_f
        -g A_{\ks}(x_{el})\big)^2 + \1_{el}\otimes H_f \;.
\eeqn
Here, $A_{\ks}(x_{el})$ denotes the (regularized) quantized vector
potential defined by
\eqn
        A_{\ks}(x_{el}):= \int \frac{dK}{|k|^{\frac12}}
        \, \ks(|k|)
        \big(\e(K)e^{i\langle k,x_{el}\rangle_{\R^3}}\otimes\,
        a_\lambda(k)+h.c.\big) \;.
	\label{Aks-def-1}
\eeqn
It couples the electron to the degrees of freedom of the quantized electromagnetic field.
Moreover, $g$ denotes a coupling constant corresponding to the electron charge.
Given an infrared regularization parameter $\sigma\geq0$,  
$\ks\in C_0^\infty([0,1];\R_+)$ is assumed to be a smooth cutoff function obeying 
\eqn
		0<\|\ks\|_\sigma+
		\|x^\sigma\partial_x\big(x^{-\sigma}\ks\big)\|_\sigma<10
		\label{kappaassump}
\eeqn
and
\eqn
		\lim_{x\rightarrow0}\frac{\ks(x)}{x^\sigma}=1\;,
		\label{kappasig-norm-def-1}
\eeqn
where
\eqn 
        \|f\|_\sigma:=\sup_{x\in\R_+}|x^{-\sigma}f(x)|\;.    
\eeqn
Most of the time, we have $\sigma>0$, although we are really interested
in the limit $\sigma\searrow0$.

By translation invariance of the system, we may write $\H$ as a direct integral 
with respect to the total momentum operator
\eqn
        P_{tot}=i\nabla_{x_{el}}\otimes\1_f+\1_{el}\otimes P_f \;,
\eeqn
given by
\eqn
        \H=\int_{\R^3}^\oplus d^3p \; \H_p \;.
\eeqn
Each fiber Hilbert space $\H_p$, labeled by a vector $p\in\R^3$
corresponding to the conserved total momentum, is isomorphic to
$\Fo$, and invariant with respect to the unitary time evolution $e^{-itH_{QED}}$.

By invariance of $\H_p$ under the unitary evolution generated by $H_{QED}$,
it suffices to study the restriction of $H_{QED}$ to the fibers $\H_p$, which
we denote by $H(p,\sigma)$ (the
fiber Hamiltonian on $\H_p$). Thus, for any fixed $p\in\R^3$,
\eqn
        \label{Hpsdef}
        H(p,\sigma) &=& H_f + \frac12\big(p - P_f - g A_{\ks} \big)^2
        \nonumber\\
        &=&\frac{|p|^2}{2}+H_f-|p|P_f^\parallel-g|p|A_{\ks}^\parallel
        + \frac12\big(P_f + g A_{\ks} \big)^2\; ,
\eeqn
where $A_{\ks}:= A_{\ks}(0)$.
We are using the notation
\eqn
        v^\parallel:=\langle v , n_p \rangle_{\R^3}
        \; \; , \; \;
        n_p:=\frac{p}{|p|}
        \; \; , \; \;
        v^\perp:=v-v^\parallel n_p
\eeqn
for $0\neq p,v\in\R^3$.


\section{Statement of the main results}

We study the spectrum of the fiber Hamiltonian $H(p,\sigma)$ in the vicinity of 
its infimum $E(p,\sigma)$. We prove that $E(p,\sigma)$
is a simple eigenvalue and construct the corresponding eigenvector
$\Psi(p,\sigma)\in\Fo\cong \H_p$
provided that either $\sigma>0$ and $0< |p|<\frac13$ or $\sigma\geq0$ if $p=0$. The vector
$\Psi(p,\sigma)$ is an {\em infraparticle state}, describing a compound
particle comprising the electron together with a cloud of low-energy (soft)
photons whose expected number diverges as $\sigma\rightarrow0$, unless $p=0$.
One of our main goals in this paper
is to determine the {\em renormalized electron mass}.
 
The renormalized electron mass can be defined 
as follows. 
Let ${\rm Hess}\,E(p,\sigma)$ denote the Hessian of $E(p,\sigma)$ at $p\in\R^3$.
We compute
\eqn
        \big({\rm Hess}\,E(p,\sigma)\big)_{ij}&=&
        \big(\delta_{ij}-\frac{p_i p_j}{|p|^2}\big)
        \frac{\partial_{|p|}E(p,\sigma)}{|p|}
        +\frac{p_i p_j}{|p|^2}
        \partial_{|p|}^2E(p,\sigma) \;.
\eeqn
As a linear map, it
has a simple radial eigenvalue $\partial_{|p|}^2E(p,\sigma)$,
and a tangential eigenvalue $\frac{\partial_{|p|}E(p,\sigma)}{|p|}$
of multiplicity two.
If $p=0$ the Hessian simplifies to a multiple of the identity,
$\partial_{|p|}^2E(0,\sigma){\bf 1}_3$.
We shall refer to the radial eigenvalue
\eqn
        m_{ren}(p,\sigma):=\frac{1}{\partial_{|p|}^2 E(p,\sigma)}\;
        \label{mren-def-1}
\eeqn
as the renormalized electron mass.

One of our main purposes in the present paper is to study $m_{ren}(p,\sigma)$.
If $p\neq0$, we shall derive bounds on $m_{ren}(p,\sigma)$ which
are {\em not}
uniform in $\sigma$. Uniform bounds are beyond the scope of this
work and are established in \cite{ch}. A corollary of the results in \cite{ch}
is that
\eqn
        \lim_{\sigma\rightarrow0}\lim_{p\rightarrow0}m_{ren}(p,\sigma)
        =\lim_{p\rightarrow0}\lim_{\sigma\rightarrow0}m_{ren}(p,\sigma)\;.
        \label{mrenlimexch}
\eeqn
In the literature, $m_{ren}(0,0)$ is the most common definition of the renormalized mass.
In the present work, we assume the commutativity of the limits in (~\ref{mrenlimexch});
see condition (~\ref{limcommhyp}) below,
and \cite{ch} for a proof.
Under this condition we prove uniform bounds in the $p=0$ case.
We emphasize that
our bounds on $m_{ren}(0,\sigma)$ are constructive and can be used to devise
an algorithm to compute the renormalized mass
to any given precision.

\begin{theorem}\label{mainthm}
Assume that $0\leq |p|<\frac13$.
Let $E(p,\sigma)$ denote the infimum of the spectrum of the fiber Hamiltonian $H(p,\sigma)$ 
defined in (~\ref{Hpsdef}), where $\sigma>0$ if $p\neq0$,
and $\sigma\geq0$ if $p=0$.

\begin{itemize}

\item{\underline{A. The case  $p\neq0$:}}
For any $\sigma>0$, there exists a constant $g_0(\sigma)>0$
such that, for all $0\leq g<g_0(\sigma)$,  $H(p,\sigma)$ has an eigenvalue
$E(p,\sigma)$ of
multiplicity one at the bottom of absolutely continuous spectrum. 
Moreover, there is a finite constant $c_0(\sigma)>0$ independent of $g$ such that
\eqnn
		\Big|E(p,\sigma)-\frac{|p|^2}{2}-
		\frac{g^2}{2}\bra\vac , A_{\ks}^2\vac\ket\Big|<
		c_0(\sigma)\,\frac{ g^2|p|^2}{2}\;,
\eeqnn 
with
\eqnn
        \Big|\partial_{|p|} E(p,\sigma)-|p|\Big|\leq c_0(\sigma)g^2|p| 
\eeqnn
and 
\eqnn
		 0<1-\partial_{|p|}^2E(p,\sigma) \leq  c_0(\sigma) g^2 \;.
\eeqnn  
Consequently, 
\eqn
        \label{derp2Ebound}
        1<m_{ren}(p,\sigma)<
        1+c_0(\sigma) g^2     \;,
\eeqn  
where
\eqn
        m_{ren}(p,\sigma)=\frac{1}{ \partial_{|p|}^2E(p,\sigma) }  
        \label{massincr-1}
\eeqn 
is the renormalized mass.

\item{\underline{B. The case $p=0$:}} 
There is a constant $g_0>0$ {\em independent} of $\sigma\geq0$ 
such that, for arbitrary $\sigma\geq0$ and for any
$0\leq g<g_0$, $H(0,\sigma)$ has an eigenvalue $E(0,\sigma)$ of
multiplicity one at the bottom of absolutely continuous spectrum.

There are finite constants $c_1, c_2>0$ {\em independent} of $g$, 
$p$ and $\sigma\geq0$ such that
\eqnn
		0<E(0,\sigma)\leq c_1 g^2 \;.
\eeqnn
Moreover
\eqnn
		(\partial_{|p|}E)(0,\sigma)=0 \;,
\eeqnn
and 
\eqnn
		1<m_{ren}(0,\sigma)\leq 1+ c_2 g^2 \;.
\eeqnn 
Assuming condition ({~\ref{mrenlimexch}}) (corresponding to (~\ref{limcommhyp}) below), 
\eqn
        \lim_{p\rightarrow0}\lim_{\sigma\rightarrow0}m_{ren}(p,\sigma)
        =1+ \tilde c_2 g^2+O(g^{\frac73})\;,
        \label{unifmassbd}
\eeqn
where
\eqn
		\tilde c_2&=&\lim_{\e\rightarrow0}\lim_{\sigma\rightarrow0}
		\Bra\vac\,,\,A_{\knull}\Big[H_f+\frac{1}{2}P_f^2 +\e\Big]^{-1}A_{\knull}\vac\Ket
		\nonumber\\
		&=& \frac{8\pi }{3}\int_{\R_+}dx\,\frac{\knull^2(x)}{1+\frac x2} \;,
		\label{unif-mass-const-1}
\eeqn
and $\knull(x):=\lim_{\sigma\rightarrow0}\ks(x)$.
\end{itemize}
\end{theorem}

\subsection{Discussion}

The renormalized mass $m_{ren}(0,0)$ can be determined perturbatively.
For a sharp ultraviolet cutoff at $\Lambda$, and 
setting the bare mass equal to 1, the ground state energy has the form
\eqn
		E(p,0)=E_0+\frac{p^2}{2m_{ren}(0,0)}+O(p^4)
\eeqn
where to leading order in $g$,
\eqn
		E_0\approx\frac{2\pi g^2\Lambda^2 }{3}
		\label{E-ren-pert-lead-1}
\eeqn
and 
\eqn
		\frac{|p|^2}{2m_{ren}(0,0)}&\approx&\frac{|p|^2}{2}
		\big(1-\frac{16\pi}{3} g^2\log(1+\frac\Lambda2)\big)
		\nonumber\\
		&\approx&\frac{|p|^2}{2}\frac{1}{1+\frac{16\pi }{3}g^2\log(1+\frac\Lambda2)} \;,
\eeqn
so that 
\eqn
		m_{ren}(0,0)\approx 1+\frac{16\pi }{3}g^2\log(1+\frac\Lambda2) \;.
		\label{m-ren-pert-lead-1}
\eeqn 
This agrees with (~\ref{unifmassbd}) and (~\ref{unif-mass-const-1}).
(It is a common convention to include a factor $\frac{1}{\sqrt2}$
in the definition (~\ref{Aks-def-1}), whereby $g$ corresponds to $\frac{\tilde g}{\sqrt2}$
and $\frac{16\pi g^2}{3}$ to $\frac{8\pi \tilde g^2}{3}$.)
A presentation of the leading order calculations producing these 
results can for instance be found in \cite{sp1}.
An important result of this paper is that the right hand side of
(~\ref{m-ren-pert-lead-1}) is the correct value of the renormalized mass, 
up to an error $o(g^2)$.

Note that (~\ref{E-ren-pert-lead-1}) and (~\ref{m-ren-pert-lead-1}) depend
on the ultraviolet cutoff $\Lambda$. Throughout this paper,
we choose $\Lambda$ to be $O(1)$. 
We do not study ultraviolet renormalization, i.e., we do not worry
about the dependence of $E_0$ and $m_{ren}(0,0)$ when $\Lambda\rightarrow\infty$.
For some preliminary results see \cite{hasei,hisp,lilo, lilo1}, and \cite{sp1} for a recent survey. 

\begin{remark}
The upper bound on $|p|$ of $\frac13$ is not optimal,
but we note that, for $E(p,\sigma)$ to be an eigenvalue, $|p|$
cannot exceed a critical value $p_c< 1$ (corresponding
to the speed of light). As $|p|\rightarrow p_c$,
it is expected that the eigenvalue at $E(p,\sigma)$ dissolves in
the continuous spectrum, while a resonance appears.
This is a manifestation of a phenomenon analogous to 
Cherenkov radiation.
\end{remark}

\begin{remark}
Our analysis is based on the operator-theoretic renormalization group
method involving the {\em smooth Feshbach map}, which is developed in \cite{bcfs1}. 
This method provides a
convergent expansion for both $E(0,0)$ and $m_{ren}(0,0)$ in powers of
the feinstructure constant $\alpha= \frac{g^2}{4\pi}$, with coefficients 
that are themselves functions of $\alpha$.
It is a subtle and significant issue that the latter diverge in the limit
$\alpha\rightarrow0$ like $\alpha^{-\e}$, where $\e>0$ is arbitrarily small. 
Note that
$E(0,0)$ and $m_{ren}(0,0)$ {\em cannot} be represented as a convergent
power series in $\alpha$.
\end{remark}

\begin{remark}
Inequality (~\ref{massincr-1}) expresses the fact that the mass of the electron is increased
by interactions with the photon field.
\end{remark}

\begin{remark}
A priori, the constant $c_2(\sigma)$ in (~\ref{derp2Ebound}) might
diverge in the limit $\sigma\rightarrow0$; but one can prove bounds that
are uniform in $\sigma$, see \cite{ch}. In this work,
we prove $c(\sigma)<C^{\frac{1}{\sigma}}$ for an explicitly computable constant $C$ which is
independent of $g$ and $\sigma$.
\end{remark}

\begin{remark}
The operator-theoretic renormalization group method
provides a {\em convergent, finite algorithm} for determining the values 
of $E(0,0)$ and $m_{ren}(0,0)$ to arbitrary precision.  
\end{remark}

\begin{remark}
The existence of the ground state at $p=0$ for $\sigma=0$, and 
renormalization of the electron mass is an important ingredient for the 
phenomenon of {\em enhanced binding}, \cite{cvv} and \cite{hisp1,hvv}. A 
Schr\"odinger operator with a non-confining potential
can exhibit a bound state when the interaction of the
electron with the quantized electromagnetic field is included. 
Binding to a shallow potential can be energetically more favorable for the electron
than forming an infraparticle through 
binding of a cloud of soft photons. 
\end{remark}

\begin{remark}
As stated above, in this paper we use an operator-theoretic renormalization group method based 
on the "smooth Feshbach map", in order to prove our main result (Theorem {~\ref{mainthm}})
and, in particular, to determine $m_{ren}(0,0)$. This method has been developed in
\cite{bcfs1}. We expect that, alternatively, A. Pizzo's method of iterated (analytic)
perturbation theory \cite{piz1} could be used to reach the same goals. In fact,
Pizzo's method has been shown to be applicable to the problem of
determining the ground state and the ground state energy of a 
hydrogen atom interacting with the quantized radiation field, see \cite{bfp}, which
is similar to the problem solved in this paper.
\end{remark}


\section{Strategy and organization of the proof}
\label{stratorg-sect-1}

The purpose of this section is to outline the key steps and analytical methods used 
in our proof of Theorem {~\ref{mainthm}}.  

\subsection{Smooth Feshbach map}

An important functional-analytic tool that we employ to
establish Theorem {~\ref{mainthm}} is the {\em smooth Feshbach map}, \cite{bcfs1}. 
This is an essential ingredient of the operator-theoretic
renormalization group method and is addressed in detail in Section
{~\ref{Fpairssubsec}}. Its main features can be summarized as follows. 

Let $\H$ denote a separable Hilbert space, let $\tau$ and $H$ be closed
operators on $\H$, and assume that the operator $\omega:=H-\tau$
is defined on the entire domain of $\tau$.
We choose a positive operator $0\leq\chi\leq 1$  on $\H$
which, together with $\bar\chi:=\sqrt{1-\chi^2}$, shall commute with $\tau$ and
leave the domain of $\tau$
invariant. The operators $(H,\tau)$ are called a {\em Feshbach pair} corresponding to $\chi$
if the bounds formulated in eq. (~\ref{Feshboundopasscond}),
below, are satisfied.

The {\em smooth Feshbach map} is defined on Feshbach pairs with values in
the linear operators on $\H$. It is defined by
\eqn
        F_\chi(H,\tau)=H_\chi-\chi\omega\bar\chi H_{\bar\chi}^{-1}
        \bar\chi\omega\chi\;,
\eeqn
where
\eqn
        H_\chi:=\tau+\chi\omega\chi\;\;,\;\;
        H_{\bar\chi}:=\tau+\bar\chi\omega\bar\chi\;.
\eeqn
A quintessential property of $F_\chi(H,\tau)$ is  
{\em (Feshbach) isospectrality} (Theorem {~\ref{Feshprop}}):
\begin{itemize}
\item  $H$ is bounded invertible on $\H$ if and only if $F_\chi(H,\tau)$
is bounded invertible on $\Ran(\chi)$.
\item  $0$ is an eigenvalue of $H$ with multiplicity $n_0$ if and only if
$0$ is an eigenvalue of $F_\chi(H,\tau)$ with the same multiplicity.
\item  If $\tau$ and $H$ are selfadjoint operators on $\H$, then $H$ and
$F_\chi(H,\tau)$ are of the same spectral type at 0; see \cite{bcfs1} for details.
\end{itemize}

In this precise sense, the smooth Feshbach map allows us to study the low-energy
spectrum of $H$ by analyzing a bounded operator, $F_\chi(H,\tau)$, on a proper 
{\em subspace}, $\Ran(\chi)$, of $\H$. The operator-theoretic renormalization group
is used to prove (in a recursive fashion) that
there is a choice of $\chi$ for which $F_\chi(H,\tau)$ is well-defined and such that
it can easily be seen that the bottom of the spectrum of $F_\chi(H,\tau)$
consists of a simple eigenvalue. By Feshbach isospectrality, one then concludes 
that the bottom of the spectrum of $H$ consists of a simple eigenvalue, too.

As compared to the version of the Feshbach
map based on sharp projectors, \cite{bfs1,bfs2}, 
the {\em smooth} Feshbach map is simpler to handle, analytically 
(differentiability of the cutoff operators), but somewhat more
complicated algebraically, since $\chi \bar\chi\neq0$.
All this is discussed in detail in Section {~\ref{overlapopsubsubsec}}.

\subsection{A Banach space of generalized Wick kernels}

Our key strategy is to relate the spectral problem outlined above to a discrete 
dynamical system in a Banach space of operators, in the spirit of K. Wilson's 
formulation of the renormalization group. The first step in this construction is to
define this Banach space, cf.  section {~\ref{Banspsubsect}}.

We consider bounded operators of the form
\eqn
        H= E+T[\opp]+W\;,
        \label{Hnroughstructure}
\eeqn
on the fixed Hilbert space $\H_{red}:=\1[H_f<1]\Fo$,
referred to as {\em effective Hamiltonians}. 
Here, we have introduced the notation
\eqn
		\opp:=(H_f,P_f) \;,
\eeqn
and we denote the spectral variable corresponding to $\opp$ by 
\eqn
		\spvar=(X_0,X)\in[0,1]\times B_1\;,
\eeqn 
where $X=(X_1,X_2,X_3)$.

\begin{itemize}
\item
In (~\ref{Hnroughstructure}), $E$ is a complex number which plays the role of a 
spectral parameter. 
\item
The operator $T[\opp]$  satisfies $T[\unull]=0$.
It is referred to as the non-interacting, or {\em free} part of $H$.
\item
The operator $W$, referred to as the interaction part of $H$, is of the form 
\eqnn
        	W=\sum_{M+N\geq1}W_{M,N} \;,
\eeqnn
where $W_{M,N}$ is a {\em generalized Wick monomial} 
in $M$ creation and $N$ annihilation operators,
\eqnn
		W_{M,N}&=&\int_{B_1^{M+N}} d\mu^{(M,N)}(K^{(M)},\tilde K^{(N)}) 
		\nonumber\\
		&&\hspace{1cm}a^*(K^{(M)})\, 
		w_{M,N}[\opp;K^{(M)},\tilde K^{(N)}] \,
		a(\tilde K^{(N)})
\eeqnn 
where $K^{(M)}:=(K_1,\dots,K_M)$, $\tilde K^{(N)}=(\tilde K_1,\dots,\tilde K_N)$,
and where $K_i=(k_i,\lambda_i)$, $\tilde K_j=(\tilde k_j,\tilde\lambda_j)$ 
are pairs of photon momenta and polarization labels.
Furthermore,
\eqnn
		d\mu(K^{(M)},\tilde K^{(N)})&:=&\prod_{i=1}^M\prod_{j=1}^N\frac{dK_i}{\sqrt{|k_i|}}
		\frac{d\tilde K_j}{\sqrt{|k_j|}}
		\nonumber\\
		a^\sharp(K^{(M)})&:=&\prod_{j=1}^M a^\sharp(K_j) \;.
\eeqnn
The precise definition is given in (~\ref{WMNwMNdef}) below. 
$W_{M,N}$ is uniquely determined by its
{\em generalized Wick kernel}
$w_{M,N}[\opp;K^{(M)},\tilde K^{(N)}]$.
We note that  the singular photon form factors $|k_j|^{-\frac12}$
are absorbed into the integration measure $d\mu^{(M,N)}$, while
the infrared regularization depending on $\sigma$ is contained in
the generalized Wick kernels $w_{M,N}$.
\end{itemize} 

Adapted to (~\ref{Hnroughstructure}), we introduce a 
{\em sequence space of generalized Wick kernels}
\eqn
        \Hspace_{\geq0}=\C\oplus\Tspace\oplus\Hspace_{\geq1}
\eeqn
that parametrizes the effective Hamiltonians.
Here
\eqn
	\Hspace_{\geq1}=\bigoplus_{M+N\geq1}\Wspace_{M,N}\;.
\eeqn

\begin{itemize}
\item
$\C$ is the range of $E$. 
\item
$\Tspace$ is the space of functions $T[\spvar]$ with $T[\underline{0}]=0$, 
equipped with the norm
$$
		\|T\|_{\Tspace}^\sharp\approx\Big(\sum_{a_0=0,1}+\sum_{ 0\leq \sum_1^3 a_i\leq2}\Big)
		|\partial_\spvar^{\ua} T| \;.
$$
The precise definition of this norm is given in (~\ref{Tnormdef}). 
\item
$\Wspace_{M,N}$ is the space of functions 
$w_{M,N}[\spvar;K^{(M)},\tilde K^{(N)}]$, equipped with the norm
\eqnn
		\;\;\;\;\;
		\|w_{M,N}\|^\sharp_\sigma&\approx&
		\sup_{k_i\atop i=1,\dots,M}\sup_{\tilde k_j\atop j=1,\dots,N}
		\sup_{|X|\leq X_0<1}\Big[ \prod_{i=1,\dots,M\atop j=1,\dots,N}
		|k_i|^{-\sigma}|\tilde k_j|^{-\sigma}\Big]
		\nonumber\\
		&&\hspace{1.5cm}\times
		\Big(|k_i|^{\sigma} |\partial_{|k_i|}|k_i|^{-\sigma} w_{M,N}|
		+||k_j|^{\sigma} \partial_{|\tilde k_j|}|k_j|^{-\sigma} w_{M,N}|
		\nonumber\\
		&&\hspace{3cm}
		+\Big(\sum_{a_0=0,1}+\sum_{1\leq \sum_1^3 a_i\leq2}\Big)|\partial_\spvar^{\ua} w_{M,N}|\Big)\;.
		\label{norm-w-sig-sharp-def-1}
\eeqnn
The precise definition of this norm is given in (~\ref{normsigmasharpdef}).
The factors $|k_i|^{-\sigma}$, $|\tilde k_j|^{-\sigma}$ implement the
infrared regularization, since finiteness of $\|w_{M,N}\|^\sharp_\sigma$ evidently implies 
\eqnn
		&&||k_i|^{\sigma} \partial_{|k_i|}|k_i|^{-\sigma} w_{M,N}|
		+||k_j|^{\sigma}\partial_{|\tilde k_j|}|k_j|^{-\sigma} w_{M,N}|
		\nonumber\\	
		&&\hspace{1.5cm}+\Big(\sum_{a_0=0,1}+\sum_{1\leq \sum_1^3 a_i\leq2}\Big)|\partial_\spvar^{\ua} w_{M,N}|
		\nonumber\\
		&&\hspace{3cm}=O\Big(\prod_{i=1,\dots,M}\prod_{ j=1,\dots,N}
		|k_i|^{\sigma}|\tilde k_j|^{\sigma}\Big)
\eeqnn
as $|k_i|,|\tilde k_j|\rightarrow0$. 
\end{itemize}

For $\h=(E ,T ,\h_1)\in\Hspace_{\geq0}$
with $\h_1:=(w_{M,N} )_{M+N\geq1}$, and a fixed number $0<\xi<1$,
we define the norm
\eqn
        \|\h\|_{\sigma,\xi}=|E|+\|T\|_{\Tspace}^\sharp+\|\h_1\|_{\sigma,\xi}
\eeqn
with
\eqn
		\|\h_1\|_{\sigma,\xi}:=\sum_{M+N\geq1}\xi^{-M-N}
        \|w_{M,N}\|_{\sigma}^\sharp\;.
\eeqn
This endows the space $\Hspace_{\geq0}$ with a Banach space structure.
There exists an injective imbedding 
\eqn
		H:\Hspace_{\geq0}\hookrightarrow\cB(\H_{red})\;,
\eeqn 
such that for every $\h\in\Hspace_{\geq0}$,
\eqn
        H[\h]=E\1+T[\opp]+W[\h]
\eeqn
with
\eqn
        W[\h]:=\sum_{M+N\geq1}W_{M,N}[w_{M,N}]\;
\eeqn
is an effective Hamiltonian of the form (~\ref{Hnroughstructure}).
Our norms are chosen such that  
\eqn
        \|H[\h]\|_{op}\leq C\|\h\|_{\sigma,\xi} \;,
\eeqn
where $\|\,\cdot\,\|_{op}$ denotes the operator norm on $\H_{red}$.

These constructions are discussed in detail in Section {~\ref{Banspsubsect}}.

\subsection{The isospectral renormalization map $\ren$}

To implement the multiscale approach outlined above, 
we define an isospectral renormalization transformation that maps a polydisc
contained in $\Hspace_{\geq0}$ into itself.

To define the polydisc, we introduce a family of non-interacting comparison
theories defined by $T^{(p;\lambda)}_0\in\Tspace$ with
\eqn
        T^{(p;\lambda)}_0[\opp] \approx H_f-|p| P_f^\parallel+
        \lTnl P_f^2\;,
\eeqn
where $P_f^\parallel$ is the projection of $P_f$ onto the
conserved momentum $p$, and $0\leq\lambda<\frac12$ is a real parameter. 
The effective Hamiltonians in our problem are compared with
the free family of theories defined by $T^{(p;\lambda)}_0\in\Tspace$.

In this spirit, we introduce a {\em polydisc} $\Polyd\Polpar\subset \Hspace_{\geq0}$ given by
\eqn
        \Polyd\Polpar\approx\Big\{\h=(E,T,\h_1)&\Big|&
        E\in\I \;,
        \nonumber\\
        &&\|T-T^{(p;\lTnl)}_0\|_{\Tspace}<\delta \;, 
        \;\;\| \h_1 \|_{\sigma,\xi}\leq\e\;\; \Big\}
        \;,
        \label{Polyd-intro-def-1}
\eeqn
where $\I:=\{z\in\C\big||z|<\frac{1}{10}\}$.
Let $0<\rho<\frac12$.
For $\e,\delta,\lTnl$ sufficiently small, and
$\h\in\Polyd\Polpar$, one can verify that $(H[\h],H_f)$ is a Feshbach pair
corresponding to $\chi_\rho[H_f]=\chi_1[\rho^{-1}H_f]$.

We let $\resc$ denote the rescaling operator,
which acts on creation- and annihilation operators by
\eqn
		\resc[ a^\sharp(K)]=\rho^{-\frac52}a^\sharp(\rho^{-1}K)
		\;\;{\rm and}
		\;\;
		\resc[\1]=\rho^{-1}\1 \;.
\eeqn 
(We note that the definition of $\resc$ includes multiplication by
$\frac1\rho$. Under unitary scaling, $a^\sharp(K)$ is mapped to 
$\rho^{-\frac32}a^\sharp(\rho^{-1}K)$).

Moreover, we denote by $E_\rho$ a {\em renormalization of the complex number} $E$.

The {\em isospectral renormalization map} is then defined by
\eqn
        \renop^H [H[\h]]=
        \Big(E_\rho\circ\resc\circ F_{\chi_\rho[H_f]}(\,\cdot\,,H_f)\Big)[H[\h]] \;,
\eeqn
by composing an application of the smooth Feshbach map with a rescaling 
transformation, cf. section {~\ref{Rentrsf}}.

We next lift $\renop^H$ to $\Hspace_{\geq0}$.
Given $\h\in\Polyd\Polpar$, we define a {\em renormalization map
acting on generalized Wick kernels}  $\ren$ whose 
domain is defined by those elements $\h\in\Hspace_{\geq0}$, for which 
$\renop^H [H[\h]]$ is well-defined and in the domain of $H^{-1}$. Accordingly,  
\eqn
		\ren:= H^{-1}\circ \renop \circ H\;,
\eeqn
where $H:\h\mapsto H[\h]$. (Note that here the injectivity of $H$ and
$\renop:\Ran(H)\rightarrow\Ran(H)$ are crucial.)

\subsection{Codimension two contraction property of $\ren$ on a polydisc}
\label{contr-intro-ssect-1}

Our main result in the context of the operator-theoretic renormalization group
states that $\ren$ is contractive on a codimension two subspace of 
$\Polyd\Polpar$. This is a consequence of the following facts.

Let $\h=(E,T,\h_1)\in\Polyd\Polpar$, and
\eqn
		\widehat\h\;:=\;\ren[\h]
		\;=\;(\widehat E,\widehat T,\widehat\h_1) \;.
\eeqn
We will show that 
\eqn
		\widehat E = \rho^{-1}(E+O(\e))\;,
\eeqn
i.e. the spectral parameter $E$
is magnified by a factor $\frac1\rho$ through the rescaling transformation.
The operator $E\1$ thus belongs to a 1-dimensional space of
{\em relevant perturbations}, but by explicit change of variables, implemented by 
the map $E_\rho$,
this 1-dimensional subspace of operators can be projected out.

We may then restrict our attention to the
spaces of marginal and irrelevant perturbations.
We shall prove that 
\eqn
		\|\widehat T - T_0^{(p;\rho\lambda)}\|_{\Tspace}\leq \delta+\e \;.
\eeqn
Thus, under application of $\ren$, the bound in (~\ref{Polyd-intro-def-1}) involving $T$
is transformed according to $\delta\rightarrow \delta+\e$ and 
$\lambda\rightarrow\rho\lambda$. The  leading terms in $T$ of the form
$\alpha H_f+\beta \Ppar$ are invariant under the rescaling transformation and hence
belong to a 2-dimensional space of {\em marginal perturbations}.

We will prove for the renormalized generalized Wick kernels that
\eqn
        \widehat w_{M,N}&=& 
        \rho^{(\sigma+1)(M+N)-1}\big(w_{M,N}+\Delta w_{M,N}\big)\;\;\;{\rm if}\;p\neq0 
\eeqn
and
\eqn
		\widehat w_{M,N}&=& 
        \rho^{\max\{M+N-1,1\}}\big(w_{M,N}+\Delta w_{M,N}\big)\;\;\;{\rm if}\;p=0 \;,
\eeqn
for $M+N\geq1$, where (with $\Delta \h_1:=(\Delta w_{M,N})_{M+N\geq1}$)
\eqn
		\|\Delta \h_1\|_{\sigma,\xi}<c\e^2 \;.
\eeqn
The powers of $\rho$ are generated by the scaling transformation (that is, by the action of 
$\resc$ on $\Wspace_{M,N}$; this is explained in Section {~\ref{resc-def-subsubsect-1}}).
The terms $\Delta w_{M,N}$ are produced by the smooth Feshbach map.
If $p\neq0$, every $w_{M,N}$ is contracted by a factor $\rho^\sigma$, or smaller, under an
application of the renormalization map. In the special case $p=0$, this factor is
given by $\rho$, {\em independently} of $\sigma$.
One obtains
\eqn
		\|\widehat\h_1\|_{\sigma,\xi}&\leq& \rho^a(\|\h_1\|_{\sigma,\xi}+\|\Delta \h_1\|_{\sigma,\xi})
		\nonumber\\
		&\leq&\rho^a(\e+O(\e^2)) \;,
\eeqn
where $a=\sigma$ if $p\neq0$, and $a=1$ if $p=0$. 
Thus,  for a suitable choice of $\rho$ (which depends on $\sigma$ if $p\neq0$, but
not if $p=0$),
\eqn
		\|\widehat\h_1\|_{\sigma,\xi}&\leq& \frac{\e}{2}\;.
\eeqn
Thus, from application of $\ren$, $\e\rightarrow\frac\e2$ in the bounds
formulated in (~\ref{Polyd-intro-def-1}).

In conclusion,
\eqn
        \ren:\Polyd\Polpar\rightarrow\Polyd(\frac{\e}{2},\delta+\e,\rho\lTnl)
        \label{codim2intro}
\eeqn
for all $0\leq \e\leq \e_0(\sigma)$, $0\leq \delta\leq \delta_0(\sigma)
+2\e_0(\sigma)$ sufficiently small if $p\neq0$, and
for all $0\leq \e\leq \e_0$, $0\leq \delta\leq \delta_0 +2\e_0$ sufficiently small  
(independently of $\sigma$)
if $p=0$. Ignoring the subspace of relevant perturbations spanned
by $E\1$, which is explicitly projected out, 
this expresses the {\em codimension two contraction property} of the
renormalization map.
The details of this analysis are presented in Section {~\ref{Codimcontrsubsect}}.

\subsection{The first Feshbach decimation step}

In the first Feshbach decimation step, the fiber Hamiltonian $H(p,\sigma)$
is mapped to an effective Hamiltonian that is used as an initial condition
for the renormalization group recursion.

To this end, we verify that for 
$E\in \frac{|p|^2}{2}+\langle\vac,A_{\ks}^2\vac\rangle+\I$,
and $g$ sufficiently small,
$(H(p,\sigma)-E,H_f)$ is a Feshbach pair corresponding to $\chi_1[H_f]$. 
We then find an element $\h^{(0)}$ in
a polydisc $\Polyd(\e_0,\delta_0,\frac12)$ such that
\eqn
		H[\h^{(0)}]=F_{\chi_1[H_f]}(H(p,\sigma)-E,H_f) \;.
\eeqn 
$H[\h^{(0)}]$ and $H(p,\sigma)-E$ are isospectral in the Feshbach sense.
The parameters $\e_0$ and $\delta_0$ are both $O(g)$.

\subsection{The isospectral renormalization group flow}

We assume that the electron charge
$g$, and therefore the parameters $\e_0,\delta_0$, are sufficiently small such that
$\ren$ is codimension 2 contractive
on $\Polyd(\e_0, \delta_0+2\e_0,\frac12)$. 
If $p\neq0$, one must assume
that $g<g_0(\sigma)$, while if $p=0$, one can assume that $g<g_0$ independently
of $\sigma$.

Repeated application of the renormalization map yields a sequence
$(\h^{(n)})_{n\geq0}$ satisfying $\h^{(n+1)}=\ren[\h^{(n)}]$. The index $n$
is referred to as the {\em scale} of the effective problem obtained after
the $n$-th recursion.
The effective Hamiltonian of the scale $n$ is given by
\eqn
	H[\h^{(n)}] = E^{(n)}\1+T^{(n)}[\opp]+W[\h^{(n)}] \;,
\eeqn
and has an operator norm bounded by 
$\|H[\h^{(n)}]\|_{op}\leq c\|\h^{(n)}\|_{\sigma,\xi}$ on $\H_{red}$.

By (~\ref{codim2intro}), the {\em scaling limit}
($n\rightarrow\infty$) is characterized by an element  
$\h^{(\infty)}\in\Polyd(0,\delta_0+2\e_0,0)$ in the vicinity of 
$T_0^{(p;0)}$ relative to the $\|\,\cdot\,\|_\Tspace$-norm. That is,    
$H[\h^{(\infty)}]=\alpha H_f+\beta \Ppar$,
where $|\alpha-1|,|\beta+|p||<\delta_0+2\e_0$.

Thus, for $\sigma>0$ and $0\leq |p|<\frac13$, or $\sigma=0$ and $p=0$, 
the fixed point set of $\ren$ (after elimination of 
the one-dimensional relevant subspace of perturbations) consists of a
2-dimensional linear center stable manifold that parametrizes a universality class
of non-interacting theories.
However, we remark that for $\sigma=0$ and $0<|p|<\frac13$, the generalized Wick kernels
$w_{1,0}$ and $w_{0,1}$ (which are hermitean conjugates of one another) 
are {\em strictly marginal} operators, as is proven in \cite{ch}. Correspondingly,
the fixed point set of $\ren$ is then given by
a 3-dimensional center stable manifold that parametrizes a universality class of theories.  

\subsection{Ground state eigenvalue and eigenvector}

In Section {~\ref{eigenvalvecsubsec}}, we use the isospectral renormalization group to
prove the existence of a simple eigenvalue $E(p,\sigma)$ at the bottom of the
spectrum of $H(p,\sigma)$, and to construct the corresponding normalized eigenvector
$\Psi(p,\sigma)$ for $\sigma>0$. In the case $p=0$, the same is achieved for $\sigma\geq0$.

Key to this analysis is the fact that the infimum of the spectrum of 
$H[\h^{(\infty)}]=\alpha H_f+\beta \Ppar$ is a simple eigenvalue
at $\{0\}\in\C$, and the corresponding eigenvector is the Fock vacuum $\vac\in\H_{red}$.
Using Feshbach isospectrality  (Theorem {~\ref{Feshprop}}), this suffices to reconstruct the ground state
eigenvalue and eigenvector of $H(p,\sigma)$.

\subsection{Soft photon sum rules}

For $\sigma>0$, it is possible to prove (~\ref{codim2intro}) 
along the lines presented in \cite{bcfs1}.
When $\sigma=0$ and $p=0$, additional, new techniques are needed,
due to the following subtle and important issue.

Our result states that the interaction is irrelevant even if $\sigma=0$,
provided that $p=0$. And indeed, based on the naive scaling of the interaction operators in the fiber
Hamiltonian $H(p=0,\sigma)$ (which is notably spherically symmetric), this might 
seem trivial. 
However, the argument based on pure scaling is unreliable.
Irrelevance of the interaction is a consequence of {\em symmetries} of the model,
but spherical symmetry alone is insufficient.
There are simple examples of spherically symmetric
models, in which marginal interactions are {\em generated} from
irrelevant ones through the renormalization map. This is a feature of its 
{\em non-linearity}.

To treat the case $p=0$, we exploit a special property of the QED model,
which is shared by the Gross transformed Nelson model. In both instances, 
there exists a hierarchy
of non-perturbative identities linking the generalized Wick kernels,
referred to as {\em soft boson sum rules} (or {\em soft photon sum rules}
for QED).

Given $n\in\R^3$, $|n|=1$, let $\e(n,\lambda)$
denote the photon polarization vector orthonormal to $n$
labeled by the polarization index $\lambda$.
For $\mu\in\R_+$.
The sequence of generalized Wick kernels $\h\in\Hspace_{\geq0}$ is said to
satisfy the  
{\bf soft photon sum rules}   \sbsr  
if the identities
\eqn
        &&g \mu  \bra\e(n,\lambda)\,,\,\partial_{X}\ket_{\R^3} \,
        w_{M,N}[\spvar;K^{(M,N)}]
        \nonumber\\
        &&\hspace{2cm}=(M+1)\lim_{\rvar\rightarrow0}\rvar^{-\sigma} w_{M+1,N}
        [\spvar;K^{(M+1)},\tilde K^{(N)}]\Big|_{K_{M+1} =(\rvar n,\lambda)}
        \\
        &&\hspace{2cm}=(N+1)\lim_{\rvar\rightarrow0}\rvar^{-\sigma} w_{M,N+1}
        [\spvar;K^{(M)},\tilde K^{(N+1)}]\Big|_{\tilde K_{N+1}=(\rvar n,\lambda)}\nonumber
\eeqn
hold for all $M,N\geq0$, and every choice of the unit vector $n$.
We recall that $X$ denotes the spectral variable corresponding to $P_f$.

Both the generalized Wick kernels of the Wick ordered fiber Hamiltonian $H(p,\sigma)$, and
$\h^{(0)}$
satisfy ${\rm \bf SR}[1]$. The value $\mu=1$ is a consequence of the normalization
condition (~\ref{kappasig-norm-def-1}). Under the action of the renormalization map
$\ren$, \sbsr is mapped to  ${\rm \bf SR}[\rho^\sigma\mu]$. This is proved in 
section {~\ref{spsrproofsssect}}. Therefore, $\h^{(n)}$
satisfies ${\rm \bf SR}[\rho^{n\sigma}]$. 

In QED, the soft photon sum rules can be viewed as a generalization of the 
differential Ward-Takahashi identities. However,
the existence of soft boson sum rules is not necessarily linked to
the presence of a gauge symmetry. The Nelson model, for instance, admits soft boson
sum rules, but does not exhibit a gauge symmetry; cf. our remarks in Section {~\ref{Nelsonremsubsect}}.
The soft photon sum rules for the QED model are discussed in detail in Section
{~\ref{spsrsect}} and proven to be preserved by $\ren$ in Section {~\ref{spsrproofsssect}}.
Then, in Sections {~\ref{codim2contrproofsubsubsect-1}}
and {~\ref{codim2contrproofsubsubsect-2}}, we establish
the codimension 2 contraction property of $\ren$ stated above.
In every application of $\ren$, the soft photon sum rules imply the
{\em precise cancellation} of all potentially {\em marginal} terms.

\subsection{Determination of the renormalized mass}
\label{det-ren-mass-intro-1}

To bound the first and second derivative of $E(p,\sigma)$ (which is a  
function only of $|p|$) with respect to $|p|$, we use the {\em Feynman-Hellman formula}
\eqn
        	\partial_{|p|}E(p,\sigma)=\frac{\bra \Psi(p,\sigma)\,,\,
        	(\partial_{|p|}H)(p,\sigma)  \Psi(p,\sigma)\ket}
		{\bra\Psi(p,\sigma)\,,\,\Psi(p,\sigma)\ket} \;,
\eeqn
from which it will follow that
\eqn\;\;\;\;\;\;\;
        \partial_{|p|}^2 E(p,\sigma)=
        1-2\frac{\bra(\partial_{|p|}\Psi)(p,\sigma)\,,\,(H(p,\sigma)
        -E(p,\sigma))(\partial_{|p|}\Psi)(p,\sigma)\ket}
        {\bra\Psi(p,\sigma)\,,\,\Psi(p,\sigma)\ket} \;.
        \label{partp2Eintro}
\eeqn
Eq. (~\ref{partp2Eintro}) makes it evident that
$m_{ren}(p,\sigma)>1$. Exploiting algebraic identities
satisfied by the smooth Feshbach map, (~\ref{partp2Eintro}) can be
directly used to derive the bounds asserted in (~\ref{derp2Ebound}), for $p\neq0$,
and $\sigma>0$.

For $p=0$, our aim is to find an estimate
$1<m_{ren}(0,\sigma)<1+cg^2$ with
$c$ {\em uniform in} $\sigma$, as $\sigma\rightarrow0$.
The key to our method is an understanding of how important
physical quantities, such as the renormalized
electron mass, can be extracted from the flow of effective Hamiltonians.

We notice that there are different ways to extract the 
renormalized mass from the renormalization flow of effective Hamiltonians.
One observes in
\eqnn
		H(p,\sigma)&=&\frac{|p|^2}{2m}+H_f-\frac{1}{m}|p|\Ppar+\frac{1}{2m}P_f^2
		-\frac{g}{m}|p|A_{\ks}^\parallel+\frac{g}{m}\bra P_f , A_{\ks}\ket_{\R^3}
		+\frac{g^2}{2m}A_{\ks}^2
\eeqnn
that the inverse of the mass appears in six terms (under the normalization condition
that the coefficient of $H_f$ is 1). Gauge invariance suggests that the inverse 
renormalized mass can be determined either through the second derivative in $|p|$ of
the ground state energy (as above), or
through the quotient between the coefficient of $P_f^2$ and the coefficient of $H_f$, 
or through the quotient between the  
coefficient of one of the other operators and the coefficient of $H_f$. 

Instead of calculating the second derivative of the ground state energy 
with respect to $|p|$, we shall, in the case $p=0$,
determine the renormalized mass from the coefficients of the operators
$P_f^2$ and $H_f$ in the Taylor expansion of the effective Hamiltonians
in $\opp$, via
\eqn
		\frac{1}{m_{ren}^*(0,\sigma)}:= 
		\lim_{n\rightarrow\infty}
		\frac{\rho^{-k}
		\bra \vac\,,\,(\partial_{P_f^\parallel}^2
		T^{(n)})\Big|_{p=0}\vac\ket}
		{\bra\vac\,,\,(\partial_{H_f}
		T^{(n)})\Big|_{p=0} \vac\ket} \;.
		\label{m-star-def-intro-1}
\eeqn 
Let
\eqn
        \Delta T^{(n)}[\opp;p]:=\rho T^{(n)}[\rho^{-1}\opp;p]-T^{(n-1)}[\opp;p]
\eeqn
denote the correction of $T^{(n)}$ due to an application of $\ren$.
For $p=0$, we define
\eqn
		\Delta \gamma_1^{(n)}:=\big(\partial_{H_f}\Delta T^{(n)}\big)
        	[\unull;0] 
		\; \; , \; \;
        	\Delta \gamma_2^{(n)}&:=&\big(\partial_{P_f}^2\Delta T^{(n)}\big)
        	[\unull;0] \;.
\eeqn
We then derive the formula
\eqn
        \frac{1}{m_{ren}^*(0,\sigma)}=
        \frac{1+\sum_{n=-1}^\infty\rho^{-n_+}\Delta\gamma_2^{(n)}}
        {1+\sum_{n=-1}^\infty \Delta\gamma_1^{(n)}}\;,
\eeqn
where $n_+=\max\{n,0\}$. The $n=-1$ term accounts for the first decimation step.
The full discussion is given in Section {~\ref{peq0masssect}}. From 
the uniform bounds on the sums in the numerator and denominator
(owing to the fact that the
operator $P_f^2$ in $T^{(n)}$ is irrelevant), we then arrive at an upper bound on
$m_{ren}^*(0,\sigma)-1>0$ which is uniform in $\sigma$.
We show that $m_{ren}^*(0,\sigma)=m_{ren}(0,\sigma)$ for $\sigma>0$.
By condition (~\ref{mren-def-1}), we then find
$\lim_{\sigma\rightarrow0}m_{ren}(0,\sigma)=
\lim_{p\rightarrow0}\lim_{\sigma\rightarrow0}m_{ren}(p,\sigma)$.
This implies a bound on the renormalized electron mass for $p=0$ which
is {\em uniform} in $\sigma$. 

\subsection{New techniques}
 
In this paper, we introduce some new techniques that can
be expected to be useful in a much broader context. Among those, we particularly
point out the following.

One of the key new methods entails the soft photon sum rules. These establish
a hierarchy of non-perturbative identities that are here used to prove the
precise cancellation of potentially marginal terms in $w_{0,1}$ and $w_{1,0}$
for $p=0$. In \cite{ch}, they are also used to prove strict marginality 
of $w_{0,1}$ and $w_{1,0}$ for $p\neq0$.

We introduce a method to determine renormalized physical parameters, such as
the renormalized mass, by following the renormalization group flow of effective Hamiltonians.
Key to the technique is a method to directly relate the fiber Hamiltonian $H(p,\sigma)$
to the effective Hamiltonian $H[\h^{(n)}]$ at the scale $n$, for arbitrarily
large, finite $n$. The corresponding identity is presented in Lemma {~\ref{first-last-scale-lemma-1}}. 
The proof is based on the recursive use of the important identity (~\ref{QHQid}).
Lemma {~\ref{first-last-scale-lemma-1}} allows us to prove equality of the 
expression (~\ref{partp2Eintro}) for the inverse
renormalized mass $m_{ren}(0,\sigma)$, obtained from the Feynman-Hellman formula, to 
the definition (~\ref{m-star-def-intro-1}) of $m_{ren}^*(0,\sigma)$, 
which only depends on $H[\h^{(n)}]$.  

The smooth Feshbach map, in the form used here, has the advantage that arbitrarily high derivatives 
with respect to $H_f$ can in principle be applied to the effective Hamiltonians. 
Moreover, derivatives in $H_f$ can simply be estimated in operator norm. In the case of the
Feshbach map based on sharp projectors, only the first derivative in $H_f$ 
could be accommodated, and the delta distributions arising therefrom 
burdened the analysis. However, the price of using the smooth Feshbach map
is that, due to the non-vanishing of overlaps $\chi\bar\chi\neq0$, the 
algebraic side of the analysis is somewhat more complicated. In  
this paper, we introduce many new methods, interspersed throughout the text,
to efficiently deal with overlap phenomena.

\subsection{Relations to Nelson's model}
\label{Nelsonremsubsect}

The analysis developed here for the QED model
is applicable to Nelson's model with minor modifications. 
The latter describes a non-relativistic, scalar particle
which interacts with a field of scalar bosons. The Hilbert space is given by
\eqn
        \H=L^2(\R^3)\otimes \Fo_{bos}\;,
\eeqn
with
$$
        \Fo_{bos}:=\bigoplus_{n\geq0} \big(L^2(\R^3)\big)^{\otimes_s n} \;
$$
denoting a Fock space of scalar bosons. Introducing creation- and annihilation
operators $a^\sharp(k)$, as above (except for the absence of polarization labels), 
the Hamiltonian of the system is then defined by
\eqn
        H_{Nelson}(\sigma)&=&\frac{1}{2}
        \big(i\nabla_{x}\otimes\1_{f}+\1\otimes
        P_{bos}\big)^2+\1\otimes H_{bos}
        \nonumber
        \\
        &&+g\int_{\R^3}d^3k\,v_\sigma(k)\big(
        e^{-i\langle k,x\rangle_{\R^3}}\otimes a^*(k)+
        e^{i\langle k,x\rangle_{\R^3}}\otimes a(k)\big)\;,
\eeqn
where $v_\sigma(k):=\frac{\ks(|k|)}{|k|^{\frac12}}$, and where
$g$ is a small coupling constant.
$H_{bos}$ and $P_{bos}$ are the Hamiltonian and momentum 
operator of the free boson field,
defined in the same manner as in QED.
By translation invariance,
it again suffices to consider the restriction of $H_{Nelson}(\sigma)$
to a fiber Hilbert space
$\H_p$, labeled by the conserved total momentum $p\in\R^3$,
\eqn
        H_{Nelson}(p,\sigma)&=&\frac{1}{2}
        \big(p-P_{bos}\big)^2+  H_{bos}+g a^*(v_\sigma)+g a(v_\sigma) \;.
\eeqn
We then apply a {\em Bogoliubov transformation},
\eqn
        H_{Nelson}(p,\sigma)\mapsto H_{BN}(p,\sigma):=
        U_{Bog,\sigma} \, H_{Nelson}(p,\sigma) \, U_{Bog,\sigma}^*\;,
\eeqn
by which
\eqn
        a^\sharp(k)\rightarrow a^\sharp(k)-|k|^{-1} v_\sigma(k) \;.
\eeqn
The operator $U_{Bog,\sigma}$ is unitary if $\sigma>0$, but
in the limit $\sigma\rightarrow0$, the image of $\Fo$ under
$U_{Bog,\sigma}$ lies in a Hilbert space carrying a representation of 
the canonical commutation relations inequivalent to the Fock representation.

The Bogoliubov-transformed Nelson Hamiltonian at fixed 
conserved total momentum $p$ is given by
\eqn
        H_{BN}(p,\sigma)=\frac{1}{2}
        \big(p-P_{bos}-g a(w_\sigma)-g a^*(
        w_\sigma)\big)^2+  H_{bos} \;,
\eeqn
where $w_\sigma(k):=v_\sigma(k)\frac{k}{|k|}$ is a vector-valued function in
the boson momentum space.

A very important issue in the context of the operator-theoretic renormalization 
group method, in which the Nelson model may differ from the QED
model, is whether it admits soft boson sum rules or not.
The answer is affirmative, even though Nelson's model
has no gauge symmetry. The soft boson sum rules
of the Bogoliubov transformed Nelson model differ from those  
presented in Section {~\ref{spsrsect}} for the QED model
only in the replacement of the photon polarization vector $\e(K)$ appearing
in the definition (~\ref{sbsr}) by the radial unit vector $\frac{k}{|k|}$.
Correspondingly, all steps in the analysis presented here
for $H_{QED}(p,\sigma)$
apply, with minor modifications, to $H_{BN}(p,\sigma)$.
Consequently, the results of the present work can be extended to
the infrared mass renormalization for the Nelson model.


\section{The smooth Feshbach map} 

We present key properties of the {\em smooth Feshbach map} in
this section. This functional analytical tool was introduced in \cite{bcfs1}, and
generalizes the standard Feshbach map based on sharp projectors in \cite{bfs1,bfs2}.
We refer to \cite{bcfs1} for a detailed exposition and for proofs.

\subsection{Feshbach pairs and smooth Feshbach map}
\label{Fpairssubsec}

Let $\H$ be a separable Hilbert space. We introduce a pair of selfadjoint
operators $0\leq\chi\leq1$ and $\bch:=\sqrt{1-\chi^2}$, acting on $\H$,
thus obtaining a partition of unity through $\chi^2+\bch^2=\1$.
Let $P_\chi$, $P_{\bar\chi}$ denote the
orthoprojectors onto the subspaces $\Ran(\chi)$, $\Ran(\bch)\subset\H$, and
$P^\perp_\chi= \1-P_\chi$, $P_{\bar\chi}^\perp=\1-P_{\bar\chi}$
their respective complements.
Clearly, $\Ran(\chi)$ and $\Ran(\bch)$ are disjoint if and only if
$\chi$ is a projector.

\begin{definition}\label{Feshbtripledef}
A pair of closed operators $(H,\tau)$ acting on $\H$
is called a Feshbach pair corresponding to $\chi$  if:
\begin{itemize}
\item   $\Dom(H)=\Dom(\tau)\subset\H$, $\chi$ and $\bch$ map $\Dom(H)$ to itself, and
$[\chi,\tau]=0=[\bch,\tau]$.

\item
Let
\eqn
        H_{\bch}&:=&\tau+\bch \omega\bch
        \nonumber\\
        \omega&:=&H -\tau \;.
\eeqn
The operators $\tau$ and $H_{\bch}$
are bounded invertible on $\Ran(\bch)$.

\item
Let
\eqn
        \bar R:=H_{\bch}^{-1}\;,
\eeqn
and let $H_{\bch}=U|H_{\bch}|$ denote the polar decomposition of
$H_{\bch}$ on $\Ran\bch$. Then,
\eqn 
    \big\|\bar{R}\big\|_{\cB(\H)}&<&\infty
    \nonumber\\
    \big\|\big|\bar{R}\big|^{\frac{1}{2}}U^{-1}
    \bch\omega\chi\big\|_{\cB(\Ran(\chi),\H)}
    & , &
    \big\|\chi \omega\
    \bch\big|\bar{R}\big|^{\frac{1}{2}}
    \big\|_{\cB(\H,\Ran(\chi))}<\infty \; .
    \label{Feshboundopasscond}
\eeqn
\end{itemize}
We denote by
\eqn
        \Fpairs(\H,\chi)
\eeqn
the set of all Feshbach pairs on $\H$ that correspond to $\chi$.
\end{definition}

The {\em smooth Feshbach map} is defined by
\eqn
        F_\chi:\Fpairs(\H,\chi)&\rightarrow&\cL(\H) \;, \nonumber\\
        (H,\tau)&\mapsto& \tau + \chi\,
      \omega \,\chi -
     \chi  \,\omega \,\bch\, \bar{R}\, \bch \, \omega \,\chi \; ,
        \label{FchiHtaudef}
\eeqn
We note that on the subspace Ran$(P_\chi^\perp)$, $F_\chi(H,\tau)$ is simply $\tau$,
and thus commutes with $\chi,\bar\chi$. On Ran$(P_\chi)$, $F_\chi(H,\tau)$
defines a bounded operator in $\cB$(Ran$(P_{\chi})$).
We also introduce {\em intertwining maps}
\eqn
        Q_\chi : \Fpairs(\H,\chi)&\rightarrow&\cB(\Ran(\chi),\H) \;,
        \nonumber\\
        (H,\tau)&\mapsto&\chi \,- \,
        \bch\, \bar{R} \,\bch \,\omega \,\chi \;,
        \label{QchiHtaudef}
\eeqn
and
\eqn
        Q^\sharp_\chi  :
        \Fpairs(\H,\chi)&\rightarrow&\cB(\H,\Ran(\chi))\;,
        \nonumber\\
        (H,\tau)&\mapsto&\chi \,- \,\chi\, \omega \,\bch\,
        \bar{R}\,\bch \;.
        \label{QschiHtaudef}
\eeqn
We shall next discuss the properties of these operators that are needed
in the present work.

The smooth Feshbach map and the intertwining operators possess very powerful
spectral properties, which we describe in the sequel. Moreover, we present a number of crucial
algebraic identities which will be of great use later.

\subsection{Feshbach isospectrality} 

The smooth Feshbach map establishes a non-linear, isospectral
map between operators on $\H$ and $\Ran(\chi)$ according to
the following theorem.

\begin{theorem}\label{Feshprop} (Feshbach isospectrality theorem)
Assume that $(H,\tau)\in\Fpairs(\H,\chi)$.  Then, the following hold.
\begin{itemize}
\item (Isospectrality)
$H$ is bounded invertible on $\H$ if and only if
$F_{\chi}(H,\tau)$ is bounded invertible on $\Ran(\chi)$.
If $H$ is invertible,  
\eqn
	F_\chi(H,\tau)^{-1}=\chi H^{-1}\chi+\bar\chi \tau^{-1} \bar\chi 
	\label{inverse-id-1}
\eeqn
and
\eqn 
	H^{-1}=Q_\chi(H,\tau) F_\chi(H,\tau)^{-1} Q^\sharp_\chi(H,\tau)
	+\bar\chi \, \bar R \, \bar\chi \;.
	\label{inverse-id-2}
\eeqn
\item 
Let $\psi\in\H$. Then, $H\psi=0$ if and only if
$F_{\chi}(H,\tau) \chi \psi = 0$ on $\Ran(\chi)$.

\item (Reconstruction of an eigenvector)
Let $\zeta\in \Ran(\chi)$. Then, $F_{\chi}(H,\tau)\zeta=0$
if and only if $H Q_{\chi}(H,\tau)\zeta=0$.
\end{itemize}
\end{theorem}

The identities formulated in the following lemma will be very important
later (to relate "effective Hamiltonians" on different scales to one another).

\begin{lemma}
Let $(H,\tau)\in\Fpairs(\H,\chi)$. Then, the following identities hold.
\eqn
        \chi F_\chi(H,\tau)&=&HQ_\chi(H,\tau)\nonumber\\
        F_\chi(H,\tau)\chi&=&Q^\sharp_\chi(H,\tau) H \;,
        \label{FHQid}
\eeqn
and
\eqn
        Q^\sharp_\chi(H,\tau)HQ_\chi(H,\tau)=F_\chi(H,\tau)
		-F_\chi(H,\tau)\bar\chi\tau^{-1}\bar\chi F_\chi(H,\tau) \;.
        \label{QHQid}
\eeqn

\end{lemma}

\prf
The proof of (~\ref{FHQid}) is given in \cite{bcfs1}. 

To prove (~\ref{QHQid}), let for brevity $\omega_A\equiv A\omega A$ for $A=\chi,\bar\chi$.
Moreover, let 
$F\equiv F_\chi(H,\tau)$ and $Q\equiv Q_\chi(H,\tau)$. Then,
\eqnn
\lefteqn{Q^\# H Q - F + F \bar\chi \tau^{-1} \bar\chi F
}
\\ & = &
F \chi Q - F + F \bar\chi \tau^{-1} \bar\chi 
\big( \tau + \omega_\chi - \chi \omega \bar\chi H_{\bar\chi}^{-1} \bar\chi \omega \chi \big)
\\ & = &
F \big( \chi^2 - \chi \bar\chi H_{\bar\chi}^{-1} \bar\chi \omega \chi - \1 
+ \bar\chi \tau^{-1} \bar\chi \tau + \bar\chi \tau^{-1} \bar\chi \omega_\chi 
\\ & & 
- \bar\chi \tau^{-1} \bar\chi \chi \omega \bar\chi H_{\bar\chi}^{-1} \bar\chi \omega \chi \big)
\\ & = &
F \big( - \chi \bar\chi H_{\bar\chi}^{-1} \bar\chi \omega \chi 
+ \chi \bar\chi \tau^{-1} \bar\chi \omega \chi - 
\chi \bar\chi \tau^{-1} \omega_{\bar\chi} H_{\bar\chi}^{-1} \bar\chi \omega \chi \big)
\\ & = &
- F \chi \bar\chi \big( H_{\bar\chi}^{-1} - \tau^{-1} 
+ \tau^{-1} \omega_{\bar\chi} H_{\bar\chi}^{-1} \big) \bar\chi \omega \chi 
\ = \ 0  \;.
\eeqnn
Thus we arrive at the assertion of the lemma.
\endprf

\subsection{Derivations}
\label{derivationssect3}

Next, we consider the action of derivations on $F_{\chi}(H,\tau)$.
Consider a Hilbert space $\H$ with a dense subspace $\mathcal{D}\subset\H$,
and let $\mathcal{L}(\mathcal{D},\H)$ denote the space of linear (not
necessarily bounded) operators from $\mathcal{D}$ to $\H$.
A derivation $\der$ is a linear map $\Dom(\der)
\rightarrow \mathcal{L}(\mathcal{D}, \H)$, defined on a subspace
$\Dom(\der) \subset \mathcal{L}(\mathcal{D}, \H)$, which obeys
Leibnitz' rule. That is, for $A, B \in \Dom(\der)$,
$\Ran( B) \subseteq \mathcal{D}$, and $A \, B \in \Dom(\der)$,
$$
        \der[A\,B]=\der[A]B + A\der[B]\;.
$$
Let $(H,\tau)\in\Fpairs(\H,\chi)$, and
assume that $H,\tau \in \mathcal{L}(\mathcal{D}, \H)$, where
$\mathcal{D} := \Dom(H) = \Dom(\tau)$ and that $H, \tau, \chi, \bch$
and the composition of operators in the definition of
$F_{\chi}(H,\tau)$ are contained in $\Dom(\der)$.

\begin{theorem}
\label{derXbarRlemma1}
\label{derXfPthm}
Assume that $\der[\bar\chi]$ and $\bar\chi$ are bounded operators which 
leave $\mathcal{D}$ invariant and commute with $\tau$ and with each other. Then,
under the above conditions,
\eqnn
            \der [F_{\chi}(H,\tau)]
            &=&
            \der[\tau] + \chi \omega \bch \bar{R}
            \der[\tau]
            \bar{R} \bch \omega \chi
            + Q^\sharp \der[\omega] Q\;\nonumber\\
            &+&\der[\chi]H Q + Q^\sharp H \der[\chi]
            \nonumber\\
	    &-&2 \chi \omega \bch\bar{R}
	    \big(\tau^{-1}\der[\bch] - \bar{R}
        \bch \omega  \tau^{-1}\der[\bch]\big)\,\tau \,
            \bar{R}\bch\omega\chi \; .
\eeqnn
If $[\der [\chi],\bch]=0=\der [\tau]$, this reduces to
\eqn
        \der [F_{\chi}(H,\tau)] = Q^\sharp \der [H] Q \;,
        \label{Feshdersimple}
\eeqn
and furthermore,
\eqn
        \der [Q_\chi(H,\tau)]&=&-\bar\chi \bar R
        \bar\chi \der [H]  Q_\chi(H,\tau)
        \nonumber\\
        \der [Q_\chi^\sharp(H,\tau)]&=&-
         Q_\chi^\sharp(H,\tau)
         \der [H] \bar\chi\bar R\bar\chi
        \label{Qdersimple}
\eeqn
holds for the intertwining operators.
\end{theorem}

\subsection{Compositions}
\label{concatlawssect}

Another key aspect of smooth Feshbach maps and intertwining
operators concerns their properties under composition.
To this end, we consider a pair of mutually commuting,
selfadjoint operators $0 \leq \chi_1 , \chi_2 \leq
1$, with $\bch_j := (\1 - \chi_j^2)^{\frac12}$. Furthermore,
let $\chi_1 \chi_2 = \chi_2
\chi_1=\chi_{2}$, such that $\Ran(\chi_2)\subseteq \Ran(\chi_1) \subset \H$.
Then, we consider Feshbach pairs
\eqnn
        (H,\tau_1)&\in&\Fpairs(\H,\chi_1)
        \\
        (H,\tau_{2})&\in&\Fpairs(\H,\chi_2)
        \\
        (F_1,\tau_{12})&\in&\Fpairs( \Ran(\chi_1),\chi_2) \;,
\eeqnn
with $F_1 := F_{\chi_1}(H,\tau_1)$, and where $\tau_1$,
and $\tau_{12}$ commute with
$\chi_j, \bch_j$.

\begin{theorem} \label{QQscomplawthm333} Under the above assumptions,
\eqn
        F_{\chi_{2}}(H,\tau_2) &=& F_{\chi_2}(F_1,\tau_{12}) \; ,
        \nonumber\\
        Q_{\chi_{2}}(H,\tau_2) &=& Q_{\chi_1}(H,\tau_1)
        \, Q_{\chi_2}(F_1,\tau_{12}) \; ,
        \nonumber\\
        Q^\#_{\chi_{2}}(H,\tau_{2}) &=& Q^\#_{\chi_2}(F_1,\tau_{12})
        \, Q^\#_{\chi_1}(H,\tau_1) \; ,
        \label{Feshcompositionrules}
\eeqn
if and only if $\tau_2=\tau_{12}$. Furthermore,
\eqn
        A \, Q_{\chi_{2}}(H,\tau_{2}) =  A \, Q_{\chi_2}(F_1,\tau_{12})
         \; \; , \; \;
        Q^\sharp_{\chi_{2}}(H,\tau_{2}) \, A
        =  Q^\sharp_{\chi_2}(F_1,\tau_{12})  \, A \; ,
        \label{AQpullthrform333}
\eeqn
for all operators $A$ on $\H$ that satisfy $A \bch_1 = \bch_1 A = 0$.
\end{theorem}

\subsection{Organizing overlap terms}
\label{overlapopsubsubsec}

For the class of Feshbach pairs $(H,\tau)\in\Fpairs(\H,\chi)$ of interest in
the present work, one can write $H=T+W$, where $W$ is a small perturbation
of (has a small relative bound
with respect to) $T$. 
Moreover, $[T,\chi]=0=[T,\tau]$,
while $[W,\chi],[W,\tau]\neq0$.   

But in contrast to the situation considered in \cite{bcfs1}, 
we shall here {\em not} make the choice $\tau=T$. 
Consequently, owing to  
$\chi\bch\neq0$, terms of the form $(T-\tau)\chi\bch$, 
which have a spectral support of small Lebesgue measure, 
may possess a large operator norm. Hence, it will in general only
be possible to study the resummation of resolvent expansions with
respect to the operator $W$, but not with respect to the operator
$\omega=H-\tau=T-\tau+W$. Consequently, the algebraic structure of the
smooth Feshbach map in the present work is more complicated than in \cite{bcfs1}. 
It is also more complicated than in \cite{bfs1,bfs2}, 
where $\chi=P$ is a sharp projector,
so that $(T-\tau)P\bar P=0$ is trivially true (the operator
defined in (~\ref{piopdef}) then simply reduces to the identity).

For future use, it will be convenient
to introduce an alternative expression for the smooth
Feshbach map that makes the separation of $\omega$ into
large and small terms manifest.

\begin{lemma}\label{piopdeflemma}
Let $(H,\tau)\in\Fpairs(\H,\chi)$, and
assume that
$H=T+W$, where $[T,\chi]=[T,\tau]=0$. Let
\eqn\label{barR0def}
        T':=T-\tau
        \; \; {\rm and} \; \;
        \bar{R}_0:=(\tau+\bch T'\bch)^{-1}
\eeqn
on $\Ran(\bch)$.
We introduce the operator
\eqn
    \piop_\chi(T,\tau) &:=& \1 - \bch T'\bch \bar{R}_0 \nonumber\\
    & =&
    P_{\bch}^\perp + P_{\bch} \tau \bar{R}_0 \;
    \label{piopdef}
\eeqn
on $\Ran(\chi)$, where
$\Ran(\piop_\chi(T,\tau)-\1)=\Ran(\chi\bch)$,
and where $\piop_\chi(T,\tau)$ commutes with $\tau,\chi,\bch$ and $T$.
Then,
\eqn
        F_{\chi}(H,\tau)\;=\;\tau &+& \chi T' \piop_\chi(T,\tau) \chi 
	\nonumber\\
        &+& \chi \piop_\chi(T,\tau) (W - W \bch\bar{R}\bch W) \piop_\chi(T,\tau)\chi\;,
\eeqn
and in particular, $\piop_\chi=\1$ if and only if $\tau=T$.
\end{lemma}

\prf
Using the second resolvent identity
\eqn
        \bar{R}&=&\bar{R}_0\,-\,\bar{R}_0\,\bch
            W\bch\bar{R}\nonumber\\
            &=&\bar{R}_0\,-\,\bar{R}\,\bch
            W\bch\bar{R}_0\;,\label{secondresident4444}\eeqn
and
$$
        \omega \;=\;T'\,+\,W\;,
$$
we find
\eqn
    F_\chi(H,\tau)&=&\tau \,+\,\chi T' \chi\,+\,
    \chi W\chi\,\nonumber\\
    &&-\,\chi\omega\bch\bar{R}\bch\omega\chi
    \nonumber\\
    &=&\tau\,+\,\chi T'\chi\,-\,\chi T'\bch\bar{R}_0\bch T'\chi
    \nonumber\\
    &&\,+\,
    \chi W\chi\,-\,\chi W\bch\bar{R}\bch W\chi\nonumber\\
    &&\,-\,
    \chi W\bch\bar{R}_0\bch T'\chi\,-\,
    \chi T'\bch\bar{R}_0\bch W\chi
    \nonumber\\
    &&\,+\,\chi T'\bch\bar{R}_0\bch W\bch\bar{R}_0\bch T'\chi
    \nonumber\\
    &&\,+\,\chi T'\bch\bar{R}_0\bch W\bch\bar{R}\bch W\chi\,+\,
    \chi W\bch\bar{R}\bch W\bch\bar{R}_0\bch T'\chi
    \nonumber\\
    &&\,-\,\chi T'\bch\bar{R}_0\bch W\bch\bar{R}\bch W
    \bch\bar{R}_0\bch T'\chi\nonumber\\
    &=&\tau\,+\,\chi T'\chi\,-\,\chi\,
    T'\bch\bar{R}_0\bch T'\chi\nonumber\\
    &&+\,\chi(\1-\bch T'\bch \bar{R}_0)W
    (\1-\bch T'\bch \bar{R}_0)\chi\nonumber\\
    &&-\,\chi(\1-\bch T'\bch \bar{R}_0)W\bar{R}W
    (\1-\bch T'\bch \bar{R}_0)\chi\;.
\eeqn 
From the second line in (~\ref{piopdef}) follows immediately that
$\piop_\chi=\1$ if and only if $\tau=T$. \endprf


\section{Isospectral renormalizaton group: Effective Hamiltonians} 
\label{Banspsubsect} 

Our first step in constructing the framework of the isospectral
operator-theoretic renormalization group is to introduce the spaces
on which we will introduce an isospectral renormalization transformation.
The central objects of interest are {\em effective Hamiltonians}, 
which are bounded operators on $\H_{red}=\1[H_f<1]\Fo\subset\Fo$
of a generalized Wick ordered normal form. 

In Section {~\ref{Banspsubsect}}, we define
a Banach space of generalized integral (Wick) kernels, which parametrize
those effective Hamiltonians.

As a first step, we consider the spectral subspace
$$
        \H_{red}:=\1[H_f<1]\Fo\subset\Fo \;.
$$
Furthermore, we choose a smooth cutoff function
\eqn
        \chi_1[x]:=\sin[\frac\pi2 \Theta(x)]
        \label{Thetadef}
\eeqn
of $[0,1)$, with
\eqn
        \Theta\in
        C_0^\infty([0,1);[0,1])
\eeqn
and
\eqn
        \Theta=1 \; \; {\rm on } \; [0,\frac34] \;.
\eeqn
Moreover, we let
$$
        \bar\chi_1[x]:=\sqrt{1-\chi_1^2[x]}\;.
$$
The selfadjoint cutoff operators $\chi_1[H_f]$ and $\bar\chi_1[H_f]$ are defined
in the sense of the functional calculus.

We use the notation
\eqn
        \opp:=(H_f,P_f)
\eeqn
and let 
\eqn
        X&=&(X_1,X_2,X_3)\in B_1 \;,
        \nonumber\\
        \spvar&=&(X_0,X)\in[0,1]\times B_1 \;,
\eeqn
denote the corresponding spectral variables, where $B_1$ is the closed unit ball in $\R^3$.

We then introduce operators, referred to as {\em effective Hamiltonians}, of the
particular form
\eqn
        H_{eff}=T[\opp;p] + E\chi_1^2[H_f]+\chi_1[H_f]W[p]\chi_1[H_f]
        \label{Heff-def-1}
\eeqn
which act on $\H_{red}$. They are parametrized
by the conserved total momentum $p\in\R^3$, and depend on
a complex parameter $E$. The function
$T[\spvar;p]$ can be written in the form
\eqn	
		T[\spvar;p]=X_0+T'[\spvar;p] 
		\;\;,\;{\rm and}\;\;T'[\spvar;p]= \chi_1^2[X_0]\tT[\spvar;p]  \;.
		\label{T-prime-def-2}
\eeqn 
It satisfies $T[\underline{0};p]=0$, is of class $C^1$ in $X_0$, and
of class $C^2$ in $X$.  
The operator $T[\opp;p]$,
defined in the context of the functional calculus, is referred to as the
{\em non-interacting}, or {\em free} (part of the effective) Hamiltonian.

The operator $W[p]$ can be written as
$$
        W[p]=\sum_{M+N\geq1}W_{M,N}[p] \;,
$$
where the operators in the sum are defined as follows.
A {\em generalized Wick monomial} of degree $(M,N)$, for $M+N\geq0$, 
is an Wick ordered polynomial in $M$ creation and $N$ annihilation
operators of the form
\eqn
        W_{M,N}[p] &\equiv&W_{M,N}[w_{M,N}]
        \nonumber\\
        &=&
        P_{red}\int_{B_1^{M+N}}\frac{ dK^{(M,N)}}{|k^{(M,N)}|^{\frac12}}
        \, a^*(K^{(M)}) \, w_{M,N}[\opp;K^{(M,N)};p] \,
        a(\tilde K^{(N)})
        P_{red} \; 
        \label{WMNwMNdef}
\eeqn
where $P_{red}\equiv\1[H_f<1]$, 
and where we introduce the notation
\eqnn
        K:=(k,\lambda) &\in& B_1\times\{+,-\}\\
        K^{(M)}:=\big(K_1, \dots, K_M \big)
        & , &
        k^{(M)}:=\big(k_1,\dots,k_M\big) \\
        K^{(M,N)}:=\big( K^{(M)},\tilde K^{(N)} \big)
        & , &
        k^{(M,N)}:=\big( k^{(M)},\tilde k^{(N)} \big)\\
        |k^{(M,N)}|:=|k^{(M)}|\cdot|\tilde k^{(N)}|
        & , &
        |k^{(M)}|:=|k_1|\cdots|k_M|  \\
        \uk:=(|k|,k) \in [0,1]\times B_1
        & , &
        \sum\big[\uk^{(m)}\big] := \uk_1+\cdots+\uk_m \\
        a^\sharp(K^{(M)})&:=&a^\sharp(K_1)\cdots a^\sharp(K_M) \\
        dK^{(M)}:= \sum_{\lambda_1,\dots,\lambda_M}
         d^3 k_1\cdots d^3k_M
        & , &
        d K^{(M,N)}:= d K^{(M)}d\tilde K^{(N)}\; .
\eeqnn
Furthermore, the notation
\eqn
        "\,k\in k^{(M,N)}\,"
\eeqn
shall imply $k=k_i$ for some $i\in\{1,\dots,M\}$, or $k=\tilde k_j$,
for some $j\in\{1,\dots,N\}$.
Through (~\ref{WMNwMNdef}), $W_{M,N}$ is uniquely determined by the 
\begin{center}
{\em generalized Wick kernel}  $w_{M,N}[\spvar;K^{(M,N)};p]$
{\em of degree $(M,N)$}. 
\end{center}
$w_{M,N}$ is separately totally symmetric with respect to the
variables $K^{(M)}$ and $\tilde K^{(N)}$.

For the problem at hand, we require that the generalized Wick kernels satisfy
\eqn
		w_{M,N}[ R\spvar;RK^{(M,N)};p]=w_{M,N}[\spvar;K^{(M,N)};R^{-1}p]
\eeqn
for all $R\in O(3)$, where
\eqn
		R\spvar&:=&(X_0,RX)
		\nonumber\\
		RK^{(M,N)}&:=&(RK^{(M)},R\tilde K^{(N)}) 
		\nonumber\\ 
		R K^{(N)}&:=&(RK_1,\dots,R K_N)  \;,
\eeqn
and $R K:=(Rk,\lambda)$.
Likewise, we require that  
\eqn
        T[X_0,RX;p]=T[\spvar;R^{-1}p] 
\eeqn
for all $R\in O(3)$.
Hence, in the special case $p=0$,
\eqn
		w_{M,N}[X_0,RX;RK^{(M,N)};0]&=&w_{M,N}[\spvar;K^{(M,N)};0] \;,
		\nonumber\\
		T[X_0,RX;0]&=&T[\spvar;0] \;,
\eeqn
for all $R\in O(3)$. That is, the effective Hamiltonian (~\ref{Heff-def-1}) is
{\em rotation and reflection symmetric} if $p=0$.

The parameter $p\in\R^3$ will in the sequel frequently be omitted from the notation.

\subsection{A Banach space of generalized Wick kernels}
\label{Banach-seq-sp-sssect-1}

We introduce the Banach space
\eqnn
        \Wspace_{M,N}=\Big\{w_{M,N}\Big|\|w_{M,N}\|^\sharp_\sigma<\infty\Big\} \;,
\eeqnn
of generalized Wick kernels of {\em degree} $(M,N)$, with $M+N\geq1$, endowed with the norm
\eqn
        \|w_{M,N}\|^\sharp_\sigma
        &:=&\sum_{a_0=0,1}\|\partial_{X_0}^{a_0}w_{M,N}\|_\sigma+
        \sum_{1\leq|\aind|\leq2\atop a_0=0}
        \|\partial_{\spvar}^{\aind}w_{M,N}\|_\sigma
        \nonumber\\
        &+&
        \sup_{k\in k^{(M,N)}}\Big\||k|^\sigma\partial_{|k|}\big(|k|^{-\sigma}w_{M,N}\big)\Big\|_\sigma
        +\1_{|p|>0}\,
        \sum_{|\aind|\leq1}\Big\|\partial_{|p|}\partial_{\spvar}^{\aind}
        w_{M,N}\Big\|_\sigma \;,
        \label{normsigmasharpdef}
\eeqn
where
\eqn
        \partial_{\spvar}^{\aind}&:=&\partial_{X_0}^{a_0}\cdots\partial_{X_3}^{a_3}
        \;\;,\;\;\aind \in \N_0^4 \;,
\eeqn
and
\eqn
        \|w_{M,N}\|_\sigma &:=&\sup_{K^{(M,N)}\in(B_1\times\{+,-\})^{M+N}}
        (2\pi^{\frac12})^{M+N}\big|k^{(M,N)}\big|^{-\sigma}
        \nonumber\\
        &&\hspace{2cm}\times
        \sup_{|X|\leq X_0<1}
        \big|w_{M,N}[\spvar;K^{(M,N)}]\big|
        \;. \; \;
        \label{normsigmadef}
\eeqn
\begin{remark}
 
We note that $\e(k,\lambda)$ is homogenous of degree zero with respect
to $|k|$, whereby $\e(\,\cdot\,,\lambda):S^2\rightarrow S^2$.
The definition of the norm (~\ref{normsigmasharpdef}) only involves the {\em radial}
derivative $\partial_{|k|}$ with respect to the photon momentum, but
no tangent derivatives. Therefore, no regularity beyond measurability
of the polarization vectors is required.  
\end{remark}
 
\begin{remark}
We will employ different methods in our analysis for the cases $p=0$ and $p\neq0$.
For this reason, the derivative with respect to $|p|$ 
does not enter the definition of the norm (~\ref{normsigmasharpdef}) if $p=0$.
\end{remark}

\begin{remark}
We use the supremum norm in $k^{(M,N)}$-space instead of a weighted $L^2$-norm 
as in \cite{bcfs1}, because the soft photon sum rules in Section {~\ref{spsrsect}} 
require a pointwise property of $w_{M,N}$ with respect to $k^{(M,N)}$.
Consequently, the norm of generalized Wick kernels (~\ref{normsigmadef}) 
is stronger than that used in
\cite{bcfs1} (there also denoted by $\|\,\cdot\,\|_\sigma$), and the spaces $\Wspace_{M,N}$
are smaller than the corresponding spaces in \cite{bcfs1}. 
Apart from this, there is no fundamental difference in comparison to \cite{bcfs1}.\index{}
Indeed, the following key theorem, which 
relates the operator norm on $\cB[\H_{red}]$ to $\|\,\cdot\,\|_\sigma$, 
follows straightforwardly
from the corresponding result proven as Theorem 3.1 in \cite{bcfs1}.
\end{remark}

\begin{theorem}\label{opL2relbound}
Let $\sigma>0$, and $M,N\in\N_0$, such that $M+N\geq1$. Assume that $w_{M,N}\in
\Wspace_{M,N}$, and $W_{M,N}:=W_{M,N}[w_{M,N}]$. Then, on $\H_{red}$,
\eqn
        \|W_{M,N}\|_{{\rm op}}&\leq& \|(H_f P_\vac^\perp)^{-M/2}
        W_{M,N}(H_f P_\vac^\perp)^{-N/2}\|_{{\rm op}}
        \nonumber\\
        & \leq&
        \frac{1}{M^{\frac M2}N^{\frac N2}}
        \|w_{M,N}\|_\sigma \;,
\eeqn
where $\|\,\cdot\,\|_{{\rm op}}$ denotes the operator norm on $\cB[\H_{red}]$.
$P_\vac:=\big|\vac\rangle\langle\vac\big|$ is the orthoprojector onto the span
of the vacuum vector in $\Fo$, and $P_\vac^\perp=\1-P_\vac$ is its complement.
\end{theorem}

For $M=N=0$, we consider the natural splitting
$$
        w_{0,0}[\spvar]=w_{0,0}[\underline{0}]
        +\big(w_{0,0}[\spvar]-w_{0,0}[\underline{0}]\big) \;,
$$
which induces the decomposition
$$
        \Wspace_{0,0}=\C\oplus\Tspace^\sharp \;,
$$
where  
\eqnn
        \Tspace^\sharp:=\Big\{T: \bigcup_{r\in[0,1)}\{r\}\times B_r\rightarrow\C
		&\Big|&\|T\|_{\Tspace}^\sharp<\infty \;,\;T[\underline{0};p]=0\;,\;\;\;\;
        \nonumber\\
        &&T[X_0,RX;p]=T[\spvar;R^{-1}p]
        \;\;\forall\; R\in O(3) \Big\} \;
\eeqnn
with  
\eqn
        \|T\|_{\Tspace}^\sharp:=\sup\Big\{\|P_{|X|\leq X_0<3/4}T\|^\sharp
        \, , \, \frac{1}{K_\Theta}\|P_{|X|\leq X_0 \in[3/4,1)}T\|^\sharp\Big\}
         \;,
        \label{Tnormdef}
\eeqn
using
\eqn
        \|T\|^\sharp:=\sup_{|X|\leq X_0<1}|\partial_{X_0}T|&+&
        \sum_{|\ua|=1,2\atop a_0=0}
        \sup_{|X|\leq X_0<1} \big|\partial_{\spvar}^{\ua}T\big|
        \nonumber\\
        &+&\sup_{|X|\leq X_0<1} \; \1_{|p|>0} \; \sum_{|\aind|\leq1}\big|
        \partial_{|p|}\partial_{\spvar}^{\aind}T\big|\;.
        \label{Tnormsharpdef}
\eeqn
The (explicitly computable) constant $1<K_\Theta<\infty$ depends only on the
smooth cutoff function $\Theta$, which was introduced in (~\ref{Thetadef}), and
is determined by the condition formulated in (~\ref{renT0part-1}) below. 
The different weights in $\|\,\cdot\,\|_{\Tspace}^\sharp$
on the spectral subintervals $[0,\frac34)$
and $[\frac34,1)$ account for the overlap effects
discussed in Section {~\ref{overlapopsubsubsec}}.
Evidently, $(\Tspace^\sharp,\|\,\cdot\,\|_{\Tspace}^\sharp)$ is a Banach space, and  
$ \|\,\cdot\,\|^\sharp$, $\|\, \cdot \,\|_{\Tspace}^\sharp$ are equivalent norms.

Correspondingly, we introduce the space of sequences of generalized Wick kernels
\eqn
        \Hspace^\sharp_{\geq0}:=\C\oplus\Tspace^\sharp\oplus
        \Hspace_{\geq1}^\sharp \;,
        \label{Hspace-0-def-1}
\eeqn
where
\eqn
        \Hspace_{\geq k}^\sharp:=
        \bigoplus_{M+N\geq k}\Wspace_{M,N} \;.
        \label{Hspace-k-def-1}
\eeqn
Elements of this space are henceforth denoted by
\eqn
        \h &=& E\oplus T \oplus
        \h_1\; \in \; \Hspace_{\geq0}^\sharp
        \label{h-Hspace-0-def-1}
\eeqn
with
\eqn
        \h_k &=& (w_{M,N})_{M+N\geq k} \; \in \; \Hspace_{\geq k}^\sharp\;.
        \label{h-Hspace-k-def-1}
\eeqn
Given $\xi\in(0,1)$, we define the norm
\eqn
        \|\h\|_{\xi,\sigma}^\sharp:=|E|+\|T\|_{\Tspace}^\sharp+\|\h_1\|^\sharp_{\xi,\sigma} \;,
\eeqn
where
\eqn
        \|\h_k\|^\sharp_{\xi,\sigma}:=\sum_{M+N\geq k}\xi^{-M-N}\|w_{M,N}\|^\sharp_\sigma\;,
\eeqn
and note that
for $A=0,1$, $(\Hspace^\sharp_{\geq A},\|\,\cdot\,\|_{\xi,\sigma})$ is a Banach space.

We will henceforth write
\eqnn
        W[\h]:=\sum_{M+N\geq1} W_{M,N}[w_{M,N}]\;.
\eeqnn
According to Theorem 3.3 in \cite{bcfs1}, the map
\eqn
        H:\Hspace_{\geq0}^\sharp&\rightarrow&\cB(\H_{red})
        \nonumber\\
        \h&\mapsto& H[\h]:=T[\opp]+
        E\chi_1^2[H_f] +\chi_1[H_f]W[\h]\chi_1[H_f]
        \label{H-Hsp-cB-map-def-1}
\eeqn
is an injective embedding of $\Hspace_{\geq0}^\sharp$
into the bounded operators on $\H_{red}$.
In particular, we have an a-priori bound
\eqn
        \|H[\h]\|\leq\|\h\|^\sharp_{\xi,\sigma} \;,
\eeqn
for $\h\in\Hspace_{\geq0}^\sharp$, and 
\eqn
        \|H[\h_1]\|\leq\xi\|\h_1\|^\sharp_{\xi,\sigma}
\eeqn
for $\h_1\in\Hspace_{\geq1}^\sharp$.


\section{Isospectral renormalizaton group: Renormalization map}
\label{Rentrsf}

We will next construct a variant of the isospectral renormalization map
presented in \cite{bcfs1} that accommodates the specific features
of the model considered in this paper.  

\subsection{Definition of the isospectral renormalizaton map}

We let $\h$ depend holomorphically on a spectral
parameter $\z\in\I:=\{\zeta\in\C\big|\,|\zeta|\leq\frac{1}{10}\}$, and consider families
of effective Hamiltonians parametrized by $\h[\z]$.
Accordingly, we introduce the Banach space $\Hspace_{\geq0}=\C\oplus \Tspace\oplus\Hspace_{\geq1}$ 
of analytic functions on the disc $\I$,
taking values in $\Hspace_{\geq0}^\sharp$, and endowed with the
norm
\eqnn
        \|\h\|_{\xi,\sigma}:=
        \sup_{\z\in\I}\|\h[\z]\|_{\xi,\sigma}^\sharp \;.
\eeqnn
Furthermore, we denote the Banach space of analytic families
$\I\rightarrow H[\Hspace_{\geq0}^\sharp]$,
$\z\mapsto H[\h[\z]]$, by $H[\Hspace_{\geq0}]$.

For $\rho\leq\frac12$, we introduce the renormalization transformation
$\ren$. Given $\h\in\Hspace_{\geq0}$, it
is defined by the composition of the following three operations:

\begin{enumerate}

\item[{\bf (F)}] The degrees of freedom in the range of photon field
energies in $[\rho,1]$ are eliminated ({\em decimated}) by use of the smooth
Feshbach map $F_{\chi_\rho[H_f]}$, applied to the Feshbach pair
$(H[\h],H_f)$.
Here,
\eqnn
        \chi_\rho[H_f]:=\sin[\frac\pi2 \Theta(H_f/\rho)]
\eeqnn
is a smooth characteristic function on $[0,\rho)$, where
$\Theta$  has been introduced in (~\ref{Thetadef}).

\item[{\bf (S)}] A unitary rescaling transformation $\resc$, under which
$\1[H_f<\rho]\mapsto\1[H_f<1]$, and $\chi_\rho[H_f]\mapsto\chi_1[H_f]$,
followed by multiplication by $\frac1\rho$.

\item[{\bf (E)}] An analytic transformation $\Ez$ of the
spectral parameter $\z\in\I$ in $\h[\z]$.
\end{enumerate}

\subsubsection{The operation {\bf (S)}} 
\label{resc-def-subsubsect-1}

The rescaling transformation $\resc$ on $\Fo$ is
given by $\resc[\1]=\frac1\rho\1$ and
\eqn
		\resc[a^\sharp(K)]=\rho^{-\frac52}a^\sharp(\rho^{-1}K)
		\label{Gammarhodef}
\eeqn
where $\rho^{-1}K:=(\rho^{-1}k,\lambda)$, and $K\in\R^3\times\{+,-\}$.
(We note that the definition of $\resc$ includes multiplication by
$\frac1\rho$. Under unitary scaling, $a^\sharp(K)$ is mapped to 
$\rho^{-\frac32}a^\sharp(\rho^{-1}K)$).

Restricted to $H[\Wspace^\sharp_{\geq0}]\subset\cB[\H_{red}]$,
it induces a rescaling map $s_\rho$ on $\Hspace_{\geq0}$ by
\eqn
        \resc[H[\h]]=:H[s_\rho[\h]]=:H[(s_\rho[w_{M,N}])_{M+N\geq0}]\;,
\eeqn
where
\eqn
        \label{rescwMNdef}
        s_\rho[w_{M,N}][\spvar;K^{(M,N)};p]=\rho^{M+N-1}
        w_{M,N}[\rho\spvar;\rho K^{(M,N)};p]\;,
\eeqn
admitting the a priori bound
\eqn
        \|s_\rho[w_{M,N}]\|_\sigma &\leq& \rho^{(1+\sigma)(M+N)-1}
        \| w_{M,N}\|_\sigma   
        \label{srhowMNaprioribd}
\eeqn
(note that the conserved momentum $p$ is {\em not} rescaled).
Thus, from application of $S_\rho$, $\|w_{M,N}\|_\sigma$ is
contracted by a factor of at least $\rho^\sigma$ for all $M,N$ with
$M+N\geq1$.
\\

\subsubsection{The operation {\bf (E)}} 

The renormalization of the spectral parameter is determined as follows.
For a given $\h\in\Hspace_{\geq0}$ with $E[\z]:=w_{0,0}[\z;\underline{0}]$,
we define
$$
        {\mathcal U}[\h]:=\Big\{
        \z\in\I\Big|\,|E[\z]|\leq\frac{\rho}{10}\Big\} \;,
$$
and consider the analytic map
\eqnn
        \Ez:{\mathcal U}[\h]&\rightarrow&\I
        \\
        \z&\mapsto& \rho^{-1} E[\z] \;.
\eeqnn
We note that $\Ez$ is a bijection, where ${\mathcal U}[\h]$ is close to the
disc of radius $\rho$ centered at 0, provided that $\h$ is close to a non-interacting
theory (defined by $\h_0^{(p;\lTnl)}$ in (~\ref{T0def}) below).
\\

\subsubsection{The operation {\bf (F)}} 

For the decimation of degrees of freedom by the smooth Feshbach map,
one must verify that given $\h\in\Polyd(\e,\delta,\lTnl)$, and
$\z\in{\mathcal U}[\h]$, $\big(H[\h[\z]],H_f\big)$
is a Feshbach pair corresponding to $\chi_\rho[H_f]$.
This is done in Proposition {~\ref{Polydlemma}} below.
\\

\subsubsection{The renormalization transformation}

Composing the rescaling transformation $\resc$, the analytic transformation
of the spectral parameter $\Ez$, and the smooth Feshbach map, we now define the
{\em renormalization transformation} $\ren$.

We note that by arguments presented in \cite{bcfs1},
the map $H:\h\mapsto H[\h]$ injectively embeds $\Hspace_{\geq0}$
into the bounded operators on $\H_{red}$. 
The domain of $\ren$, $\Dom(\ren)$, is defined by those elements $\h\in\Hspace_{\geq0}$ for which 
\eqn
        \renop^H\big[\, H[\h] \,\big] [\zeta]  &:=& \resc\Big[\;F_{\chi_\rho[H_f]}
        \Big(\;H\big[\,\h[\,\Ez^{-1}[\zeta]\,]\,\big] \,,\, H_f\;\Big) \;\Big]  
        \label{renHh-def-1}
\eeqn
is well-defined and in the domain of $H^{-1}$, where $\zeta\in\I$.
The map $\renop^H$ is referred
to as the {\em renormalization map acting on operators}.

Accordingly, we define the {\em renormalization map (acting on generalized Wick
kernels)}
\eqn
		\ren:=H^{-1}\circ \renop^H \circ H 
\eeqn
on $\Dom(\ren)$. We shall prove below that the intersection of the 
domain and range of $\ren$ contains a family of polydiscs.

\subsection{Choice of a free comparison theory}

An essential part of our analysis is based on the comparison of
$\h\in\Dom(\ren)$ to a family of {\em non}-interacting
theories parametrized by $\h_0^{(p;\lTnl)}[z]\in\Dom(\ren)$, which converges to a fixed point of the
renormalization transformation $\ren$ in the limit $\z,\lTnl\rightarrow0$. 
We shall here, as a first step, construct this family of free comparison theories,
and discuss the structure of fixed points of $\ren$ that
parametrize non-interacting theories.

As the free comparison kernel, we consider
\eqn
        \h_0^{(p;\lTnl)}[\z]=\z\oplus T_0^{(p;\lTnl)}[\z;\spvar]
        \oplus\underline{0}_1 \;.
        \label{h0plTnl-def-1}
\eeqn
$\h_0^{(p;\lTnl)}[\z]$ is chosen in a manner that on the subspace
$\Ran(\bar P_1^\perp)\subset\H_{red}$ (on which $\chi_1[H_f]\equiv1$), the operator
$T_0^{(p;\lTnl)}[\z;\opp]=H[\h_0^{(p;\lTnl)}]$ equals 
$H_f-|p|\Ppar +\lTnl P_f^2 + \z$.
The latter corresponds to the rescaled non-interacting part of the
fiber Hamiltonian $H(p,\sigma)$.
On the complementary subspace $\Ran(\bar P_1)\subset\H_{red}$, the operator
$T_0^{(p;\lTnl)}[\z;\opp]$ is more complicated due to overlap effects from 
$\chi_1[H_f]\bar\chi_1[H_f]\neq0$, as we show below.

For the precise discussion, we first of all observe that 
\eqn
        \Big(H_f-|p|\Ppar+
        \lTnl \rho^{-1}P_f^2 +\z\rho\chi_1^2[H_f] \, ,
        \, H_f\Big)\in\Fpairs(\H_{red},\chi_\rho[H_f])  
\eeqn
is a Feshbach pair corresponding to $\chi_\rho[H_f]$. 

We define $\h_0^{(p;\lTnl)}[\z]$ by
\eqn
        H[\h_0^{(p;\lTnl)}[\z]]
        =\renop^H\Big[ H_f-|p|\Ppar  +\lTnl
        \rho^{-1}P_f^2 +\z\rho\chi_1^2[H_f] \Big] \;,  
        \label{h0plambdadef}
\eeqn
see (~\ref{renHh-def-1}), so that
\eqn
        H[\h_0^{(p;\lTnl)}[\z]]
        &=&H_f+\chi_1^2[H_f]\big(\z-|p|\Ppar+\lTnl P_f^2\big)
        \nonumber\\
        &-&  \frac{\big(|p|\Ppar -\lTnl P_f^2\big)^2
        \chi_1^2[H_f]\bar\chi_1^2[H_f]}{H_f+\bar\chi_1^2[H_f]
        \big(\z-|p|\Ppar +\lTnl P_f^2\big)}
        \Big|_{\Ran(\bar\chi_1[H_f])} \;. \;\;\;\;\;\;
        \label{fixedpt-1}
\eeqn
We observe that in the limit $\z,\lTnl\rightarrow0$, the operator
\eqnn
        \lim_{\z\rightarrow0}\lim_{\lambda\rightarrow0}H[\h_0^{(p;\lTnl)}[\z]]
        =H_f-\chi_1^2[H_f] |p|\Ppar
        - \left.\frac{(|p|\Ppar)^2
        \chi_1^2[H_f]\bar\chi_1^2[H_f]}{H_f-\bar\chi_1^2[H_f]
        |p|\Ppar }\right|_{\Ran(\bar\chi_1[H_f])}
\eeqnn
defines a {\em fixed point} of the renormalization transformation $\ren$.
Due to the non-linear nature of $\ren$, and $\chi_1[H_f]\bar\chi_1[H_f]\neq0$,
it notably {\em differs} on the subspace 
$\Ran(\chi_1[H_f])\cap\Ran(\bar\chi_1[H_f])$
from the operator $H_f-|p|\Ppar$, which is
a fixed point of the (linear) {\em rescaling} transformation $\resc$ on $\cB(\H_{red})$.

Let $P_{\Tspace}$ denote the projection
\eqn
        P_{\Tspace}:\left\{\begin{array}{rcl}
        \Hspace_{\geq0}&\rightarrow&\Tspace
        \\
        \h=(E,T,\h_{1} )&\mapsto&T
        \;.
        \end{array}\right.
        \label{PTspacedef}
\eeqn
Then, the function $T_0^{(p;\lTnl)}$ in (~\ref{h0plTnl-def-1}) is given by
\eqn
        T_0^{(p;\lTnl)}[\z;\spvar]
        &=&P_{\Tspace}\ren\Big[ \big(\rho \z\big) \oplus\big( X_0-|p|X^\parallel
        +\lTnl\rho^{-1} X^2\big) 
        \oplus\unull_1\Big] \; .
        \label{T0def}
\eeqn 
$T_0^{(p;\lTnl)}$ is used for the definition of polydiscs in the next section.

\subsection{The domain of $\ren$}
\label{polyd-subsect-1-1}

We shall next prove that the domain of $\ren$ contains a {\em polydisc} of the form
\eqn
        \Polyd\Polpar :=\Big\{\h\in\Hspace_{\geq 0}\Big|
        \sup_{\z\in\I}\|T[\z;\,\cdot\,]- T_0^{(p;\lTnl)}[\z;\,\cdot\,] \|_{\Tspace}^\sharp
        &<&\delta \;,
        \nonumber\\
        \sup_{\z\in\I}|E[\z]-\z|&<&\e \; ,
        \nonumber\\
        \sup_{\z\in\I}|\partial_{|p|}E[\z]|&<&\e \; ,
        \nonumber\\
        \sup_{\z\in\I}\|\h_{1}[\z] \|_{\xi,\sigma}^\sharp&<&\e  \;
        \Big\} \;, \;\;\;\;
        \label{Polyddef}
\eeqn
for $0<|p|<\frac13$,  $\e,\delta>0$, and $0\leq\lTnl<\frac12$.
As noted above, the function $T_0^{(p;\lTnl)}[\z;\spvar]$, defined in (~\ref{T0def}),
is chosen close to a fixed point of the renormalization
transformation. One can verify along the lines demonstrated in \cite{bcfs1} that
\eqnn
        \Big\{\h\in\Hspace_{\geq0}\Big|\,\|\h-\h_0^{(p;\lTnl)}\|_{\xi,\sigma}\leq\e\Big\}
        &\subseteq&\Polyd(\e,\delta,\lTnl)
        \\
        &\subseteq&
        \Big\{\h\in\Hspace_{\geq0}\Big|\,\|\h-\h_0^{(p;\lTnl)}\|_{\xi,\sigma}\leq
        2\delta+2\e\Big\} \;.
\eeqnn
Thus, $\Polyd\Polpar$
is comparable to an $(\e,\delta)$-ball about $\h_0^{(p;\lTnl)}[\z]$.

\begin{lemma}\label{Ezlemma}
Let $0<\xi<1$, $\sigma>0$ and $0<\rho\leq\frac12$. Then,
\eqn
        D_{\frac{\rho}{20}}\subseteq {\mathcal U}[\h]\subseteq D_{\frac{3\rho}{20}}
\eeqn
for all $\h\in\Polydrho$, and
\eqn
        |\rho\partial_{\z}E[\z]-1|\leq 1 \;,
\eeqn
for all $\z\in{\mathcal U}[\h]$. Then, $\Ez:{\mathcal U}[\h]\rightarrow
\I$ is a bijection, for all $\h\in\Polydrho$.
\end{lemma}

This Lemma corresponds to Lemma 3.4 in \cite{bcfs1}, where we refer for the proof.

\begin{proposition}\label{Polydlemma}
(Domain of the smooth Feshbach map)
For any fixed choice of $0\leq|p|<\frac13$,
$0< \rho<\frac12$, $\sigma>0$, and $0<\xi<1$,
$(H[\h[\z]],H_f)$ defines a Feshbach pair corresponding to $\chi_\rho[H_f]$,
for all $\h\in\Polydrho$
and all $\z\in{\mathcal U}[\h]$.
\end{proposition}

\prf
Since $H[\h[\z]]$ and $H_f$ define bounded operators on $\H_{red}$,
it suffices to verify the invertibility of $T[\z;\opp]+E[\z] $
on Ran$(\bar\chi_1[H_f])$.

Besides
$0\leq|p|<\frac13$ and $0\leq\lTnl\leq\frac12$, we have $|E[\z]|<\frac{\rho}{10}$. 
Using $|P_f|\leq H_f$, $\chi_1^2[H_f]\bar\chi_1^2[H_f]\leq\frac14$,
and $H_f\geq\frac34$ on Ran$(\bar\chi_1[H_f])$, we find
\eqn
        |T_0^{(p;\lTnl)}[\z;\opp]|&\geq&
        H_f(1-|p|)-\frac{H_f}{4}\frac{(|p|+\lTnl)^2}{(1-|p|)-4|E[\z]|/3}
        \nonumber\\
        &\geq& \frac{H_f}{5} \; ,
        \label{HToPlTnllowbd}
\eeqn
respectively,
\eqn
        \inf_{|X|\leq X_0\leq1}|T_0^{(p;\lTnl)}[\z;\spvar]|\geq \frac{X_0}{5} \;.
        \label{T0ptTnllowbd}
\eeqn
Thus, we observe that
\eqn
        &&\inf_{|X|\leq X_0\leq1}|T[\z;\spvar]-E[\z]|
        \nonumber\\
        &&\hspace{2cm}\geq\,
        \inf_{|X|\leq X_0\leq1}\Big\{|T_0^{(p;\lTnl)}[\z;\spvar]|
        -|T[\z;\spvar]-T_0^{(p;\lTnl)}[\z;\spvar]|\Big\}-|E[\z]|
        \nonumber\\
        &&\hspace{2cm}\geq\, X_0\Big(\frac15-\|T[\z;\spvar]-T_0^{(p;\lTnl)}
        [\z;\spvar]\|_{\Tspace}\Big)
        -\sup_{\z\in{\mathcal U}[\h]}|E[\z]|
        \nonumber\\
        &&\hspace{2cm}\geq\,\frac{3\rho}{4}\Big(\frac13-\frac{\rho}{10}\Big)-\frac{\rho}{10}
        >
        \frac{\rho}{100} \;.
\eeqn
The claim follows.
\endprf

\subsection{Generalized Wick ordering}

The next step in the construction of $\widehat\h=\ren[\h]$
consists of determining the {\em generalized Wick ordered} 
form of the right hand side of (~\ref{renHh-def-1}).
To this end, we let
\eqn
        W[\z]:=\sum_{M+N\geq1}W_{M,N}[w_{M,N}[\z]] \;,
\eeqn
and we observe that by Proposition {~\ref{Polydlemma}} and Theorem {~\ref{opL2relbound}},
\eqn
        \|\bar\chi_\rho[H_f]\big|T[\z;\opp]+E[\z] \big|\Big|_{\Ran(\bar\chi_\rho[H_f])}^{-1}
        \bar\chi_\rho[H_f]\|_{op}
        &\leq& \frac{100}{\rho} \;,
        \nonumber\\
        \|W[\z]\|_{op}&\leq&\frac{\xi\rho}{10} \;,
        \label{TWbounds}
\eeqn
for $\h\in\Polydrho$.
We introduce the notation
\eqn
        \piop_{\rho}[\z;\opp]:=\piop_{\chi_\rho[H_f]}\big(T[\z;\opp]+E[\z]\chi_1^2[H_f] \, , \,H_f\big) \;,
        \label{piop-def-1}
\eeqn
see (~\ref{piopdef}),
and write
\eqn
        \bar R_0[\h[\z]]:=\Big[H_f+(\bar\chi_\rho^2\chi_1^2)[H_f]
        \big(\tT[\z;\opp]+E[\z] \big) \Big]^{-1}
        \label{bar-R-0-def-1}
\eeqn
for the free resolvent on $\Ran(\bar\chi_\rho)$, with
\eqn
        T[\z;\opp]=H_f+\chi_1^2[H_f]\tT[\z;\opp]\;.
\eeqn 
Then, Lemma {~\ref{piopdeflemma}} implies that
\eqn
        F_{\chi_\rho[H_f]}\big(H[\h[\z]],H_f\big) 
        &=&E[\z]\chi_\rho^2[H_f]
        +H_f+\tT[\z;\opp]\piop_{\rho}[\z;\opp]\chi_\rho^2[H_f] 
        \nonumber\\
        &+&\sum_{L=1}^\infty(-1)^{L-1}\chi_\rho[H_f]
        \piop_{\rho}[\z;\opp] W[\h[\z]]
        \nonumber\\
        &&\hspace{0.5cm} \times\,
        \Big[ (\bar\chi_\rho^2\chi_1^2)[H_f]
        \bar R_0[\h[\z]] W[\h[\z]]\Big]^{L-1}
        \piop_{\rho}[\z;\opp]  \chi_\rho[H_f] \;,
        \label{resolvexp}
\eeqn
which is norm convergent due to (~\ref{TWbounds}).

Next, we introduce the operators
\eqn
        &&W_{p,q}^{m,n}[\h\big|\spvar;K^{(m+p,n+q)}]
        \;:=\; P_{red}\int_{B_1^{p+q}}
        \frac{dQ^{(p,q)}}{|Q^{p,q}|^{\frac12}}a^*(Q^{(p)})
        \nonumber\\ 
        &&\hspace{3cm}\times\,
        w_{m+p,n+q}[\opp+\spvar;Q^{(p)},K^{(m)};\tilde Q^{(q)},\tilde K^{(n)}]
        a(\tilde Q^{(q)})P_{red} \;.
        \label{Wmnpqdef}
\eeqn 
The (generalized) Wick ordering of the
resolvent expansion (~\ref{resolvexp}) is governed by the following theorem.

\begin{theorem}
\label{Wickorderthm}
Let $\h=(w_{M,N})_{M+N\geq1}\in\Hspace_{\geq1}^\sharp$. Writing
$W_{M,N}:=W_{M,N}[w_{M,N}]$, $W=\sum_{M+N\geq1}W_{M,N}$, and $F_0,\dots,F_L\in
\Wspace_{0,0}$, let $S_M$ denote the $M$-th symmetric group. Then,
\eqnn
        F_0 W F_1 W\cdots W F_{L-1} W F_L = H[\tilde\h] \;,
\eeqnn
where $\tilde \h \in (\widehat w^{(sym)}_{M,N})_{M+N\geq0}\in \Hspace_{\geq0}^\sharp$
is determined by the symmetrization with respect to $K^{(M)}$ and $\tilde K^{(N)}$,
\eqnn
        \tilde w_{M,N}^{(sym)}[\spvar;K^{(M,N)}]
        =
        \frac{1}{M!N!}\sum_{\pi\in S_M}\sum_{\tilde\pi\in S_N}
        \tilde w_{M,N}[\spvar;K_{\pi(1)},\dots,K_{\pi(M)};\tilde K_{\tilde\pi(1)},
        \dots,\tilde K_{\tilde\pi(N)}]\;,
        \nonumber
\eeqnn
of
\eqnn
        \tilde w_{M,N}[\spvar;K^{(M,N)}]&=&\sum_{ m_1+\cdots+m_L=M \atop
        n_1+\cdots+n_L=N}
        \sum_{ p_1,q_1,\dots,p_L,q_L \atop
        m_\ell+p_\ell+n_\ell+q_\ell\geq1}
        \nonumber\\ 
        &\times&
        \Big[\prod_{\ell=1}^L 
        \Big(\begin{array}{c}
        m_\ell+p_\ell\\
        p_\ell
        \end{array}\Big)
        \Big(\begin{array}{c}
        n_\ell+q_\ell\\
        q_\ell
        \end{array}\Big)\Big] F_0[\spvar+\tilde\spvar_0]\, 
        F_L[\spvar+\tilde\spvar_L]
        \nonumber\\
        &\times&\Bra \, \vac \, , \,
        \tilde W[\spvar+\spvar_1;K_1^{(m_1,n_1)}]F_1[\opp+\spvar+\tilde\spvar_1;
        K_2^{(m_2,n_2)}]
        \nonumber\\ 
        &&
        \hspace{0.5cm}\cdots F_{L-1}[\opp+\spvar+\tilde\spvar_{L-1}]
        \tilde W[\spvar+\spvar_L;K_L^{(m_l,n_L)}]\, \vac\,\Ket \, 
\eeqnn
with the definitions
\eqn
        \tilde W_\ell[\spvar+\spvar_\ell;K^{(m_\ell,n_\ell)}_\ell]
        &:=&W^{m_\ell,n_\ell}_{p_\ell,q_\ell}[\h\big|\spvar+\spvar_\ell;
        K^{(m_\ell,n_\ell)}_\ell] \;,
\eeqn
\eqnn
        K^{(M,N)}=(K^{(m_1,n_1)},\dots,K^{(m_L,n_L)})
        & , &
        K_\ell^{(m_\ell,n_\ell)}:=(K_\ell^{(m_\ell)},\tilde K_\ell^{(n_\ell)}) \;,
\eeqnn
and
\eqnn
        \spvar_\ell&:=& \sum\big[\tuk_1^{(n_1)}\big]+\cdots+
        \sum\big[\tuk_{\ell-1}^{(n_{\ell-1})}\big]
        +\sum\big[\uk_{\ell+1}^{(m_{\ell+1})}\big]+\cdots
        +\sum\big[\uk_L^{(m_L)}\big]\nonumber\\
        \tilde \spvar_\ell&:=&\sum\big[\tuk_1^{(n_1)}\big]+\cdots+
        \sum\big[\tuk_\ell^{(n_\ell)}\big]
        +\sum\big[\uk_{\ell+1}^{(m_{\ell+1})}\big]+\cdots+\sum\big[\uk_L^{(m_L)}\big]
        \;.
\eeqnn
\end{theorem}

Applying the rescaling transformation, and transforming
the spectral parameter, we obtain  
\eqn
        H[\widehat\h[\zeta]]&=&\ren^H[H[\h][\Ez^{-1}[\zeta]]]
	\nonumber\\
	&=&
        \resc(F_{\chi_\rho}[H_f](H[\h][\Ez^{-1}[\zeta]],H_f)) \;,
        \label{hath-renopH-id-1}
\eeqn
(see (~\ref{renHh-def-1})).
$\widehat \h[\zeta]$ is characterized in the following theorem.

\begin{theorem}
\label{hatwMNformalseriesthm}
Let $\zeta\in \I$, and $z:=\Ez^{-1}[\zeta]\in{\mathcal U}[\h]$.
(~\ref{hath-renopH-id-1}) determines $\widehat\h=(\widehat\h^{(sym)}_{M,N})_{M+N\geq0}$
as the symmetrization with respect to $K^{(M)}$ and $\tilde K^{(N)}$ of
\eqn
        &&\widehat w_{M,N}[\zeta;\spvar;K^{(M,N)}]\; =\;\rho^{M+N-1}
        \sum_{L=1}^\infty
        (-1)^{L-1}
        \nonumber\\
        &&\hspace{1cm}\times\, \sum_{m_1+\cdots+m_L=M \atop n_1+\cdots n_L=N}
        \sum_{p_1,q_1,\dots,p_L,q_L: \atop m_\ell+p_\ell+n_\ell+q_\ell\geq1}\Big[\prod_{\ell=1}^L
        \Big(\begin{array}{c}
        m_\ell+p_\ell\\
        p_\ell
        \end{array}\Big)
        \Big(\begin{array}{c}
        n_\ell+q_\ell\\
        q_\ell
        \end{array}\Big)\Big]
        \nonumber\\
        &&\hspace{1cm}\times\,
        \Bra\,\vac\,,\,
        \piop_{\rho}[\z;\opp+\rho (\spvar+\tilde\spvar_0)]
        \tilde W_1[\z;\rho(\spvar+\spvar_1);\rho K^{(m_1,n_1)}_1]
        \nonumber\\
        &&\hspace{2cm} \;\;\;\;
        (\bar\chi_\rho^2\chi_1^2 \bar R_0)[\z;\opp+\rho(\spvar+\tilde\spvar_1)]
        \tilde W_2[\z;\rho(\spvar+\spvar_2);\rho K^{(m_2,n_2)}_2]
        \\
        &&\hspace{2cm} \;\;\;\;
        (\bar\chi_\rho^2\chi_1^2\bar R_0)[\z;\opp+\rho(\spvar+\tilde\spvar_2)]
        \cdots\cdots
        \nonumber\\
        &&\hspace{2cm} \hspace{3.5cm}\cdots\cdots
        (\bar\chi_\rho^2\chi_1^2\bar R_0)
        [\z;\opp+\rho(\spvar+\tilde\spvar_{L-1})]
        \nonumber\\
        &&\hspace{2cm} \;\;\;\;
        \tilde W_L[\z;\rho(\spvar+\spvar_L);\rho K_L^{(m_L,n_L)}]
        \piop_{\rho}[\z;\opp+\rho (\spvar+\tilde\spvar_L)]\, \vac\,\Ket
        \nonumber
\eeqn
in the case of $M+N\geq1$, and
\eqn
        \widehat w_{0,0}&=& \ren[-E[\,\cdot\,]\oplus w_{0,0}\oplus\unull_1]
        \nonumber\\
        &+&\rho^{-1}\sum_{L=2}^\infty(-1)^{L-1}
        \sum_{p_1+q_1\geq1}\cdots\sum_{p_L+q_L\geq1}\piop_{\rho}^2[\z;\spvar]
        \nonumber\\
        && 
        \times\,\Bra \,\vac\,,\,\tilde W_{p_1,q_1}[\h[\z]\big|\rho\spvar]
        (\bar\chi_\rho^2\chi_1^2\bar R_0)[\z;\opp+\rho\spvar]
        \tilde W_{p_2,q_2}[\h[\z]\big|\rho\spvar]
        \nonumber\\
        &&\hspace{3cm}
        \cdots (\bar\chi_\rho^2\chi_1^2\bar R_0)[\z;\opp+\rho\spvar]
        \tilde W_{p_L,q_L}[\h[\z]\big|\rho\spvar]
        \, \vac\,\Ket
\eeqn
\end{theorem}

It remains to verify that $\widehat\h$ is again an element of $\Hspace_{\geq0}$.
We will in fact establish a much stronger result, and prove that $\ren$
is contractive on a codimension-2 subspace of $\Hspace_{\geq0}$.
One of the key tools in this analysis are soft photon sum rules, 
which we address next.

\subsection{Soft photon sum rules}
\label{spsrsect}

It is an important and deep property of the model analyzed in this work
that the generalized Wick kernels $w_{M,N}$ generated by repeated applications
of the renormalization map are, for different values of $M,N$, all mutually linked.
We shall here prove the existence of a hierarchy of non-perturbative identities,
which we refer to as the {\em soft photon sum rules},
that interrelate all $w_{M,N}$ in the small photon momentum regime.

The key point is that the soft photon sum rules are {\em preserved
by the renormalization map}.
The soft photon sum rules are in the present analysis used  
to prove that for the value $p=0$ of the conserved total momentum, all
interaction operators of the effective Hamiltonians are irrelevant 
even if the infrared regularization is removed, that is, when $\sigma\rightarrow0$.

\begin{definition}
Let $g$ denote the electron charge, cf. (~\ref{Hpsdef}).
Given $n\in\R^3$, $|n|=1$, let $\e(n,\lambda)$
denote the photon polarization vector orthonormal to $n$
labeled by the polarization index $\lambda$.
Let $\mu\in\R_+$.

The sequence of generalized Wick kernels $\h\in\Hspace_{\geq0}$ is said to
satisfy the 
\begin{center}
{\bf soft photon sum rules} $\;$  \sbsr 
\end{center}
if  the identity
\eqn
        &&g \mu  \bra\e(n,\lambda),\partial_{X}\ket_{\R^3} \,
        w_{M,N}[\spvar;K^{(M,N)}] 
        \nonumber\\
        &&\hspace{3cm}=(M+1)\lim_{\rvar\rightarrow0}\rvar^{-\sigma} w_{M+1,N}
        [\spvar;K^{(M+1,N)}]\Big|_{K_{M+1} =(\rvar n,\lambda)}
        \nonumber\\
        &&\hspace{3cm}=(N+1)\lim_{\rvar\rightarrow0}\rvar^{-\sigma} w_{M,N+1}
        [\spvar;K^{(M,N+1)}]\Big|_{\tilde K_{N+1}=(\rvar n,\lambda)}
        \label{sbsr}
\eeqn
is satisfied for all $M,N\geq0$, and every choice of the unit vector $n$ (we recall that
$X$ is the spectral variable corresponding to the photon momentum operator $P_f$).
\end{definition}

Applying (~\ref{sbsr}) inductively, beginning with $M,N=0$, in the order indicated by
\eqn
        \begin{array}{llllllll}
        &&&&&&&\nearrow\cdots\\
        &&&&&&w_{3,0}&\\
        &&&&&\nearrow&&\searrow\cdots\\
        &&&&w_{2,0}&&&\\
        &&&\nearrow&&\searrow&&\nearrow\cdots\\
        &&w_{1,0}&&&&w_{2,1}&\\
        &\nearrow&&\searrow&&\nearrow&&\searrow\cdots\\
        w_{0,0}&&&&w_{1,1}&&&\\
        &\searrow&&\nearrow&&\searrow&&\nearrow\cdots\\
        &&w_{0,1}&&&&w_{1,2}&\\
        &&&\searrow&&\nearrow&&\searrow\cdots\\
        &&&&w_{0,2}&&&\\
        &&&&&\searrow&&\nearrow\cdots\\
        &&&&&&w_{0,3}&\\
        &&&&&&&\searrow\cdots
        \end{array}
\eeqn
all generalized Wick kernels are recursively linked to one another in the vicinity of
the origin in photon momentum space.

As remarked in Section {~\ref{stratorg-sect-1}}, these identities are due
to $U(1)$ gauge invariance in the special case of
non-relativistic quantum electrodynamics. But since in general, the existence of
soft boson sum rules is independent from gauge invariance, 
such as the Gross transformed,
translation-invariant 1-particle Nelson model
described in Section {~\ref{Nelsonremsubsect}} shows,
we shall here prefer the notion of soft photon sum rules instead
of Ward-Takahashi identities (which would be equally appropriate in
non-relativistic quantum electrodynamics).
 
It is interesting to ask whether one can avoid the use of
soft photon sum rules, and achieve this result 
by a more insightful use of $O(3)$-invariance in the case $p=0$.
It may appear surprising that rotational and reflection 
symmetry alone do in fact {\em not} suffice
to prove irrelevance of the interaction, even if $p=0$. There exist
simple rotationally invariant
models for which the interaction operators of the bare Hamiltonian
(corresponding to the fiber Hamiltonian $H(p=0,\sigma)$ in our case)
scale naively like irrelevant operators, but which are in fact {\em marginal}.
This is an artifact of the non-linear nature of the renormalization map.
$\ren$ can
generate marginal operators from irrelevant operators, unless this is suppressed by
symmetries of the model beyond rotation and reflection invariance.


\subsection{Codimension two contractivity of $\ren$ on a polydisc}
\label{Codimcontrsubsect}

A quintessential property of the renormalization map $\ren$ is that it is contractive on
a codimension 2 subspace of a polydisc of the form (~\ref{Polyddef}) (after projecting
out the relevant direction corresponding to $E$ in $\h=(E,T,\h_1)$, cf. our introductory
remarks in Section {~\ref{contr-intro-ssect-1}}). 

Let
\eqnn
        \Polyd^{(\mu)}\Polpar:=\Big\{\h\in\Polyd\Polpar\Big|
        \h\;{\rm satisfies \; the \; soft \; photon
        \; sum \; rules\;}\sbsrm\Big\} \;
\eeqnn
denote the subset of the polydisc $\Polyd\Polpar$ of generalized
Wick kernels in the Banach space $(\Hspace_{\geq 0},\|\,\cdot\,\|_{\sigma,\xi})$,
for given $\xi$,
which {\em satisfy the soft photon
sum rules}. The definitions were given in (~\ref{Polyddef}) and (~\ref{sbsr}).
We shall prove
that $\ren$ is a {\em contraction} on
a codimension 2 subspace of
$\Polyd^{(\mu)}\Polpar$.

\begin{theorem}\label{codim2contrthm}
The renormalization map $\ren$
is codimension-2 contractive on the polydisc
$\Polyd^{(\mu)}\Polpar$ in the following sense.

\begin{itemize}

\item{\underline{A. The case $0<|p|<\frac13$:}}
For $\sigma>0$,
there exist constants $\rho=\rho(\sigma)$, $\xi$, $\e_0(\sigma)$,
such that for
all $0\leq\e  \leq \e_0(\sigma)$ and $0\leq\delta \leq \e_0(\sigma)$,
\eqn
        \ren: \Polyd^{(\mu)}(\e ,\delta,\lTnl)\rightarrow
        \Polyd^{(\rho^\sigma\mu)}(\frac{\e}{2},\delta+\e,\rho\lTnl) \;.
        \label{codim2contr}
\eeqn

\item{\underline{B. The case $|p|=0$:}}
For $\sigma\geq0$, there are constants $\rho,\xi,\e_0$,
all independent of $\sigma$, and a constant $c$  
independent of $\rho, \e_0$ and $\sigma$ such that for
arbitrary $0\leq  \e \leq \e_0$ and $0\leq\delta \leq \e_0$,
\eqn
        \ren: \Polyd^{(\mu)}(\e,\delta,\lTnl)\rightarrow
        \Polyd^{(\rho^\sigma\mu)}(\ccontr\rho\e ,\delta+\e,\rho\lTnl) \;,
        \label{codim2contrp0}
\eeqn
where the constant $\ccontr$ is defined in (~\ref{ccontr-def-1}) below.
The renormalization map is therefore contractive on a codimension 2
subspace of $\Polyd(\e,\delta,\lTnl)$, even without
infrared regularization. 

\end{itemize}
\end{theorem}

\section{Proof of Theorem {~\ref{codim2contrthm}}: Generalized Wick ordering}
\label{gen-wick-prf-subsect-1}

We introduce the following notation.
For fixed $L\in\N$, let
\eqn
        \underline{m,p,n,q}:=(m_1,p_1,n_1,q_1,\dots,m_L,p_L,n_L,q_L)\in\N_0^{4L} \;
\eeqn
and
\eqn
        M:=m_1+\dots+m_L \; \; , \; \;
        N:=n_1+\dots+n_L \;.
\eeqn
We let
\eqn 
        &&V^{(L)}_{\underline{m,p,n,q}}\big[\h\,\big|\,\spvar;K^{(M,N)}\big]
        \nonumber\\
        &&\hspace{2cm}:=\;\Bra\, \vac\,,\,
        F_0[\spvar]\prod_{\ell=1}^L\Big\{\tilde W_\ell
        [\z;\rho(\spvar+\spvar_\ell);\rho K^{(m_\ell,n_\ell)}_\ell]
        F_\ell[\spvar]\Big\} \vac\,\Ket \;, 
        \label{VLmpnq-def-1}
\eeqn
where
\eqn
        F_0[\spvar]:=\piop_{\rho}[\z;\opp+\rho(\spvar+\tilde \spvar_{0})]
        \; \; , \; \;
        F_L[\spvar]:=\piop_{\rho}[\z;\opp+\rho(\spvar+\tilde \spvar_L)]
        \label{F0FLdef}
\eeqn
(see (~\ref{piop-def-1}))  and
\eqn
        F_\ell[\spvar]:=\frac{(\bar\chi_\rho^2\chi_1^2)[H_f+\rho(X_0+\tilde X_{\ell,0})]}
        {H_f+\rho(X_0+\tilde X_{\ell,0})
        +(\bar\chi_\rho^2\chi_1^2)[H_f+\rho(X_0+\tilde X_{\ell,0})]
        \big(\tT[\z;\opp+\rho(\spvar+\tilde \spvar_\ell)]
        +E[\z]\big)}
        \nonumber\\
        \label{Felldef}
\eeqn
for $\ell=1,\dots,L-1$, with $T[\z;\spvar]=X_0+\chi_1^2[X_0]T'[\z;\spvar]$. 
We note that the definition of the operators $F_\ell[\spvar]$ 
in \cite{bcfs1} contains a misprint.

Then,
\eqn
        &&\widehat w_{M,N}[\zeta;\spvar;K^{(M,N)}]=
        \sum_{L=1}^\infty (-1)^{L-1}\rho^{M+N-1}
        \sum_{m_1+\cdots+m_L=M \atop
        n_1+\cdots +    n_L=N}
        \label{hatwMNdefform}
        \\
        &&\hspace{1cm}
        \sum_{p_1,q_1,\dots,p_L,q_L: \atop
        m_\ell+p_\ell+n_\ell+q_\ell\geq1}
        \Big[\prod_{\ell=1}^L 
        \Big(\begin{array}{c}
        m_\ell+p_\ell\\
        p_\ell
        \end{array}\Big)
        \Big(\begin{array}{c}
        n_\ell+q_\ell\\
        q_\ell
        \end{array}\Big)\Big]
        V^{(L)}_{\underline{m,p,n,q}}\big[\h\,\big|\,\spvar;K^{(M,N)}\big] \;,
        \nonumber
\eeqn
where the scaling factors $\rho^{M+N-1}$ were
explained in connection with (~\ref{rescwMNdef}).

\begin{lemma}
\label{VLboundlm}
For $L\geq1$ fixed, and $\underline{m,p,n,q}\in\N_0^{4L}$, one has
$V^{(L)}_{\underline{m,p,n,q}}\in\Wspace_{M,N}^\sharp$. In particular,
\eqn
        &&\rho^{M+N-1}\max\Big\{
        \|\partial_{X_0}V^{(L)}_{\underline{m,p,n,q}}\|_\sigma \,,\,
        \||k|^\sigma\partial_{|k|}\big(|k|^{-\sigma}V^{(L)}_{\underline{m,p,n,q}}\big)\|_\sigma
        \Big\}
        \nonumber\\
        &&\hspace{0.5cm}\leq 2 (L+1)  C_\Theta^{L+1}\rho^{(1+\sigma)(M+N) -L}
        \prod_{l=1}^L\frac{\| w_{m_l+p_l,n_l+q_l}[\z]\|_\sigma^\sharp}
        {p_l^{p_l/2}q_l^{q_l/2}} \;,\;\;\;
\eeqn
for $\spvar=(X_0,X)$ and any $k\in k^{(M,N)}$. Furthermore,
\eqn
        &&\rho^{M+N-1}
        \|\partial_{\spvar}^{\aind} V^{(L)}_{\underline{m,p,n,q}}\|_\sigma
        \nonumber\\
        &&\hspace{0.5cm}\leq 10 (L+1)^2 C_\Theta^{L+2}
        \rho^{(1+\sigma)(M+N) -L}
        \prod_{l=1}^L\frac{\| w_{m_l+p_l,n_l+q_l}[\z]\|_\sigma^\sharp}
        {p_l^{p_l/2}q_l^{q_l/2}} \;,\;\;\;
        \label{partspvaraindVL}
\eeqn
for $0\leq|\aind|\leq2$, $a_0=0$.
For $|p|>0$, and $|\aind|\leq1$,
\eqn
        &&\rho^{M+N-1}
        \|\partial_{|p|}
        \partial_{\spvar}^{\aind} V^{(L)}_{\underline{m,p,n,q}}\|_\sigma
        \nonumber\\
        &&\hspace{0.5cm}\leq 10 (L+1)^2 C_\Theta^{L+2}
        \rho^{(1+\sigma)(M+N) -L}
        \prod_{l=1}^L\frac{\| w_{m_l+p_l,n_l+q_l}[\z]\|_\sigma^\sharp}
        {p_l^{p_l/2}q_l^{q_l/2}} \;.\;\;\;
        \label{partppartspvaraindVL}
\eeqn
Consequently,
\eqn
        &&\rho^{M+N-1}\|V^{(L)}_{\underline{m,p,n,q}}\|_\sigma^\sharp
        \nonumber\\
        &&\hspace{0.5cm}\leq 10 (L+1)^2 C_\Theta^{L+2}\rho^{(1+\sigma)(M+N)-L}
        \prod_{l=1}^L\frac{\| w_{m_l+p_l,n_l+q_l}[\z]\|_\sigma^\sharp}
        {p_l^{p_l/2}q_l^{q_l/2}} \;,\;\;\;
\eeqn
using the convention $p^p=1$ for $p=0$. The constant $C_\Theta$ only
depends on the choice of the smooth cutoff function $\chi_\rho$.
\end{lemma}

\subsection{Proof of Lemma {~\ref{VLboundlm}}}

There exists a constant $1\leq C_\Theta<\infty$ that only depends on the
choice of the smooth cutoff function $\Theta$, such that
\eqn
        \;\;\;\;\;\;
        \|\partial_{X_0}F_\ell[\spvar]\|_{op}
        +\sum_{0\leq|\aind|\leq2\atop a_0=0}\|\partial_{\spvar}^{\aind} F_\ell[\spvar]\|_{op}
        +\sum_{|\aind|\leq1}\|\partial_{|p|}\partial_{\spvar}^{\aind} F_\ell[\spvar]\|_{op}
        \leq\frac{C_\Theta}{\rho^{\eta(\ell)}}\;,
        \label{CThetadef}
\eeqn
uniformly in $\z\in\I$, where
\eqn
        \eta(\ell):=\left\{\begin{array}{rl}
        0&\;{\rm if}\; \ell=0,L\\
        1&\;{\rm if}\; \ell=1,\dots,L-1 \,,
        \end{array}
        \right.
\eeqn
cf. ({~\ref{F0FLdef}}) and ({~\ref{Felldef}}). Here and in the sequel,
$\|\,\cdot\,\|_{op}$ denotes the operator norm on $\H_{red}=\1[H_f<1]\Fo$.

Let us to begin with consider (~\ref{partspvaraindVL}), for
the case $\aind=\underline{0}$, where we have
\eqn
        |V_{\underline{m,p,n,q}}^{(L)}[\h\,\big|\,\spvar;K^{(M,N)}]|
        &\leq& \prod_{\ell=0}^L \|F_\ell[\spvar_\ell]\|_{op}
        \nonumber\\
        &&\times
        \prod_{\ell=1}^L\|\tilde W_\ell[z;\rho(\spvar+\spvar_\ell);
        \rho K_\ell^{(m_\ell,n_\ell)}]\|_{op}
        \label{VLmpnqbounds}\\
        &\leq&C_\Theta^{L+1}\rho^{-L+1}\prod_{\ell=1}^L\|
        \tilde W_\ell[z;\rho(\spvar+\spvar_\ell);
        \rho K_\ell^{(m_\ell,n_\ell)}]\|_{op} \;.
        \nonumber
\eeqn
$\|\,\cdot\,\|_{op}$ denotes the operator norm on $\cB[\H_{red}]$.

Next, we discuss the various terms corresponding to $1\leq|\aind|\leq2$.

\subsubsection{The case $|\aind|=1$} 

We have
\eqnn
        \partial_{\spvar}^{\aind}V_{\underline{m,p,n,q}}^{(L)}[\h\,\big|\,\spvar;K^{(M,N)}]
        \;=\;V_{\underline{m,p,n,q}}^{(L,i)}[\h\,\big|\,\spvar;K^{(M,N)}]+
        V_{\underline{m,p,n,q}}^{(L,ii)}[\h\,\big|\,\spvar;K^{(M,N)}] \;,
\eeqnn
where
\eqn
        &&V_{\underline{m,p,n,q}}^{(L,i)}[\h\,\big|\,\spvar;K^{(M,N)}]
        \nonumber\\
        &&\hspace{2cm}:=\;\sum_{j=0}^L
        \Bra   \vac \,,\, 
        \Big[\prod_{\ell=1}^{j-1}
        F_{\ell-1}[\spvar]\tilde W_{\ell}[z;\rho(\spvar+\spvar_\ell);
        \rho K_\ell^{(m_\ell,n_\ell)}]
        \Big]
        \nonumber\\
        &&\hspace{2cm}
        \times\;\Big(\partial_{\spvar}^{\aind}F_j[\spvar]\Big) 
        \Big[\prod_{\ell=j+1}^L  \tilde W_{\ell}[z;\rho(\spvar+\spvar_\ell);
        \rho K_\ell^{(m_\ell,n_\ell)}]
        F_\ell[\spvar]\Big] \vac \, \Ket 
\eeqn
and
\eqn
        &&V_{\underline{m,p,n,q}}^{(L,ii)}[\h\,\big|\,\spvar;K^{(M,N)}]
        \nonumber\\
        &&\hspace{2cm}:=\;\sum_{j=1}^L
        \Bra \,  \vac \,,\, F_0[\spvar]
        \Big[\prod_{\ell=1}^{j-1}
        \tilde W_{\ell}[z;\rho(\spvar+\spvar_\ell);
        \rho K_\ell^{(m_\ell,n_\ell)}] F_{\ell}[\spvar]
        \Big]
        \nonumber\\
        &&\hspace{2cm}
        \times\;\rho\Big(\partial_{\spvar}^{\aind}
        \tilde W_{j}[z;\rho(\spvar+\spvar_\ell);
        \rho K_j^{(m_j,n_j)}]\Big)
        \nonumber\\
        &&\hspace{2cm}
        \times\;
        \Big[\prod_{\ell=j+1}^L  
        F_\ell[\spvar]\tilde W_{\ell}[z;\rho(\spvar+\spvar_\ell);
        \rho K_\ell^{(m_\ell,n_\ell)}]\Big] 
        F_L[\spvar] \,\vac \, \Ket \;.
\eeqn
It follows that
\eqn
        &&|V_{\underline{m,p,n,q}}^{(L,i)}[\h\,\big|\,\spvar;K^{(M,N)}]|
        \nonumber\\
        &&\hspace{2cm}\leq\;
        \sum_{j=0}^L \|\partial_{\spvar}^{\aind} F_j\|_{op}
        \Big[\prod_{\ell=0\atop \ell\neq j}^L
        \|F_\ell[\spvar]\|_{op} \Big]
        \nonumber\\
        &&\hspace{2cm}\times\;
        \prod_{\ell=1}^L
        \|\tilde W_\ell[\rho(\spvar+\spvar_\ell);\rho K_\ell^{(m_\ell,n_\ell)}]\|_{op}
        \nonumber\\
        &&\hspace{2cm}\leq\;(L+1)C_\Theta^{L+1}\rho^{-L+1}\prod_{\ell=1}^L
        \|\tilde W_\ell[z;\rho(\spvar+\spvar_\ell);\rho K_\ell^{(m_\ell,n_\ell)}]\|_{op} 
\eeqn
and
\eqn
        &&|V_{\underline{m,p,n,q}}^{(L,ii)}[\h\,\big|\,\spvar;K^{(M,N)}]| 
        \nonumber\\
        &&\hspace{2cm}\leq\;
		\Big[\prod_{\ell=0}^L \|F_\ell[\spvar]\|_{op}\Big]
        \Big\{ \sum_{j=1}^L \|\partial_{\spvar}^{\aind}\tilde W_j
        [z;\rho(\spvar+\spvar_j);\rho K_j^{(m_j,n_j)}]\|_{op}
        \nonumber\\
        &&\hspace{2cm}\times\;\prod_{\ell=1 \atop \ell\neq j}^L
        \|\tilde W_\ell[z;\rho(\spvar+\spvar_j);\rho K_j^{(m_j,n_j)}]\|_{op}\Big\}
        \nonumber\\
        &&\hspace{2cm}\leq\; L C_\Theta^{L+1} \rho^{-L+2}
        \Big\{ \sum_{j=1}^L \|W_{p_j,q_j}^{m_j,n_j}
        \Big[\partial_{\spvar}^{\aind}\h[z]\Big|\rho(\spvar+\spvar_j);
        \rho K_j^{(m_j,n_j)}\Big]\|_{op}
        \nonumber\\
        &&\hspace{2cm}\times\;\prod_{\ell=1 \atop \ell\neq j}^L
        \|\tilde W_\ell[z;\rho(\spvar+\spvar_\ell);
        \rho K_\ell^{(m_\ell,n_\ell)}]\|_{op}\Big\} \;.
\eeqn

\subsubsection{The case $|\aind|=2$ and $a_0=0$} 

Due to $a_0=0$, no derivatives
with respect to $X_0$ (the spectral variable corresponding to the
operator $H_f$) appear here.
We have
\eqn
        \partial_{\spvar}^{\aind}V_{\underline{m,p,n,q}}^{(L)}[\h\,\big|\,\spvar;K^{(M,N)}]
        &=&V_{\underline{m,p,n,q}}^{(L,iii)}[\spvar;K^{(M,N)}]
        \nonumber\\
        &+&\cdots\cdots\;+\;
        V_{\underline{m,p,n,q}}^{(L,viii)}[\spvar;K^{(M,N)}] \;,
\eeqn
where
\eqn
        &&V_{\underline{m,p,n,q}}^{(L,iii)}[\h\,\big|\,\spvar;K^{(M,N)}] 
        \;:=\;\sum_{\aind_1+\aind_2=\aind}\sum_{j_2>j_1=0}^L
        \nonumber\\
        &&\hspace{2cm}
        \Bra   \vac \,,\, \Big[\prod_{\ell=1}^{j_1-1}
        F_{\ell-1}[\spvar]\tilde W_{\ell}[z;\rho(\spvar+\spvar_\ell);
        \rho K_\ell^{(m_\ell,n_\ell)}]
        \Big]
        \nonumber\\
        &&\hspace{2cm} 
        \times\;\Big(\partial_{\spvar}^{\aind_1}F_{j_1}[\spvar]\Big) 
        \Big[\prod_{\ell=j_1}^{j_2-1}
        \tilde W_{\ell}[z;\rho(\spvar+\spvar_\ell);
        \rho K_\ell^{(m_\ell,n_\ell)}]F_{\ell-1}[\spvar]\Big]
        \nonumber\\
        &&\hspace{2cm} 
        \times\;
        \tilde W_{j_2-1}[z;\rho(\spvar+\spvar_\ell);
        \rho K_{j_2-1}^{(m_{j_2-1},n_{j_2-1})}]
        \Big(\partial_{\spvar}^{\aind_2}F_{j_2}[\spvar]\Big)
        \nonumber\\
        &&\hspace{2cm} 
        \times\;
        \Big[\prod_{\ell=j_2+1}^L  \tilde W_{\ell}[z;\rho(\spvar+\spvar_\ell);
        \rho K_\ell^{(m_\ell,n_\ell)}]
        F_\ell[\spvar]\Big] \vac \, \Ket \;,
\eeqn
and
\eqn
        &&V_{\underline{m,p,n,q}}^{(L,iv)}[\h\,\big|\,\spvar;K^{(M,N)}]
        \;:=\;\sum_{\aind_1+\aind_2=\aind}\sum_{j_2>j_1=0}^L      
        \nonumber\\
        &&\hspace{2cm}
        \Bra \,  \vac \,,\, \Big[\prod_{\ell=1}^{j_1}
        F_{\ell-1}[\spvar] \tilde W_{\ell}[z;\rho(\spvar+\spvar_\ell);
        \rho K_\ell^{(m_\ell,n_\ell)}]
        \Big]
        \nonumber\\
        &&\hspace{2cm}
        \times\;\Big(\partial_{\spvar}^{\aind_1}F_{j_1}[\spvar]\Big) 
        \Big[\prod_{\ell=j_1+1}^{j_2-1}
        \tilde W_{\ell}[z;\rho(\spvar+\spvar_\ell);
        \rho K_\ell^{(m_\ell,n_\ell)}]
        F_{\ell}[\spvar]\Big]
        \nonumber\\
        &&\hspace{2cm} 
        \times\;\rho\Big(\partial_{\spvar}^{\aind_2}
        \tilde W_{j_2}[z;\rho(\spvar+\spvar_\ell);
        \rho K_{j_2}^{(m_{j_2},n_{j_2})}]\Big)
        \nonumber\\
        &&\hspace{2cm} 
        \times\;
        \Big[\prod_{\ell=j_2+1}^L  
        F_{\ell-1}[\spvar]\tilde W_{\ell}[z;\rho(\spvar+\spvar_\ell);
        \rho K_\ell^{(m_\ell,n_\ell)}]\Big] 
        F_L[\spvar]\, \vac \, \Ket \;.
\eeqn
$V_{\underline{m,p,n,q}}^{(L,v)}$ is defined similarly as
$V_{\underline{m,p,n,q}}^{(L,iv)}$, but applies to the case $j_2<j_1$.
\eqn
        &&V_{\underline{m,p,n,q}}^{(L,v)}[\h\,\big|\,\spvar;K^{(M,N)}]
        \;:=\;\sum_{\aind_1+\aind_2=\aind}\sum_{j_2>j_1=0}^L
        \nonumber\\
        &&\hspace{2cm}
        \Bra \, \vac \,,\, F_0[\spvar]
        \Big[\prod_{\ell=1}^{j_1-1}
        \tilde W_{\ell}[z;\rho(\spvar+\spvar_\ell);
        \rho K_\ell^{(m_\ell,n_\ell)}] F_{\ell}[\spvar]
        \Big]
        \nonumber\\
        &&\hspace{2cm}
        \times\;\rho\Big(\partial_{\spvar}^{\aind_1}
        \tilde W_{j_1}[z;\rho(\spvar+\spvar_\ell);
        \rho K_{j_1}^{(m_{j_1},n_{j_1})}]\Big)
        \nonumber\\  
        &&\hspace{2cm}\times \; F_{j_1}[\spvar]
        \Big[\prod_{\ell=j_1+1}^{j_2-1}
        \tilde W_{\ell}[z;\rho(\spvar+\spvar_\ell);
        \rho K_\ell^{(m_\ell,n_\ell)}]
        F_{\ell}[\spvar]\Big]
        \nonumber\\
        &&\hspace{2cm} 
        \times\;\rho\Big(\partial_{\spvar}^{\aind_2}
        \tilde W_{j_2}[z;\rho(\spvar+\spvar_\ell);
        \rho K_{j_2}^{(m_{j_2},n_{j_2})}]\Big)
        \nonumber\\
        &&\hspace{2cm}
        \times\;
        \Big[\prod_{\ell=j_2+1}^L  
        F_\ell[\spvar]\tilde W_{\ell}[z;\rho(\spvar+\spvar_\ell);
        \rho K_\ell^{(m_\ell,n_\ell)}]\Big] 
        F_L[\spvar] \, \vac \, \Ket \;.
\eeqn
Finally,
\eqn
        &&V_{\underline{m,p,n,q}}^{(L,vii)}[\h\,\big|\,\spvar;K^{(M,N)}]
        \;:=\;\sum_{j=0}^L
        \Bra \, \vac \,,\, \Big[\prod_{\ell=1}^{j-1}
        F_{\ell-1}[\spvar]\tilde W_{\ell}[z;\rho(\spvar+\spvar_\ell);
        \rho K_\ell^{(m_\ell,n_\ell)}]
        \Big]
        \nonumber\\
        &&\hspace{2cm} 
        \times\;\Big(\partial_{\spvar}^{\aind}F_j[\spvar]\Big) 
        \Big[\prod_{\ell=j+1}^L 
        \tilde W_{\ell}[z;\rho(\spvar+\spvar_\ell);
        \rho K_\ell^{(m_\ell,n_\ell)}]
        F_\ell[\spvar]\Big]\, \vac \, \Ket
        \nonumber
\eeqn
and
\eqn
        &&V_{\underline{m,p,n,q}}^{(L,viii)}[\h\,\big|\,\spvar;K^{(M,N)}]
        :=\;\sum_{j=0}^L
        \Bra  \, \vac \,,\, F_0[\spvar] 
        \nonumber\\
        &&\hspace{2cm}\times\;
        \Big[\prod_{\ell=1}^{j-1}
        \tilde W_{\ell}[z;\rho(\spvar+\spvar_\ell);
        \rho K_\ell^{(m_\ell,n_\ell)}] F_{\ell}[\spvar]
        \Big]
        \nonumber\\
        &&\hspace{2cm} 
        \times\;\rho^2\Big(\partial_{\spvar}^{\aind}
        \tilde W_{j}[z;\rho(\spvar+\spvar_\ell);
        \rho K_j^{(m_j,n_j)}]\Big)
        \nonumber\\
        &&\hspace{2cm} 
        \times\;
        \Big[\prod_{\ell=j+1}^L  
        F_{\ell-1}[\spvar]\tilde W_{\ell}[z;\rho(\spvar+\spvar_\ell);
        \rho K_\ell^{(m_\ell,n_\ell)}]\Big] 
        F_L[\spvar] \, \vac \, \Ket \;.
\eeqn
We then conclude that
\eqn
        &&|V_{\underline{m,p,n,q}}^{(L,iii)}[\h\,\big|\,\spvar;K^{(M,N)}]|
        \;\leq\;\sum_{\aind_1+\aind_2=\aind}
        \sum_{j_2>j_1=0}^L \|\partial_{\spvar}^{\aind_1} F_{j_1}\|_{op}
        \|\partial_{\spvar}^{\aind_2} F_{j_2}\|_{op}        \nonumber\\
        &&\hspace{2cm}\times\;
        \prod_{\ell=0\atop \ell\neq j_1,j_2}^L
        \|F_\ell[\spvar]\|_{op}\prod_{\ell=1}^L
        \|\tilde W_\ell[\rho(\spvar+\spvar_\ell);\rho K_\ell^{(m_\ell,n_\ell)}]\|_{op}
        \\
        &&\hspace{2cm}\leq\; L(L+1)C_\Theta^{L+1}\rho^{-L+1}\prod_{\ell=1}^L
        \|\tilde W_\ell[z;\rho(\spvar+\spvar_\ell);\rho K_\ell^{(m_\ell,n_\ell)}]\|_{op}
        \;, \nonumber
\eeqn
and
\eqn
        &&|V_{\underline{m,p,n,q}}^{(L,iv)}[\h\,\big|\,\spvar;K^{(M,N)}]|
        \;,\;  |V_{\underline{m,p,n,q}}^{(L,v)}[\spvar;K^{(M,N)}]|
        \;\leq\;\sum_{\aind_1+\aind_2=\aind} \rho^{|\ua_2|}
        \sum_{j_2>j_1=0}^L 
        \nonumber\\
        &&\hspace{2cm} \times \;\|\partial_{\spvar}^{\aind_1} F_{j_1}\|_{op}
        \|W_{p_{j_2},q_{j_2}}^{(m_{j_2},n_{j_2})}
        \Big[\partial_{\spvar}^{\aind_2}\h[z]\Big|\rho(\spvar+\spvar_{j_2});
        \rho K_{j_2}^{(m_{j_2},n_{j_2})}\Big]\|_{op}
        \nonumber\\
        &&\hspace{2cm} \times\;
        \Big[\prod_{\ell=0\atop \ell\neq j_1 }^L  \|F_\ell[\spvar]\|_{op}\Big]
        \prod_{\ell=0\atop \ell\neq j_2 }^L
        \|\tilde W_\ell[\rho(\spvar+\spvar_\ell);\rho K_\ell^{(m_\ell,n_\ell)}]\|_{op}
        \\
        &&\hspace{2cm}\leq\; L(L+1)C_\Theta^{L+1}\rho^{-L+1}\prod_{\ell=1}^L
        \|\tilde W_\ell[z;\rho(\spvar+\spvar_\ell);\rho K_\ell^{(m_\ell,n_\ell)}]\|_{op}
        \;. \nonumber
\eeqn
Similarly,
\eqn
        &&|V_{\underline{m,p,n,q}}^{(L,vi)}[\h\,\big|\,\spvar;K^{(M,N)}]|
        \;\leq\;   C_\Theta^{L+1} \rho^{-L+2}\sum_{\aind_1+\aind_2=\aind}
        \nonumber\\
        &&\hspace{2cm}\times\; 
        \Big\{ \sum_{j_2>j_1=1}^L \|W_{p_{j_1},q_{j_1}}^{(m_{j_1},n_{j_1})}
        \Big[\partial_{\spvar}^{\aind_1}\h[z]\Big|\rho(\spvar+\spvar_{j_1});
        \rho K_{j_1}^{(m_{j_1},n_{j_1})}\Big]\|_{op}
        \nonumber\\
        &&\hspace{2cm}\times\;
        \|W_{p_{j_2},q_{j_2}}^{(m_{j_2},n_{j_2})}
        \Big[\partial_{\spvar}^{\aind_2}\h[z]\Big|\rho(\spvar+\spvar_{j_2});
        \rho K_{j_2}^{(m_{j_2},n_{j_2})}\Big]\|_{op}
        \nonumber\\
        &&\hspace{2cm}\times\,\prod_{\ell=1 \atop \ell\neq j_1,j_2}^L
        \|\tilde W_\ell[z;\rho(\spvar+\spvar_\ell);
        \rho K_\ell^{(m_\ell,n_\ell)}]\|_{op}\Big\} \;,
\eeqn
\eqn
        &&|V_{\underline{m,p,n,q}}^{(L,vii)}[\h\,\big|\,\spvar;K^{(M,N)}]|
        \;\leq\;
        \sum_{j=0}^L \|\partial_{\spvar}^{\aind} F_j\|_{op}
        \prod_{\ell=0\atop \ell\neq j_1,j_2}^L
        \|F_\ell[\spvar]\|_{op}
        \nonumber\\
        &&\hspace{2cm} \times\;\prod_{\ell=1}^L
        \|\tilde W_\ell[\rho(\spvar+\spvar_\ell);\rho K_\ell^{(m_\ell,n_\ell)}]\|_{op}
        \\
        &&\hspace{2cm}\leq\;(L+1)C_\Theta^{L+1}\rho^{-L+1}\prod_{\ell=1}^L
        \|\tilde W_\ell[z;\rho(\spvar+\spvar_\ell);\rho K_\ell^{(m_\ell,n_\ell)}]\|_{op}
        \nonumber
\eeqn
and
\eqn
        &&|V_{\underline{m,p,n,q}}^{(L,viii)}[\h\,\big|\,\spvar;K^{(M,N)}]|
        \;
        \leq\; \rho^2 \prod_{\ell=0}^L \|F_\ell[\spvar]\|_{op}
        \nonumber\\
        &&\hspace{2cm}\times\;
        \Big\{ \sum_{j=1}^L \|\partial_{\spvar}^{\aind}\tilde W_j
        [z;\rho(\spvar+\spvar_j);\rho K_j^{(m_j,n_j)}]\|_{op}
        \nonumber\\
        &&\hspace{2cm} \times\;\prod_{\ell=1 \atop \ell\neq j}^L
        \|\tilde W_\ell[z;\rho(\spvar+\spvar_j);\rho K_j^{(m_j,n_j)}]\|_{op}\Big\}
        \nonumber\\
        &&\hspace{2cm}\leq \; L C_\Theta^{L+1} \rho^{-L+3}
        \nonumber\\
        &&\hspace{2cm}\times\;
        \Big\{ \sum_{j=1}^L \|W_{p_j,q_j}^{(m_j,n_j)}
        \Big[\partial_{\spvar}^{\aind}\h[z]\Big|\rho(\spvar+\spvar_j);
        \rho K_j^{(m_j,n_j)}\Big]\|_{op}
        \nonumber\\
        &&\hspace{2cm} \times\;\prod_{\ell=1 \atop \ell\neq j}^L
        \|\tilde W_\ell[z;\rho(\spvar+\spvar_\ell);
        \rho K_\ell^{(m_\ell,n_\ell)}]\|_{op}\Big\} \;.
\eeqn

\subsubsection{Radial $C^1$-bound in $k$} 

The discussion of the derivative 
$|k|^\sigma\partial_{|k|} \big(|k|^{-\sigma}V_{\underline{m,p,n,q}}^{(L)}\big)$
is analogous to the case for $\partial_{\spvar}^{\ua}$ with $|\aind|=1$ treated above, 
and will here not be reiterated.
One finds
\eqn
        &&||k|^\sigma\partial_{|k|}
        \big(|k|^{-\sigma}V_{\underline{m,p,n,q}}^{(L)}[\h\,\big|\,\spvar;K^{(M,N)}]\big)|
        \nonumber\\
        &&\hspace{2cm}\leq\;
        (L+1)C_\Theta^{L+1}\rho^{-L+1}
        \prod_{\ell=1}^L
        \|\tilde W_\ell[z;\rho(\spvar+\spvar_\ell);
        \rho K_\ell^{(m_\ell,n_\ell)}]\|_{op}
        \nonumber\\
        &&\hspace{2cm}+\;  L C_\Theta^{L+1} \rho^{-L+2}
        \nonumber\\
        &&\hspace{2cm}\times\;
        \Big\{ \sum_{j=1}^L \|W_{p_j,q_j}^{(m_j,n_j)}
        \Big[|k|^\sigma\partial_{|k|}
        \big(|k|^{-\sigma}\h[z]\big)\Big|\spvar+\spvar_j;
        K_j^{(m_j,n_j)}\Big]\|_{op}
        \nonumber\\
        &&\hspace{2cm} \times\;\prod_{\ell=1 \atop \ell\neq j}^L
        \|\tilde W_\ell[z;\rho(\spvar+\spvar_\ell);
        \rho K_\ell^{(m_\ell,n_\ell)}]\|_{op}\Big\} \;,
        \label{der-k-VLmpnq-1}
\eeqn
for any $k\in k^{(M,N)}$.

\subsubsection{Bounds involving derivatives in $|p|$ for $|p|>0$} 

The asserted estimate for $|p|>0$ on 
$|\partial_{|p|}V_{\underline{m,p,n,q}}^{(L)}|$
is obtained precisely in the
same way as in the case $|\aind|=1$, or (~\ref{der-k-VLmpnq-1}).
However, one does not obtain a factor $\rho$ from differentiating $\tilde W_\ell$, 
because $\partial_{|p|}$ is invariant under
rescaling of photon momenta.

The bound on $|\partial_{|p|}\partial_{\spvar}^{\aind}
V_{\underline{m,p,n,q}}^{(L)}|$ is derived in a manner very similar to
the case discussed under \underline{2.}, for $a_0=0$ and $|\aind|=2$.
We shall not reiterate the calculations explicitly, but only again note
that there is no factor $\rho$ involved in the derivative of $\tilde W_\ell$ with
respect to $|p|$, in contrast to derivatives with respect to  $\spvar$.

\subsubsection{Completing the proof}

Collecting the above estimates,
\eqn
        \sup_{\spvar}|V_{\underline{m,p,n,q}}^{(L,i)}
        [\h\,\big|\,\spvar;K^{(M,N)}]|
        &+&\cdots\;+\;
        \sup_{\spvar} |V_{\underline{m,p,n,q}}^{(L,viii)}
        [\h\,\big|\,\spvar;K^{(M,N)}]|
        \nonumber\\
        &+& 
        \sup_{\spvar}||k|^\sigma\partial_{|k|}
        \big(|k|^{-\sigma}V_{\underline{m,p,n,q}}^{(L)}[\h\,\big|\,\spvar;K^{(M,N)}]\big)|
        \nonumber\\
        &+& 
        \sup_{\spvar}\; 
        \1_{|p|>0} \; \sum_{|\aind|\leq1}
        |\partial_{|p|}\partial_{\spvar}^{\aind}
        V_{\underline{m,p,n,q}}^{(L)}[\h\,\big|\,\spvar;K^{(M,N)}]|
        \nonumber 
\eeqn
is bounded by
\eqn
		&&(L+1)^2
        C_\Theta^{L+2}\rho^{-L+1}\prod_{\ell=1}^L
        \Big\{\sup_{|X|\leq X_0<1} \; \rho \;
        \| \partial_{X_0} \tilde W_\ell
        [z;\rho\spvar;\rho K_\ell^{(m_\ell,n_\ell)}]\|_{op}
        \nonumber\\
        &&\hspace{2cm} +\;\sum_{0\leq|\aind|\leq2 \atop a_0=0} 
        \sup_{|X|\leq X_0<1}
        \; \rho^{|\aind|} \; \|\partial_{\spvar}^{\aind} \tilde W_\ell
        [z;\rho\spvar;\rho K_\ell^{(m_\ell,n_\ell)}]\|_{op}
        \\
        &&\hspace{2cm} +\; \sup_{k\in k^{(M,N)}}
        \sup_{|X|\leq X_0<1}\rho \; \||k|^\sigma\partial_{|k|}
        \big(|k|^{-\sigma}\tilde W_\ell
        [z;\rho\spvar;\rho K_\ell^{(m_\ell,n_\ell)}\big)]\|_{op}
        \nonumber\\
        &&\hspace{2cm} +\;
        \sup_{|X|\leq X_0<1}
        \; \1_{|p|>0}\;
        \sum_{|\aind|\leq1} \rho^{|\aind|} \;
        \|\partial_{|p|}\partial_{\spvar}^{\aind} \tilde W_\ell
        [z;\rho\spvar;\rho K_\ell^{(m_\ell,n_\ell)}]\|_{op}
        \Big\}\;. \nonumber
\eeqn
Thus,
\eqn
        &&(2\pi^{\frac12})^{M+N}\sup_{K^{(M,N)}}|k^{(M,N)}|^{-\sigma}
        \sup_{|X|\leq X_0<1}
        |V_{\underline{m,p,n,q}}^{(L)}|
        \nonumber\\
        &&\hspace{2cm}\leq\;
        (L+1)^2
        C_\Theta^{L+2}\rho^{-L+1}
        \prod_{\ell=1}^L
        \Big\{\sup_{K_\ell^{(m_\ell,n_\ell)}}(2\pi^{\frac12})^{m_\ell+n_\ell}
        |k_\ell^{(m_\ell,n_\ell)}|^{-\sigma}        \nonumber\\
        &&\hspace{2cm}\times\;
        \Big[\sup_{\spvar} \; \rho \;  \| \partial_{X_0} \tilde W_\ell
        [z;\rho\spvar;\rho K_\ell^{(m_\ell,n_\ell)}]\|_{op}
        \nonumber\\
        && \hspace{2cm} 
        + \;\sum_{0\leq|\aind|\leq2 \atop a_0=0} 
        \sup_{|X|\leq X_0<1}
        \; \rho^{|\aind|} \; \|\partial_{\spvar}^{\aind} \tilde W_\ell
        [z;\rho\spvar;\rho K_\ell^{(m_\ell,n_\ell)}]\|_{op} 
        \nonumber\\
        &&\hspace{2cm}
        +  \; \sup_{k\in k^{(M,N)}}
        \sup_{|X|\leq X_0<1}\;\rho\;\||k|^\sigma\partial_{|k|}
        \big(|k|^{-\sigma}\tilde W_\ell
        [z;\rho\spvar;\rho K_\ell^{(m_\ell,n_\ell)}]\big)\|_{op}
        \nonumber\\
        &&\hspace{2cm}
        +  \; 
        \sup_{|X|\leq X_0<1}
        \; \1_{|p|>0} \; 
        \sum_{|\aind|\leq1} \rho^{|\aind|} \;        
        \|\partial_{|p|}\partial_{\spvar}^{\aind} \tilde W_\ell
        [z;\rho\spvar;\rho K_\ell^{(m_\ell,n_\ell)}]\|_{op}\Big]\Big\} \;.
\eeqn
Using the coordinate change $k^{(m_\ell,n_\ell)}
\rightarrow\rho^{-1}k^{(m_\ell,n_\ell)}$ from rescaling, this is bounded by
\eqn
        &&(L+1)^2
        C_\Theta^{L+2}\rho^{\sigma(M+N)-L+1}\prod_{\ell=1}^L
        \Big\{\sup_{K_\ell^{(m_\ell,n_\ell)}}(2\pi^{\frac12})^{m_\ell+n_\ell}
        |k_\ell^{(m_\ell,n_\ell)}|^{-\sigma}
        \nonumber\\
        && \hspace{2cm} \times\;
        \Big[\sup_{|X|\leq X_0<1} \; \rho \;
        \| \partial_{X_0} \tilde W_\ell
        [z; \spvar;  K_\ell^{(m_\ell,n_\ell)}]\|_{op}^2
        \nonumber\\
        && \hspace{4cm}
        + \;
        \sum_{0\leq|\aind|\leq2 \atop a_0=0} 
        \sup_{|X|\leq X_0<1} \; \rho^{|\aind|} \;
        \|\partial_{\spvar}^{\aind} \tilde W_\ell
        [z; \spvar;   K_\ell^{(m_\ell,n_\ell)}]\|_{op}^2
        \nonumber\\
        &&\hspace{2cm} 
        + \; \sup_{k\in k^{(M,N)}}
        \sup_{|X|\leq X_0<1}
        \; \rho  \; \||k|^\sigma\partial_{|k|}
        \big(|k|^{-\sigma}\tilde W_\ell
        [z; \spvar; K_\ell^{(m_\ell,n_\ell)}]\big)\|_{op}
        \nonumber\\
        && \hspace{2cm}
        + \;
        \sup_{|X|\leq X_0<1} \; 
        \1_{|p|>0}\; 
        \sum_{|\aind|\leq1}\rho^{|\aind|} \; 
        \|\partial_{|p|}\partial_{\spvar}^{\aind} \tilde W_\ell
        [z;\spvar;K_\ell^{(m_\ell,n_\ell)}]\|_{op}\Big]
        \Big\}\;.\nonumber
\eeqn
Using Theorem {~\ref{opL2relbound}}, we find
\eqn
        &&(2\pi^{\frac12})^{m_\ell+n_\ell}
        \sup_{K^{(m_\ell,n_\ell)}\in B_1^{m_\ell+n_\ell}} 
        |k^{(m_\ell,n_\ell)}|^{-\sigma}
        \sup_{\spvar}
        \|\tilde W_\ell [z;\spvar; K^{(m_\ell, n_\ell)}]\|_{op}
        \nonumber\\
        &&\hspace{2cm}\leq\,\frac{1}{p_\ell^{p_\ell/2}q_\ell^{q_\ell/2}}
        \sup_{K^{(m_\ell+p_\ell,n_\ell+q_\ell)}\in B_1^{m_\ell+p_\ell+n_\ell+q_\ell}}
        |k^{(m_\ell+p_\ell,n_\ell+q_\ell)}|^{-\sigma}
        \nonumber\\
        &&\hspace{2cm}\times\,
        (2\pi^{\frac12})^{m_\ell+p_\ell+n_\ell+q_\ell}
        \sup_{\spvar}
        |w_{m_\ell+p_\ell,n_\ell+q_\ell}[z; X;K^{(m_\ell+p_\ell,n_\ell+q_\ell)}]|
        \nonumber\\
        &&\hspace{2cm}\leq\,\frac{1}{p_\ell^{p_\ell/2}q_\ell^{q_\ell/2}}
        \| w_{m_\ell+p_\ell, n_\ell+q_\ell}[z]\|_\sigma
        \;.
\eeqn
Consequently,
\eqnn 
        &&\rho^{(M+N)-1}\|V_{\underline{m,p,n,q}}^{(L)}\|_\sigma^\sharp
        \nonumber\\
        &&\hspace{2cm}\leq \; 10 \rho^{(1+\sigma)(M+N)-L} (L+1)^2
        C_\Theta^{L+2}\prod_{\ell=1}^L
        \frac{\| w_{m_\ell+p_\ell, n_\ell+q_\ell}[z]\|_\sigma^\sharp}
        {p_\ell^{p_\ell/2}q_\ell^{q_\ell/2}} \;.
\eeqnn
This proves Lemma {~\ref{VLboundlm}}.

\section{Proof of Theorem {~\ref{codim2contrthm}}: Soft photon sum rules}
\label{spsrproofsssect}

We will next prove that the renormalization map $\ren$ preserves
the soft photon sum rules. More precisely, we shall demonstrate that the soft photon
sum rules \sbsr are transformed according to 
\eqn
		{\rm{\bf SR}}[\mu]\rightarrow{\rm{\bf SR}}[\mu \rho^\sigma]
\eeqn
under $\ren$.

Let us first formulate some remarks regarding how the soft photon sum rules have influenced
on our construction of the isospectral renormalization group.
We recall from (~\ref{sbsr}) that \sbsr involves a derivative 
in $P_f$, and a limit $|k|\rightarrow0$, where $k$ is a photon momentum.
In order to accommodate the latter, we did not use photon momentum space integral norms
for the definition of the Banach space $\Hspace_{\geq0}$ in Section
{~\ref{Banach-seq-sp-sssect-1}}, in contrast to \cite{bcfs1}.

As one will see from the proof, the issue is to verify essentially a purely 
algebraic property
of the renormalization map. However, we emphasize that this is so only because
no derivatives of the cutoff operators $\chi_\rho[H_f]$, $\bar\chi_\rho[H_f]$
enter, since they do not depend on $P_f$. 
For this reason, we have used $H_f$, and not a combination of $H_f$ and 
$P_f$ as the reference operator to 
slice the Fock space. 

Moreover, the soft photon sum rules have also influenced our choice
of $\tau$ (see Definition {~\ref{Feshbtripledef}}) 
in the application of the smooth Feshbach map (as an ingredient of $\ren$). We are using
$\tau=H_f$, which is also independent of $P_f$. The issue here is that
$\tau=H_f$ appears without a factor $(\bar\chi_\rho\chi_1)^2[H_f]$ in the free resolvent 
\eqn
        \bar R_0[\z;\opp]=\Big[H_f+(\bar\chi_\rho^2\chi_1^2)[H_f]
        \big(\tT[\z;\opp]+E[\z]\big) \Big]^{-1}
\eeqn
(cf. (~\ref{bar-R-0-def-1}), and  
we recall that $\tT$ is defined by $T=H_f+\chi_1^2\tT$) in (~\ref{eqnsbsrR0}) below 
(where $\spvar=(X_0,X)$ is the spectral 
variable acounting for $\opp=(H_f,P_f)$). This is in contrast to the 
$P_f$-dependent operators (that is, $X$-dependent terms below).
The renormalization map is here so defined that
differentiating the resolvents in (~\ref{eqnderPfDeltwMN}) with respect to $P_f$ (that is, $X$)  
produces a factor $(\bar\chi_\rho\chi_1)^2$, which is necessary for $\ren$ to preserve the algebraic structure 
of the soft photon sum rules \sbsr.

For brevity, let
\eqn
        \piop[\z;p;\spvar]:=\Bra\vac\,,\,\piop_\rho[\z;\opp+\spvar]\,\vac\Ket \;,
\eeqn
cf. (~\ref{piopdef}) and (~\ref{piop-def-1}).
Recalling the definition of
$W_{p,q}^{m,n}[\spvar;z;K^{(m,n)}]$ from (~\ref{Wmnpqdef}), the soft photon sum rules 
\sbsr state that for some $\mu\in\R_+$, and every unit vector
$n\in\R^3$, $|n|=1$,
\eqnn
        &&g \mu \langle\eps(n,\lambda),\partial_{X}\rangle_{\R^3}
        W_{p_r,q_r}^{m_r,n_r}
        [\spvar;\z;K^{(m_r,n_r)}]
        \\
        &&\hspace{2cm}=(n_r+q_r+1)\lim_{\rvar\rightarrow0}\rvar^{-\sigma}
        W_{p_r,q_r}^{m_r,n_r+1}[\spvar;\z;K^{(m_r,n_r+1)}]
        \Big|_{\tilde K_{n_r+1}=(\rvar n,\lambda)}\\
        \\
        &&\hspace{2cm}=(m_r+p_r+1)\lim_{\rvar\rightarrow0}\rvar^{-\sigma}
        W_{p_r,q_r}^{m_r+1,n_r}[\spvar;\z;K^{(m_r+1,n_r)}]
        \Big|_{\tilde K_{m_r+1}=(\rvar n,\lambda)}  \;.
\eeqnn
Assuming \sbsr, one easily checks that
\eqn
        &&g \mu \langle \eps(n,\lambda),\partial_{X}\rangle_{\R^3}\bar R_0[\z;p;\spvar]
        \nonumber\\
        &&\hspace{0.5cm}=- \bar R_0[\z;p;\spvar](\bar\chi_\rho^2\chi_1^2)[X_0]
        \Big(g \mu \langle \eps(n,\lambda),\partial_{X}\rangle_{\R^3}  \tT[\spvar;\z]\Big)
        \bar R_0[\z;p;\spvar] 
        \label{eqnsbsrR0}\\
        &&\hspace{0.5cm}=
        -\lim_{\rvar\rightarrow0}\rvar^{-\sigma}\bar R_0[\z;p;\spvar](\bar\chi_\rho^2\chi_1^2)[X_0]
        W_{0,0}^{0,1}[\spvar;\z;\rvar n, \lambda](\bar\chi_\rho^2\chi_1^2)[X_0]\bar R_0[\z;p;\spvar]  \; ,
        \nonumber
\eeqn
and
\eqn
        &&g \mu \langle \eps(n,\lambda),\partial_{X}\rangle_{\R^3}\piop[\z;X]
        \nonumber\\
        &&\hspace{1cm}=
        -\lim_{\rvar\rightarrow0}\rvar^{-\sigma} \piop[\z;X] W_{0,0}^{0,1}[\spvar;\z;\rvar n,\lambda]
        (\bar\chi_\rho^2\chi_1^2)[X_0]
        \bar R_0[\z;p;\spvar]\bch[X_0]  \;.
        \label{eqnsbsrpiop}
\eeqn
In both (~\ref{eqnsbsrR0}) and (~\ref{eqnsbsrpiop}), $W^{0,1}_{0,0}$ can be
exchanged with $W^{1,0}_{0,0}$ without any effect.

We next consider $g \langle\eps(K),\partial_{X}\rangle_{\R^3}\widehat w_{M,N}$, which is
given by the symmetrization in $K^{(M)}$ and $\tilde K^{(N)}$ of
\eqn
        &-& \, g\mu  \rho^{ M+N -1}
        \sum_{L\geq1}(-1)^L
        \sum_{m_j+n_j+p_j+q_j\geq1 \atop 1\leq j\leq L}
        \delta_{N,|\underline{n}|}\delta_{M,|\underline{m}|}
        \nonumber\\
        &\times&
         \prod_{j=1}^L
        \left(\begin{array}{c} m_j+p_j \\ p_j\end{array}\right)
        \left(\begin{array}{c} n_j+q_j \\ q_j\end{array}\right)
        \nonumber\\
        &\times&\Big[\;\;
        \sum_{r=1}^{L-1}\;
        \Bra \,\vac\,,\,\Big[\cdots
        \langle \eps(K),\partial_{X}\rangle_{\R^3} (\bar\chi_\rho^2\chi_1^2 R_0)
        [\Ez^{-1}[\widehat \z];\opp+\rho(\spvar+\tilde \spvar_r) ]\cdots\Big]\,\vac\, \Ket 
        \nonumber\\
        && +
        \sum_{r=1}^L
        \Bra\,\vac\,,\,\Big[  \cdots
        \langle \eps(K),\partial_{X}\rangle_{\R^3} W_{p_r,q_r}^{m_r,n_r}
        [\rho(\spvar+\spvar_r) ;\Ez^{-1}[\widehat \z];\rho K_r^{(m_r,n_r)}]
        \cdots \Big]\,\vac\,\Ket
        \nonumber
        \\
        &&\hspace{1.5cm} + \,
        \Bra\,\vac\,,\,\Big[  \langle \eps(K),\partial_{X}\rangle_{\R^3}
        \piop[\Ez^{-1}[\widehat \z];\opp+\rho(\spvar+\tilde \spvar_0) ]
        \cdots\cdots\Big]\,\vac\,\Ket\hspace{0.3cm}
        \label{eqnderPfDeltwMN}\\
        &&\hspace{1.5cm} + \,
        \Bra \,\vac\,,\,\Big[\cdots\cdots
        \langle \eps(K),\partial_{X}\rangle_{\R^3}
        \piop[\Ez^{-1}[\widehat \z];\opp+\rho(\spvar+\tilde \spvar_L) ]
        \Big]\,\vac\,\Ket\;\;\Big]
        \; .\nonumber
\eeqn
Substituting (~\ref{eqnsbsrR0}) and (~\ref{eqnsbsrpiop}), the
first expectation in $[\,\dots\,]$ can be written as
\eqn
        &&-\lim_{\rvar\rightarrow0}\rho (\rho\rvar )^{-\sigma}\sum_{r=1}^{L}
        (m_r+p_r+1)
        \\
        &&\hspace{2cm}\times\;V^{(L+1)}_{\underline{m+e_r,p,n,q}}
        \left[\h[\Ez^{-1}[\widehat \z]]\Big|
        \rho\spvar;\rho K^{(M+1,N)}\right]\Big|_{
        K_{M+1}=( R n,\lambda)}\nonumber
\eeqn
or
\eqn
        &&-\lim_{\rvar\rightarrow0}\rho (\rho\rvar )^{-\sigma}\sum_{r=1}^{L}
        (n_r+1)
        \\
        &&\hspace{2cm}\times\;V^{(L+1)}_{\underline{m+e_r,p,n,q}}
        \left[\h[\Ez^{-1}[\widehat \z]]\Big|
        \rho\spvar;\rho K^{(M+1,N)}\right]\Big|_{
        \tilde K_{N+1}=( R n,\lambda)} \;, \nonumber
\eeqn
where $\underline{e_r}:=(0,\dots,0,1,0,\dots,0)$ is the $L$-dimensional
unit vector with 1 at the $r$-th entry.
Let us next discuss the expectations in $[\,\dots\,]$ which involve derivatives of
$\bar R_0$ and $\piop$. Substituting (~\ref{eqnsbsrR0}) and (~\ref{eqnsbsrpiop}),
the number of interaction operators is increased from $L$ to $L+1$ in every case.
Relabelling all operators according to their product order from
1 to $L+1$, the terms on the three last lines in (~\ref{eqnderPfDeltwMN})
can be combined into
\eqn\;\;\;\;\;
        \lim_{\rvar\rightarrow0}\rho (\rho\rvar )^{-\sigma}\sum_{r=1}^{L+1}
        V^{(L+1)}_{\underline{m,p,n,q}}
        \left[\h[\Ez^{-1}[\widehat \z]]\Big|
        \rho\spvar;\rho K^{(M+1,N)}\right]\Big|_{q_r=p_r=n_r=0 \atop
        K_r^{(1,0)}=( R n,\lambda)}
\eeqn
or
\eqn\;\;\;\;\;
        \lim_{\rvar\rightarrow0}\rho (\rho\rvar )^{-\sigma}\sum_{r=1}^{L+1}
        V^{(L+1)}_{\underline{m,p,n,q}}
        \left[\h[\Ez^{-1}[\widehat \z]]\Big|
        \rho\spvar;\rho K^{(M,N+1)}\right]\Big|_{q_r=p_r=m_r=0 \atop
        K_r^{(1,0)}=( R n,\lambda)}
\eeqn
The terms corresponding to $r=1$ and $r=L+1$ are obtained from derivatives of $\piop$.
Since this expectation value contains $L+1$
interaction operators, we rearrange the sum over $L$ accordingly.

In the special case $M+N=1$, there notably is an additional contribution
\eqn
        &&g \mu \langle\eps(n,\lambda),\partial_{X}\rangle_{\R^3} 
        (\piop \tT)[\z;\rho\spvar]
        \nonumber\\
        &&\hspace{0.5cm}
        = - \lim_{\rvar\rightarrow0}\rho (\rho\rvar )^{-\sigma}
        \piop[\z;p;\spvar] W^{0,1}_{0,0}[\rho\spvar;\z;\rvar n, \lambda]\piop[\z;p;\spvar]
        \,,\;\;
\eeqn
where $W^{0,1}_{0,0}$ can be exchanged with $W^{1,0}_{0,0}$ without any effect.

Thus, after rearranging the terms in the sum over $L$, and using
\eqn
        (n_r+q_r+1)\left(\begin{array}{c} n_r+q_r   \\
        q_r\end{array}\right)
        =(n_r+1)\left(\begin{array}{c} n_r+q_r+1  \\
        q_r\end{array}\right) \; ,
\eeqn
we obtain
\eqn
        &-& \lim_{\rvar\rightarrow0}
        \rho^{M+N} (\rho\rvar )^{-\sigma}
        \sum_{L\geq1}(-1)^L
        \sum_{m_j+n_j+p_j+q_j\geq1 \atop {0<j\leq L \atop m_r+n_r+p_r+q_r\geq0}}
        \delta_{N,|\underline{n}|}\delta_{M,|\underline{m}|}
        (n_r+1)
        \nonumber\\
        &&\hspace{2cm}\times\;
        \sum_{r=1}^L
        \left(\begin{array}{c} n_r+q_r+1  \\ q_r\end{array}\right)
        \Big[\prod_{j=1}^L
        \left(\begin{array}{c} m_j+p_j \\ p_j\end{array}\right)\Big]
        \Big[
        \prod_{j=1 \atop j\neq r}^L
        \left(\begin{array}{c} n_j+q_j  \\ q_j\end{array}\right)\Big]
        \nonumber\\
        &&\hspace{2cm}\times\;
        V^{(L)}_{\underline{m,p,n+e_r,q}}
        \left[\h[\Ez^{-1}[\widehat \z]]\Big|
        \rho\spvar;\rho K^{(M+1,N)}\right]
        \Big|_{\tilde K_{N+1}=(\rvar n,\lambda)} \; , 
\eeqn
or the corresponding expression from interchanging $(n_r,q_r,N)$ and $(m_r,p_r,M)$.
Relabelling the index $n_r\rightarrow n_r+1$, we have
$\delta_{N,\sum n_i}\rightarrow \delta_{N+1,\sum n_i}$, and
thus indeed find
\eqnn
        &&g \mu\rho^\sigma \langle\eps(n,\lambda),\partial_{X}\rangle_{\R^3}
        \widehat w_{M,N}[\widehat \z;\opp;K^{(M,N)}]\\
        &&\hspace{2cm}=  (N+1) \lim_{\rvar\rightarrow0}  \rvar^{-\sigma}
        \widehat w_{M,N+1}[\opp ;\z;K^{(M)},\tilde K^{(N+1)}]
        \Big|_{\tilde K_{N+1}=(\rvar n,\lambda)} \; ,
\eeqnn
or likewise the corresponding expression obtained from
interchanging $N$ and $M$.

In particular, we observe that the parameter $\mu$ is rescaled by a factor
$\rho^\sigma$ so that
\eqn
		 {\rm \bf SR}[\mu]\rightarrow {\rm \bf SR}[\mu\rho^\sigma] 
\eeqn 
under the action of $\ren$, as claimed.

Moreover, we observe that in the limit $\sigma\rightarrow0$, $\mu$ becomes invariant
under $\ren$, and has the constant value $\mu=1$ (the value
1 is connected to the normalization condition (~\ref{kappasig-norm-def-1})).

\section{Proof of Theorem {~\ref{codim2contrthm}}: Codimension two contractivity for $p\neq 0$}
\label{codim2contrproofsubsubsect-1}
 
Let us first address the case of non-vanishing conserved total momentum
$0<|p|<\frac13$. Our aim is to prove that the renormalization
transformation diminishes the interaction essentially by a factor $\rho^\sigma$.

Using Theorem {~\ref{hatwMNformalseriesthm}}, Lemma {~\ref{VLboundlm}}, and 
$\left( m +p \atop p  \right)\leq 2^{m+p}$,
we find
\eqn
        \;\;\;\;\;
        \|\widehat w_{M,N}[\zeta]\|_\sigma^\sharp&\leq&
        \sum_{L=1}^\infty 10 \, C_\Theta^2 (L+1)^2
        \Big(\frac{C_\Theta}{\rho}\Big)^L(2\rho^{1+\sigma})^{M+N}
        \nonumber\\
        && \times\,\sum_{m_1+\cdots+m_l=M,\atop n_1+\cdots+n_L=N}
        \sum_{p_1,q_1,\dots,p_L,q_L:\atop
        m_\ell+p_\ell+n_\ell+q_\ell\geq1}
        \nonumber\\
        &&\times\;\;
        \prod_{\ell=1}^L\Big\{\Big(\frac{2}{\sqrt{p_\ell}}\Big)^{p_\ell}
        \Big(\frac{2}{\sqrt{q_\ell}}\Big)^{q_\ell}
        \|w_{m_\ell+p_\ell,n_\ell+q_\ell}[\z]\|_\sigma^\sharp
        \Big\}
        \label{hatwMNbd-1}
\eeqn
The sum over $\underline{m,p,n,q}$ yields
\eqn
        \|\widehat \h_1 [\zeta] \|^\sharp_{\sigma,\xi}
        &\leq& 20 \, C_\Theta^2\rho^{1+\sigma}\sum_{M+N\geq1}\xi^{-M-N} \sum_{L=1}^\infty (L+1)^2
        \Big(\frac{C_\Theta}{\rho}\Big)^L
        \nonumber\\
        &&\times\,
        \sum_{m_1+\cdots+m_l=M,\atop n_1+\cdots+n_L=N}
        \sum_{p_1,q_1,\dots,p_L,q_L:\atop
        m_\ell+p_\ell+n_\ell+q_\ell\geq1}\prod_{\ell=1}^L\xi^{m_\ell+n_\ell}
        \nonumber\\
        && \times\,\prod_{\ell=1}^L
        \Big\{
        \Big(\frac{2\xi}{\sqrt{p_\ell}}\Big)^{p_\ell}
        \Big(\frac{2\xi}{\sqrt{q_\ell}}\Big)^{q_\ell}
        \xi^{-(m_\ell+p_\ell+n_\ell+q_\ell)}
        \|w_{m_\ell+p_\ell,n_\ell+q_\ell}[\z]\|_\sigma^\sharp
        \Big\}
        \nonumber\\
        &\leq& 20 \, C_\Theta^2\rho^{1+\sigma} \sum_{L=1}^\infty (L+1)^2
        \Big(\frac{C_\Theta}{\rho}\Big)^L
        \label{hatwMNgeq1est1}
        \\
        && \times\,\Big\{\sum_{M+N\geq1}
        \Big(\sum_{p=0}^M\Big(\frac{2\xi}{\sqrt{p}}\Big)^{p}\Big)
        \Big(\sum_{q=0}^N\Big(\frac{2\xi}{\sqrt{q}}\Big)^{q}\Big)
        \xi^{-(M+N)}
        \|w_{M,N}[\z]\|_{\sigma}^\sharp
        \Big\}^L
        \nonumber\\
        &\leq& 20 \, C_\Theta^2\rho^{1+\sigma} \sum_{L=1}^\infty (L+1)^2
        \Big(\frac{C_\Theta}{\rho}\Big)^L
        A^{2L}(\|\h_1[\z] \|_{\xi,\sigma}^\sharp)^L\nonumber
\eeqn
with
\eqn
        A:=\sum_{p=0}^\infty\Big(\frac{2\xi}{\sqrt{p}}\Big)^{p}
        \leq\sum_{p=0}^\infty(2\xi)^p =\frac{1}{1-2\xi}
        \leq\frac12\;,
        \label{Aconstdef}
\eeqn
assuming that $\xi\leq\frac14$.
Furthermore,
\eqn
        B:=\frac{C_\Theta}{\rho(1-2\xi)^2 }\|\h_1\|_{\xi,\sigma}
        \leq\frac{4C_\Theta }{\rho}\|\h_1\|_{\xi,\sigma}\;.
\eeqn
Hence,
\eqn
        \sum_{L=1}^\infty (L+1)^2 B^L&=&\Big(B\frac{d^2}{dB^2}+\frac{d}{dB}\Big)
        \sum_{L=0}^\infty B^L-1
        \nonumber\\
        &=&\frac{1}{(1-B)^2}-1- \frac{B}{(1-B)^3}
        \leq12B\;.
\eeqn
We consequently find
\eqn
        \|\widehat \h_1[\zeta] \|^\sharp_{\xi,\sigma}
        &\leq&240 \, C_\Theta^2\rho^{1+\sigma}B
        \nonumber\\
        &=&960 C_\Theta^3\rho^\sigma\|\h_1\|_{\xi,\sigma}^\sharp
        \leq\frac{\e}{2}
        \label{hatwMNbd-2}
\eeqn
by choosing $\rho=(2\cob)^{-1/\sigma}$ with $\cob:=960 C_\Theta^3$.

We next discuss the case $M+N=0$, where
\eqn
        \widehat w_{0,0}[\zeta;\spvar]&=&H\big[\ren[E[\z]\oplus T\oplus
        \unull_1]\big]
        \nonumber\\
        &&\hspace{0.5cm}
        +\,\rho^{-1}\sum_{L=2}^\infty(-1)^{L-1}\sum_{p_1,q_1,\dots,p_L,q_L:\atop
        p_\ell+q_\ell\geq1} V^{(L)}_{\underline{0,p,0,q}}[\spvar] \;,
        \label{hatw00def}
\eeqn
with
\eqn
        \rho^{-1}\|V^{(L)}_{\underline{0,p,0,q}}\|_{\sigma}^\sharp\leq10\,(L+1)^2
        C_\Theta^2 \rho^{-L}\prod_{\ell=1}^\infty \frac{\|w_{p_\ell,q_\ell}[\z]\|}
        {p_\ell^{p_\ell/2}q_\ell^{q_\ell/2}} \;,
        \label{VL0p0qest}
\eeqn
and
\eqnn
        &&\ren\Big[E[\z]\oplus T\oplus
        \unull_1\Big]
        \nonumber\\
        &&\hspace{2cm}
        =\,E[\z]\oplus\big(X_0 +\big(\rho^{-1}
        w_{0,0}[z;\rho\spvar]-X_0\big)
        \piop_{\rho}[z;\rho\spvar]\chi_1^2[X_0]\big)
        \oplus\unull_1\;.
\eeqnn
Writing $P_{\bar\chi_1}:=P_{\bar\chi_1[|X|\leq X_0]}$, we have
\eqn
        &&\Big\|P_{\Tspace}\Big(\ren[E[\z]\oplus T\oplus
        \unull_1]-T_0^{(p;\rho\lTnl)}\Big)\Big\|_{\Tspace}
        \nonumber\\
        &&\hspace{2cm}= \, \sup\Big\{
        \Big\|P_{\bar\chi_1}^\perp \Big(P_{\Tspace}\ren[E[\z]\oplus T\oplus
        \unull_1]-T_0^{(p;\rho\lTnl)}\Big)\Big\|^\sharp \;,
        \nonumber\\
        &&\hspace{4cm}\frac{1}{K_\Theta}\Big\|P_{\bar\chi_1}
        \Big(P_{\Tspace}\ren[E[\z]\oplus T\oplus
        \unull_1]-T_0^{(p;\rho\lTnl)}\Big)\Big\|^\sharp
        \Big\}
        \\
        &&\hspace{2cm}\leq\,\sup\Big\{
        \delta\,,\,
        \frac{1}{K_\Theta}\Big\|P_{\bar\chi_1}
        \Big( P_{\Tspace}\ren[E[\z]\oplus T\oplus
        \unull_1]-T_0^{(p;\rho\lTnl)}\Big)\Big\|^\sharp
        \Big\}
        \nonumber
\eeqn
(the projection $P_{\Tspace}:\Hspace_{\geq0}\rightarrow\Tspace$
is defined in (~\ref{PTspacedef})). A straightforward calculation shows that
\eqn
        \Big\|P_{\bar\chi_1[|X|\leq X_0]}
        \Big( P_{\Tspace}\ren[E[\z]\oplus T\oplus
        \unull_1]-T_0^{(p;\rho\lTnl)}\Big)\Big\|^\sharp
        &\leq & \tilde K_\Theta \|P_{|X|\leq X_0\leq3/4}
        \big(T-T_0^{(p;\lTnl)}\big)\|^\sharp \nonumber\\
        &\leq & \tilde K_\Theta \delta \;,
        \label{renT0part-1}
\eeqn
for a constant $1\leq\tilde K_\Theta<\infty$, which only depends on $C_\Theta$ and numerical factors
independent of $\e,\delta,\lTnl$.
Hence, choosing $K_\Theta:=\tilde K_\Theta$ in (~\ref{Tnormdef}),
we find
\eqn
        \|\widehat T-T_0^{(p;\rho\lTnl)}\|_{\Tspace}&\leq&\|P_{\Tspace}
        \ren[E[\z]\oplus T\oplus
        \unull_1]-T_0^{(p;\rho\lTnl)}\|_{\Tspace}
        \nonumber\\
        &+&\frac1\rho\sum_{L\geq2}\sum_{p_1,q_1,\dots,p_L,q_L:\atop
        p_\ell+q_\ell\geq1}
        \Big[\sup_{\spvar}|\partial_{X_0}
        V^{(L)}_{\underline{0,p,0,q}}[\spvar]|
        \nonumber\\
        &+&\sum_{|\aind|=1}\sup_{\spvar}|\partial_{|p|}\partial_{\spvar}^{\aind}
        V^{(L)}_{\underline{0,p,0,q}}[\spvar]| 
        +\sum_{1\leq|\aind|\leq2\atop a_0=0}
        \sup_{\spvar}|\partial_{\spvar}^{\aind}
        V^{(L)}_{\underline{0,p,0,q}}[\spvar]|
        \Big]
        \nonumber\\
        &\leq&\delta+10\,C_\Theta^2\sum_{L=2}^\infty (L+1)^2
        \Big(\frac{C_\Theta}{\rho}\Big)^L
        \Big(\sum_{p+q\geq1} \|w_{p,q}[\z]\|_\sigma^\sharp\Big)^L
        \nonumber\\
        &\leq&\delta+10\,C_\Theta^2\sum_{L=2}^\infty (L+1)^2
        \Big(\frac{C_\Theta}{\rho}\Big)^L
        \Big(\xi\sum_{p+q\geq1}\xi^{-p-q}\|w_{p,q}[\z]\|_\sigma^\sharp\Big)^L
        \nonumber\\
        &\leq&
        \delta+10\,C_\Theta^2\sum_{L=2}^\infty (L+1)^2
        \Big(\frac{ C_\Theta\xi}{\rho}\Big)^L
        \Big(\|\h_1\|_{\xi,\sigma}^\sharp\Big)^L
        \nonumber\\
        &\leq&\delta+ 120\, C_\Theta \Big(\frac{C_\Theta\xi}{\rho}
        \|\h_1\|_{\xi,\sigma}^\sharp\Big)^2
        \nonumber\\
        &\leq&\delta+ 120\,  \frac{C_\Theta^3 \xi^2 }{\rho^2}\e^2
        \nonumber\\
        &\leq&\delta+\e \;,
        \label{DiffThatTest}
\eeqn
with   
\eqn
		\e&<&\e_0\;<\;\rho^2
		\nonumber\\
		\xi&:=& \coe^{\frac12}
		\;\;{\rm where}\;\;
		\coe\;:=\;\frac{1}{120 C_\Theta^3 }\;,
		\nonumber\\
		\rho&=&(2\cob)^{-1/\sigma}
		\;\;{\rm where} \;\;
		\cob=960 C_\Theta^3\;,
		\label{parm-choice-1}
\eeqn 
as our choice of constants.

Finally, let $\widehat E[\zeta]:=\widehat w_{0,0}[\zeta;\underline{0}]$. With 
$E[z]=w_{0,0}[z;\underline{0}]$ and $\zeta=E_\rho[z]=\frac{1}{\rho}E[z]$,
we derive from (~\ref{hatw00def}) that
\eqn
        \widehat E[\zeta]=\zeta+\rho^{-1}\sum_{L=2}^\infty
        \sum_{p_1,q_1,\dots,p_L,q_L:\atop
        p_\ell+q_\ell\geq1} V^{(L)}_{\underline{0,p,0,q}}[\underline{0}]\;.
\eeqn 
We can again use the bounds (~\ref{VL0p0qest}) and (~\ref{DiffThatTest}), whereby
\eqn
        |\partial_{|p|}^a (\widehat E[\zeta]-\zeta)|&\leq& 120C_\Theta\Big(
        \frac{C_\Theta\xi}{\rho}
        \|\partial_{|p|}^a \h_1\|^\sharp\Big)^2
        \nonumber\\
        &\leq&120\,  \frac{C_\Theta^3 \xi^2}{\rho^2}\e^2 
        \;<\;\e\;,
        \label{partpEest}
\eeqn
for $a=0,1$ and (~\ref{parm-choice-1}).
Thus,
\eqn
		\ren:\Polyd^{(\mu)}\Polpar\rightarrow
		\Polyd^{(\rho^\sigma\mu)}(\widehat\e,\widehat\delta,\widehat\lTnl)\;,
\eeqn
with
\eqn
		\widehat\e&\leq&\frac\e2
		\nonumber\\
		\widehat\delta&\leq&\delta+\e
		\nonumber\\
		\widehat\lTnl&=&\rho\lTnl
\eeqn
given the ($\sigma$-dependent) choice of parameters (~\ref{parm-choice-1}).

This proves the assertion of Theorem  {~\ref{codim2contrthm}} for the case $0<|p|<\frac13$.

\section{Proof of Theorem {~\ref{codim2contrthm}}: Codimension two contractivity for $p=0$}
\label{codim2contrproofsubsubsect-2}

Next, we address the case of vanishing conserved total momentum $p=0$.
Our aim is to prove that instead of only $\rho^\sigma$,
the bounds on the interaction gain a full factor $\rho$
from the renormalization transformation.  
This holds, in particular, {\em independently} of the infrared regularization 
$\sigma$.

For $M+N\geq2$, (~\ref{hatwMNbd-1}) and the subsequent calculations leading to
(~\ref{hatwMNbd-2}) imply that (using the same notation)
\eqn
        \|\widehat\h_2\|^\sharp_{\xi,\sigma}
        &\leq&40 C_\Theta^2\rho^{2(1+\sigma)}  \sum_{L=1}^\infty
        (L+1)^2\Big(\frac{4C_\Theta}{\rho}\Big)^L  (\|\h_1[z]\|_{\xi,\sigma}^\sharp)^L
        \nonumber\\
        &\leq&480 C_\Theta^3\rho^{1+\sigma}B
        \nonumber\\
        &\leq&1920 C_\Theta^3\rho \|\h_2\|_{\xi,\sigma} 
\eeqn
($\h_2$ is defined in (~\ref{h-Hspace-k-def-1})).

For $M+N=1$, we use the soft photon sum rules \sbsr, which state that for any
arbitrary unit vector $n\subset\R^3$, and some $\mu\in\R_+$,
\eqn
        \lim_{\rvar\rightarrow0}\rvar^{-\sigma} w_{1,0}[z;\spvar;\rvar n, \lambda]
        =g\mu\bra\e(n,\lambda),\partial_{X}\ket_{\R^3} T[z;\spvar] \;,
\eeqn
and likewise for $w_{0,1}$.
If $p=0$, $T\in\Tspace$ is $O(3)$-invariant,
and is a function only of $X_0$ and $X^2$. Therefore,
\eqn
        \partial_{X}\Big|_{X=0}  T[z;\spvar;p=0]=0\;,
\eeqn
so that
\eqn
        \lim_{X\rightarrow0}\lim_{\rvar\rightarrow0}
        \rvar^{-\sigma} w_{1,0}[z;\spvar;\rvar n, \lambda] =0\;,
\eeqn
and hence
\eqn
        \lim_{X\rightarrow0}\lim_{\rvar\rightarrow0}
        w_{1,0}[z;\spvar;\rvar n, \lambda] =0 \;,
        \label{w-deg1-zero-1}
\eeqn
and likewise for $w_{0,1}$.
This is the crucial step where the soft photon sum rules are used to prove 
irrelevance of the generalized Wick kernels of degree 1. 
The consequence of (~\ref{w-deg1-zero-1}) is that  $w_{0,1}$ and $w_{1,0}$ 
(which superficially scale marginally, that is, with a factor $\rho^\sigma\rightarrow1$
as $\sigma\rightarrow0$) have a Taylor expansion 
in $|k|$ and $\spvar$ with zero constant coefficient, and thus scale with
a factor $\rho$. 

The leading term in $\widehat w_{1,0}$ corresponding to $L=1$ (where $\underline p$ and
$\underline q$ are necessarily zero) is given by
\eqn
		V^{(L=1)}_{\underline{1,0,0,0}}[\h|\spvar;K]=\Bra\vac\,,\,
		F_0[\spvar+\uk]\tilde W_{1}[z;\rho\spvar;\rho K]F_1[\spvar]\vac\Ket\;,
\eeqn 
which can be estimated by 
\eqn
		\|V^{(L=1)}_{\underline{1,0,0,0}}[z]\|_{\sigma}^\sharp
		&\leq& 2\|w_{1,0}[z;\rho \spvar;\rho K]\|_{\sigma}^\sharp \|F_0\|^\sharp
		\|F_1 \|^\sharp  \;.
\eeqn
This is obtained from applying the definition of the norms $\|\,\cdot\,\|_{\sigma}^\sharp$ in
(~\ref{norm-w-sig-sharp-def-1}) and $\|\,\cdot\,\|^\sharp$ in (~\ref{Tnormsharpdef}),
and using the Leibnitz rule repeatedly. 
An additional derivative with respect to $|k|$ in $F_0$ is accounted for by the factor 2,
where we have used
\eqn
		\||k|^\sigma\partial_{|k|}(|k|^\sigma F_0 w_{1,0} F_1)\|_\sigma
		&=&\||k|^\sigma\partial_{|k|}( F_0 (|k|^\sigma w_{1,0}) F_1)\|_\sigma
		\nonumber\\
		&\leq&(\sup_{|X|\leq X_0<1}\|\partial_{|k|}F_0[\spvar+\uk]\|_{op}) \|w_{1,0}\|_\sigma^\sharp \|F_1\|^\sharp
		\nonumber\\
		&&\hspace{1cm}+\|F_0\|^\sharp \||k|^\sigma\partial_{|k|}
		(|k|^\sigma w_{1,0})\|_\sigma^\sharp\|F_1\|^\sharp \;,
\eeqn
since 
\eqn
		\sup_{|X|\leq X_0<1}\|\partial_{|k|}F_0[\spvar+\uk]\|_{op}\leq 
		\sum_{|\aind|=1}\sup_{|X|\leq X_0<1}\|\partial_{\spvar}^\aind F_0\|_{op}<\|F_0\|^\sharp\;.
\eeqn
Use of (~\ref {w-deg1-zero-1}) implies  
\eqn
        \| w_{1,0}[z;\rho \spvar;\rho K] \|_\sigma &\leq&
        \rho\sum_{|\aind|=1\atop a_0=0}\|\partial_{\spvar}^{\aind} w_{1,0}\|_\sigma
        +\rho\||k|^\sigma\partial_{|k|}\big(|k|^{-\sigma} w_{1,0}\big)\|_\sigma
        \nonumber\\
        &\leq&\rho\|w_{1,0}\|_\sigma^\sharp
        \;,
\eeqn
so that
\eqn
        \| w_{1,0}[z;\rho \spvar;\rho K]\|_\sigma^\sharp &=&
		\sum_{a_0=0,1}
        \|\partial_{X_0}^{a_0} w_{1,0}[z;\rho \spvar;\rho K]\|_\sigma
        \nonumber\\
        &&\hspace{1cm}+\sum_{1\leq|\aind|\leq2\atop a_0=0}
        \|\partial_{\spvar}^{\aind}  w_{1,0}[z;\rho \spvar;\rho K]\|_\sigma
        \nonumber\\
        &&\hspace{1cm}+\||k|^\sigma\partial_{|k|}\big(|k|^{-\sigma} w_{1,0}[z;\rho \spvar;\rho K]\big)\|_\sigma
        \nonumber\\
        &\leq&
        \rho\|\partial_{X_0} w_{1,0}\|_\sigma
        +2\rho\sum_{1\leq|\aind|\leq2\atop a_0=0}
        \|\partial_{\spvar}^{\aind}   w_{1,0}\|_\sigma
        \nonumber\\
        &&\hspace{1cm}+2\rho\||k|^\sigma\partial_{|k|}\big(|k|^{-\sigma} w_{1,0}\big)\|_\sigma 
		\nonumber\\
		&\leq& 2\rho \| w_{1,0}\|_\sigma^\sharp 
        \;.
\eeqn
We note that since the norm $\|\,\cdot\,\|_{\xi,\sigma}$ does not
involve any derivative with respect to $|p|$ if $p=0$, all derivatives
scale with a factor $\rho$.
Moreover, 
\eqn
		\|F_0 \|^\sharp\,,\,\|F_1 \|^\sharp&\leq& C_\Theta  \;.
\eeqn
Consequently,
\eqn
		\|V^{(L=1)}_{\underline{1,0,0,0}}[z]\|_{\sigma}^\sharp
		\leq 10\rho C_\Theta^2 \| w_{1,0}\|_\sigma^\sharp  \;.
\eeqn
The case for $\widehat w_{0,1}$ is identical.

The sum of terms contributing to $\widehat w_{1,0}$  for $L\geq2$ can be bounded by
\eqn
	    \lefteqn{
	    20\,C_\Theta^2 \sum_{L=2}^\infty    (L+1)^2
        \Big(\frac{C_\Theta}{\rho}\Big)^L(2\rho^{1+\sigma})^{M+N}
        }
        \nonumber\\
        &&\hspace{0.5cm}\times\,\sum_{m_1+\cdots+m_l=M,\atop n_1+\cdots+n_L=N}
        \sum_{p_1,q_1,\dots,p_L,q_L:\atop
        m_\ell+p_\ell+n_\ell+q_\ell\geq1}
        \nonumber\\
        &&\hspace{0.5cm}\times\,
        \prod_{\ell=1}^L\Big\{\Big(\frac{2}{\sqrt{p_\ell}}\Big)^{p_\ell}
        \Big(\frac{2}{\sqrt{q_\ell}}\Big)^{q_\ell}
        \|w_{m_\ell+p_\ell,n_\ell+q_\ell}[\z]\|_\sigma 
        \Big\}
        \nonumber\\
        &\leq&20\,C_\Theta^2\rho^{1+\sigma}\xi\sum_{L=2}^\infty    (L+1)^2 B^L
        \nonumber\\
        &\leq&3840 \frac{C_\Theta^4}{\rho} \, \xi\,\|\h_1\|_{\xi,\sigma}^2 \;,
        \label{hat-w-pnull-bd-1}
\eeqn 
by use of similar arguments as in the derivation of
(~\ref{hatwMNbd-2}), and $\sum_{L=2}^\infty    (L+1)^2 B^L<12 B^2$ for 
$B=\frac{4C_\Theta}{\rho}\|\h_1\|_{\xi,\sigma}<\frac{1}{10}$.

In conclusion,
\eqn
        \|\widehat \h_1\|_{\xi,\sigma}^\sharp&\leq& \xi^{-1} \|\widehat w_{1,0}\|_\sigma^\sharp
        +\xi^{-1}\|\widehat w_{0,1}\|_\sigma^\sharp
        +\|\widehat w_2\|_{\sigma,\xi}
        \nonumber\\
        &\leq&
        (20 C_\Theta^2 +7680 \frac{C_\Theta^4\e}{\rho^2} + 1920 C_\Theta^3 )  \rho \,
        \|\h_1\|_{\xi,\sigma}^\sharp
		\nonumber\\
        &<&9620 C_\Theta^4 \rho\,\e\;,
        \label{hatwMNest}
\eeqn
independently of $\sigma$, for $\e<\e_0<\rho^2$.

In the case $M+N=0$, we again choose $K_\Theta:=\tilde K_\Theta$, see (~\ref{Tnormdef}),
as in (~\ref{DiffThatTest}), and
we find
\eqn
        \|\widehat T-T_0^{(p=0;\rho\lTnl)}\|_{\Tspace}&\leq&
        \delta+ 120\,C_\Theta \Big(\frac{C_\Theta\xi}{\rho}
        \|\h_1\|_{\xi,\sigma}^\sharp\Big)^2
        \nonumber\\
        &\leq&\delta+120\frac{C_\Theta^3 \xi^2}{\rho^2}\e^2
		\nonumber\\ 
        &<&\delta+\e 
        \label{T00p0est}
\eeqn
using  $\e<\e_0< \rho^2$,
and $\xi= (120 C_\Theta^3)^{-\frac12}=\coe^{\frac12}$, all independent of $\sigma$.

Therefore,
\eqn
		\ren:\Polyd^{(\mu)}\Polpar\rightarrow \Polyd^{(\rho^\sigma\mu)}(\widehat\e,\widehat\delta,\widehat\lTnl)
\eeqn
with
\eqn
		\widehat\e&<&\ccontr\rho\,\e\;\;{\rm where}\;\;\ccontr:=9620 C_\Theta^4
		\nonumber\\
		\widehat\delta&<&\delta+\e 
		\nonumber\\
		\widehat\lTnl&=&\rho\lTnl\;,
		\label{ccontr-def-1}
\eeqn
independently of $\sigma$.
Thus, $\ren$ is codimension 2 contractive with $\widehat\e<\frac\e2$ for $\rho<\frac{1}{2\ccontr}$.
\\

This concludes the proof of Theorem {~\ref{codim2contrthm}}.


\section{The first Feshbach decimation step}
\label{ground-state-sect-1} 
 
One of our key aims is to find the ground state eigenvalue and
eigenvector of the fiber Hamiltonian $H(p,\sigma)$ on $\Fo$,
using the isospectral, operator-theoretic renormalization group constructed
in our previous analysis.  
$H(p,\sigma)$ has been defined in (~\ref{Hpsdef}),
and is written in Wick ordered normal form in (~\ref{Hpsdef2}) below.

We use the following strategy.

In this section, we construct an initial condition for the 
renormalization group recursion
by isospectrally mapping the fiber Hamiltonian $H(p,\sigma)$
on $\Fo$ to
an effective Hamiltonian on $\H_{red}=\1[H_f<1]\Fo$. The latter is
parametrized by an element $\h^{(0)}$ inside a polydisc $\Polyd(\e_0,\delta_0,\frac12)$
(defined in  (~\ref{Polyddef})).
We refer to this procedure as the "first Feshbach decimation step".
For sufficiently small values of the electron charge $g$,
the constants $\e_0,\delta_0$ are sufficiently small that 
$\ren$ is codimension 2 contractive on $\Polyd(\e_0,\delta_0,\frac12)$
according to Theorem {~\ref{codim2contrthm}}. 

Thus, iteration of $\ren$ generates a convergent
sequence $\{\h^{(n)}\}_{n\in\N_0}$ in $\Hspace_{\geq0}$. 
In Section {~\ref{eigenvalvecsubsec}}, we show that in the limit $n\rightarrow\infty$, 
the spectral problem of
finding the ground state for $H[\h^{(n)}]$ can be explicitly solved. 
By a recursive use of the Feshbach isospectrality
theorem, Theorem {~\ref{Feshprop}}, this result allows us to 
reconstruct the ground state of $H(p,\sigma)$.

\subsection{The result}
\label{firstdecsubsect}

In this section, we prove the following theorem, which provides an initial
condition $\h^{(0)}$ for the isospectral renormalization group.

We recall here that $\Fpairs(\Fo,\chi_1[H_f])$ denotes the set of Feshbach pairs corresponding 
to $\chi_1[H_f]$, cf. Definition {~\ref{Feshbtripledef}}, and
$\chi_1$ is the smooth cutoff function defined in ({~\ref{Thetadef}}). 
Moreover,   we recall the definition of the injective embedding  
$H:\Hspace_{\geq0}\rightarrow\cB(\H_{red})$ (see (~\ref{H-Hsp-cB-map-def-1})) 
from Section {~\ref{Banspsubsect}},
where $(\Hspace_{\geq0},\|\,\cdot\,\|_{\sigma,\xi})$ 
is a Banach space of generalized Wick kernels.

\begin{theorem}
\label{firstdecthm}
Assume that $0\leq |p|<\frac13$, $\sigma\geq0$, and 
$\zeta\in D_{\frac{1}{9}}$. 
Then, for sufficiently small values of the electron charge $g$,
\eqn
        (H(p,\sigma)-\frac{|p|^2}{2}-\frac{g^2}{2}\bra\vac,A_{\kappa_\sigma}^2\vac\ket+\zeta\,,\,H_f)\,
        \in\,\Fpairs(\Fo,\chi_1[H_f]) \;.
\eeqn
For sufficiently small values of the electron charge $g$,
there exist small constants $\e_0,\delta_0,\xi>0$ independent of $\sigma$, 
and 
\eqn
        \h^{(0)}=(E^{(0)},T^{(0)},\h_1^{(0)})\in 
        \Polyd(\e_0,\delta_0,2^{-1}) \subset \Hspace_{\geq0}
\eeqn
such that
\eqn
        H[\h^{(0)}[z]] = 
        F_{\chi_1[H_f]}(H(p,\sigma)-\frac{|p|^2}{2}
        -\frac{g^2}{2}\bra\vac,A_{\kappa_\sigma}^2\vac\ket+\zeta\,,\,H_f) \;.
\eeqn
The spectral parameters $z$ and $\zeta$ are related by
\eqn
        z =\zeta-E^{(0)}[\zeta]=:J_{(-1)}[\zeta]\;,
\eeqn
and $z\in\I$. The parameters $\e_0,\delta_0$ are $O(g)$.
\end{theorem}

\prf
We closely follow the proof of Theorem {~\ref{codim2contrthm}}. 
However, while in Theorem  {~\ref{codim2contrthm}}, we were concerned with
bounded operators on $\H_{red}$, we are now studying unbounded
operators on $\Fo$. Thus, instead of operator bounds, we use relative
operator bounds of the interaction operator with respect to the free Hamiltonian.
Otherwise, the arguments are identical; for the algebraic part of the proof
related to Wick ordering, essentially no modification is required.

We recall the notation
$$
		\opp=(H_f,P_f) \;,
$$
and 
$$
		\spvar=(X_0,X)\;\in\R_+\times\R^3\;,
$$
which denotes the spectral variable corresponding to $\opp$.

\subsubsection{Generalized Wick normal form of $H(p,\sigma)$}

We rewrite the fiber Hamiltonian $H(\vp,\sigma)$, defined
in  (~\ref{Hpsdef}), in Wick ordered form
\eqn
        H(p,\sigma) &=& \frac{p^2}{2}+\frac{g^2}{2}
        \bra\vac, A_{\ks}^2 \vac\ket +
        T[\opp] + W \; ,
        \label{Hpsdef2}
\eeqn
where
\eqn
        T[\opp]&=& H_f -|p|\Ppar+ \frac{1}{2} P_f^2
\eeqn
is the free Hamiltonian. Since $0\leq |p|<\frac13$, and $|P_f|\leq H_f$,  
\eqn
        \frac{H_f}{2}<T[\opp]< 2 H_f 
\eeqn
is immediately clear.

The interaction Hamiltonian is given by
\eqn
        W &=& \sum_{1\leq M+N\leq 2} W_{M,N} \;,
        \label{Tmin1-def-1}
\eeqn
where
\eqn 
	\sum_{M+N=1} W_{M,N}&=&  -g\bra (p-P_f),A_{\ks}\ket_{\R^3}
        \\
        \sum_{M+N=2}
        W_{M,N}&=&\frac{1}{2}g^2\,:\,A_{\ks}^2\,: 
\eeqn
with $M,N\in\{0,1,2\}$. Here, $:\,(\cdot)\,:$ denotes Wick ordering.

The generalized Wick kernels $w_{M,N}$ corresponding to the generalized Wick monomials
$W_{M,N}$ (for definitions, see Section {~\ref{Banspsubsect}}) are given by
\eqn
		w_{1,0}[\spvar;K]&=& -g\bra (p-X),\e(K)\ket_{\R^3}
        	\ks(|k|)\;=\;
		w_{0,1}^*[\spvar;K]  
        \label{appendwMNdef1}
\eeqn
for $M+N=1$.
Moreover,
\eqn
		w_{1,1}[\spvar;K,\tilde K] &=&g^2
        \bra\e(K),\e(\tilde K)\ket_{\R^3}
        \ks(|k|)\ks(|\tilde k|)
        \label{appendwMNdef3} 
\eeqn
and
\eqn
		\;\;\;\;\;
		w_{2,0}[\spvar;K,\tilde K]&=&
 		\frac{1}{2}g^2
        \bra\e(K),\e(\tilde K)\ket_{\R^3}\ks (|k|) 
        \ks(|\tilde k|)
        \;=\;
        w_{0,2}^*[\spvar;K,\tilde K]  
        \label{appendwMNdef2} 
\eeqn 
for $M+N=2$.
The smooth cutoff function  
$\ks$, which implements the ultraviolet and infrared regularization,
has been defined in (~\ref{kappaassump}).

\subsubsection{Soft photon sum rules for $H(p,\sigma)$}

The generalized Wick kernels $w_{M,N}$ together with $T$ satisfy the soft photon
sum rules ${\rm\bf SR}[\mu=1]$ in (~\ref{sbsr}). Indeed,
\eqn
        g \bra\e(n,\lambda),\partial_{X}\ket_{\R^3}  
        T[\spvar]  
        &=&g  \bra\e(n,\lambda),\partial_{X}\ket_{\R^3} \,
        \Big[X_0-|p|X^\parallel+\frac{X^2}{2}\Big]
        \nonumber\\
        &=&-g  \bra(p-X),\e(n,\lambda) \ket_{\R^3} \,
        \nonumber\\
        &=&\lim_{\rvar\rightarrow0}\rvar^{-\sigma} w_{1,0}
        [\spvar;K]\Big|_{k =\rvar n }
        \\
        &=&\lim_{\rvar\rightarrow0}\rvar^{-\sigma} w_{0,1}
        [\spvar;K ]\Big|_{k=\rvar n}\nonumber \;,
\eeqn 
since 
\eqn
		\lim_{\rvar\rightarrow0}\rvar^{-\sigma}\ks(\rvar)=1\;,
\eeqn
see (~\ref{kappasig-norm-def-1}). We note that the value of $\mu$ depends
on this normalization condition.
Moreover,
\eqn
        g  \bra\e(n,\lambda),\partial_{X}\ket_{\R^3} \,
        w_{1,0}[\spvar;\tilde K ] 
        &=&-g^2  \Big[ \bra\e(n,\lambda),\partial_{X}\ket_{\R^3} \,
        \bra(p-X),\e(\tilde K) \ket_{\R^3} \Big] \ks(|\tilde k|)
        \nonumber\\
        &=&g^2   \bra\e(n,\lambda)\, ,\,\e(\tilde K) \ket_{\R^3} \ks(|\tilde k|)
        \nonumber\\
        &=& 2\lim_{\rvar\rightarrow0}\rvar^{-\sigma} w_{2,0}
        [\spvar;K,\tilde K]\Big|_{  K  =(\rvar n,\lambda)}
        \\
        &=& 2\lim_{\rvar\rightarrow0}\rvar^{-\sigma} w_{1,1}
        [\spvar;K,\tilde K]\Big|_{ K =(\rvar n,\lambda)}\;.
        \nonumber
\eeqn
The case for $w_{0,1}$ is completely analogous. This establishes
that the generalized Wick kernels of $H(p,\sigma)$ satsify the soft
photon sum rules ${\rm\bf SR}[1]$.

\subsubsection{Basic estimates}

From (~\ref{appendwMNdef1}) $\sim$ (~\ref{appendwMNdef2}), 
one can straightforwardly read off the following estimates.
For $M+N=1$, one obtains
\eqn
    \big|w_{M,N}[\spvar;K]\big| &\leq&
    	g (|p| + |X|) \ks(|k|)  
	\label{W1physhamest1-1}
\eeqn
and
\eqn
        \big|\partial_{X_0}  w_{M,N}[\spvar;K ]\big|
        \, , \,
        \big|\partial_{X}^2 w_{M,N}[\spvar;K ]\big|  &=& 0 \;
        \nonumber\\
		\big|\partial_{|p|} w_{M,N}[\spvar;K ]\big|&\leq& c g \ks(|k|) 
        \nonumber\\
        \big|\partial_{X} w_{M,N}[\spvar;K ]\big|  &\leq& c g \ks(|k|) 
        \nonumber\\
        \big||k|^\sigma\partial_{|k|}\big(|k|^{-\sigma}w_{M,N}[\spvar;K]\big)\big|
        &\leq& cg \;.
        \label{W1physhamest1-2}
\eeqn
For $M+N=2$, one gets
\eqn
        \big|w_{M,N}[\spvar;K,\tilde{K} ] \big| &\leq& g^2 \ks(|k|)
        \ks(|\tilde k|)
        \label{W2physhamest1-1}
\eeqn
and
\eqn
        \big|\partial_{|p|} w_{M,N}[\spvar;K]\big|&=&0
        \nonumber\\
        \big|\partial_{\spvar} w_{M,N}[\spvar;K,\tilde{K} ] \big| \,,\,
        \big|\partial_{X}^2 w_{M,N}[\spvar;K ,\tilde{K} ] \big| &=& 0 
        \nonumber\\
        \big||k|^\sigma\partial_{|k|}\big(|k|^{-\sigma}w_{M,N}[\spvar;K,\tilde{K}]\big)\big|
        &\leq&cg       
        \label{W2physhamest1-2}\\
        \big||\tilde k|^\sigma\partial_{|\tilde k|}\big(|\tilde k|^{-\sigma}
        w_{M,N}[\spvar;K,\tilde{K}]\big)\big|&\leq& cg \;.\nonumber
\eeqn
We shall divide the rest of the proof into similar segments as in
Section {~\ref{Codimcontrsubsect}}.

\subsection{Proof of Theorem {~\ref{firstdecthm}}: Domain of $F_{\chi_1[H_f]}$}

We shall first verify that for $g$ and $|\zeta|$ small, $(T[\opp]+W+\zeta,H_f)$ lies in the
domain of the smooth Feshbach map.

\begin{lemma} (Domain of the smooth Feshbach map)
\label{dom-Fesh-lemma-0-1}
For sufficiently small electron charge $g$ and $\zeta\in D_{\frac{1}{9}}$,
\eqn
        (T[\opp]+W+\zeta\,,\, H_f) 
        \in \Fpairs(\Fo,\chi_1[H_f]) \;
\eeqn
is a Feshbach pair corresponding to $\chi_1[H_f]$.
\end{lemma}

\prf
The task is to verify properties
(~\ref{Feshboundopasscond}) that define Feshbach pairs.

To this end, let $P_1$ denote the orthoprojector onto
$\Ran(\chi_1[H_f])=\H_{red}=\1[H_f<1]\Fo$, and let $\bar P_1$ be its complementary
projection. According to the definition given in (~\ref{Thetadef}),  
\eqn
        \bar P_1=\1[H_f>\frac34]\;.
\eeqn
We introduce the free resolvent
\eqn
        \bar R_0 :=\Big[H_f+\bar\chi_1[H_f]\big(T'[\opp]+\zeta  \big)\bar\chi_1[H_f]\Big]^{-1}
		\label{barR0-init-def-1}
\eeqn
on $\Ran(\bar P_1)$, where $T'[\opp]:=T[\opp]-H_f=-|p|\Ppar+\frac{P_f^2}{2}$. 
$\bar R_0$ is well-defined since
$H_f$ is invertible on Ran$(\bar P_1)$, and
\eqn
        |\bar{R}_0|\leq c\bar P_1 \big[ H_f+P_f^2\big]^{-1} \bar P_1 \;.
        \label{barR0bound-1}
\eeqn
We shall first verify that 
\eqn
        \big\||\bar{R}_0|^{\frac{1}{2}} \bch_1[H_f]W \bch_1[H_f]
        |\bar{R}_0|^{\frac{1}{2}} \big\|_{op}  &\leq& cg 
        \label{R0-min1-est-1}
\eeqn
and
\eqn
	   	\big\||\bar{R}_0|^{\frac{1}{2}}\bch_1[H_f]W 
        \chi_1[H_f] \big\|_{op}&\leq& c g  \;.
        \label{R0-min1-est-2}
\eeqn
To this end, we estimate the contributions from $W=\sum_{M+N=1,2}W_{M,N}$ separately.

For $M+N=1$, the Schwarz inequality yields
\eqn
        \big\||\bar R_0|^{\frac12}\bar\chi_1[H_f]W_{0,1}\bar\chi_1[H_f]
        |\bar R_0|^{\frac12}\phi\big\| 
        &\leq& g\big\|
        |\bar R_0|^{\frac12} (|p|+|P_f|)\big\|_{op} 
        \big\|A_{\ks}^-\bar\chi_1[H_f]|\bar R_0|^{\frac12}\phi\big\|
        \nonumber\\
        &\leq& cg\int dK |k|^{-1}\ks(|k|)\big\|a(K)\phi'\big\|
        \nonumber\\
        &\leq& cg\Big[\int d^3k |k|^{-2}\ks^2(|k|)\Big]^{\frac12}
        \|H_f^{\frac12}\phi'\|
        \nonumber\\
        &\leq&cg\|H_f^{\frac12} |\bar R_0| \phi\|
        \nonumber\\
        &\leq& cg\|\phi\| \;,
        \label{appendest3}
\eeqn
for arbitrary $\phi\in\Fo$ and $\phi':=|\bar R_0|^{\frac12}\phi$.
Here, $A_{\ks}^-$ denotes the term in the quantized electromagnetic
vector potential $A_{\ks}$ that contains annihilation operators.
The case for $W_{1,0}$ is analogous.

Moreover, 
\eqn
    		&&\big\||\bar{R}_0|^{\frac{1}{2}}
    		\bch_1[H_f]W_{0,1} \chi_1[H_f]
    		\phi\big\|
		\; \leq \; c g\int dK |\vk|^{-\frac{1}{2}}\ks(|k|)
    		\big\| a(K)  \chi_1[H_f] \phi  
    		\nonumber\\
    		&&\hspace{2cm}\times \; \big\| (|p|+|P_f|)\Big[H_f-|\zeta|-|p|\Ppar+
    		\frac{P_f^2}{2} \Big]\Big|^{-\frac{1}{2}}_{\Ran\bar\chi_1[H_f]}\bar\chi_1[H_f] \big\|_{op}
    		\nonumber\\ 
    		&&\hspace{2cm} \leq \; c g
    		\Big[\int_{|\vk| <1} d^3 \vk |\vk|^{-2} \ks^2(|k|)
    		\Big]^{\frac{1}{2}}\big\|H_f^{\frac12}\chi_1[H_f]\phi\big\|
    		\label{appendest4} 
    		\nonumber\\
		&&\hspace{2cm} \leq \; cg\|\phi\|  
\eeqn
for arbitrary $\phi\in\Fo$.
The case for $W_{1,0}$ is analogous.

For $M+N=2$, one obtains 
\eqn
        \big\||\bar{R}_0|^{\frac{1}{2}} \bch_1[H_f]W_{M,N} \bch_1[H_f]
        |\bar{R}_0|^{\frac{1}{2}} \big\|_{op} &\leq& c g^{2} 
        \label{appendest1}
\eeqn
and
\eqn
        \big\||\bar{R}_0|^{\frac{1}{2}}\bch_1[H_f]W_{M,N}
        \chi_1[H_f] \big\|_{op}&\leq& c g^{2} \;.
        \label{appendest2}
\eeqn
The proof can be straightforwardly adapted from \cite{bfs1}.

This establishes (~\ref{R0-min1-est-1}) and (~\ref{R0-min1-est-2}).

Let
\eqn
        \bar R :=\Big[H_f+\bar\chi_1[H_f]\big(T'[\opp]+W+\zeta \big)\bar\chi_1[H_f]\Big]^{-1}
\eeqn
on $\Ran(\bar P_1)$. 
Applying a resolvent expansion in powers of $W$, and using the estimates
(~\ref{R0-min1-est-1}), (~\ref{R0-min1-est-2}), 
one finds for $g$ sufficiently small
\eqn
        \|\bar R\|_{op}&<&\sum_{L=0}^\infty\big\||\bar R_0|^{\frac12}\big\|_{op}
	\big\||\bar R_0|^{\frac12}W|\bar R_0|^{\frac12}\big\|_{op}^L
	\big\||\bar R_0|^{\frac12}\big\|_{op}
	\nonumber\\
	&<&c\sum_{L=0}^\infty(cg)^L\;<\; c 
	\label{Fpair-ver-init-1}
\eeqn
and similarly,
\eqn
        &&\big\||\bar R|^{\frac12}\bar\chi_1[H_f](T'[\opp]+W+\zeta)\chi_1[H_f]\big\|_{op}^2
        \nonumber\\
        &&\hspace{1cm}=\;\big\|\chi_1[H_f](T'[\opp]+W+\zeta)\bar\chi_1[H_f] 
        \nonumber\\
        &&\hspace{4cm}
        \times\,\bar R \bar\chi_1[H_f](T'[\opp]+W+\zeta)\chi_1[H_f]\big\|_{op}
        \nonumber\\
        &&\hspace{1cm}\leq\;  
        \big\||\bar R_0|^{\frac12}\bar\chi_1[H_f](T'[\opp]+W+\zeta)\chi_1[H_f]\big\|_{op}^2
        \sum_{L=0}^\infty
        \big\||\bar R_0|^{\frac12}W|\bar R_0|^{\frac12}\big\|_{op}^L
        \nonumber\\
        &&\hspace{1cm}<\;c \;.
	\label{Fpair-ver-init-2}
\eeqn
We note that $(T'+\zeta)\chi_1[H_f]=\chi_1[H_f](T'+\zeta)$ is bounded.

By (~\ref{Feshboundopasscond}), this establishes the assertion of the lemma.
\endprf

\subsection{Proof of Theorem {~\ref{firstdecthm}}: Generalized Wick ordering}

Application of the smooth Feshbach map yields
\eqn
        F_{\chi_1[H_f]}(T[\opp]+W+\zeta\,,\,H_f)  \;,
\eeqn
which, by Lemma {~\ref{dom-Fesh-lemma-0-1}}, 
restricts to a bounded operator on $\Ran(\chi_1[H_f])$.
The corresponding intertwining operators are given by
\eqn
        Q_{\chi_1[H_f]}(T[\opp]+W+\zeta\,,\,H_f)&\in&\cB(\Ran(\chi_1[H_f])\rightarrow\Fo)
        \nonumber\\
        Q^\sharp_{\chi_1[H_f]}(T[\opp]+W+\zeta\,,\,H_f)
        &\in&\cB(\Fo\rightarrow\Ran(\chi_1[H_f]))  \;,
\eeqn
see (~\ref{FchiHtaudef}),
(~\ref{QchiHtaudef}), and (~\ref{QschiHtaudef}).

Our next goal is to prove that there exist $\e_0,\delta_0=O(g)$,
and $\h^{(0)}\in\Polyd(\e_0,\delta_0,\frac12)$ such that
\eqn
        H[\h^{(0)}]=F_{\chi_1[H_f]}(T[\opp]+W+\zeta\,,\,H_f) \;.
\eeqn
The first step is to prove an analogue of the statement of 
Theorems {~\ref{Wickorderthm}} and {~\ref{hatwMNformalseriesthm}}.
That is, we
determine the entries of $\h^{(0)}[z]$ as integrals of formal power series with 
respect to the generalized Wick kernels $w_{M,N}$ in (~\ref{appendwMNdef1} )
$\sim$ (~\ref{appendwMNdef2} ). The result is given in (~\ref{apphatwMNdef}) 
below.  (~\ref{apphatwMNdef})  is  
obtained by Neumann series expansion of $F_{\chi_1[H_f]}(T[\opp]+W+\zeta,H_f)$
with respect to the operator $W$, and Wick ordering.
These operations are purely algebraic, and the proofs of
Theorems {~\ref{Wickorderthm}} and {~\ref{hatwMNformalseriesthm}} 
apply to the present case with with trivial modifications, 
where $\chi_\rho[H_f]$ is replaced by $\chi_1[H_f]$.

To prove the convergence of  (~\ref{apphatwMNdef}), 
we adapt the proof of Theorem
{~\ref{codim2contrthm}}. The necessary modifications are due
to the fact that the operators $T$ and $W$ are unbounded, in contrast to
the cases studied in Section {~\ref{Codimcontrsubsect}}.

We closely follow our discussion in Section {~\ref{Codimcontrsubsect}},
and suitably modify the elements of the proof of Theorem
{~\ref{codim2contrthm}} from Section {~\ref{gen-wick-prf-subsect-1}}
$\sim$ Section {~\ref{codim2contrproofsubsubsect-1}} step by step.

We first adapt Lemma {~\ref{VLboundlm}} 
in Section {~\ref{gen-wick-prf-subsect-1}}, which governs
the Wick ordering operation, to the first Feshbach decimation step.

Recalling
\eqn 
	T'[ \spvar]=T[ \spvar]-X_0=-|p|X^\parallel
	+\frac12 X^2 
\eeqn 
and the definition of the operators
$W_{p,q}^{m,n}[\spvar;K^{(m+p,n+q)}]$ from (~\ref{Wmnpqdef}), 
and of $\tilde W_\ell$ from Theorem {~\ref{Wickorderthm}}.

For fixed $L\in\N$, we recall that
\eqn
        \underline{m,p,n,q}:=(m_1,p_1,n_1,q_1,\dots,m_L,p_L,n_L,q_L)\in\N_0^{4L} \;
\eeqn
and
\eqn
        M=m_1+\dots+m_L \; \; , \; \;
        N=n_1+\dots+n_L \;.
\eeqn
We let, for $1\leq m_\ell+p_\ell+n_\ell+q_\ell\leq2$,
\eqn 
        &&V^{(L)}_{\underline{m,p,n,q}}[\spvar;K^{(M,N)}]
        \nonumber\\
        &&\hspace{2cm}:= \; 
        \Bra\,\vac\,,\,
        F_0[\spvar]
        \Big[\prod_{\ell=1}^L
        \tilde W_\ell
        [\z;\rho(\spvar+\spvar_\ell);\rho K^{(m_\ell,n_\ell)}_\ell]
        F_\ell[\spvar]\Big]  \,\vac\,\Ket \;, 
        \label{firstdecVLdef}
\eeqn
where
\eqn
        F_0[\spvar]:=\piop_{\chi_1}[\z;\opp+ \spvar+\tilde \spvar_{0} ]
        \; \; , \; \;
        F_L[\spvar]:=\piop_{\chi_1}[\z;\opp+ \spvar+\tilde \spvar_L ]
        \label{F0FLdefn0}
\eeqn
and
\eqn
        F_\ell[\spvar]:=\frac{\bar\chi_1^2[H_f+X_0+\tilde X_{\ell,0}]}
        {H_f+ X_0+\tilde X_{\ell,0}
        +\bar\chi^2_1[H_f]\big(T'[\z; \opp+\spvar+\tilde \spvar_\ell ]
        +E[\z]\big)}
        \label{Felldefn0}
\eeqn
for $\ell=1,\dots,L-1$.

Then,
\eqn
		\widehat w_{M,N}[\zeta;\spvar;K^{(M,N)}]&=&
        \sum_{L=1}^\infty (-1)^{L-1}
        \sum_{m_1+\cdots+m_L=M \atop
        n_1+\cdots +    n_L=N}
        \sum_{p_1,q_1,\dots,p_L,q_L: \atop
        1\leq m_\ell+p_\ell+n_\ell+q_\ell\leq2}
        \nonumber\\
        &\times& \prod_{\ell=1}^L
        \Big\{
        \Big(\begin{array}{c}
        m_\ell+p_\ell\\
        p_\ell
        \end{array}\Big)
        \Big(\begin{array}{c}
        n_\ell+q_\ell\\
        q_\ell
        \end{array}\Big)\Big\}
        V^{(L)}_{\underline{m,p,n,q}}[\spvar;K^{(M,N)}] \;.
        \label{apphatwMNdef}
\eeqn
We claim that the statement of
Lemma {~\ref{VLboundlm}} also holds for the definition of
$F_\ell$, $W_{M,N}$, and
$V^{(L)}_{\underline{m,p,n,q}}[\spvar;K^{(M,N)}]$
given here, but with $\rho$ replaced by $1$. 

\begin{lemma}
\label{VLboundlm-0}
For any $L\geq1$  and $\underline{m,p,n,q}\in\N_0^{4L}$, one has
$V^{(L)}_{\underline{m,p,n,q}}\in\Wspace_{M,N}^\sharp$ with
\eqn
        \max\Big\{
        \|\partial_{X_0}V^{(L)}_{\underline{m,p,n,q}}\|_\sigma 
        &,&
        \||k|^\sigma\partial_{|k|}\big(|k|^{-\sigma}V^{(L)}_{\underline{m,p,n,q}}\big)\|_\sigma
        \Big\}
        \nonumber\\
        &\leq& 2 (L+1)  C_\Theta^{L+1}
        \prod_{l=1}^L\frac{\| w_{m_l+p_l,n_l+q_l}[\z]\|_\sigma^\sharp}
        {p_l^{p_l/2}q_l^{q_l/2}} \;,\;\;\;
\eeqn
for any $k\in k^{(M,N)}$. Furthermore,
\eqn    
        \|\partial_{\spvar}^{\aind} V^{(L)}_{\underline{m,p,n,q}}\|_\sigma
        &\leq& 10 (L+1)^2 C_\Theta^{L+2}
        \prod_{l=1}^L\frac{\| w_{m_l+p_l,n_l+q_l}[\z]\|_\sigma^\sharp}
        {p_l^{p_l/2}q_l^{q_l/2}} \;,\;\;\;
        \label{partspvaraindVL-0}
\eeqn
for $0\leq|\aind|\leq2$, $a_0=0$.
For $|p|>0$, and $|\aind|\leq1$,
\eqn 
        \|\partial_{|p|}
        \partial_{\spvar}^{\aind} V^{(L)}_{\underline{m,p,n,q}}\|_\sigma
        &\leq& 10 (L+1)^2 C_\Theta^{L+2}
        \prod_{l=1}^L\frac{\| w_{m_l+p_l,n_l+q_l}[\z]\|_\sigma^\sharp}
        {p_l^{p_l/2}q_l^{q_l/2}} \;.\;\;\;
        \label{partppartspvaraindVL-0}
\eeqn
Consequently,
\eqn
        \|V^{(L)}_{\underline{m,p,n,q}}\|_\sigma^\sharp
        &\leq& 10 (L+1)^2 C_\Theta^{L+2}
        \prod_{l=1}^L\frac{\| w_{m_l+p_l,n_l+q_l}[\z]\|_\sigma^\sharp}
        {p_l^{p_l/2}q_l^{q_l/2}} \;,\;\;\;
\eeqn
using the convention $p^p=1$ for $p=0$. The constant $C_\Theta$ only
depends on the choice of the smooth cutoff function $\chi_1$.
\end{lemma}

\prf
We follow the proof of Lemma {~\ref{VLboundlm}}
step by step, and replace arguments wherever necessary.

First of all, one can straightforwardly verify that
there is a constant $1\leq C_\Theta<\infty$ that only depends on the
choice of the smooth cutoff function $\Theta$ (used in the definition 
of $\chi_1[H_f]$, see (~\ref{Thetadef})) such that
\eqn
		|\partial_{X_0}F_\ell[\spvar]|
        &+&\sum_{0\leq|\aind|\leq2\atop a_0=0}|\partial_{\spvar}^{\aind} F_\ell[\spvar]|
        +\sum_{|\aind|\leq1}|\partial_{|p|}\partial_{\spvar}^{\aind} F_\ell[\spvar]|
        \nonumber\\
        &\leq& \frac{C_\Theta}{H_f+X_0+(P_f+X)^2}
        \1\Big[|X|\leq X_0\big|H_f+X_0\geq\frac34\Big] \;,
        \label{CThetadefn0}
\eeqn
uniformly in $\z\in\I$. The characteristic function on the right hand side
accommodates the fact that $|P_f|\leq H_f$, and
supp$(\bar \chi_1)=[\frac34,\infty)$.

Next, we bound $|V_{\underline{m,p,n,q}}^{(L)}[\spvar;K^{(M,N)}]|$.
The operator norm estimates in (~\ref{VLmpnqbounds}) cannot be applied here, because
the interaction operators $\tilde W_\ell$ are not bounded on
Ran$(\bar\chi_1[H_f])\subset\Fo$. However, with
\eqn
        |V_{\underline{m,p,n,q}}^{(L)}[\spvar;K^{(M,N)}]|
        &\leq&  \Big\|\,|F_0[\spvar_0]|^{\frac12}\Big\|_{op}\,
        \Big\|\,|F_L[\spvar_L]|^{\frac12}\Big\|_{op} 
        \nonumber\\
        & \times&
        \prod_{\ell=1}^L\Big\||F_{\ell-1}[\spvar_{\ell-1}]|^{\frac12}
        \tilde W_\ell[z; \spvar+\spvar_\ell ;
        K_\ell^{(m_\ell,n_\ell)}]|F_\ell[\spvar_\ell]|^{\frac12}\Big\|_{op}
\eeqn
we may use (~\ref{CThetadefn0}) together with the relative norm bounds
\eqnn
        \Big\||F_{\ell-1}[\spvar_{\ell-1}]|^{\frac12}
        \tilde W_\ell[z; \spvar+\spvar_\ell ;
        K_\ell^{(m_\ell,n_\ell)}]|F_\ell[\spvar_\ell]|^{\frac12}\Big\|_{op}\leq
        c g \;,
\eeqnn
for $\ell=1,\dots,L$.
The proof of the latter is a straightforward adaptation of the arguments used
to prove (~\ref{appendest1}), (~\ref{appendest2}), (~\ref{appendest3}),
and (~\ref{appendest4}).

To estimate $|\partial_{\spvar}^{\aind}
V_{\underline{m,p,n,q}}^{(L)}[\spvar;K^{(M,N)}]|$, for
$1\leq|\aind|\leq2$ with $a_0\in\{0,1\}$, we use
(~\ref{W1physhamest1-1}), (~\ref{W1physhamest1-2}),
(~\ref{W2physhamest1-1}), (~\ref{W2physhamest1-1}), and (~\ref{CThetadefn0}),
finding
\eqn
        \Big\||\partial_{\spvar}^{\aind}F_{\ell-1}[\spvar_{\ell-1}]|^{\frac12}
        \tilde W_\ell[z; \spvar+\spvar_\ell ;
        K_\ell^{(m_\ell,n_\ell)}]|F_\ell[\spvar_\ell]|^{\frac12}\Big\|_{op}&,&
        \nonumber\\
        \Big\||F_{\ell-1}[\spvar_{\ell-1}]|^{\frac12}
        \partial_{\spvar}^{\aind}\tilde W_\ell[z; \spvar+\spvar_\ell ;
        K_\ell^{(m_\ell,n_\ell)}]|F_\ell[\spvar_\ell]|^{\frac12}\Big\|_{op}&\leq&
        c g  \;,
        \nonumber
\eeqn
etc.,
as well as
\eqn
        \Big\||\partial_{\spvar}^{\aind_1}F_{\ell-1}[\spvar_{\ell-1}]|^{\frac12}
        \tilde W_\ell[z; \spvar+\spvar_\ell ;
        K_\ell^{(m_\ell,n_\ell)}]|
        \partial_{\spvar}^{\aind_2} F_\ell[\spvar_\ell]|^{\frac12}\Big\|_{op}&,&
        \nonumber\\
        \Big\||\partial_{\spvar}^{\aind_1}F_{\ell-1}[\spvar_{\ell-1}]|^{\frac12}
        \partial_{\spvar}^{\aind_2} \tilde W_\ell[z; \spvar+\spvar_\ell ;
        K_\ell^{(m_\ell,n_\ell)}]|
        F_\ell[\spvar_\ell]|^{\frac12}\Big\|_{op}&\leq&
        c g \;,
        \nonumber
\eeqn
etc., for $\aind_1+\aind_2=\aind$, and $\ell=1,\dots,L$.
 
We estimate derivatives of $V_{\underline{m,p,n,q}}^{(L)}$ by separately bounding
the terms  $V_{\underline{m,p,n,q}}^{(L,j)}$,  
$j=i,ii,\dots,viii$,  introduced in the proof of Lemma {~\ref{VLboundlm}} in
Section {~\ref{gen-wick-prf-subsect-1}}.
Only now, we use the relative norm bounds derived above, and replace
$\rho$ by $1$. Then, the statement of Lemma {~\ref{VLboundlm}} can straightforwardly
be verified to hold also for the present case.  
\endprf

\subsection{Proof of Theorem {~\ref{firstdecthm}}: Mapping into a polydisc}

\begin{lemma}
For $g$ sufficiently small 
and under the assumptions of Theorem {~\ref{firstdecthm}}, there are 
small, positive constants $\e_0,\delta_0 = O(g)$, and $\xi$, such that
\eqn
        H[\h^{(0)}]=F_{\chi_1[H_f]}(T[\opp]+W+\zeta\,,\,H_f)
\eeqn
for an element $\h^{(0)}$ contained in the polydisc $\Polyd(\e_0,\delta_0,\frac12)$,
which is endowed with the norm $\|\,\cdot\,\|_{\xi,\sigma}$.
\end{lemma}

\prf
We may here adapt the segment of the proof of Theorem {~\ref{codim2contrthm}} 
presented in Section {~\ref{codim2contrproofsubsubsect-1}} line by line.
The only difference here again is that $\rho$ is replaced by 1 in all estimates.

We note that a detailed account on this part of the discussion 
for a similar model can be found in \cite{bfs1}.
\endprf

\subsection{Proof of Theorem {~\ref{firstdecthm}}: Soft photon sum rules}

\begin{lemma}
\label{sbsr-sF-init-lemma-1}
Under the assumptions of Theorem {~\ref{firstdecthm}}, 
$\h^{(0)}\in\Polyd(\e_0,\delta_0,\frac12)$ satisfies the
soft photon sum rules ${\rm{\bf SR}}[1]$.
\end{lemma}

\prf
Again, we can straightforwardly adapt the arguments presented
in Section {~\ref{spsrproofsssect}} line by line to the present case.
The algebraic structure of $V_{\underline{m,p,n,q}}^{(L)}$
is equal to that of the corresponding terms discussed there,
while $\rho$ is replaced by $1$.
One then concludes that $\h^{(0)}$ satisfies the soft photon sum rules
{\bf SR}$[1]$,  for the value $\mu=1$ in (~\ref{sbsr}) (the value of $\mu$
depends on the 
normalization condition (~\ref{kappasig-norm-def-1})).
\endprf


\section{Reconstruction of the ground state eigenvalue and eigenvector}
\label{eigenvalvecsubsec}

In this section, we prove that the infimum of the
spectrum of $H(p,\sigma)$ is a simple eigenvalue at the bottom of
essential spectrum, and to construct the corresponding ground state
eigenvector. 

The strategy is based on combining the isospectral renormalization
group from Sections {~\ref{Banspsubsect}}, {~\ref{Rentrsf}} and {~\ref{Codimcontrsubsect}} 
with recursive applications of the reconstruction part
of the Feshbach theorem, Theorem {~\ref{Feshprop}}.

\begin{theorem}
\label{gs-const-thm-1}
If $0<|p|<\frac13$, assume for $\sigma>0$ that $\delta_0=\delta_0(\sigma)$,
$\rho=\rho(\sigma)$, $\xi$, $\e_0=\e_0(\sigma)>0$
are sufficiently small such that the renormalization
map $\ren$ has the codimension 2 contractivity property
(~\ref{codim2contr}) on $\Polyd(\e_0,\delta_0+2\e_0,1/2)$, endowed with 
the norm $\|\,\cdot\,\|_{\xi,\sigma}$. 
If $p=0$, assume that independently of $\sigma\geq0$, 
$\rho$, $\xi$, $\e_0$, $\delta_0$ are sufficiently small  
such that the latter holds.

Suppose that the electron charge $g$ is sufficiently small that
$\h^{(0)}\in \Polyd(\e_0,\delta_0,\frac12)$.

Then, with $e_{(0,\infty)}\in \I$ defined in (~\ref{eninftydef1})
below,  
$H[\h^{(0)}[e_{(0,\infty)}]]$ has a simple ground state eigenvalue at $0$, 
with eigenvector 
$\Psi_{(0,\infty)}$ given in (~\ref{Psi-lim-def-1}).

Moreover, $E(p,\sigma)$, as defined in (~\ref{Epsigmaemin1def}), 
is the simple ground state eigenvalue
of $H(p,\sigma)$, and the corresponding eigenvector is given by 
$\Psi_{(-1,\infty)}$, as defined in (~\ref{Psigrdst}).
\end{theorem}

\prf
We recall from Section {~\ref{Codimcontrsubsect}} that 
$\Polyd^{(\mu)}\Polpar\subset\Hspace_{\geq0}$ denotes the subset of elements of 
$\Polyd\Polpar$ satisfying the soft photon sum rules \sbsr. By Lemma {~\ref{sbsr-sF-init-lemma-1}}
and Theorem {~\ref{codim2contrthm}}, we have 
\eqnn
        \ren:\Polyd^{(1)}(\e_0,\delta_0,2^{-1})&\rightarrow&
        \Polyd^{(\rho^\sigma)}(2^{-1}\e_0,\delta_0+\e_0,2^{-1}\rho) \;,
\eeqnn
so that
\eqnn
        \ren^n:\Polyd^{(1)}(\e_0,\delta_0,2^{-1})\rightarrow
        \Polyd^{(\rho^{n\sigma})}(\e_n,\delta_n,\lTnl_n) \;,
\eeqnn
where
\eqn
		\e_n&\leq&2^{-n}\e_0
		\nonumber\\
		\delta_n&\leq&\delta_0+[1+2^{-1}+
        \dots+2^{-n+1}]\e_0\leq \delta_0+2\e_0
        \nonumber\\
        \lTnl_n&=&\frac{\rho^n }{2} \;.
        \label{en-dn-ln-def-1}
\eeqn
It hence follows that
\eqn
        \h^{(n)}&:=&\ren^n[\h^{(0)}]\label{rgflowparambds}
        \\
        &\in&\Polyd^{(\rho^{n\sigma})}
        (\e_n,\delta_n,\lTnl_n)
        \nonumber
\eeqn
for $n\in\N_0$.

Let now
\eqn
        E_{(n)}[z]:=w_{0,0}^{(n)}[z;\underline{0}]\;,
\eeqn
and
\eqn
        {\mathcal U}_{(n)}:={\mathcal U}[\h^{(n)}]=\Big\{z\in \I\Big|
        |E_{(n)}[z]|\leq\frac{\rho}{10}\Big\}\;.
\eeqn
From Lemma {~\ref{Ezlemma}}, we recall that
\eqn
        J_{(n)} :{\mathcal U}_{(n)}\rightarrow \I
        \; \; , \; \; z\mapsto \rho^{-1}E_{(n)}[z]\;,
        \label{Jn-maps-def-1}
\eeqn
is an analytic bijection. which satisfies
\eqn
        \frac{\rho}{20}|\zeta-\zeta'|\leq|J_{(n)}^{-1}[\zeta]-J_{(n)}^{-1}[\zeta']|
        \leq\frac{3\rho}{20}|\zeta-\zeta'| \;,
        \label{Jninvbd}
\eeqn
cf. \cite{bcfs1}.

We then define, for $0\leq n\leq m$,
\eqn
        e_{(n,m)}:=J_{(n)}^{-1}\circ\cdots\circ J_{(m)}^{-1}[0]\;.
        \label{Jn-concat-1}
\eeqn
By (~\ref{Jninvbd}),
\eqn
        |e_{(n,m)}-e_{(n,m+1)}|\leq\Big(\frac{3\rho}{20}\Big)^{m-n}|e_{(m,m)}-e_{(m,m+1)}|
        \leq\Big(\frac{3\rho}{20}\Big)^{m-n}\;,
\eeqn
since $e_{m,m},e_{m,m+1}\in \I$, \cite{bcfs1}. As $\frac{3\rho}{20}<1$,
the limit
\eqn
        e_{(n,\infty)}:=\lim_{m\rightarrow\infty}e_{(n,m)}\in{\mathcal U}_{(n)}
        \label{eninftydef1}
\eeqn
exists for all $n\in\N_0$, and by construction,
\eqn
        \rho^{-1}E_{(n)}[e_{(n,\infty)}]=J_{(n)}[e_{(n,\infty)}]=e_{(n+1,\infty)}\;.
        \label{JnenEnid}
\eeqn
This, together with $|E_{(n)}[z]-z|\leq 2^{-n}\e_0$, implies that
\eqn
        |e_{(n,\infty)}-\rho e_{(n+1,\infty)}|=
        |e_{(n,\infty)}-E_{(n)}[e_{(\alpha,\infty)}]|
        \leq2^{-n}\e_0\;,
\eeqn
and consequently,
\eqn
        |e_{(n,\infty)}|\leq2^{-n+1}\e_0
        \label{eninftybds}
\eeqn
tends to zero in the limit $n\rightarrow\infty$.

Next, we let
\eqn
        e_{(-1,\infty)}&:=&J_{(-1)}^{-1}[e_{(0,\infty)}]\;, 
        \nonumber\\
        E(p,\sigma)&:=&e_{(-1,\infty)}+\frac{p^2}{2}+
        \frac{g^2}{2}\bra\vac , A_{\ks}^2 \vac\ket 
        \label{Epsigmaemin1def}
\eeqn
and
\eqn
        H_{(-1)}&:=&e_{(-1,\infty)}\1+T_{(-1)} +W_{(-1)}
        \label{H-min1-def-1}
\eeqn
on $\Fo$, where
\eqn
        T_{(-1)}&=&H_f-|p|\Ppar+P_f^2\\
        W_{(-1)}&=&g\bra (p-P_f),A_{\ks}\ket_{\R^3}+\frac{g^2}{2}:
        \,A_{\ks}^2\,: \;.
        \label{H-min1-def-2}
\eeqn
Then,
\eqn
        H_{(-1)}=H(p,\sigma)-E(p,\sigma) \;,
\eeqn
where $H(p,\sigma)$ is the fiber Hamiltonian, see (~\ref{Hpsdef}) and (~\ref{Hpsdef2}).

Moreover, we introduce the notation 
\eqn
        H_{(n)}&:=&H[\h^{(n)}[e_{(n,\infty)}]]
        \nonumber\\
        &=&T_{(n)} +e_{(n,\infty)}\chi_1^2[H_f]+W_{(n)} 
        \label{H-n-def-1}
\eeqn
for $n\geq0$,  where
\eqn
        T_{(n)}&=&w_{0,0}^{(n)}[e_{(n,\infty)};\opp]-
        w_{0,0}^{(n)}[e_{(n,\infty)};\underline{0}] \;,
        \label{H-n-def-2}
\eeqn
and
\eqn
        W_{(n)}&=&\sum_{M+N\geq1}\chi_1[H_f]W_{M,N}[w_{M,N}^{(n)}[e_{(n,\infty)}]]
        \chi_1[H_f]\;.
        \label{H-n-def-3}
\eeqn
Since by construction (~\ref{JnenEnid}) is satisfied, we have for $n\geq0$
\eqn
		H_{(n)}&=&\resc\big[F_{\chi_\rho}\big(H_{(n-1)} , H_f\big)\big] \;
		\nonumber\\
		&=&\renop^H[H_{(n-1)}]
		\nonumber\\
		&=&(\renop^H)^n[H_{(0)}]\;,
		\label{HnRen-def-eq-1}
\eeqn
and
\eqn
		H_{(0)}=F_{\chi_1[H_1]}(H(p,\sigma)-E(p,\sigma),H_f)\;.
		\label{HnRen-def-eq-2}
\eeqn
$H_{(0)}$ has been constructed in the first decimation step.

We shall now demonstrate that for $\sigma>0$ and $0\leq|p|<\frac13$,
\eqn
        E(p,\sigma)&=& \inf{\rm spec}\{H(p,\sigma)\} \;. 
\eeqn
Furthermore, we shall prove that $E(p,\sigma)$ is a non-degenerate eigenvalue,
and construct the corresponding eigenvector.

Let
\eqn
        Q_{(-1)}&:=&Q_{\chi_1[H_f]}\big(H_{(-1)},
        H_f\big)\;,
        \label{Qn-it-def-1}
\eeqn
which is a map $\Ran(\chi_1[H_f])\rightarrow\Fo$,
and for $n\geq0$,
\eqn
        Q_{(n)}&:=&Q_{\chi_\rho[H_f]}\big(H_{(n)},
        H_f\big)\;,
        \label{Qn-it-def-2}
\eeqn
which is a map $\Ran(\chi_\rho)\rightarrow\H_{red}$.
Let $\Gamma_\rho$ denote the unitary dilatation operator, 
defined by $\Gamma_\rho\vac=\vac$, so that $\resc=\frac1\rho\Gamma_\rho(\,\cdot\,)\Gamma_\rho^*$, 
see (~\ref{Gammarhodef}).
Then, the first equation in (~\ref{FHQid}), Eq. (~\ref{HnRen-def-eq-1}), 
and $\resc[\chi_\rho[H_f]]=\frac1\rho\chi_1[H_f]$
together imply the key {\em intertwining property}
\eqn
        H_{(-1)}Q_{(-1)}=\chi_1[H_f]H_{(0)}\;,
		\label{HQ-intertw-id-0}
\eeqn
and
\eqn
        H_{(n-1)}Q_{(n-1)}\Gamma^*_\rho&=&
        \chi_\rho[H_f]F_{\chi_\rho[H_f]}(H_{(n-1)},H_f)\Gamma^*_\rho
        \nonumber\\
        &=&\rho\chi_\rho[H_f]\Gamma_\rho^* H_{(n)}
        \nonumber\\
        &=&
        \rho\Gamma^*_\rho\chi_1[H_f] H_{(n)}
		\label{HQ-intertw-id-n}
\eeqn
for $n\geq0$.

Next, we define vectors
\eqn
        \Psi_{(n,m)}:=Q_{(n)}\Gamma_\rho^* Q_{(n+1)}\Gamma_\rho^*
        \cdots Q_{(m-1)}\vac \;\;\in\;\H_{red} \;,
\eeqn
for $0\leq n<m$. In the case $n=-1$,
\eqn
        \Psi_{(-1,m)}=Q_{(-1)}\Psi_{(0,m)}
\eeqn
is an element not of $\H_{red}=\1[H_f<1]\Fo$, but of $\Fo$.

Noting that $\vac=\Gamma_\rho^*\chi_\rho[H_f]\vac$,
\eqn
        \Psi_{(n,m+1)}-\Psi_{(n,m)} =Q_{(n)}\Gamma^*_\rho Q_{(n+1)}\Gamma_\rho^*
        \cdots Q_{(m-1)}\Gamma_\rho^*(Q_{(m)}-\chi_\rho[H_f])\vac\;.
        \nonumber\\
        \label{Psimndiff}
\eeqn
We shall next estimate (~\ref{Psimndiff}). 
The discussion here is
more complicated than in \cite{bcfs1},
due to overlap terms of the form $T'_{(j)}[\opp]\chi_\rho[H_f]\bar\chi_\rho[H_f]$,
which can be large
(cf. Section {~\ref{overlapopsubsubsec}}), where $T_{(j)}'$ is defined in (~\ref{Tprime-def-3}).

\begin{lemma} 
\label{QnQnpl1-lemma-1}
Let $j\in\N_0$. Then,
\eqn
        \|(Q_{(j)}-\chi_\rho[H_f])
        \Gamma_\rho^* Q_{(j+1)}\|_{op}\leq c\,\frac{2^{-j}\e_0}{\rho^3} \;.
        \label{QjQjp1est}
\eeqn
\end{lemma}

\prf
Writing out (~\ref{Qn-it-def-2}),  
\eqn
        Q_{(j)}&=&\chi_\rho[H_f]-\bar\chi_\rho[H_f]\bar R_{(j)}
        \bar\chi_\rho[H_f]
        \nonumber\\
        &&\hspace{2cm}
        \Big( T'_{(j)}+e_{(j,\infty)}\chi_1^2[H_f] +W_{(j)}  \Big)
        \chi_\rho[H_f] \;,
        \label{Qjdefrecall}
\eeqn
for $j\geq0$, where
\eqn
        \bar R_{(j)}&:=&\Big[H_f+\bar\chi_\rho[H_f]\Big( T'_{(j)}+ e_{(j,\infty)}\chi_1^2[H_f]  
        + W_{(j)}\Big) 
        \bar\chi_\rho[H_f]\Big]^{-1}
\eeqn
on Ran$(\bar\chi_\rho[H_f])$,
and 
\eqn
		T'_{(j)} = T_{(j)} -H_f  \;.
		\label{Tprime-def-3}
\eeqn
We define
\eqn
        \tilde Q_{(j)}&:=&\chi_\rho[H_f]-\bar\chi_\rho[H_f]\bar R_{(j)}
        \bar\chi_\rho[H_f]  W_{(j)}
        \chi_\rho[H_f]
        \label{tildeQjdef}\\
        \bar Q_{(j)}&:=&\chi_\rho[H_f]-\bar\chi_\rho[H_f]\bar R_{(j)}
        \bar\chi_\rho[H_f] \Big(T'_{(j)}  +
        e_{(j,\infty)}\Big) 
        \chi_\rho[H_f] \;.
        \nonumber
\eeqn
We note that  
\eqn
        Q_{(m)}\vac=\tilde Q_{(m)}\vac\;,
\eeqn
since $T'_{(m)}$ and $e_{(m,\infty)}$ commute with
$\bar\chi_\rho[H_f]$, and $\bar\chi_\rho[H_f]\vac=0$.

The first term on the right hand side of
\eqn
        (Q_{(j)}-\chi_\rho[H_f])
        \Gamma_\rho^* Q_{(j+1)}&=&(\tilde Q_{(j)}-\chi_\rho[H_f])
        \Gamma_\rho^* Q_{(j+1)}
        \nonumber\\
        &+&(\bar Q_{(j)}-\chi_\rho[H_f])
        \Gamma_\rho^* Q_{(j+1)}\;
        \label{QjtildeQjkeyid}
\eeqn
is in operator norm bounded by
\eqn
        \|(\tilde Q_{(j)}-\chi_\rho[H_f])
        \Gamma_\rho^* Q_{(j+1)}\|_{op}&\leq&
        \|\tilde Q_{(j)}-\chi_\rho[H_f]\|_{op}\| Q_{(j+1)}\|_{op}
        \nonumber\\
        &\leq&\|\bar R_{(j)}\|_{op}\|W_{(j)}\|_{op}
        \|\bar R_{(j)}\|_{op}\|H_{(j+1)}-H_f\|_{op}
        \nonumber\\
        &\leq&c\,\frac{2^{-j}\e_j}{\rho^2}\;,
\eeqn
where we used the estimate
\eqn
        &&\Big\|\bar\chi_\rho[H_f]\Big(H_f+\bar\chi_\rho [H_f](T'_{(j)}[\opp]+
        e_{(j,\infty)}\chi_1^2[H_f]+ W_{(j)})\bar\chi_\rho[H_f]\Big)^{-1}\bar\chi_\rho [H_f]
        \Big\|_{op}
        \nonumber\\
        &&\hspace{4cm}\leq(c\rho-\|W_{(j)}\|_{op})^{-1} \;,
\eeqn
cf. the proof of Proposition {~\ref{Polydlemma}}, and
\eqn
        \|W_{(j)}\|_{op}\leq2^{-j}\e_0 \;,
\eeqn
for general $j\in\N_0$. Furthermore, $\|H_{(j+1)}-H_f\|_{op}<c$, where
the bound is independent of $j$.

The second term on the right hand side of (~\ref{QjtildeQjkeyid})
can be written as
\eqn
        (\bar Q_{(j)}-\chi_\rho[H_f])
        \Gamma_\rho^* Q_{(j+1)}=(I)+(II) \;,
\eeqn
where
\eqn
        (I):=(\bar Q_{(j)}-\chi_\rho[H_f])
        \Gamma_\rho^*  \tilde Q_{(j+1)}
\eeqn
and
\eqn
        (II):=(\bar Q_{(j)}-\chi_\rho[H_f])
        \Gamma_\rho^*(\bar Q_{(j+1)}-\chi_\rho[H_f]) \;.
\eeqn
We have
\eqn
        \|(I)\|_{op}&\leq&\|\bar Q_{(j)}-\chi_\rho[H_f]\|_{op}
        \|\bar R_{(j+1)}\|_{op}\|W_{(j+1)}\|_{op}
        \nonumber\\
        &\leq&
        c\,\frac{2^{-j}\e_0}{\rho^2}\;,
\eeqn
since
\eqn
        \|\bar Q_{(j)}-\chi_\rho[H_f]\|_{op}\leq
        \|\bar R_{(j)}\|_{op}\|T'_{(j)}+e_{(j,\infty)}\|_{op}
        \leq \frac{c}{\rho}\;,
        \label{barQkeyest}
\eeqn
where $\|T'_{(j)}+e_{(j,\infty)}\|_{op}<c$, independently of $j$.
Furthermore, by expanding the resolvent $\bar R_{(j)}$ once (with
$\bar R_{0,(j+1)}$ corresponding to $\bar R_{(j)}$ with $W_{(j)}$
set equal to zero),
\eqnn
        (II)&=&(\bar Q_{(j)}-\chi_\rho[H_f])
        \Gamma_\rho^*\bar\chi_\rho[H_f]\bar R_{0,(j+1)}\bar\chi_\rho[H_f]
        (T'_{(j+1)}+e_{(j+1,\infty)})\chi_\rho[H_f]
        \nonumber\\
        &-&(\bar Q_{(j)}-\chi_\rho[H_f])
        \Gamma_\rho^*\bar\chi_\rho[H_f]\bar R_{0,(j+1)}\bar\chi_\rho[H_f]
        W_{(j+1)} (\bar Q_{(j+1)}-\chi_\rho[H_f]) \;.
\eeqnn
The first product of operators on the right hand side of the equality
sign is identically zero, as one
easily sees by commuting the cutoff operator $\chi_\rho[H_f]$ to the left,
and by noting that
\eqn
        (\bar Q_{(j)}-\chi_\rho[H_f])
        \Gamma_\rho^*\chi_\rho[H_f]=
        (\bar Q_{(j)}-\chi_\rho[H_f])\chi_{\rho^2}[H_f]
        \Gamma_\rho^*
        =0\;,
\eeqn
since $\bar\chi_\rho[H_f]\chi_{\rho^2}[H_f]=0$.
Thus,
\eqn
        \|(II)\|_{op}&\leq&\|\bar Q_{(j)}-\chi_\rho[H_f]\|_{op}
        \|\bar R_{0,(j+1)}\|_{op}\|W_{(j+1)}\|_{op}
        \|\bar Q_{(j+1)}-\chi_\rho[H_f]\|_{op}
        \nonumber\\
        &\leq&c\,\frac{ 2^{-j}\e_0}{\rho^3}\;,
\eeqn
by (~\ref{barQkeyest}). This proves (~\ref{QjQjp1est}). 
\endprf

Hence, we find for
\eqn
        \Psi_{(n,m)}= Q_{(n)}\Gamma_\rho^* \ Q_{(n+1)}\Gamma_\rho^*
        \cdots Q_{m-2}\Gamma_\rho^*\tilde Q_{(m-1)}\vac \;,
        \label{Psinmdef}
\eeqn
that
\eqnn
        \|\Psi_{(n,m+1)}-\Psi_{(n,m)}\|&\leq&
        \|\tilde Q_{(m-1)}-\chi_\rho[H_f]\|_{op}
        \nonumber\\
        &\times&
        \prod_{k=0}^{\frac{m-n-2}{2}}
        \Big(1+\|( Q_{(n+2k)}-\chi_\rho[H_f])\Gamma_\rho^*
        Q_{(n+2k+1)}\|_{op}\Big)\;,
\eeqnn
if $m-n$ is even. The modification for $m-n$ odd is evident, and will
not be elaborated on separately.
Thus,
\eqn
        \|\Psi_{(n,m+1)}-\Psi_{(n,m)}\|\leq c\, 2^{-m} \frac{ \e_0}{\rho}
        \exp[c'\, 2^{-n}\e_0\rho^{-3}]
\eeqn
for constants which are independent of $\e_0,\rho,\sigma$, and $m,n$.
It thus follows that for each fixed $n\geq0$, the sequence of vectors
$\{\Psi_{(n,m)}\}_{m=0}^\infty$ in $\H_{red}$
\eqn
        \Psi_{(n,\infty)}:=\lim_{m\rightarrow\infty}\Psi_{(n,m)}
		\label{Psi-lim-def-1}
\eeqn
exists. For $n=-1$, $\{\Psi_{(-1,m)}\}_{m=0}^\infty$ is a convergent sequence
of vectors in $\Fo$. In particular,
\eqn
        \|\Psi_{(n,\infty)}-\vac\|=\|\Psi_{(n,\infty)}-\Psi_{(n,n)}\|
        \leq c\,2^{-n+1}\frac{ \e_0}{\rho}\exp[c'\,2^{-n}\e_0\rho^{-3}]
        \label{Psininftybound}
\eeqn
which implies that there is $n^*$, such that $\Psi_{(n,\infty)}$ is non-zero for
all $n>n^*$.

For every $n\geq-1$, the vector $\Psi_{(n,\infty)}$ is an element of the kernel of
$H_{(n)}$. To prove this, we use
\eqn
        H_{(n)}\Psi_{(n,m)}&=&(H_{(n)}Q_{(n)}\Gamma_\rho^*)Q_{(n+1)}\Gamma_\rho^*
        \cdots  Q_{(m-1)}\vac\nonumber\\
        &=&\rho\Gamma_\rho^*\chi_1[H_f](H_{(n+1)}Q_{(n+1)}\Gamma^*)Q_{n+2}\Gamma_\rho^*
        \cdots Q_{(m-1)}\vac\nonumber\\
        &=&\cdots\;=\rho^{m-n}(\Gamma_\rho^*\chi_1[H_f])^{m-n}H_{(m)}\vac\;.
\eeqn
Using $T_{(m)}\vac=0$,
\eqn
        \|\chi_1[H_f]H_{(m)}\vac\| &=&\|\chi_1[H_f]W_{(m)}\vac+E_{(m)}\vac\|
        \nonumber\\
        &\leq&2^{-m}\e_0+\rho|e_{(m+1,\infty)}|
        \nonumber\\
        &\leq&2^{-m+1}\e_0\;.
\eeqn
Hence,
\eqn
        \|H_{(n)}\Psi_{(n,m)}\|\leq2^{-m+1}\e_0\rightarrow0\;(m\rightarrow\infty)\;.
\eeqn
A more detailed discussion is given in \cite{bcfs1}.

By continuity of $H_{(n)}$ on $\H_{red}$,
\eqn
        H_{(n)}\Psi_{(n,\infty)}=\lim_{m\rightarrow\infty}H_{(n)}\Psi_{(n,m)}=0 \;,
        \label{HnPsininftyeq}
\eeqn
for all $n\geq0$. This implies that $H_{(0,\infty)}\Psi_{(0,\infty)}=0$.
In particular,
\eqn
        \Psi_{(-1,\infty)}=Q_{(-1)}\Psi_{(0,\infty)}
        \label{Psigrdst}
\eeqn
satisfies
\eqn
        H_{(-1,\infty)}\Psi_{(-1,\infty)}=(H(p,\sigma)-E(p,\sigma))\Psi_{(-1,\infty)}=0\;
\eeqn
on $\Fo$. This proves the theorem.
\endprf


\section{The renormalized mass at non-vanishing conserved momentum}
\label{massrensubsec}

Starting with this section, we focus on the renormalized mass of the electron,
and prove the main result of this paper. 
The treatment of the cases $|p|>0$ (in this section) and $p=0$ 
(in Sections {~\ref{peq0masssect}} and {~\ref{it-masses-sect-1}}) 
will differ substantially.

A key input is the main result of Section {~\ref{eigenvalvecsubsec}}, 
where we have determined 
\eqn 
	E(p,\sigma)={\rm infspec}\{H(p,\sigma)\}  
\eeqn 
for $0\leq p<\frac13$ and $\sigma>0$.
Moreover, we established
that $E(p,\sigma)$ is a simple eigenvalue, and have
constructed the corresponding eigenvector
\eqn 
	\Psi(p,\sigma):=\Psi_{(-1,\infty)}\in\Fo \;.
\eeqn 
We will also use many of the intermediate steps and results
presented in the proof of Theorem {~\ref{gs-const-thm-1}}.

Our discussion is structured as follows.

In this section, we study the case $p\neq 0$,
and prove bounds on  
\eqn
        m_{ren}(p,\sigma)=\frac{1}{\partial_{|p|}^2E(p,\sigma)} \;
\eeqn
for $\sigma>0$ and $0<|p|<\frac13$ which are
not uniform in $\sigma$. Uniform
bounds for $p\neq0$ are beyond the scope of the present
work, and addressed elsewhere, \cite{ch}. 

Section {~\ref{peq0masssect}} addresses the case $p=0$.
We shall use a different definition of the renormalized mass than in the case $p\neq0$,
and refer to the corresponding quantity as $m_{ren}^*(p=0,\sigma)$.
It is determined by the ratio of the coefficients of certain operators 
appearing in the effective Hamiltonians that are produced by the isospectral 
renormalization group.
Our bound on $m_{ren}^*(p=0,\sigma)$ is {\em uniform}
in $\sigma$. In Section {~\ref{it-masses-sect-1}}, we prove that for $\sigma>0$, 
both definitions of the renormalized mass agree. 
Invoking condition (~\ref{limcommhyp}), we arrive at a uniform bound on the
renormalized mass at $p=0$ in the limit $\sigma\rightarrow0$
(for the proof of condition (~\ref{limcommhyp}), cf.
\cite{ch}). 

Our constructive proof, based on the
operator-theoretic renormalization group, provides an explicit algorithm
to compute the renormalized mass to any desired precision.
This is discussed in Section {~\ref{ren-mass-comp-sect-1}}.

\subsection{The main theorem}
\label{pneq0masssect}

We shall first address the case $|p|>0$ and $\sigma>0$. 
We prove the estimates on
the derivatives of $E(p,\sigma)$ with respect to $|p|$ asserted
in Theorem {~\ref{mainthm}}, which are summarized in the following theorem.

\begin{theorem}
\label{mren-pnonzero-thm-1}
For $0<|p|<\frac13$, there exist finite constants $g_0(\sigma),c_0(\sigma)>0$ for every $\sigma>0$ 
such that for all $0\leq g<g_0(\sigma)$,  
\eqn
		\Big|E(p,\sigma)-\frac{|p|^2}{2}-\frac{g^2}{2}\bra\vac ,  A_{\ks}^2\vac\ket
		\Big|\leq c_0(\sigma)\,\frac{g^2|p|^2}{2} \;.
		\label{E-min-p2-Og2-bd-1}
\eeqn
Furthermore, 
\eqn
        \Big|\partial_{|p|}  E(p,\sigma)-|p|\Big|\leq c_0(\sigma)g^2|p|     
        \label{partpE-apriori-1}
\eeqn
and
\eqn
        \Big|\partial_{|p|}^2  E(p,\sigma) -1\Big|\leq
        c_0(\sigma) g^2     \;.
        \label{part-2-E-min-1-est-1}
\eeqn 
In particular,  
$\partial_{|p|}^2  E(p,\sigma)<1$.
\end{theorem}

\prf
To begin with, we remark that 
\eqn
		\lim_{p\rightarrow0}E(p,\sigma)&\leq& \bra\vac\,,\,H(0,\sigma)\vac\ket
		\nonumber\\
		&=&\frac{g^2}{2}\bra \vac ,A_{\ks}^2\vac\ket
		\;=\;O(g^2)\;,		
		\label{E-0-prf-1}
\eeqn
uniformly in $\sigma$.
Thus, (~\ref{E-min-p2-Og2-bd-1}) follows from (~\ref{partpE-apriori-1}).

Applying $\partial_{|p|}$ to
\eqn
        (H(p,\sigma)-E(p,\sigma))\Psi(p,\sigma)=0 \;,
		\label{H-E-Psi-eigeneq-1}
\eeqn
we get
\eqn 
	(\partial_{|p|}H(p,\sigma)&-&\partial_{|p|}E(p,\sigma))\Psi(p,\sigma)
	\nonumber\\
	&=& -
	(H(p,\sigma)-E(p,\sigma))\partial_{|p|}\Psi(p,\sigma)
	\label{derp-H-E-Psi-eigeneq-1}
\eeqn
and taking the inner product with $\Psi(p,\sigma)$,
we get the Feynman-Hellman formula
\eqn
        \partial_{|p|}E(p,\sigma)=\frac{\bra \Psi(p,\sigma)\,,\,
        (\partial_{|p|}H)(p,\sigma)
        \Psi(p,\sigma)\ket}{\bra\Psi(p,\sigma)\,,\,\Psi(p,\sigma)\ket } \;.
        \label{FeynHell}
\eeqn
The second derivative of the ground state energy is given by 
\eqn\;\;\;\;\;\;\;
        \partial_{|p|}^2 E(p,\sigma)=
        1-2\frac{\bra(\partial_{|p|}\Psi)(p,\sigma)\,,\,(H(p,\sigma)
        -E(p,\sigma) )(\partial_{|p|}\Psi)(p,\sigma)\ket}
        {\bra\Psi(p,\sigma)\,,\,\Psi(p,\sigma)\ket} \;.
        \label{part2pEform}
\eeqn
Since    
\eqn
        H(p,\sigma)-E(p,\sigma)\geq0 \;,
        \label{HminEpos}
\eeqn 
(~\ref{part2pEform}) immediately implies $\partial_{|p|}^2  E(p,\sigma)<1$.

To verify (~\ref{part2pEform}), let us momentarily suppress the arguments $(p,\sigma)$, and write $f'$
for $\partial_{|p|}f$. Then,
\eqn
		\partial_{|p|}^2E(p,\sigma)&=&\frac{\bra\Psi,H''\Psi\ket+
		\bra\Psi',H'\Psi\ket +\bra H'\Psi,\Psi'\ket}{\bra\Psi,\Psi\ket}
		\nonumber\\
		&&\hspace{2cm}
		-\bra\Psi,H'\Psi \ket\frac{\bra\Psi,\Psi'\ket+\bra\Psi',\Psi\ket}{\bra\Psi,\Psi\ket^2}
		\nonumber\\
		&=&\frac{\bra\Psi,H''\Psi\ket-2\bra\Psi',(H-E)\Psi'\ket}{\bra\Psi,\Psi\ket}
		\nonumber\\
		&&\hspace{2cm}+\Big(E'\bra\Psi,\Psi\ket- \bra\Psi,H'\Psi\ket\Big)
		\frac{\bra\Psi',\Psi\ket+\bra\Psi,\Psi'\ket }{\bra\Psi,\Psi\ket^2}\;.
		\label{derp2-E-calc-1}
\eeqn
Here, we have applied $\partial_{|p|}$ to (~\ref{FeynHell}), and used 
\eqn
		H'\Psi=E'\Psi-(H-E)\Psi'
\eeqn
from  (~\ref{derp-H-E-Psi-eigeneq-1}). 
The last line in (~\ref{derp2-E-calc-1}) vanishes, due to  (~\ref{FeynHell}), and we note that $H''=1$.
This establishes (~\ref{part2pEform}).

To prove (~\ref{partpE-apriori-1}), we first derive an a priori bound from 
(~\ref{FeynHell}), which implies that $\partial_{|p|}E(p,\sigma)$ exists for 
every $0\leq|p|<\frac13$, and uniformly in $\sigma>0$. To this end, we observe that
\eqn
        H(p,\sigma)=H_f+\frac12(\partial_{|p|}H(p,\sigma))^2 \;.
\eeqn
Therefore,
\eqn
        |\partial_{|p|}E(p,\sigma)| &\leq&  
        \left(\frac{\bra\Psi(p,\sigma)\,,\,(\partial_{|p|}H(p,\sigma))^2
        \Psi(p,\sigma)\ket}
        {\bra\Psi(p,\sigma)\,,\,\Psi(p,\sigma)\ket}\right)^{\frac12}
        \nonumber\\
        &\leq& 
        \left(\frac{2\bra\Psi(p,\sigma)\,,\,H(p,\sigma)
        \Psi(p,\sigma)\ket}
        {\bra\Psi(p,\sigma)\,,\,\Psi(p,\sigma)\ket}\right)^{\frac12}
        \nonumber\\
        &=&\Big(2E(p,\sigma)\Big)^{\frac12}\;,
\eeqn
by the Schwarz inequality and positivity of $H_f$. However,
\eqn
		0\;\leq\;
		E(p,\sigma)&\leq& \Bra\vac\,,\,H(p,\sigma)\vac\Ket
		\nonumber\\
		&=&\frac{|p|^2}{2}+\frac{g^2}{2}\bra\vac , A_{\ks}^2\vac\ket \;.
\eeqn
Thus, 
\eqn
		|\partial_{|p|}E(p,\sigma)|\leq 
		\big(|p|^2+g^2\bra\vac , A_{\ks}^2\vac\ket\big)^{\frac12}
		=|p|+O(g^2)
\eeqn
is bounded for any $0\leq|p|<\frac13$, and uniformly in $\sigma>0$.
By rotation symmetry, 
\eqn
		\lim_{p\rightarrow0}\partial_{|p|}E(p,\sigma)=0 \;.
		\label{der-E-0-prf-1}
\eeqn
Therefore, (~\ref{partpE-apriori-1}) follows from (~\ref{part-2-E-min-1-est-1}).

To prove  (~\ref{part-2-E-min-1-est-1}), we recall the following
definitions from the proof of Theorem {~\ref{gs-const-thm-1}}:
\eqn
		H_{(-1)}&=&H(p,\sigma)-E(p,\sigma)
		\nonumber\\
		&=&e_{(-1,\infty)} +T_{(-1)}+W_{(-1)} 
		\nonumber\\
		\bar R_{(-1)}&=&\Big[H_f -\bar\chi_1[H_f]
		\big(H_{(-1)}-H_f\big)\bar\chi_1[H_f]\Big]^{-1}
		\;\;{\rm on}\;\Ran(\bar\chi_1[H_f])
		\nonumber\\
		Q_{(-1)}&=&\chi_1[H_f]-\bar\chi_1[H_f]\bar R_{(-1)}\chi_1[H_f](H_{(0)}-H_f)\chi_1[H_f]
		\;,
\eeqn
see (~\ref{Epsigmaemin1def}) $\sim$ (~\ref{H-min1-def-2}),
and
\eqn 
		H_{(0)}&=&F_{\chi_1[H_f]}(H_{(-1)},H_f) 
		\;=\;H[\h[e_{(0,\infty)}]]
		\nonumber\\
		H_{(n)}&=&(\renop^H)^n[H_{(0)}]
		\nonumber\\
		&=&H[\h[e_{(n,\infty)}]]
		\nonumber\\
		&=&e_{(n,\infty)}\chi_1^2[H_f]+T_{(n)}+\chi_1[H_f]W_{(n)}\chi_1[H_f] 
		\nonumber\\
		\bar R_{(n)}&=&\Big[H_f+\bar\chi_\rho[H_f]\big(H_{(n)}-H_f \big)
		\bar\chi_\rho[H_f]\Big]^{-1}
		\;\;
		{\rm on}\;\Ran(\bar\chi_\rho[H_f])
		\nonumber\\
		Q_{(n)}&=&Q_{\chi_\rho[H_f]}(H_{(n)},H_f)
		\nonumber\\
		&=&\chi_\rho[H_f]-\bar\chi_\rho[H_f]\bar R_{(n)}
		\bar\chi_\rho[H_f](H_{(n)}-H_f)\chi_\rho[H_f]
		\;,
		\label{H-n-formulas-recall-1}	
\eeqn
see  (~\ref{HnRen-def-eq-2}), (~\ref{H-n-def-1}), (~\ref{H-n-def-2}), (~\ref{H-n-def-3}), 
(~\ref{HnRen-def-eq-1}).

To bound (~\ref{part2pEform}), we recall from (~\ref{Epsigmaemin1def}) that
\eqn
        E(p,\sigma)=\frac{p^2}{2}+\frac{g^2}{2}
        \bra\vac , A_{\ks}^2 \vac\ket
        +e_{(-1,\infty)}
\eeqn
where $e_{(-1,\infty)}$ is obtained from  
\eqn
        e_{(-1,\infty)}=\lim_{n\rightarrow\infty}
        J_{(-1)}^{-1}\circ\cdots\circ J_{(n)}^{-1}[0]\;,
\eeqn
see (~\ref{Jn-maps-def-1}), (~\ref{Jn-concat-1}) and  (~\ref{Epsigmaemin1def}).
More generally,
\eqn
        e_{(j,\infty)}= J_{(j-1)}\circ\cdots\circ J_{(-1)}[e_{(-1,\infty)}]\;,
        \label{ejinftyJform}
\eeqn
for $j\geq0$, cf. (~\ref{Jn-concat-1}) and the subsequent discussion.

We recall from (~\ref{Psigrdst}) and  (~\ref{Psi-lim-def-1}) 
that the ground state eigenvector is obtained from
\eqn
        \Psi(p,\sigma)&=&\Psi_{(-1,\infty)}
        \nonumber\\
        &=& Q_{(-1)}  Q_{(0)}\Gamma_\rho^* Q_{(1)}\cdots
        \Gamma_\rho^*  Q_{(n)}\Gamma^*_\rho\Psi_{(n+1,\infty)} \;.
        \label{psiprodform}
\eeqn 
We observe that due to
\eqn 
		\bra\vac,Q_{(0)}\Gamma_\rho^* Q_{(1)}\cdots
        \Gamma_\rho^*  Q_{(n)}\vac\ket=\bra\vac,\vac\ket=1\;
\eeqn
for all $n\geq0$, where $\vac$ is the Fock vacuum, $\Psi(p,\sigma)$ is normalized by
\eqn
		\bra\vac,\Psi(p,\sigma)\ket &=&\lim_{n\rightarrow\infty}
		\bra\vac,Q_{(0)}\Gamma_\rho^* Q_{(1)}\cdots
        \Gamma_\rho^*  Q_{(n)}\vac\ket
        \nonumber\\
        &=&1 \;.
        \label{psi-vac-norm-1}
\eeqn
Consequently, we obtain
\eqn
		\bra\Psi(p,\sigma),\Psi(p,\sigma)\ket\geq1
		\label{Psi-norm-low-bd-1}
\eeqn
as a trivial lower bound for the denominator in (~\ref{part2pEform}).

Furthermore, it follows from Feshbach isospectrality, Theorem {~\ref{Feshprop}}, that
\eqn
		H_{(-1)}=H(p,\sigma)-E(p,\sigma)\geq0
\eeqn 
implies
\eqn
		H_{(0)} = F_{\chi_1[H_f]}(H_{(-1)},H_f) \geq 0 \;,
\eeqn
and by iteration, 
\eqn
		H_{(n)}=(\renop^H)^n[H_{(0)}] \geq0 \;\;, \;\;\;n\geq0\;.
		\label{QHQ-pos-F-bd-2}
\eeqn
For the definition of the renormalization map $\renop^H$ acting on 
operators on $\H_{red}$, see (~\ref{renHh-def-1}).

An important ingredient in our argument is that by (~\ref{QHQid}),  
\eqn
        Q^\sharp_{(-1)}H_{(-1)} Q_{(-1)}&\leq& F_{\chi_1[H_f]}(H_{(-1)},H_f)
        \nonumber\\
        &=& H_{(0)}  \;,
        \label{QHQ-pos-F-bd-0}
\eeqn
and
\eqn
        Q^\sharp_{(n)}H_{(n)} Q_{(n)}&\leq& F_{\chi_\rho[H_f]}(H_{(n)},H_f)
        \nonumber\\
        &=&\rho \Gamma_\rho^* H_{(n+1)} \Gamma_\rho \; \; , \; \; \;n\geq0
        \label{QHQ-pos-F-bd-1}
\eeqn
as the last term in  (~\ref{QHQid}) is always non-positive for $\tau=H_f$.

The numerator in  (~\ref{part2pEform}) can be estimated recursively using (~\ref{psiprodform}).
Due to (~\ref{HminEpos}), we have
\eqn
        A_{(-1)}&:=&\| H_{(-1)}^{\frac12}
        \partial_{|p|}\Psi_{(-1,\infty)}\|
        \nonumber\\
        &\leq&\|H_{(-1)}^{\frac12}
        (\partial_{|p|} Q_{(-1)})\Psi_{(0,\infty)}\|
        +\|H_{(-1)}^{\frac12}
        Q_{(-1)}\partial_{|p|}\Psi_{(0,\infty)}\|\;.
        \nonumber
\eeqn
The first term after the inequality sign is bounded by
\eqn
        \|H_{(-1)}^{\frac12}
        (\partial_{|p|} Q_{(-1)})\Psi_{(0,\infty)}\|
        \leq a_{(-1)}\|\Psi_{(1,\infty)}\| \;,
\eeqn
with
\eqn
        a_{(-1)}:=\| 
        H_{(-1)}^{\frac12} (\partial_{|p|} Q_{(-1)})
        Q_{(0)}\|_{op} \;.
\eeqn
Next, we use (~\ref{QHQ-pos-F-bd-0}) and get 
\eqn
        \| H_{(-1)}^{\frac12}
        Q_{(-1)}\partial_{|p|}\Psi_{(0,\infty)}\|
        \leq \|H_{(0)}^{\frac12}\partial_{|p|}\Psi_{(0,\infty)}\| \;.
\eeqn
We extend this argument to $n\in\N_0$ by induction.

For $n\in\N_0$, we use
\eqn
        H_{(n)}\geq0 
        \label{Hnpos}
\eeqn
from (~\ref{QHQ-pos-F-bd-2}), and find
\eqn
        A_{(n)}&:=&\|H_{(n)}^{\frac12}\partial_{|p|}\Psi_{(n,\infty)}\|
        \nonumber\\
        &\leq& a_{(n)}\|\Psi_{(n+2,\infty)}\|+
        \|H_{(n)}^{\frac12} Q_{(n)}\Gamma_\rho^*\partial_{|p|}\Psi_{(n+1,\infty)}\|\;,
\eeqn
where
\eqn
        a_{(n)}:=\|H_{(n)}^{\frac12}(\partial_{|p|} Q_{(n)})
        \Gamma_\rho^*Q_{(n+1)}\|_{op} \;.
\eeqn
From (~\ref{QHQ-pos-F-bd-1}) and (~\ref{QHQ-pos-F-bd-2}) follows that
\eqn
        A_{(n)}\leq a_{(n)}\|\Psi_{(n+2,\infty)}\| + \rho^{\frac12}A_{(n+1)}\;.
\eeqn
We therefore find
\eqn
        A_{(-1)}&\leq& a_{(-1)}\|\Psi_{(1,\infty)}\| 
        +\sum_{n=0}^\infty \rho^{\frac n2}a_{(n)}\|\Psi_{(n+2,\infty)}\|\;.
        \label{A0bound}
\eeqn
Our main task is to bound the real numbers $a_{(n)}$ for $n\geq-1$.

The necessary estimates are summarized in the following lemma, 
whose proof we postpone until Section {~\ref{aux-rel-bds-lemma-1-proof}}.

\begin{lemma}
\label{aux-rel-bds-lemma-1}
We assume that $0<|p|<\frac13$.
For $n=-1$ and $a=0,1$, the bounds 
\eqn 
        \| \bar R_{(-1)}^{\frac12}\bar\chi_1[H_f]
        (\partial_{|p|}^a H_{(-1)})\bar\chi_1[H_f]
         \bar R_{(-1)}^{\frac12}\|_{op}&\leq& c
        \nonumber\\
        \|\bar R_{(-1)}^{\frac12}\bar\chi_1[H_f]
        (\partial_{|p|}W_{(-1)})
        \chi_1[H_f] \|_{op}&\leq& cg
        \nonumber\\
        \|\bar R_{(-1)}^{\frac12}\bar\chi_1[H_f]
        W_{(-1)}
        \chi_1[H_f] \|_{op}&\leq& cg
        \label{RWHmin1est}
\eeqn
hold for explicitly computable constants which are independent of the coupling
constant $g$ and of the infrared regularization $\sigma$.

For $n\geq0$, the bounds
\eqn
        \|\bar R_{(n)}\|_{op} \,,\,\|Q_{(n)}\|_{op} &\leq& \frac{c}{\rho}
        \nonumber\\
        \| T_{(n)} \|_{op} \, ,\,
        \| \partial_{|p|}T_{(n)} \|_{op}&\leq& c
        \nonumber\\
        |e_{(n,\infty)}| \,,\,
        \| W_{(n)}  \|_{op} \,,\,
        \| \partial_{|p|}W_{(n)} \|_{op}&\leq& c\,\e_n  
         \label{TRWjbounds}
\eeqn
hold where the constants are independent of $n$ and $\sigma$.

Moreover, the bounds
\eqn
	\|\partial_{H_f}  Q_{(n)}\vac\|&<&c\,\frac{\e_n}{\rho^{3}}
	\nonumber\\
	\|\partial^{a}_{P_f} Q_{(n)}\vac\|&<&c\,\frac{\e_n}{\rho^{1+a}}
	\nonumber\\
	\|\partial_{|p|}  Q_{(n)}\vac\|&<&c\,\frac{\e_n}{\rho^{2}}
	\label{der-Qn-est-1}
\eeqn
are satisfied for $a=0,1,2$
(these are used in Section {~\ref{it-masses-sect-1}}).
\end{lemma}

\subsubsection{Bounds on $a_{(-1)}$}

Let us first estimate $a_{(-1)}$.
We recall
\eqnn
        Q_{(-1)}=\chi_1[H_f]-\bar\chi_1[H_f]\bar R_{(-1)}\bar\chi_1[H_f]
        \Big(T'_{(-1)}+e_{(-1,\infty)}+W_{(-1)}\Big)\chi_1[H_f]
\eeqnn
($W_{(-1)}$ denotes the interaction operator in $H(p,\sigma)$, and $T'_{(-1)}=
T_{(-1)}-H_f$, where $T_{(-1)}$ is the non-interacting Hamiltonian, of
vacuum expectation value $0$),
and
\eqn
        \bar R_{(-1)}=\Big[H_f+\bar\chi_1[H_f]
        \Big(H_{(-1)}-H_f\Big)\bar\chi_{1}[H_f]\Big]^{-1}
\eeqn
on Ran$(\bar\chi_1[H_f])\subset\Fo$, with $H_{(-1)}=H(p,\sigma)-E(p,\sigma)$.

From (~\ref{Qdersimple}),
\eqn
        H_{(-1)}^{\frac12}(\partial_{|p|} Q_{(-1)})Q_{(0)}&=&-
        H_{(-1)}^{\frac12}\bar\chi_1[H_f]\bar R_{(-1)}\bar\chi_1[H_f]
        (\partial_{|p|}H_{(-1)}) Q_{(-1)}Q_{(0)}
        \nonumber\\
        &=&-(I)-(II)
        \label{H-min1-0-decomp-I-II-1}
\eeqn
where
\eqn
        (I)&:=&H_{(-1)}^{\frac12}\bar\chi_1[H_f]\bar R_{(-1)}\bar\chi_1[H_f]
        (\partial_{|p|}H_{(-1)}) \chi_1[H_f]Q_{(0)} \;,
        \nonumber\\
        (II)&:=&H_{(-1)}^{\frac12}\bar\chi_1[H_f]\bar R_{(-1)}\bar\chi_1[H_f]
        (\partial_{|p|}H_{(-1)})
        ( Q_{(-1)}-\chi_1[H_f] )Q_{(0)}\;.
        \label{H-min1-I-II-def-1}
\eeqn
From $(I)=(I_1)+(I_2)$ with
\eqn
        (I_1)&:=&H_{(-1)}^{\frac12}\bar\chi_1[H_f]\bar R_{(-1)}\bar\chi_1[H_f]
        (\partial_{|p|}(T_{(-1)}+e_{(-1,\infty)})) \chi_1[H_f]Q_{(0)}
        \nonumber\\
        (I_2)&:=&H_{(-1)}^{\frac12}\bar\chi_1[H_f]\bar R_{(-1)}\bar\chi_1[H_f]
        (\partial_{|p|}W_{(-1)}) \chi_1[H_f]Q_{(0)} \;,
\eeqn
we find, by expanding the resolvent in $Q_{(0)}$ once, 
\eqn
        (I_1)&=&H_{(-1)}^{\frac12}\bar\chi_1[H_f]\bar R_{(-1)}\bar\chi_1[H_f]
        (\partial_{|p|}(T_{(-1)}+e_{(-1,\infty)})) \chi_1[H_f]
        \nonumber\\
        &&\hspace{1cm}\times\,\Big(\chi_\rho[H_f]
        -\bar\chi_\rho[H_f]\bar R_{(0)}\bar\chi_\rho[H_f](T_{(0)}+e_{(0,\infty)})
        \chi_\rho[H_f]\Big)
        \nonumber\\
        &+&H_{(-1)}^{\frac12}\bar\chi_1[H_f]\bar R_{(-1)}\bar\chi_1[H_f]
        (\partial_{|p|}(T_{(-1)}+e_{(-1,\infty)})) \chi_1[H_f]
        \label{I1decomp}\\
        &&\hspace{1cm}\times\,\Big(\bar\chi_\rho[H_f]
        \bar R_{(0)}\bar\chi_\rho[H_f]W_{(0)}(Q_{(0)}-\chi_\rho[H_f]))
        \chi_\rho[H_f]\Big) \,. \;\;\nonumber
\eeqn
The first product of operators on the right hand side of the equality sign
equals zero because the cutoff operator $\chi_\rho[H_f]$ on the far right
can be commuted to the left, and $\bar\chi_1[H_f]\chi_\rho[H_f]=0$.
Therefore,
\eqn
        \|(I_1)\|_{op}&\leq&\|H_{(-1)}^{\frac12}\bar R_{(-1)}^{\frac12}\|_{op}
        \|\bar R_{(-1)}^{\frac12}(\partial_{|p|}(T_{(-1)}+e_{(-1,\infty)})) \chi_1[H_f]\|_{op}
        \nonumber\\
        &&\hspace{1cm}\times\,
        \|\bar R_{(0)}\|_{op} \|W_{(0)}\|_{op}
        \|Q_{(0)}-\chi_\rho[H_f]\|_{op} \;.
        \label{I-1-bound-1}
\eeqn
Consequently, we find that
\eqn
        \|(I_1)\|_{op}\leq c\,\frac{\e_0}{\rho} \;,
\eeqn
and
\eqn
        \|(I_2)\|_{op}&\leq&\|H_{(-1)}^{\frac12} \bar R_{(-1)}^{\frac12}\|_{op}
        \|\bar R_{(-1)}^{\frac12}(\partial_{|p|}W_{(-1)})\chi_1[H_f]\|_{op}
        \|Q_{(0)}\|_{op}
        \nonumber\\
        &\leq&c\,\frac{g}{\rho} \;,
        \label{I-2-bound-1}
\eeqn
for some constants $c$ which are independent of $g,\sigma$.

Likewise, $(II)=(II_1)+(II_2)$ with
\eqn
        (II_1)&:=&H_{(-1)}^{\frac12}\bar\chi_1[H_f]\bar R_{(-1)}\bar\chi_1[H_f]
        (\partial_{|p|}H_{(-1)})\bar\chi_1[H_f]\bar R_{(-1)}
        \nonumber\\
        &&\hspace{2cm}\times\,\bar\chi_1[H_f]
        (T_{(-1)}+e_{(-1,\infty)})\chi_1[H_f] Q_{(0)}\nonumber\\
        (II_2)&:=&H_{(-1)}^{\frac12}\bar\chi_1[H_f]\bar R_{(-1)}\bar\chi_1[H_f]
        (\partial_{|p|}H_{(-1)})\bar\chi_1[H_f]\bar R_{(-1)}
        \nonumber\\
        &&\hspace{2cm}\times\,\bar\chi_1[H_f]
        W_{(-1)}\chi_1[H_f] Q_{(0)} \;.
\eeqn
We have
\eqnn
        \|(II_1)\|_{op}&\leq&\|H_{(-1)}^{\frac12}\bar R_{(-1)}^{\frac12}\|_{op}
        \|\bar R_{(-1)}^{\frac12} (\partial_{|p|}H_{(-1)})\bar R_{(-1)}^{\frac12}\|_{op}
        \nonumber\\
        &&\hspace{1.5cm}\times\,
        \|\bar R_{(-1)}^{\frac12} 
        (T_{(-1)}+e_{(-1,\infty)})\chi_1[H_f] Q_{(0)}\|_{op} \;.
        \label{II-1-bound-1}
\eeqnn
Expanding the resolvent in $Q_{(0)}$ once,  similarly as
in (~\ref{I1decomp}), one can see that
the term on the last line is bounded by $\frac{c\e_0}{\rho}$. Hence,
\eqn
        \|(II_1)\|_{op}\leq c\,\frac{\e_0}{\rho}\;.
\eeqn
Furthermore,
\eqn
        \|(II_2)\|_{op}&\leq&\|H_{(-1)}^{\frac12}\bar R_{(-1)}^{\frac12}\|_{op}
        \|\bar R_{(-1)}^{\frac12}(\partial_{|p|}H_{(-1)})\bar R_{(-1)}^{\frac12}\|_{op}
        \nonumber\\
        &&\hspace{0.5cm}\times\,
        \|\bar R_{(-1)}^{\frac12}W_{(-1)}\chi_1[H_f]\|_{op}
        \|Q_{(0)}\|_{op}
        \nonumber\\
        &\leq&c\,\frac{g}{\rho}\;,
        \label{II-2-bound-1}
\eeqn
for constants $c$ which  are independent of $g,\sigma$.

It follows that
\eqn
        a_{(-1)}\leq c\,\frac{\e_0}{\rho}\, ,
\eeqn
since $g=c\e_0$.

\subsubsection{Bounds on $a_{(n)}$ for $n\geq0$}

We have
\eqn
        a_{(n)}\leq \|H_{(n)}\|_{op}^{\frac12}
        \|(\partial_{|p|} Q_{(n)})\Gamma_\rho^* Q_{(n+1)}\|_{op} \;.
\eeqn
To bound $a_{(n)}$,  we can straightforwardly adapt the steps 
between (~\ref{H-min1-0-decomp-I-II-1}) and (~\ref{II-2-bound-1}) 
in our discussion of the case $n=-1$.

To this end, we observe that in all of these expressions, 
the indices $-1$ and $0$ can be simultaneously replaced by $n\geq0$ and $n+1$, 
provided that the operators $\chi_1[H_f]$ and $\bar\chi_1[H_f]$
are replaced by $\chi_\rho[H_f]$ and $\bar\chi_\rho[H_f]$.
Correspondingly, we arrive at the bounds given in (~\ref{I-1-bound-1}), 
(~\ref{I-2-bound-1}), (~\ref{II-1-bound-1}), (~\ref{II-2-bound-1}), but
with the indices $-1$ and $0$ replaced by $n\geq0$ and $n+1$, and with
$\chi_1[H_f]$ and $\bar\chi_1[H_f]$ replaced by $\chi_\rho[H_f]$ and $\bar\chi_\rho[H_f]$. 

Using (~\ref{TRWjbounds}) in Lemma {~\ref{aux-rel-bds-lemma-1}}, we thereby obtain
\eqn
        \|H_{(n)}^{\frac12}(\partial_{|p|}\tilde Q_{(n)})\Gamma_\rho^*Q_{(n+1)}\|_{op}
        \leq c\,\frac{\e_0}{\rho^3} \;,
\eeqn
and
\eqn
         a_{(n)} <c\,\frac{\e_0}{\rho^3}\;,
\eeqn
where the constants are independent of $n$.

\subsubsection{Completing the proof}

Collecting the above estimates, we are now in the position to finish the proof
of Theorem {~\ref{mren-pnonzero-thm-1}}. 
Using $\delta_n\leq \delta_0+2\e_0$ and $\e_n\leq \e_0$
by (~\ref{rgflowparambds}), and recalling (~\ref{A0bound}),
we conclude that
\eqn
        A_{(0)}&\leq& a_{(-1)}\|\Psi_{(1,\infty)}\| 
        +\sum_{n=0}^\infty \rho^{\frac n2}a_{(n)}\|\Psi_{(n+2,\infty)}\|
        \nonumber\\
        &\leq&c\,\frac{\e_0}{\rho^3}
        \Big(1+c'\,\frac{\e_0}{\rho}\exp\Big[\frac{c''\e_0}{ \rho^{3} }\Big]\Big)
        \;,
\eeqn
using (~\ref{Psininftybound}).
We thus find
\eqn
        |\partial_{|p|}^2E(p,\sigma)-1|&\leq&
        2   A_{(0)}^2\,
        \;\leq\;c\,\frac{\e_0^2}{\rho^6}
        \Big(1+c'\,\frac{\e_0}{\rho}\exp\Big[\frac{c''\e_0}{ \rho^{3} }\Big]\Big)^2
        \nonumber\\
        &\leq & c_0(\sigma) g^2\;,
\eeqn
where we have bounded the denominator in (~\ref{part2pEform}) 
from below by 1 (see (~\ref{Psi-norm-low-bd-1})),
and recalled from (~\ref{parm-choice-1}) that $\rho=\rho(\sigma)=\cob^{1/\sigma}$ 
for $p\neq0$. For $\e_0=O(g)<\rho^3$ (which is compatible with  (~\ref{parm-choice-1})), 
we have $c_0(\sigma)<C^{1/\sigma}$ with $C$ 
independent of $g$ and $\sigma$. 
\endprf

\subsection{Proof of Lemma {~\ref{aux-rel-bds-lemma-1}}}
\label{aux-rel-bds-lemma-1-proof}

We shall here establish the bounds asserted
in Lemma  {~\ref{aux-rel-bds-lemma-1}}.

\subsubsection{Bounds for $n=-1$}

To prove the estimates (~\ref{RWHmin1est}) for the case $n=-1$, we observe that 
\eqn 
	\bar\chi_1[H_f]H_{(-1)}\bar\chi_1[H_f]\geq H_f+
	\bar\chi_1[H_f](H_{(-1)}-H_f)\bar\chi_1[H_f] \;,
	\label{H-min1-rel-bd-1}
\eeqn
and recall that
\eqn 
	\bar R_{(-1)}=\Big[H_f+
	\bar\chi_1[H_f](H_{(-1)}-H_f)\bar\chi_1[H_f]\Big]^{-1}
\eeqn
on $\Ran(\bar\chi_1[H_f])$.
Thus,
\eqn 
	\Big\|H_{(-1)}^{\frac12}\bar\chi_1[H_f]\bar R_{(-1)}^{\frac12}\Big\|_{op}  
	&<&c
\eeqn 
follows immediately.

For the second inequality in  (~\ref{RWHmin1est}), we note that
\eqn 
	\partial_{|p|}H_{(-1)}&=&-\partial_{|p|}E(p,\sigma)
	+\partial_{|p|}H(p,\sigma) \;,
\eeqn
and recall   
\eqn 
	H(p,\sigma)=H_f+(\partial_{|p|}H(p,\sigma))^2 = E(p,\sigma)+H_{(-1)}\;,
\eeqn
where $E(p,\sigma)>0$ and $H_{(-1)}\geq0$.
Thus, 
\eqn
	&&\Big\|\bar R_{(-1)}^{\frac12}\bar\chi_1[H_f](\partial_{|p|}H_{(-1)} )
	\bar\chi_1[H_f]\bar R_{(-1)}^{\frac12}\Big\|_{op}
	\nonumber\\
	&\leq&|\partial_{|p|}E(p,\sigma)|\,\|\bar R_{(-1)} \| +
	\Big\|\bar R_{(-1)}^{\frac12}\bar\chi_1[H_f]\sqrt{H(p,\sigma)}
	\bar\chi_1[H_f]\bar R_{(-1)}^{\frac12}\Big\|_{op}
	\nonumber\\
	&\leq&c +
	\Big\|\bar R_{(-1)}^{\frac12}\bar\chi_1[H_f]\sqrt{H(p,\sigma)}
	\bar\chi_1[H_f]\bar R_{(-1)}^{\frac12}\Big\|_{op}
	\nonumber\\
	&\leq&c +
	\Big\|\bar R_{(-1)}^{\frac12}\sqrt{\bar\chi_1[H_f]H_{(-1)}\bar\chi_1[H_f]}
	\bar R_{(-1)}^{\frac12}\Big\|_{op}
	\nonumber\\
	&\leq&c \;,
\eeqn
using (~\ref{H-min1-rel-bd-1}), $0\leq\bar\chi_1[H_f]\leq1$, 
and $\|\bar R_{(-1)}\|<c$ (see (~\ref{Fpair-ver-init-1})).

Next, we have
\eqn 
	&&\Big\| \chi_1[H_f](\partial_{|p|}W_{(-1)} )\bar\chi_1[H_f]
	\bar R_{(-1),0}^{\frac12}\bar\chi_1[H_f](\partial_{|p|}W_{(-1)} )\chi_1[H_f] \Big\|_{op}
	\nonumber\\
	&\leq&
	\Big\|\bar R_{(-1)}\bar\chi_1[H_f](\partial_{|p|}W_{(-1)} )\chi_1[H_f] \Big\|_{op}
	\nonumber\\
	&&\hspace{2cm}\times\,
	\sum_{L\geq0}\Big\| \bar R_{(-1),0}^{\frac12}\bar\chi_1[H_f] W_{(-1)}\bar\chi_1[H_f]
	\bar R_{(-1),0}^{\frac12} \Big\|_{op}^L 
	\nonumber\\
	&\leq&cg\sum_{L\geq0}(cg)^L
	\;\leq \;cg
\eeqn
where
\eqn 
	\bar R_{(-1),0}=\Big[H_f+
	\bar\chi_1[H_f](e_{(-1,\infty)}+T_{(-1)}-H_f)\bar\chi_1[H_f]\Big]^{-1}
\eeqn
on $\Ran(\bar\chi_1[H_f])$. Here, we have used
\eqn 
	\Big\| \bar R_{(-1),0}^{\frac12}\bar\chi_1[H_f]W_{(-1)}\bar\chi_1[H_f]
	\bar R_{(-1),0}^{\frac12} \Big\|_{op}<cg
\eeqn
which was proved in  (~\ref{Fpair-ver-init-2}), and 
\eqn 
	\Big\|\bar R_{(-1),0}^{\frac12}\bar\chi_1[H_f](\partial_{|p|}W_{(-1)} )\chi_1[H_f] \Big\|_{op}
	<cg
\eeqn
which is obtained in the same way as (~\ref{appendest1}) 
(since $\partial_{|p|}W_{(-1)}=gA_{\ks}$).

The last estimate in  (~\ref{RWHmin1est}) was already proven in (~\ref{Fpair-ver-init-2}).

\subsubsection{Bounds for $n\geq0$}

We recall that 
\eqn 
	H_{(n)}=H[\h^{(n)}[e_{(n,\infty)}]]=e_{(n,\infty)}\chi_1^2[H_f]
	+T_{(n)}+W_{(n)}
\eeqn
with
$\h^{(n)}[ e_{(n,\infty)}]\in\Polyd(\e_n,\delta_n,\lambda_n)$. 
According to the definition of the polydiscs $\Polyd\Polpar$, 
in (~\ref{Polyddef}), and recalling (~\ref{eninftybds}), we have
\eqn
		\| T_{(n)} \|_{op} &\leq&c
		\nonumber\\
		|e_{(n,\infty)}|\,,\,\| W_{(n)}  \|_{op}&\leq& c\e_n 
		\label{barR-Q-bounds-2-1}
\eeqn
The bounds 
\eqn
		\|\bar R_{(n)}\|_{op} \,,\,\|Q_{(n)}\|_{op} &\leq&\frac{c}{\rho}
		\label{barR-Q-bounds-2-2}
\eeqn
were established in the proof of Lemma {~\ref{QnQnpl1-lemma-1}}.

For the derivatives in $|p|$, we recall that 
\eqn
		\h^{(n)}[z]=(E[z],T^{(n)}[z],\h_1[z])\in\Polyd(\e_n,\delta_n,\lTnl_n) 
\eeqn 
is {\em analytic} for $z\in\I$, where $\e_n\leq 2^{-n}\e_0$, $\delta_n\leq\delta_0+2\e_0$, 
and $\lTnl_n=\rho^n/2$ (see (~\ref{en-dn-ln-def-1})). Moreover, we observe that by
\eqn
	H_{(n)}&=&H[\h^{(n)}[z]]\Big|_{z\rightarrow e_{(n,\infty)}}
	\nonumber\\
	&=&e_{(n,\infty)}\chi_1^2[H_f]+\big( \, T[z;\opp] +
	\chi_1[H_f] \, W[\h[z]] \, \chi_1[H_f] \, \big)\Big|_{z\rightarrow e_{(n,\infty)}}\;,
\eeqn 
we have
\eqn
        \partial_{|p|}H_{(n)} 
        &=& H[\partial_{|p|}\h^{(n)}[z]]
        \Big|_{z\rightarrow e_{(n,\infty)}}  
        \nonumber\\
        &+& H[\partial_{z}\h^{(n)}[z]] \Big|_{z\rightarrow e_{(n,\infty)}} 
         \partial_{|p|}e_{(n,\infty)}  \;.
\eeqn
Let us first address the operator $T^{(n)}[z;\opp]$ on $\H_{red}$, and estimate
$\|\partial_{|p|}T^{(n)}[z;\opp]\|_{op}$.

For the free comparison operator
$T_0^{(p,\lambda)}[z;\opp]$ defined in (~\ref{h0plambdadef}), it is easy to verify that
\eqn
        \| (\partial_{|p|}T_0^{(p,\lambda_n)}[z;\opp])
        \Big|_{z\rightarrow e_{(n,\infty)}}\|_{op}\leq c \;,
\eeqn
given $0<|p|<\frac13$. Recalling
(~\ref{TRWjbounds}), we have 
\eqn
        \|(\partial_{|p|}T^{(n)}[z;\opp])
        \Big|_{z\rightarrow e_{(n,\infty)}}\|_{op}
        &\leq&\| (\partial_{|p|}T_0^{(p,\lambda_n)}[z;\opp] )
        \Big|_{z\rightarrow e_{(n,\infty)}}\|_{op}
        \nonumber\\
        &&+\| (\partial_{|p|} T^{(n)}[z;\opp]- \partial_{|p|} T_0^{(p,\lambda_n)}[z;\opp])
        \Big|_{z\rightarrow e_{(n,\infty)}}\|_{op}
        \nonumber\\
        &\leq& c+\|T^{(n)}-T_0^{(p,\lambda_n)}\|_{\sigma} 
        \nonumber\\
        &\leq&c+K_\Theta\delta_n \;,
\eeqn
for a constant $c$ that is independent of $n$. 
$K_\Theta$ is the constant that appears in the definition of 
the norm $\|\,\cdot\,\|_{\Tspace}$ in (~\ref{Tnormdef}).

Next, we consider   
\eqn
		W_{(n)}=H[\h_1^{(n)}[z]]
        \Big|_{z\rightarrow e_{(n,\infty)}} \;,
\eeqn
where we find
\eqn
        \|(\partial_{|p|}H[\h_1^{(n)}[z]])
        \Big|_{z\rightarrow e_{(n,\infty)}}\|_{op}
        &\leq& 
        \|(\h^{(n)})_{M+N\geq1}\|_{\xi,\sigma}
        \;\leq\;\e_n\;.
\eeqn
Note that a partial derivative with respect to $|p|$,
for $z$ fixed,
is contained in the Banach space norm $\|\,\cdot\,\|_{\sigma,\xi}$ 
on the polydisc $\Polyd(\e_n,\delta_n,\lambda_n)$,
see Section {~\ref{Banspsubsect}}.

To bound the partial derivatives with respect to the spectral parameter $z$, 
we use the fact that $\h^{(n)}[z]$ depends analytically on $z\in \I$. 
By (~\ref{eninftybds}), we know that $|e_{(n,\infty)}|<\frac{1}{40}$.
Hence, the derivative in $z$ at $e_{(n,\infty)}$ can be represented
by the Cauchy integral
\eqn
        \;\;\;\;\;\;
        \partial_{z}w_{M,N}^{(n)}[z;\spvar;K^{(M,N)}] 
        =\frac{1}{2\pi i}\oint_{|\zeta|=\frac{1}{10}}
        \frac{d\zeta}{(\zeta-z)^2} w_{M,N}^{(n)}[z;\spvar;K^{(M,N)}]\;.
\eeqn
Thus,
\eqn
        \big\|(\partial_{z}H[\h_1^{(n)}[z])\Big|_{z\rightarrow e_{(n,\infty)}}\big\|_{op}
        &\leq&
        \sup_{|z|<\e_n}\| (\partial_{z}\h^{(n)}) \|_{\xi,\sigma}^\sharp
        \nonumber\\
        &\leq&\frac{1}{2\pi(\frac{1}{20})^2}  
        \sup_{|z|<\frac{1}{10}}\| \h_1^{(n)}[z] \|_{\xi,\sigma}^\sharp  
		\;\leq\;c\,\e_n \;,
        \label{partzHanalytbd}
\eeqn
where $c$ does not depend of $n$. 

Likewise,
\eqn 
		\big\|(\partial_{z}T^{(n)}[z])\Big|_{z\rightarrow e_{(n,\infty)}}\big\|_{op}
		\;<\; c
\eeqn
follows from the same argument.

To bound $|\partial_{|p|}e_{n,\infty}|$, we recall from (~\ref{JnenEnid}) that
\eqn
        e_{(n+1,\infty)}=J_{(n)}[e_{(n,\infty)}]=\rho^{-1}E_{(n)}[e_{(n,\infty)}]\;,
\eeqn
with $|E_{(n)}[z]-z|\leq 2^{-j}\e_0$, and
$E_{(n)}[z]$ analytic in $\I$.
We infer
\eqn
        \partial_{|p|}e_{(n,\infty)}-\rho\partial_{|p|}e_{(n+1,\infty)} 
        &=&\partial_{|p|}e_{(n,\infty)}-\partial_{|p|} E_{(n)}[e_{(n,\infty)}]
       	\nonumber\\
        &=&\Big[
        \partial_{|p|}e_{(n,\infty)}\partial_z(z-E_{(n)}[z])
        +
        \partial_{|p|}E_{(n)}[z]\Big]\Big|_{z\rightarrow e_{(n,\infty)}}\;.
         \label{partpenid}
\eeqn
Representing the derivative in $z$ at $e_{(n,\infty)}$ as a Cauchy integral, 
the argument used in (~\ref{partzHanalytbd}) yields
\eqn
        b_n:=|\partial_z(z-E_{(n)}[z])|\Big|_{z\rightarrow e_{(n,\infty)}}<c\,2^{-n}\e_0 \;,
\eeqn
for a constant independent of $n$. Moreover,
\eqn
        |\partial_{|p|}E_{(n)}[z]|<c\,2^{-n}\e_0\;.
\eeqn
Thus, we find
\eqn
        |\partial_{|p|}e_{(n,\infty)}|&\leq&\frac{1}{1-b_n}\Big[
        |\partial_{|p|}E_{(n)}[z]|
        \Big|_{z\rightarrow e_{(n,\infty)}}+\rho|\partial_{|p|}e_{(n+1,\infty)}|
        \Big]
        \nonumber\\
        &\leq&\frac{1}{1-b_n}\Big[
        |\partial_{|p|}E_{(n)}[z]|
        \Big|_{z\rightarrow e_{(n,\infty)}}
        \nonumber\\
        &&\hspace{1.5cm}+
        \frac{\rho}{1-b_{n+1}} \Big(
        |\partial_{|p|}E_{(n+1)}[z]|
        \Big|_{z\rightarrow e_{(n+1,\infty)}}
        \nonumber\\
        &&\hspace{3cm}+\frac{\rho}{1-b_{n+2}}\Big[\cdots\cdots\cdots
        \Big]\,\Big)\,\Big]
        \nonumber\\
        &\leq&\Big[\prod_{j=0}^\infty \frac{1}{1-b_{n+j}}\Big]
        \sum_{j=0}^\infty \rho^j|\partial_{|p|}E_{(n+j)}[z]|
        \Big|_{z\rightarrow e_{(n+j,\infty)}}
        \nonumber\\
        &\leq&c\e_0\exp\Big(c'\sum_{j=0}^\infty 2^{-n-j}\e_0\Big) \leq c \e_0\;,
\eeqn
by iteration of the identity (~\ref{partpenid}).

Next, we prove the estimates (~\ref{der-Qn-est-1}). We recall that
\eqn
		\;\;\;\;
		Q_{(n)}&=&\chi_\rho[H_f]-\bar\chi_\rho[H_f]\bar R_{(n)}\bar\chi_\rho[H_f]
		(T'_{(n)}+e_{(n,\infty)}+W_{(n)})\chi_\rho[H_f]\;.
\eeqn
Thus,  
\eqn
		\partial_{H_f}Q_{(n)}\vac&=&\partial_{H_f}
		\Big[\vac-\bar\chi_\rho[H_f]\bar R_{(n)}\bar\chi_\rho[H_f]
		(T'_{(n)}+e_{(n,\infty)}+W_{(n)})\vac\Big]
		\nonumber\\
		&=&-\partial_{H_f}
		\Big[\bar\chi_\rho[H_f]\bar R_{(n)}\bar\chi_\rho[H_f]
		(e_{(n,\infty)}+W_{(n)})\Big]\vac
		\nonumber\\
		&=&\bar R_{(n)}\Big(\partial_{H_f}\Big[H_f+\bar\chi_\rho[H_f]
		\big(H_{(n)}-H_f\big)\bar\chi_\rho[H_f]\Big]
		\Big)
		\nonumber\\
		&=&-\Big[(\partial_{H_f}\bar\chi_\rho[H_f])\bar R_{(n)}\bar\chi_\rho[H_f]
		\nonumber\\
		&&\hspace{2cm}+
		\bar\chi_\rho[H_f]\bar R_{(n)}(\partial_{H_f}\bar\chi_\rho[H_f])\Big] 
		(e_{(n,\infty)}+W_{(n)})\vac
		\nonumber\\
		&=&\bar R_{(n)}\Big(\partial_{H_f}\Big[H_f+\bar\chi_\rho[H_f]
		\big(H_{(n)}-H_f\big)\bar\chi_\rho[H_f]\Big]
		\Big)
		\nonumber\\
		&&\hspace{2cm}\times\bar R_{(n)}\bar\chi_\rho[H_f]
		(e_{(n,\infty)}+W_{(n)}) \vac
		\nonumber\\
		&&- \bar\chi_\rho[H_f]\bar R_{(n)}\bar\chi_\rho[H_f]
		(\partial_{H_f}W_{(n)}) \vac\;,
\eeqn
so that
\eqn
		\Big\|\partial_{H_f}Q_{(n)}\vac\Big\|&\leq&\|\bar R_{(n)}\|_{op}^2
		\Big[1+\|\partial_{H_f}\bar\chi_\rho[H_f]\|_{op}
		\|H_{(n)}-H_f\|_{op}
		\nonumber\\
		&&\hspace{2cm}+\|\partial_{H_f}(H_{(n)}-H_f)\|_{op}\Big]
		\Big(|e_{(n,\infty)}|+\|W_{(n)}\|_{op}\Big)		
		\nonumber\\
		&&+2\|\partial_{H_f}\bar\chi_\rho[H_f]\|_{op}\|\bar R_{(n)}\|_{op}
		\Big(|e_{(n,\infty)}|+\|W_{(n)}\|_{op}\Big)
		\nonumber\\
		&&+\|\bar R_{(n)}\|_{op}\|\partial_{H_f}W_{(n)}\|_{op}
		\nonumber\\
		&<& \frac{c}{\rho^3}\;,
\eeqn
using 
\eqn
		\|\partial_{H_f}H_{(n)}\|_{op}+\|\partial_{H_f}H_{(n)}\|_{op}
		+\sum_{a=1,2}\|\partial_{P_f}^a H_{(n)}\|_{op}&\leq&\|\h^{(n)}\|_{\sigma,\xi}
		\;<\;c
		\nonumber\\
		\|\partial_{H_f}\bar\chi_\rho[H_f]\|_{op}&<& \frac{c}{\rho}
		\label{H-n-op-norm-bds-1}
\eeqn
and (~\ref{barR-Q-bounds-2-1}), (~\ref{barR-Q-bounds-2-2}). 

Along the same lines, one obtains
\eqn
		\Big\|\partial_{P_f}^aQ_{(n)}\vac\Big\|&\leq& \frac{c}{\rho^{1+a}}
\eeqn
for $a=1,2$, using (~\ref{H-n-op-norm-bds-1}), and noting that
$\partial_{P_f}\bar\chi_\rho[H_f]=0$ (thus, there is an
inverse factor of $\rho$ less in comparison to the derivative in $H_f$).

This establishes Lemma  {~\ref{aux-rel-bds-lemma-1}}.
\endprf


\section{Uniform bounds for vanishing conserved momentum}
\label{peq0masssect}

In this section, we study the renormalized electron mass for vanishing
conserved momentum $p=0$. We shall introduce
a quantity $m_{ren}^*(0,\sigma)$, and prove bounds on $m_{ren}^*(0,\sigma)$ 
that are {\em uniform}
in the infrared regularization $\sigma$.
We identify $m_{ren}^*(0,\sigma)$ with the renormalized mass later, 
in Theorem {~\ref{derp2Eidthm}}.

\subsection{A second definition of the renormalized mass}

From Theorem {~\ref{codim2contrthm}} and the discussion in Section {~\ref{ground-state-sect-1}}, 
we know that there is a small constant $g_0>0$  (independent of $\sigma$) 
such that for all values of the electron charge $g<g_0$, there is an element
\eqn
		\tilde\h_{(0)}:=\h^{(0)}[e_{(0,\infty)}]\Big|_{p=0}
\eeqn 
inside a polydisc $\Polyd(\e_0,\delta_0,2^{-1})$ such that
\eqn
		H[\tilde\h_{(0) }] =F_{\chi_1[H_f]}(H(0,\sigma)-E(0,\sigma),H_f) \;.
\eeqn
This was accomplished in the first Feshbach decimation step.
Moreover, for $g_0$ sufficiently small (independently of $\sigma$),
the parameters $\e_0$ and $\delta_0$ are sufficiently small that
the renormalization map $\ren$ has the codimension 2 contraction property
(~\ref{codim2contrp0}) on $\Polyd(\e_0,\delta_0+2\e_0,2^{-1})$.
Then,
\eqn
		\tilde\h_{(n) }:=\ren^n[\tilde\h_{(0) }]=\h^{(n)}[e_{(n,\infty)}]\Big|_{p=0}
\eeqn
is an element of $\Polyd(\e_n,\delta_n,2^{-1}\rho^n)$, where
we recall from Theorem {~\ref{codim2contrthm}} that 
\eqn
		\e_n&\leq&\ccontr\rho\,\e_{n-1}\;\leq\;(\ccontr\rho)^n\e_0
		\nonumber\\
		\delta_n&\leq&\e_0+2\delta_0 
\eeqn
for a constant $\ccontr$ is independent of $\rho$, $n$, and $\e_0$.

Let
\eqn
		\tilde H_{(n) }:=H[\tilde\h_{(n)}]
		\;=\;H_{(n)}\Big|_{p=0}\;.
\eeqn
We may then define 
\eqn 		 
		\frac{1}{m_{ren}^*(0,\sigma)}:= \lim_{n\rightarrow\infty}
		\frac{\rho^{-n}
		\bra\,\vac\,,\,(\partial_{P_f^\parallel}^2 \tilde H_{(n) }) \vac\ket}
		{\bra\,\vac\,,\,(\partial_{H_f}\tilde H_{(n) })\vac\ket}\;,
		\label{mrenstar-def-1}
\eeqn
for an arbitrary choice of $n\in\R^3$, $|n|=1$, and $P_f^\parallel:=P_f\cdot n$.

We prove in Theorem {~\ref{derp2Eidthm}} below that
\eqn
		\frac{1}{m_{ren}^*(0,\sigma)}=\partial_{p}^2E(0,\sigma)
		\label{mrenstar-derE-1}
\eeqn
for $\sigma>0$. 
 
\subsection{The main theorem}

We assume the following condition.

\begin{hypothesis}
We assume that  
\eqn
        	\lim_{\sigma\rightarrow0}\lim_{p\rightarrow0}  \partial_{|p|}^2 E(p,\sigma) =
        	\lim_{p\rightarrow0}\lim_{\sigma\rightarrow0}
        	\partial_{|p|}^2 E(p,\sigma) \;.
        	\label{limcommhyp}
\eeqn
\end{hypothesis}

Condition  (~\ref{limcommhyp}) relates the definition of the renormalized mass
$m^*_{ren}(0,\sigma)$ to the definition of the renormalized electron mass $m_{ren}(p,\sigma)$
given in (~\ref{mren-def-1}):
\eqn
		\lim_{\sigma\rightarrow0}m_{ren}^*(0,\sigma)=\lim_{p\rightarrow0}
		\lim_{\sigma\rightarrow0}m_{ren}(p,\sigma) \;.
\eeqn  
The proof of condition (~\ref{limcommhyp}) is given in \cite{ch}. 
Condition (~\ref{limcommhyp}) and Theorem {~\ref{derp2Eidthm}} combined thus imply 
the following
{\em uniform} bound on the infrared renormalized mass, which is the main result
of this section.

\begin{theorem}
\label{partP2Eunifbdp0}
For $p=0$, 
\eqn
		0<E(0,\sigma)\leq c_1 g^2
		\label{E-0-bd-1}
\eeqn
and 
\eqn
		(\partial_{|p|}E)(0,\sigma)=0 \;,
		\label{der-E-0-bd-1}
\eeqn
for a finite constant $c_1>0$ that is independent of $\sigma$ and $g$.
There is a constant $g_0>0$ independent of $\sigma\geq0$ such that, for arbitrary $\sigma\geq0$ and for
any $0\leq g<g_0$, such that 
\eqn
        	1< m_{ren}^*(0,\sigma)  < 1 + c_2 g^2 \;,
        	\label{renmassp0}
\eeqn
for a finite constant $c_2>0$ that is independent of $g$ and $\sigma$.

Moreover,
\eqn
		1<\lim_{p\rightarrow0}\lim_{\sigma\rightarrow0}m_{ren}(p,\sigma)< 1 + c_2 g^2 
		\label{mren-est-2}
\eeqn
under the assumption that Condition (~\ref{limcommhyp}) holds.
\end{theorem}

\begin{lemma}
\label{ren-mass-thm-unif-1}
For $n\geq0$, we define the difference kernels
\eqn 
        	\nonumber\
        	\Delta \gamma_{(n)}[\spvar]
        	&:=&\rho\gamma_{(n+1)}[\rho^{-1}\spvar]
        	-
        	\gamma_{(n)}[\spvar]
        	\tilde \piop_{\rho}^{(n)}[\spvar]  \;,
\eeqn
where
\eqn
		\gamma_{(n)}[\spvar]&:=&w_{0,0}^{(n)}[p;e_{(n,\infty)}[p];\spvar]\Big|_{p=0} 
		\nonumber\\
		\tilde \piop_\rho^{(n)}[\spvar]&:=&
		\piop_{\rho}^{(n)}[p;e_{(n,\infty)}[p];\spvar]\Big|_{p=0}
\eeqn
see also Theorem {~\ref{hatwMNformalseriesthm}}.
The following identity then holds independently of $\sigma$.
\eqn
        	\label{part2Eidenty}
        	\\
        	\nonumber
        	\frac{1}{m_{ren}^*(0,\sigma)}
        	&=&\frac{1+\sum_{n=-1}^\infty \rho^{-n_+}
        	(\partial_{X^\parallel}^2\Delta \gamma_{(n)}) [\unull]}
        	{1+\sum_{n=-1}^\infty  
        	(\partial_{X_0} \Delta \gamma_{(n)}) [\unull]} \;,
\eeqn
where $n_+=\max\{n,0\}$. The series in the numerator
and denominator both converge absolutely, and uniformly in $\sigma$.
\end{lemma}

\prf
Let
\eqn
		\tilde T_{(n)}&:=&T_{(n)}\Big|_{p=0}
		\nonumber\\
		\tilde W_{(n)}&:=&W_{(n)}\Big|_{p=0}\;.
\eeqn
Since $\langle\vac,\tilde W_{(k)}\vac\rangle=0$, we find
\eqn
		\nonumber
		(~\ref{mrenstar-def-1})= 
		\lim_{k\rightarrow\infty}
		\frac{\rho^{-k}
		\bra \vac\,,\,(\partial_{P_f^\parallel}^2\tilde T_{(k)}) \vac\ket}
		{\bra\,\vac\,,\,(\partial_{H_f} \tilde T_{(k)}) \vac\ket}\;.
\eeqn 
Recalling that
\eqn
		\tilde T_{(n)}[\spvar]=\gamma_{(n)}[\spvar]-
		\gamma_{(n)}[ \unull] \;,
\eeqn
we write $\tilde T_{(n)}$ in the form
\eqn
		\tilde T_{(n)}[\spvar]&=&
		\rho^{-n}\tilde T_{(-1)}[\rho^{n}\spvar]
		\nonumber\\
		&+&\sum_{j=-1}^{n-1}
		\rho^{-(n-j_+)}
		\Big[\Delta \gamma_{(n)}[\rho^{j_+}\spvar] 
		-\Delta \gamma_{(n)}[\underline{0}]\Big]\;,
\eeqn
where $j_+=\max\{0,j\}$. 

Here, $\tilde T_{(-1)} [\rho^{n}\opp]$ 
denotes the non-interacting part of
$H(p,\sigma)$, cf. (~\ref{Tmin1-def-1}), which satisfies
\eqn
		\rho^{-n}\partial_{H_f}\tilde T_{(-1)} [\rho^{n}\opp]=1
		=\rho^{-2n}\partial_{\Ppar}^2\tilde T_{(-1)} [\rho^{n}\opp]  \;.
\eeqn 
Recalling Theorem {~\ref{piopdef}},
we note that $\tilde \piop_{\rho}^{(n)}$ is identical to $\1$ in
an open vicinity of $\spvar=\underline{0}$,
hence all of its derivatives with respect to
$\spvar$ or $p$ are zero at $\spvar=\underline{0}$.
Taking derivatives with respect to the spectral variable $\spvar$
for the vector-operator $\opp=(H_f,P_f)$, and evaluating
at $\spvar=\underline 0$, we thus find
\eqn
        (\partial_{\spvar}^{\ua}\tilde T_{(n)}) 
        &=&\rho^{-n}(\partial_{\spvar}^{\ua}\tilde T_{(-1)})[\unull] 
		\nonumber\\
        &+&\sum_{j=-1}^{n-1}
		\rho^{(|\ua|-1)(n-j_+)}
		(\partial_{\spvar}^{\ua}\Delta \gamma_{(n)})[\unull]
        \label{partaX-T-1}
\eeqn
for $0\leq|\ua|\leq2$. The $H_f$-part of (~\ref{partaX-T-1}) gives the numerator 
of (~\ref{part2Eidenty}), and the $P_f$-part the denominator.

We shall next prove absolute convergence of the series in the numerator and denominator.
The estimate
\eqn
        \|\partial_{\spvar}^{\aind}\Delta \gamma_{(n)} \|_\sigma\leq \cm\e^2_n
        \label{partaX-T-2}
\eeqn
was established as part of the derivation of (~\ref{T00p0est}),
for $0\leq|\aind|\leq2$ with $a_0=0$,
where the constant $\cm$ is independent of $\e,\delta, \lTnl,
\rho$, and $\sigma$.

Based on this estimate, we therefore obtain
\eqn
        &&\sum_{j\geq-1} \rho^{-j_+}
        \big|\partial_{\Ppar}^2 
        \Delta \gamma_{(j)}[\unull]  \big|  
        \nonumber\\
		&&\hspace{2cm}\leq\;\sum_{j=0}^\infty\cm\rho^{-j}\e^2_j\; 
        \leq\;  \cm  \sum_{j=0}^\infty (\ccontr^2\rho)^{j}\e^2_0 
		\;<\;2\cm\e_0^2\;,
        \label{renmassstar-ren-1}
\eeqn
for $\e_0$ sufficiently small and
$\rho\leq(2 \ccontr^2)^{-1}\leq\frac12$.

Moreover,
\eqn
		&&\sum_{j\geq-1} 
        \big|(\partial_{H_f} 
        \Delta \gamma_{(j)})[\unull] \big|  \leq \sum_{j=0}^\infty\cm \e^2_j  
	    <2\cm\e_0^2\;.
        \label{renmassstar-ren-2}
\eeqn
Both (~\ref{renmassstar-ren-1}) and (~\ref{renmassstar-ren-2})
are uniform  with respect to $\sigma$. 
\endprf

\subsubsection{Proof of Theorem {~\ref{partP2Eunifbdp0}}}

(~\ref{E-0-bd-1}) and (~\ref{der-E-0-bd-1}) were already proved in (~\ref{E-0-prf-1}) 
and (~\ref{der-E-0-prf-1}).

The estimate (~\ref{renmassp0}) follows from (~\ref{renmassstar-ren-1}), 
(~\ref{renmassstar-ren-2}), and (~\ref{part2Eidenty}),
for a constant $c$ that is independent of $\e_0$ and $\sigma$. 
Finally, under the assumption that Condition  (~\ref{limcommhyp}) holds,
(~\ref{mren-est-2}) is a consequence of Theorem {~\ref{derp2Eidthm}} below.


\section{Identification of two definitions of the renormalized mass}
\label{it-masses-sect-1}
 
For $\sigma>0$ and $g<g_0(\sigma)$, we shall next verify that
the inverse of $m_{ren}^*(0,\sigma)$ in Lemma {~\ref{ren-mass-thm-unif-1}}
agrees with $\partial_{|p|}^2E(p=0,\sigma)$.

\begin{theorem}
\label{derp2Eidthm}
For $\sigma>0$ and $|p|\geq0$, there exists $g_0(\sigma)>0$ so that
for all $g<g_0(\sigma)$, 
\eqn
        \lim_{p\rightarrow0}\partial_{|p|}^2 E (p,\sigma)=\frac{1}{m_{ren}^*(0,\sigma)}\;,
\eeqn
where $g$ is the electron charge.
\end{theorem}
 
We make the following choice of constants.

For  $|p|\geq0$ and $\sigma>0$, we choose constants 
$\rho=\rho(\sigma)$, $\delta_0=\delta_0(\sigma)$, $\e_0=\e_0(\sigma)>0$  
sufficiently small so that the renormalization map
$\ren$ is codimension 2 contractive in the sense of (~\ref{codim2contr})
for all $0\leq\e\leq\e_0(\sigma)$ and
$0\leq\delta\leq\delta_0(\sigma)+2\e_0(\sigma)$.

We choose the electron charge $g$ sufficiently small so
that in the first decimation step,
$\h^{(0)}$ defined in Section {~\ref{firstdecsubsect}} 
belongs to $\Polyd(\e_0(\sigma),\delta_0(\sigma),\frac12)$.

Then, as in (~\ref{H-n-formulas-recall-1}), we recursively define
\eqn
		\h^{(n)} [ e_{(n,\infty)}] = \ren[\h^{(n-1)}[ e_{(n-1,\infty)}]] \;\;,\;\;n\geq1\;,
\eeqn
i.e.  $\{\h^{(n)}[ e_{(n,\infty)}] \}_{n\in\N_0}$ is the orbit generated by $\ren$ 
with initial condition $\h^{(0)}[ e_{(0,\infty)} ]$. Here,
\eqn
		e_{(n,\infty)}&\equiv&e_{(n,\infty)}[p]
		\nonumber\\
		\h^{(n)}[ e_{(n,\infty)}]&\equiv&\h^{(n)}[p; e_{(n,\infty)}[p]]\;.
\eeqn
The dependence on $p$ is not explicitly accounted for in the notation; 
we are only interested in small $|p|$,
and will eventually let $|p|\rightarrow0$.

By Theorem {~\ref{codim2contrthm}}, $\h^{(n)}$ is an element
of the polydisc $\Polyd(\e_n,\delta_n,\lTnl_n)$, for any $n\in\N_0$, where
we recall that
\eqn
		\e_n&\leq&2^{-n}\e_0(\sigma)
		\nonumber\\
		\delta_n&\leq&\delta_0(\sigma)+2\e_0(\sigma)
        \nonumber\\
        \lTnl_n&=&\frac{1}{2}\rho^n \;.
\eeqn
Correspondingly, $H[\h^{(n)}[ e_{(n,\infty)} ]$
is a bounded, selfadjoint (since $e_{(n,\infty)}$ is real) 
operator on $\H_{red}$ for every $n\in\N_0$. 

To prove Theorem {~\ref{derp2Eidthm}}, we shall need the following identity.

\begin{lemma}
\label{first-last-scale-lemma-1}
Let, for $n> m\geq0$,
\eqn
        Q_{(m,n)}&:=&Q_{(m)}\Gamma_\rho^* Q_{(m+1)}\Gamma_\rho^*
        \cdots Q_{(n-1)}\Gamma_\rho^*\;,
        \nonumber\\
        Q_{(m,n)}^\sharp&:=&\Gamma_\rho 
        Q_{(n-1)}^\sharp\Gamma_\rho \cdots Q_{(m+1)}^\sharp\Gamma_\rho
        Q_{(m)}^\sharp
\eeqn
and 
\eqn
		Q_{(-1,n)}&:=&Q_{(-1)}Q_{(0,n)}\;,
		\nonumber\\
		Q_{(-1,n)}^\sharp&:=& Q_{(0,n)}^\sharp Q_{(-1)}^\sharp\;,
\eeqn
where for $k\geq0$,
\eqn
		Q_{(k)}^{(\sharp)}=Q_{\chi_{\rho}[H_f]}^{(\sharp)}(H_{(k)},H_f)\;,
\eeqn
and 
\eqn
		Q_{(-1)}^{(\sharp)}=Q_{\chi_{1}[H_f]}^{(\sharp)}(H_{(-1)},H_f)
\eeqn
(see (~\ref{Qjdefrecall})).  
Then, the identities
\eqn
		H_{(-1)}Q_{(-1,n)}&=&\rho^n (\Gamma_\rho^*)^n \chi_1[H_f] H_{(n)} \;,
		\nonumber\\
		Q_{(-1,n)}^\sharp H_{(-1)}&=&\rho^n  H_{(n)}\chi_1[H_f] (\Gamma_\rho)^n
		\label{HQ-id-min1-n-1}
\eeqn
and the "collapsing identity" 
\eqn
		Q^\sharp_{(-1,n)}H_{(-1)}Q_{(-1,n)} 
		&=&\rho^{n} 
		\Big[H_{(n)} - H_{(n)}\bar\chi_1[H_f]H_f^{-1}\bar\chi_1[H_f]H_{(n)}\Big] 
		\label{QHQ-id-basic-1}
\eeqn
(which directly relates the fiber Hamiltonian $H_{(-1)}$ to the effective
Hamiltonian $H_{(n)}$ of scale $n$)
are satisfied for $n\geq0$.
\end{lemma}

\prf
The identity (~\ref{HQ-id-min1-n-1}) follows from recursively applying 
the intertwining identities
(~\ref{HQ-intertw-id-0}) and (~\ref{HQ-intertw-id-n}).

To prove (~\ref{QHQ-id-basic-1}), we first verify that whenever $m-n\geq1$ and $m\geq0$,
\eqn
		Q^\sharp_{(m,n)}H_{(m)}Q_{(m,n)}=\rho \,Q^\sharp_{(m+1,n)}H_{(m+1)}Q_{(m+1,n)}\;.
\eeqn
This follows from
\eqn
		Q^\sharp_{(m,n)}H_{(m)}Q_{(m,n)}&=&Q^\sharp_{(m+1,n)}\Gamma_\rho 
		Q_{(m)}^\sharp H_{(m)}Q_{(m)}\Gamma_\rho^*Q_{(m+1,n)} 
		\nonumber\\
		&=&\rho \, Q^\sharp_{(m+1,n)} H_{(m+1)} Q_{(m+1,n)}
		\label{QmHmQm-id-1}
		\\
		&&
		- \rho \, Q^\sharp_{(m+1,n)} 
		H_{(m+1)}\bar\chi_1[H_f]H_f^{-1}\bar\chi_1[H_f] H_{(m+1)} Q_{(m+1,n)}\;,
		\nonumber
\eeqn
since
\eqn
		&&\Gamma_\rho 
		Q_{(m)}^\sharp H_{(m)}Q_{(m)}\Gamma_\rho^*
		\nonumber\\
		&=&
		\Gamma_\rho \Big[
		F_{\chi_\rho[H_f]}( H_{(m)},H_f)
		\nonumber\\
		&&-
		F_{\chi_\rho[H_f]}( H_{(m)},H_f)\bar\chi_\rho[H_f]H_f^{-1}\bar\chi_\rho[H_f] 
		F_{\chi_\rho[H_f]}( H_{(m)},H_f)\Big]\Gamma_\rho^*
		\nonumber\\
		&=&\rho \Big[H_{(m+1)}-H_{(m+1)}\bar\chi_1[H_f]H_f^{-1}\bar\chi_1[H_f]
		H_{(m+1)}\Big] \;,
\eeqn
using (~\ref{QHQid}).
Next, we observe that the last term in (~\ref{QmHmQm-id-1}) yields
\eqn		
		\label{QHCHQ-canc-1}&&
		\\
		&&\rho \, Q^\sharp_{(m+1,n)} 
		H_{(m+1)}\bar\chi_1[H_f]H_f^{-1}\bar\chi_1[H_f] H_{(m+1)} Q_{(m+1,n)}
		\nonumber\\
		&=&\rho \, Q^\sharp_{(m+2,n)}\Gamma_\rho Q^\sharp_{(m+1)} 
		H_{(m+1)}\bar\chi_1[H_f]H_f^{-1}\bar\chi_1[H_f] H_{(m+1)} Q_{(m+1)}\Gamma_\rho^*Q_{(m+1,n)}
		\nonumber\\
		&=&\rho^3 \, Q^\sharp_{(m+2,n)} H_{(m+2)}\chi_1[H_f]\Gamma_\rho
		\bar\chi_1[H_f] H_f^{-1} \bar\chi_1[H_f]
		\Gamma_\rho^* \chi_1[H_f] H_{(m+2)} Q_{(m+2,n)}
		\nonumber\\
		&=&\rho^2 \, Q^\sharp_{(m+2,n)} H_{(m+2)}\chi_1[H_f]\Gamma_\rho
		\bar\chi_{\rho^{-1}}[H_f] H_f^{-1} \bar\chi_{\rho^{-1}}[H_f]
		\chi_{1}[H_f] H_{(m+2)} Q_{(m+2,n)}
		\nonumber\\
		&=&0\;,
		\nonumber
\eeqn
since the supports of $\chi_1$ and $\bar\chi_{\rho^{-1}}$ do not intersect.
Thus, we conclude by recursion that
\eqn
		Q^\sharp_{(m,n)}H_{(m)}Q_{(m,n)}&=&\cdots\;=\;\rho^{n-m-1} \,Q^\sharp_{(n-1)}H_{(n-1)}Q_{(n-1)}
		\nonumber\\
		&=&\rho^{n-m}\Big[H_{(n)}-H_{(n)}\bar\chi_1[H_f]H_f^{-1}\bar\chi_1[H_f]H_{(n)}\Big]
\eeqn
for $m\geq0$. For $m=-1$, we have
\eqn
		Q^\sharp_{(-1,n)}H_{(-1)}Q_{(-1,n)} &=& 
		Q^\sharp_{(0,n)}Q^\sharp_{(-1)}H_{(-1)}Q_{(-1)}Q_{(0,n)} 
		\nonumber\\
		&=&Q^\sharp_{(0,n)} H_{(0)} Q_{(0,n)}
		\\
		&&-Q^\sharp_{(0,n)} H_{(0)}\bar\chi_1[H_f]H_f^{-1}\bar\chi_1[H_f]H_{(0)} Q_{(0,n)}\;,
		\nonumber
\eeqn
and the expression on the last line vanishes for $n\geq1$ by the same arguments as in 
(~\ref{QHCHQ-canc-1}).
We thus obtain (~\ref{QHQ-id-basic-1}).
\endprf

\begin{lemma}
For $n\geq1$, and $|p|\geq0$ sufficiently small,
\eqn
		\Bra\vac\,,\,\partial_{H_f}H_{(n)}\vac\Ket
		\;=\;\Big\|\,Q_{(-1,n)}\vac\,\Big\|^2+\err_n \;,
		\label{derHf-Hn-id-1}
\eeqn
where $\vac$ is the Fock vacuum, and
\eqn
		|\err_n|<c\,\frac{\e_n^2}{\rho^3}
\eeqn
for some finite constant $c$ which is independent of $n$ and $\rho$.
\end{lemma}

\prf
By the previous lemma,
\eqn
		\Bra\vac\,,\,\partial_{H_f}H_{(n)}\vac\Ket&=&(I)+(II) \;,
\eeqn
where
\eqn
		(I)&:=&\rho^{-n}
		\Bra\vac\,,\,\partial_{H_f}\Big[Q_{(-1,n)}^\sharp H_{(-1)}Q_{(-1,n)}\Big]\vac\Ket
		\nonumber\\
		(II)&:=&\Bra\vac\,,\,\partial_{H_f}\Big[H_{(n)}\bar\chi_1[H_f]H_f^{-1}\bar\chi_1[H_f]
		H_{(n)}\Big]\vac\Ket \;.
\eeqn
The term $(II)$ can be easily estimated,
\eqn
		(II)&=&\Bra\vac\,,\,\partial_{H_f}\Big[W_{(n)}\bar\chi_1[H_f]H_f^{-1}\bar\chi_1[H_f]
		W_{(n)}\Big]\vac\Ket
		\nonumber\\
		&\leq&c\|W_{(n)}\|_{op}\|\partial_{H_f}W_{(n)}\|_{op}
		+c\|W_{(n)}\|_{op}^2
		\nonumber\\
		&<&c\|w^{(n)}\|_\sigma^2
		\nonumber\\
		&<& c\e_n^2   \;,
\eeqn
as it only depends on operators on the scale $n$.

To study the term $(I)$, we note that
\eqn
		\partial_{H_f}\Gamma_\rho=\rho\Gamma_\rho\partial_{H_f}
		\;\; , \;\;
		\partial_{H_f}\Gamma_\rho^*=\rho^{-1}\Gamma_\rho^*\partial_{H_f}\;.
\eeqn
Accordingly,
\eqn
		(I)=(I_1)+(I_2)+(I_3) \;,
\eeqn
where
\eqn		\;\;\;\;
		(I_1)&:=&\sum_{j=-1}^{n-1} \rho^{\min\{0,-j\}}
		\Bra\vac\,,\,Q_{(-1,n)}^\sharp H_{(-1)}Q_{(-1,j-1)} 
		(\partial_{H_f}Q_{(j)})\Gamma_\rho^* Q_{(j+1,n)}\vac\Ket \;,
		\nonumber\\
		&&
\eeqn
and
\eqn
		(I_2)&:=&(I_1)^*
		\nonumber\\
		(I_3)&:=&\Bra\vac\,,\,Q_{(-1,n)}^\sharp (\partial_{H_f}H_{(-1)})Q_{(-1,n)}\vac \Ket\;.
\eeqn
The factor $\rho^{-j}$ in the $j$-th term of $(I_1)$ stems from pulling the operator $\partial_{H_f}$
through the $n$ dilatation operators $\Gamma_\rho$ contained in $Q_{(-1,n)}^\sharp$,
which produces a factor $\rho^n$, and through the $j$ inverse dilatation operators
$\Gamma_\rho^*$ in $Q_{(-1,j-1)}\Gamma_\rho^*$, which generates a factor $\rho^{-j}$.
Together with the overall factor $\rho^{-n}$ contained in $(I)$, we obtain the factor $\rho^{-j}$
for $j\geq0$. 
In the case of $(I_3)$, the overall factor $\rho^{-n}$ has been cancelled by a factor $\rho^n$
obtained from pulling the operator $\partial_{H_f}$ through the $n$ dilatation operators $\Gamma_\rho$
contained in $Q_{(-1,n)}^\sharp$.

Since $\partial_{H_f}H_{(-1)}=1$, it is clear that $(I_3)$ is the desired main
term in (~\ref{derHf-Hn-id-1}).

It remains to show that $(I_{i})$, $i=1,2$, contribute only to the error
$\err_n$.
To this end, we note that
\eqn
		&& \rho^{-j}Q_{(-1,n)}^\sharp H_{(-1)}Q_{(-1,j-1)} 
		(\partial_{H_f}Q_{(j)})\Gamma_\rho^* Q_{(j+1,n)}
		\nonumber\\
		&=& \rho^{n-j} H_{(n)}\chi_1[H_f](\Gamma_\rho)^n Q_{(-1,j-1)} 
		(\partial_{H_f}Q_{(j)})\Gamma_\rho^* Q_{(j+1,n)}
		\nonumber\\
		&=& \rho^{n-j} H_{(n)}(\Gamma_\rho)^n \chi_1[\rho^{-n}H_f] Q_{(-1,j-1)} 
		(\partial_{H_f}Q_{(j)})\Gamma_\rho^* Q_{(j+1,n)}
		\nonumber\\
		&=& \rho^{n-j} H_{(n)}(\Gamma_\rho)^{n-j} \chi_1[\rho^{-n+j}H_f]  
		(\partial_{H_f}Q_{(j)})\Gamma_\rho^* Q_{(j+1,n)}
		\nonumber\\
		&=& \left\{
		\begin{array}{rl}
		0 &{\rm if}\;j<n-1\\
		\rho H_{(n)}\Gamma_\rho\chi_\rho[H_f] \partial_{H_f} Q_{(n-1)}\Gamma_\rho^*
		&{\rm if}\;j=n-1
		\end{array}
		\right.\;,
		\label{err-jscale-1}
\eeqn
where the 'collapse' to the scale $n$ is a consequence of
\eqn
		\chi_1[\rho^{-n}H_f]Q_{(-1,j-1)}&=&\chi_1[\rho^{-n}H_f]Q_{(-1)}Q_{(0)}\Gamma_\rho^*
		Q_{(1)}\Gamma_\rho^*\cdots Q_{(j-1)}\Gamma_\rho^*
		\nonumber\\
		&=&\chi_1[\rho^{-n}H_f] Q_{(0)}\Gamma_\rho^*
		Q_{(1)}\Gamma_\rho^*\cdots Q_{(j-1)}\Gamma_\rho^*
		\nonumber\\
		&=&\Gamma_\rho^*\chi_1[\rho^{-n+1}H_f]  
		Q_{(1)}\Gamma_\rho^*\cdots Q_{(j-1)}\Gamma_\rho^*
		\nonumber\\
		&=&
		\cdots = (\Gamma_\rho^*)^k\chi_1[\rho^{-n+k}H_f] 
		Q_{(k)}\Gamma_\rho^*\cdots Q_{(j-1)}\Gamma_\rho^*
		\nonumber\\
		&=&(\Gamma_\rho^*)^{j}\chi_1[\rho^{-n+j}H_f]\;,
\eeqn
since for all $m>1$,
\eqn
		\chi_1[\rho^{-m}H_f]Q_{(k)}&=&
		\chi_1[\rho^{-m}H_f](\chi_\rho[H_f]
		\nonumber\\
		&&\hspace{2cm}-\bar\chi_\rho[H_f]\bar R_{(k)}\bar\chi_\rho[H_f]
		(H_{(k)}-H_f)\chi_\rho[H_f])
		\nonumber\\
		&=&
		\chi_1[\rho^{-m}H_f]\chi_\rho[H_f]
		\nonumber\\
		&=&\chi_1[\rho^{-m}H_f]  \;,
\eeqn
see also (~\ref{AQpullthrform333}).
Furthermore,
\eqn
		\chi_1[\rho^{-n+j}H_f]\partial_{H_f}Q_{(j)}=0
\eeqn
for $j<n-1$, because
\eqn
		\partial_{H_f}Q_{(j)}&=&\partial_{H_f}\chi_\rho[H_f]
		-P_{\Ran(\bar\chi_\rho[H_f])}\partial_{H_f}\Big[\bar\chi_\rho[H_f]\bar R_{(n)}
		\nonumber\\
		&&\hspace{2cm}
		\bar\chi_\rho[H_f](H_{(n)}-H_f)\chi_\rho[H_f]\Big]\;,
\eeqn
and
\eqn
		\chi_1[\rho^{-n+j}H_f]\partial_{H_f}\chi_\rho[H_f]&=&0
		\nonumber\\
		\chi_1[\rho^{-n+j}H_f]P_{\Ran(\bar\chi_\rho[H_f])}&=&0
\eeqn
for $j<n-1$. Here, we have used that for $\rho\leq\frac12$, 
${\rm supp}(\partial_{x}\chi_\rho[x])\subseteq[\frac{3\rho}{4},\rho]$
does not intersect ${\rm supp}(\chi_1[\rho^{-n+j}x])\subseteq[0,\rho^2]$ if $j<n-1$
(see the definition of $\chi_1$ in (~\ref{Thetadef}) and below).
Moreover, $P_{\Ran(\bar\chi_\rho[H_f])}$ denotes the orthogonal projection 
onto the range of $\bar\chi_\rho[H_f]$.
One finds
\eqn
		\rho \Big|\Bra\vac\,,\,H_{(n)}\Gamma_\rho\chi_\rho[H_f] \partial_{H_f}  Q_{(n-1)}\vac\Ket\Big|
		&\leq&\rho \Big\|\chi_1[H_f] W_{(n)}\vac\Big\|\,\Big\|  \partial_{H_f}  Q_{(n-1)}\vac\Big\|
		\nonumber\\
		&<&c\rho^{n-j-3}\e_n^2
\eeqn
for a constant $c$ which is independent of $\rho$, $n$, and $j$. Here, we have used (~\ref{der-Qn-est-1}).
We get
\eqn
		|\,(I_1)\,|&<&c\sum_{j=0}^{n-1}\rho^{n-j-3}\e_n^2
		\nonumber\\
		&<& c\,\frac{\e_n^2}{\rho^3} \;.
\eeqn 
Thus,
\eqn
		\err_n:=(I_1)+(I_2)+(II)
\eeqn
is bounded by $\frac{c\e_n^2}{\rho^3}$ as claimed.
\endprf

\begin{proposition}
\label{partHf-Hn-prop-1}
Let $\Psi(0,\sigma)$ denote the ground state of the fiber Hamiltonian $H(0,\sigma)$
for conserved momentum $p=0$, normalized by
\eqn
		\Bra\vac\,,\,\Psi(0,\sigma)\Ket=1 \;.
		\label{gspsi-normalization-1}
\eeqn
Then,
\eqn
		\lim_{n\rightarrow\infty}\Bra\vac\,,\,
		(\partial_{H_f}H_{(n)})\Big|_{p=0}\,\vac\Ket
		=\Bra\Psi(0,\sigma),\Psi(0,\sigma)\Ket \;.
\eeqn
\end{proposition}

\prf
We recall that the ground state of $H(p,\sigma)$, derived from
\eqn
		\Psi(p,\sigma)=\lim_{n\rightarrow\infty}Q_{(-1,n)}\vac \;,
		\label{Psi-recall-Qvac-1}
\eeqn
satsifies the normalization condition 
\eqn
		\Bra\vac\,,\,\Psi(p,\sigma)\Ket=1 \;,
\eeqn
independently of $p$ and $\sigma>0$, see
(~\ref{psi-vac-norm-1}). Furthermore, $\lim_{n\rightarrow\infty}\err_n=0$,
since $\e_n\rightarrow0$ as $n\rightarrow\infty$. 
Thus, the assertion of this proposition
is an immediate corollary of the previous lemma.
\endprf

Next, we shall address derivatives in $\Ppar:=\langle P_f,n\rangle_{\R^3}$, where
$n\in\R^3$, $|n|=1$ is an arbitrary unit vector.

\begin{lemma}
\label{Q-intertw-id-scales-1}
The identity 
\eqn 
	\Bra\phi\,,\,(\partial_{\Ppar}Q_{(-1,n)}^\sharp) \psi\Ket= 
	\rho^n \Bra (\partial_{\Ppar}Q_{(-1,n)}) \phi\,,\,\psi\Ket
	\label{part-Ppar-QQ-1}
\eeqn
holds for $n\geq0$, and any $\psi\in\Fo$, $\phi\in\H_{red}$.
\end{lemma}

\prf
The intertwining relations between the operators $\partial_{\Ppar}$, $\Gamma_\rho$
and $\Gamma_\rho^*$ are given by
\eqn 
	\partial_{\Ppar}\Gamma_\rho^* = \frac1\rho \Gamma_\rho^*\partial_{\Ppar}
	\; \; \; , \; \; \;
	\partial_{\Ppar}\Gamma_\rho = \rho \Gamma_\rho\partial_{\Ppar} \;.
\eeqn
We thus find
\eqn 
	\partial_{\Ppar}Q_{(0,n)}^\sharp=
	\sum_{j=n}^{-1}\rho^{\min\{n,n-j\}}Q_{(j+1,n)}^\sharp\Gamma_\rho^*  (\partial_{\Ppar}Q_{(j)})
	Q_{(-1,j-1)}^\sharp \;.
	\label{derPpar-Q-scales-1}
\eeqn
There is a factor $\rho^{ n-j }$ in the $j$-th term of the sum if $0\leq j\leq n$,
because $\partial_{\Ppar}$ is pulled to the right
through $n-j$ dilatation operators $\Gamma_\rho$ until it acts on $Q_{(j)}$.
Only for $j=-1$, no additional factor $\rho$ is introduced.

On the other hand,
\eqn 
	\partial_{\Ppar}Q_{(0,n)}=\sum_{j=-1}^{n}\rho^{\min\{0,-j\}}
	Q_{(-1,j-1)}  (\partial_{\Ppar}Q_{(j)})
	\Gamma_\rho^* Q_{(j+1,n)} \;.
	\label{derPpar-Q-scales-2}
\eeqn
The factor $\rho^{-j}$ in the $j$-th term of the sum arises because 
$\partial_{\Ppar}$ is pulled to the right
through $j$ dilatation operators $\Gamma_\rho^*$ to act on $Q_{(j)}$.
Only for $j=-1$, there is no additional factor $\rho$.

Direct comparison of (~\ref{derPpar-Q-scales-1}) and (~\ref{derPpar-Q-scales-2})
establishes (~\ref{part-Ppar-QQ-1}).
\endprf

Next, we link derivatives in $|p|$ at $p=0$ with derivatives in $\Ppar$.

\begin{lemma}
\label{derp-derPpar-psi-lm-1}
Let $\Psi(p,\sigma)$ denote the ground state of $H(p,\sigma)$ as defined in 
(~\ref{Psi-recall-Qvac-1}).
Then,
\eqn 
	(\partial_{|p|}\Psi)(0,\sigma)=-\lim_{n\rightarrow\infty}
	\partial_{\Ppar}Q_{(-1,n)}\Big|_{p=0}\vac
\eeqn
\end{lemma}

\prf
We note that from
\eqn
		H_{(-1)}\Psi_{(-1,\infty)}=0
\eeqn
clearly follows that
\eqn 
	\partial_{|p|}H_{(-1)}\Big|_{p=0}\Psi(0,\sigma)
	=- H_{(-1)}\Big|_{p=0}(\partial_{|p|}\Psi)(0,\sigma) \;.
	\label{derp-H-id-eq-1}
\eeqn
By $O(3)$-symmetry, and differentiability of $E(p,\sigma)$ at $p=0$,
\eqn
	(\partial_{|p|}E)(0,\sigma)=0 \;.
\eeqn
Thus, one immediately verifies that
\eqn 
	\partial_{|p|}H_{(-1)}\Big|_{p=0}
	&=&-\partial_{\Ppar}H_{(-1)}\Big|_{p=0}\;.
\eeqn
On the other hand,
\eqn 
	\partial_{\Ppar}H_{(-1)}\Big|_{p=0}\Psi(0,\sigma)
	= - H_{(-1)}\Big|_{p=0}(\partial_{\Ppar}Q_{(-1,\infty)}\Big|_{p=0})\vac\;,
	\label{derp-H-id-eq-2}
\eeqn
where we recall that $\Psi(0,\sigma)=Q_{(-1,\infty)}\Big|_{p=0}\vac$.
For brevity, let 
\eqn 
	\phi&:=&(\partial_{|p|}\Psi)(0,\sigma)
	\nonumber\\
	\zeta&:=&\partial_{\Ppar}Q_{(-1,\infty)}\Big|_{p=0} \vac \;.
\eeqn
Comparing (~\ref{derp-H-id-eq-1}) and (~\ref{derp-H-id-eq-2}), we find that
\eqn 
	H_{(-1)}\Big|_{p=0}(\phi+\zeta)=0\;.
\eeqn
Thus,  
\eqn 
	\phi+\zeta=\lambda \Psi(0,\sigma)
\eeqn
for some $\lambda\in\C$. We shall next show that $\lambda=0$. To this end, we prove that
\eqn 
	\Bra\vac\,,\,\phi\Ket=0=\Bra\vac\,,\,\zeta\Ket \;.
	\label{phi-zeta-vac-1}
\eeqn
Since
\eqn 
	\Bra\vac\,,\,\Psi(0,\sigma)\Ket=1\;,
\eeqn
cf. (~\ref{psi-vac-norm-1}),
this immediately implies $\lambda=0$.

To prove (~\ref{phi-zeta-vac-1}), we observe that  
\eqn 
	\Bra\vac\,,\,\phi\Ket&=&\lim_{n\rightarrow\infty}
	\sum_{j=-1}^n
	\Bra\vac\,,\,Q_{(-1,j-1)} (\partial_{|p|}Q_{(j)})
	\Gamma_\rho^* Q_{(j+1,n)}\Big|_{p=0}\vac\Ket
	\nonumber\\
	&=&
	\lim_{n\rightarrow\infty}
	\sum_{j=-1}^n
	\Bra\vac\,,\, (\partial_{|p|}Q_{(j)})
	\Gamma_\rho^* Q_{(j+1,n)}\Big|_{p=0}\vac\Ket
	\nonumber\\
	&=&0 
\eeqn
(passing to the second line, we have used (~\ref{AQpullthrform333})), since
\eqn 
	&&\Bra\vac\,,\,(\partial_{|p|}Q_{(j)})\phi'\Ket 
	\nonumber\\
	&=&
	\Bra\vac\,,\,\Big[\partial_{|p|}\Big(\chi_\rho[H_f]-\bar\chi_\rho[H_f]
	\bar R_{(j)}\bar\chi_\rho[H_f](H_{(j)}-H_f)\chi_\rho[H_f]\Big)\Big]\phi'\Ket
	\nonumber\\
	&=&
	-\Bra\vac\,,\,\Big[\partial_{|p|}\Big( \bar\chi_\rho[H_f]
	\bar R_{(j)}\bar\chi_\rho[H_f](H_{(j)}-H_f)\chi_\rho[H_f]\Big)\Big]\phi'\Ket
	\nonumber\\
	&=&
	-\Bra\bar\chi_\rho[H_f]\vac\,,\,\Big[\partial_{|p|}\Big( 
	\bar R_{(j)}\bar\chi_\rho[H_f](H_{(j)}-H_f)\chi_\rho[H_f]\Big)\Big]\phi'\Ket
	\nonumber\\
	&=&0
\eeqn
for any vector $\phi'\in\Ran(\chi_\rho[H_f])$.

Similarly,
\eqn 
	\Bra\vac\,,\,\zeta\Ket&=&\lim_{n\rightarrow\infty}
	\sum_{j=-1}^n \rho^{-\min\{0,j\}}
	\Bra\vac\,,\,Q_{(-1,j-1)} (\partial_{\Ppar}Q_{(j)})
	\Gamma_\rho^* Q_{(j+1,n)}\Big|_{p=0}\vac\Ket
	\nonumber\\
	&=&
	\lim_{n\rightarrow\infty}
	\sum_{j=-1}^n \rho^{-\min\{0,j\}}
	\Bra\vac\,,\, (\partial_{\Ppar}Q_{(j)})
	\Gamma_\rho^* Q_{(j+1,n)}\Big|_{p=0}\vac\Ket
	\nonumber\\
	&=&0 \;.
\eeqn
This concludes the proof.
\endprf

\begin{proposition}
\label{partPpar2-Hn-prop-1}
We have
\eqn 
	\lim_{n\rightarrow\infty}\rho^{-n}
	\Bra\vac\,,\,\partial_{\Ppar}^2H_{(n)}\Big|_{p=0}\vac\Ket
	&=&\Bra\Psi(0,\sigma)\,,\,\Psi(0,\sigma)\Ket
	\nonumber\\
	&-&
	2\Bra(\partial_{|p|}\Psi)(0,\sigma) \,,\,
	H_{(-1)}\Big|_{p=0}
	(\partial_{|p|}\Psi)(0,\sigma)\Ket 
\eeqn
\end{proposition}

\prf
(~\ref{QHQ-id-basic-1}) yields
\eqn 
	\partial_{\Ppar}^2H_{(n)}\Big|_{p=0} &=&\rho^{-n}\partial_{\Ppar}^2\Big[
	Q_{(-1,n)}^\sharp H_{(-1)} Q_{(-1,n)}\Big|_{p=0}\Big]
	\nonumber\\
	&&\hspace{3cm}
	+\partial_{\Ppar}^2\Big[H_{(n)}\bar\chi_1[H_f]H_f^{-1}\bar\chi_1[H_f]H_{(n)}\Big|_{p=0}\Big] 
\eeqn
We note for the first term after the equality sign 
that for every operator $\partial_{\Ppar}$ which is pulled through $Q^\sharp_{(-1,n)}$
from the left, we obtain a factor $\rho^n$. Therefore,
\eqn
	\partial_{\Ppar}^2H_{(n)}\Big|_{p=0} &=&\rho^{n}
	Q_{(-1,n)}^\sharp(\partial_{\Ppar}^2 H_{(-1)}) Q_{(-1,n)}\Big|_{p=0}
	\nonumber\\
	&&\hspace{3cm}
	+2 (\partial_{\Ppar}Q_{(-1,n)}^\sharp) H_{(-1)} (\partial_{\Ppar}Q_{(-1,n)})\Big|_{p=0}
	\nonumber\\
	&&+2\rho^{n}
	Q_{(-1,n)}^\sharp(\partial_{\Ppar} H_{(-1)})(\partial_{\Ppar} Q_{(-1,n)})\Big|_{p=0}
	\nonumber\\
	&&\hspace{3cm}+2(\partial_{\Ppar}
	Q_{(-1,n)}^\sharp)(\partial_{\Ppar} H_{(-1)}) Q_{(-1,n)}\Big|_{p=0} 
	\nonumber\\
	&&+\rho^{-n}(\partial_{\Ppar}^2 
	Q_{(-1,n)}^\sharp) H_{(-1)} Q_{(-1,n)}\Big|_{p=0} 
	\nonumber\\
	&&\hspace{3cm}+\rho^{n}
	Q_{(-1,n)}^\sharp H_{(-1)} (\partial_{\Ppar}^2 Q_{(-1,n)})\Big|_{p=0} 
	\nonumber\\ 
	&&+\rho^n\partial_{\Ppar}^2\Big[H_{(n)}\bar\chi_1[H_f]
	H_f^{-1}\bar\chi_1[H_f]H_{(n)}\Big|_{p=0}\Big] \;.
\eeqn
Using
\eqn
	(\partial_{\Ppar} H_{(-1)}) Q_{(-1,n)}\Big|_{p=0} 
	&=&- H_{(-1)}(\partial_{\Ppar} Q_{(-1,n)} )\Big|_{p=0}
	\nonumber\\
	&&
	+(\Gamma_\rho^*)^n\chi_1[H_f](\partial_{\Ppar}H_{(n)})\Big|_{p=0}
\eeqn
and
\eqn
	Q_{(-1,n)}^\sharp(\partial_{\Ppar} H_{(-1)}) \Big|_{p=0} 
	&=&-\rho^{-n}  (\partial_{\Ppar} Q_{(-1,n)} )H_{(-1)}\Big|_{p=0}
	\nonumber\\
	&&
	+(\partial_{\Ppar}H_{(n)})\Big|_{p=0} \chi_1[H_f](\Gamma_\rho)^n\;,
\eeqn
we obtain
\eqn 
	&&\Bra\vac\,,\,(\partial_{\Ppar}^2H_{(n)})\Big|_{p=0} \vac\Ket
	\nonumber\\
	&=&\rho^{n}
	\Bra\vac \,,\, Q_{(-1,n)}^\sharp(\partial_{\Ppar}^2H_{(-1)})Q_{(-1,n)}\Big|_{p=0}\vac\Ket
	\nonumber\\
	&-&
	2\Bra\vac\,,\,
	(\partial_{\Ppar}Q_{(-1,n)}^\sharp)H_{(-1)}(\partial_{\Ppar}Q_{(-1,n)} )\Big|_{p=0}\vac\Ket
	+\tilde\err_n
	\nonumber\\ 
	&=&\rho^{n}\Bra\vac\,,\,
	Q_{(-1,n)}^\sharp Q_{(-1,n)}\Big|_{p=0}\vac\Ket
	\label{parialPpar-Hn-id-err-2}\\
	&-& 2\rho^{n}\Bra
	(\partial_{\Ppar}Q_{(-1,n)})\Big|_{p=0}\vac\,,\, 
	H_{(-1)}(\partial_{\Ppar}Q_{(-1,n)} )\Big|_{p=0}\vac\Ket 
	+\tilde\err_n
	\nonumber
\eeqn
where we have used Lemma {~\ref{Q-intertw-id-scales-1}}.
The error term is defined by
\eqn
	\tilde\err_n=\tilde\err_n^{(1)}+\tilde\err_n^{(2)}+\tilde\err_n^{(3)}
\eeqn
where
\eqn
	\tilde\err_n^{(1)}&:=&\rho^{n}\Bra\vac\,,\,
	Q_{(-1,n)}^\sharp  H_{(-1)} (\partial_{\Ppar}^2 Q_{(-1,n)})\Big|_{p=0}\vac\Ket
	+h.c.
	\nonumber\\
	\tilde\err_n^{(2)}&:=&\rho^n\Bra\vac\,,\,
	(\partial_{\Ppar} H_{(n)})\chi_1[H_f](\Gamma_\rho)^n(\partial_{\Ppar}Q_{(-1,n)})\vac\Ket
	+h.c.
	\nonumber\\
	\tilde\err_n^{(3)}&:=&\rho^n\Bra\vac\,,\,
	\partial_{\Ppar}^2\Big[H_{(n)}\bar\chi_\rho[H_f]H_f^{-1}\bar\chi_\rho[H_f]H_{(n)}\Big]\Big|_{p=0}
	\vac\Ket \;.
\eeqn
We find the following estimates.
\eqn
	|\tilde\err_n^{(1)}|&\leq&2\rho^{2n}\Big|\Bra\vac\,,\,
	H_{(n)}\chi_1[H_f](\Gamma_\rho)^n (\partial_{\Ppar}^2 Q_{(-1,n)})\Big|_{p=0}\vac\Ket\Big|
	\nonumber\\
	&=&2\rho^{n}\Big|\Bra\vac\,,\,
	H_{(n)} \chi_1[H_f](\Gamma_\rho)^n(\partial_{\Ppar}^2 Q_{(n-1)})\Big|_{p=0}\vac\Ket\Big|
	\nonumber\\
	&\leq&2\rho^{n}\Big\| \chi_1[H_f]
	W_{(n)}\vac\Big\|\, \Big\|(\partial_{\Ppar}^2 Q_{(n-1)})\Big|_{p=0}\vac \Big\|
	\nonumber\\
	&\leq& c\,\rho^{n-3}\,\e_n^2 \;,
\eeqn
using (~\ref{der-Qn-est-1}).
Furthermore,
\eqn
	|\tilde\err_n^{(2)}|&\leq&2\rho^n\Big|\Bra\vac\,,\,
	(\partial_{\Ppar} H_{(n)})\chi_1[H_f](\Gamma_\rho)^n(\partial_{\Ppar}Q_{(-1,n)})\vac\Ket\Big|
	\nonumber\\
	&\leq&2\rho^n\Big\| \chi_1[H_f](\partial_{\Ppar}Q_{(n-1)}^\sharp)\vac\Big\|\,
	\Big\|\chi_1[H_f](\partial_{\Ppar}W_{(n)})\vac \Big\|
	\nonumber\\
	&\leq& c\,\rho^{n-2}\,\e_n^2  \;,
\eeqn
using (~\ref{der-Qn-est-1}), and
\eqn
	|\tilde\err_n^{(3)}|&=&\rho^n\Bra\vac\,,\,
	\partial_{\Ppar}^2\Big[W_{(n)}\bar\chi_\rho[H_f]H_f^{-1}\bar\chi_\rho[H_f]W_{(n)}\Big]\Big|_{p=0}
	\vac\Ket
	\nonumber\\
	&<&\rho^n\|\bar\chi_\rho[H_f]H_f^{-1}\bar\chi_\rho[H_f]\|_{op}
	\Big(\sum_{a=0}^2\|\partial_{\Ppar}^a W_{(n)}\Big|_{p=0}\|_{op}\Big)^2
	\nonumber\\
	&<&c\,\rho^{n-1}\|\h_1^{(n)}\|_\sigma^2
	\nonumber\\
	&<&c\,\rho^{n-1}\,\e_n^2  \;.
\eeqn
Thus, $\lim_{n\rightarrow\infty} \rho^{-n}\tilde\err_n =0$, 
since $\e_n\rightarrow0$ as $n\rightarrow\infty$.

The claim follows from (~\ref{parialPpar-Hn-id-err-2}) and 
Lemma {~\ref{derp-derPpar-psi-lm-1}}.
\endprf

\noindent{\bf Proof of Theorem {~\ref{derp2Eidthm}}:}
From Propositions {~\ref{partHf-Hn-prop-1}} and {~\ref{partPpar2-Hn-prop-1}}, we find
\eqn
	\frac{1}{m_{ren}^*(0,\sigma)}&=&\lim_{n\rightarrow\infty}
	\frac{\rho^{-n}\bra\vac\,,\,(\partial_{\Ppar}^2H_{(n)}\Big|_{p=0})\vac\ket}
	{\bra\vac\,,\,(\partial_{H_f}H_{(n)}\Big|_{p=0})\vac\ket}
	\nonumber\\
	&=&1-
	2\frac{\bra(\partial_{|p|}\Psi)(0,\sigma)\,,\,H_{(-1)}\Big|_{p=0}
	(\partial_{|p|}\Psi)(0,\sigma)\ket}{\bra \Psi(0,\sigma)\,,\,\Psi(0,\sigma)\ket}
	\;, 
\eeqn
which agrees with the definition of the renormalized mass
\eqn
	\frac{1}{m_{ren}(0,\sigma)}&=&(\partial_{|p|}^2E)(0,\sigma)
	\nonumber
	\\
	&=&1-
	2\frac{\bra(\partial_{|p|}\Psi)(0,\sigma)\,,\,(H(0,\sigma)-E(0,\sigma))
	(\partial_{|p|}\Psi)(0,\sigma)\ket}{\bra \Psi(0,\sigma)\,,\,\Psi(0,\sigma)\ket}
	\;
\eeqn
from (~\ref{part2pEform}). This establishes the equality of the two notions $m_{ren}(0,\sigma)$
and $m_{ren}^*(0,\sigma)$ of the renormalized electron mass.

 
\section{Computation of the renormalized mass}
\label{ren-mass-comp-sect-1}

The renormalized mass can be determined to any arbitrary level of precision by use 
the identity
\eqn
		m_{ren}(0,0)&=&
		\frac{1+\sum_{n=0}^\infty \partial_{X_0}\Big|_{\spvar\rightarrow\unull}\Delta\gamma_{(n)}}
		{1+\sum_{n=0}^\infty \rho^{-n}
		\partial_{X^\parallel}^2\Big|_{\spvar\rightarrow\unull}\Delta\gamma_{(n)}}
		\label{part2Eidenty-2}
\eeqn
obtained from formula  (~\ref{part2Eidenty}).

This follows from the fact that for $n\geq-1$, 
\eqn
        \;\;\;\;\;
        \Delta \gamma_{(n)} [\spvar]=\sum_{L=2}^\infty(-1)^{L-1}
        \sum_{p_1,q_1,\dots,p_L,q_L: \atop p_\ell+q_\ell\geq1}
        V_{\underline{0,p,0,q}}^{(L)}
        \Big[\h^{(n)}[e_{(n,\infty)}]\,\Big|\,\spvar\Big]\Big|_{p=0}\;,
        \label{Deltw00-VL-def-1}
\eeqn
where the quantities $V_{\underline{0,p,0,q}}^{(L)}$, defined in (~\ref{VLmpnq-def-1}),
are explicitly computable. Indeed, the isospectral renormalization group provides
a constructive, finite, and convergent algorithm.

We recall here that $n=-1$ accounts for  
the first decimation step, which is discussed in Section {~\ref{firstdecsubsect}}. 
Through (\ref{Deltw00-VL-def-1}), the numerator and denominator in (~\ref{part2Eidenty}),
and thus
the renormalized mass $m_{ren}^*(0,0)$, can be determined up to any given level of precision.

\subsection{Determination of the renormalized mass in leading order}

As an application of our analysis, we shall here calculate the term of leading order 
$O(g^2)$ for the renormalized mass $m_{ren}(0,0)$, and rigorously bound the errors by $o(g^2)$.

\begin{theorem}
The renormalized mass at conserved momentum $p=0$, and in the limit $\sigma\rightarrow0$, 
is given by
\eqn
		m_{ren}(0,0)=1+\frac{8\pi }{3}g^2\tilde c_2 + O(g^{\frac73})\;.
\eeqn
where
\eqn
		\tilde c_2:=
		\int_{\R_+}dx\,\frac{\knull^2(x)}{1+\frac x2} \;,
\eeqn
and $\knull(x):=\lim_{\sigma\rightarrow0}\ks(x)$.
\end{theorem}

\prf
We set $\sigma=0$, and let 
\eqn
		\tilde H_{(n)}&=&H[\h^{(n)}[e_{(n,\infty)}]\Big|_{p=0}]
		\;=\;\tilde e_{(n,\infty)}\chi_1^2[H_f]+\tilde T_{(n)}+\tilde W_{(n)} 
\eeqn
denote the effective Hamiltonian on scale $n$, where
here and in the sequel, the tilde shall notationally account for evaluation at $p=0$.

We have
\eqn
		\h^{(n)}\in\Polyd^{(\mu=1)}(\e_n,\delta_n,\lTnl_n)
\eeqn
with 
\eqn
		\e_n&<&(\ccontr\rho)^n\e_0\;<\;2^{-n}\e_0
		\nonumber\\
		\delta_n&\leq&\delta_0+2\e_0
		\nonumber\\
		\lTnl_n&=&\frac{1}{2}\rho^n
\eeqn
($\ccontr$ is defined in (~\ref{ccontr-def-1})) and
\eqn 
		|e_{(n,\infty)}|&<&2^{-n}\e_0 \;,
\eeqn
where
\eqn
		\e_0,\delta_0\;=\;O(g)\;.
\eeqn
Codimension 2 contractivity of $\ren$ on $\Polyd(\e_0,\delta_0+2\e_0,1/2)$ is ensured by 
the requirement
\eqn
		\rho&<&\frac{1}{2\ccontr^2}  \;,
\eeqn 
see Section {~\ref{codim2contrproofsubsubsect-2}}. 

\subsubsection{Calculation of the leading term}
\label{calc-main-term-subsect-1}

We determine the leading order contribution $V_{\underline{0,1,0,1}}^{(L=2;n)}[\spvar]$ 
in (~\ref{Deltw00-VL-def-1}) explicitly for $n=-1$ and $n=0$.

We have
\eqn
		V_{\underline{0,1,0,1}}^{(L=2;-1)}[\spvar]&=&- 
		\Bra\vac\,,\, \tilde W_{0,1}^{(-1)}[\opp+\spvar]
		\bar\chi_1[H_f+ X_0]
		\nonumber\\
		&&\hspace{2cm}\bar R_0^{(-1)}[\opp+\spvar]\bar\chi_1[H_f+ X_0]
		\tilde W_{1,0}^{(-1)}[\opp+ \spvar]\vac\Ket  \;,
		\label{V-lead-min1-1}
\eeqn
and
\eqn
		V_{\underline{0,1,0,1}}^{(L=2;0)}[\spvar]&=&-\frac1\rho
		\Bra\vac\,,\, \tilde W_{0,1}^{(0)}[\opp+\rho\spvar]
		\bar\chi_\rho[H_f+\rho X_0]
		\nonumber\\
		&&\hspace{2cm}\bar R_0^{(0)}[\opp+\rho\spvar]\bar\chi_\rho[H_f+\rho X_0]
		\tilde W_{1,0}^{(0)}[\opp+\rho \spvar]\vac\Ket  \;.
		\label{V-lead-0-1}
\eeqn
We recall the definition of the operators appearing here.

In the case $n=-1$, we have
\eqn
		\tilde T_{(-1)}=e_{(-1,\infty)}\1+H_f+\frac{P_f^2}{2} \;,
\eeqn
so that 
\eqn
		\bar R_0^{(-1)}[\opp]&=&\Big[ H_f+\big(e_{(-1,\infty)}+\frac{1}{2}P_f^2\big)\bar\chi_1^2[H_f]\Big]^{-1} 
\eeqn
on the range of $\Ran(\bar\chi_1[H_f])$. Furthermore,
\eqn
		W_{1,0}^{(-1)}=\int_{\R^3}\frac{dK}{|k|^{\frac12}} a^*(K)w_{1,0}^{(-1)}[\opp;K]
		=(W_{0,1}^{(-1)})^*
\eeqn
with 
\eqn
		w_{1,0}^{(-1)}[\opp;K]=g\bra\eps(K)\,,\,P_f\ket_{\R^3}\knull(|k|) \;,
\eeqn
where $\knull(|k|)=\lim_{\sigma\rightarrow0}\ks(|k|)$.

In the case $n=0$,  our analysis of the first Feshbach decimation step in 
Section {~\ref{firstdecsubsect}} has yielded
\eqn
		\tilde T_{(0)}[\opp]&=&H_f+ \frac{1}{2}P_f^2\chi_1^2[H_f]
		-\frac14 P_f^4\frac{\bar\chi_1^2[H_f]\chi_1^2[H_f]}
		{H_f+ \frac{1}{2}P_f^2\bar\chi_1^2[H_f]}
		\nonumber\\
		&&\hspace{2cm}+\Delta \tilde T_{(0)}[\opp] \;,
\eeqn
where $\|\Delta \tilde T_{(0)}\|_{\Tspace}\leq O(g^2)$ contains all terms depending on
$e_{(-1,\infty)}=O(g^2)$. We therefore have
\eqn
		\bar R_0^{(0)}[\opp]&=&\Big[ H_f+\frac12 P_f^2
		\bar\chi_\rho^2[H_f]\chi_1^2[H_f]
		\nonumber\\
		&&\hspace{2cm}
		-\frac14 P_f^4\frac{\bar\chi_\rho^2[H_f]\chi_1^2[H_f]}
		{H_f+ \frac{1}{2}P_f^2\bar\chi_1^2[H_f]}\Big]^{-1} 
		+O(g^2)
\eeqn
on the range of $\Ran(\bar\chi_\rho[H_f])$, where the term of order $O(g^2)$
is small with respect to $\|\,\cdot\,\|_{\Tspace}$. Moreover, we recall that
\eqn
		\piop_1^{(-1)}[e_{(-1,\infty)};\opp]&=&\1-\frac{1}{2}P_f^2\frac{\bar\chi_1^2[H_f]}
		{H_f+\frac{1}{2}P_f^2\bar\chi_1^2[H_f]}
		\nonumber\\
		&&\hspace{2cm}+\Delta\piop_1^{(-1)}[e_{(-1,\infty)};\opp] \;,
\eeqn
where the terms depending on $e_{(-1,\infty)}=O(g^2)$ have been absorbed into
the error term with $\|\Delta\piop_1^{(-1)}\|_{\Tspace}\leq O(g^2)$.
Then,
\eqn
		W_{1,0}^{(0)}=\int_{B_1}\frac{dK}{|k|^{\frac12}} a^*(K)w_{1,0}^{(0)}[\opp;K]
		=(W_{0,1}^{(0)})^*
\eeqn
with  
\eqn
		w_{1,0}^{(0)}[\opp;K]=
		\piop_1^{(-1)}[\opp+\underline{k}]w_{1,0}^{(-1)}[\opp]\piop_1^{(-1)}[\opp]
		+\Delta w_{1,0}^{(0)}[\opp;K]\;,
\eeqn
where the analysis in Section {~\ref{codim2contrproofsubsubsect-2}} has shown that 
$\|\Delta w_{1,0}^{(0)}\|_\sigma^\sharp\leq O(\e_0^2)=O(g^2)$.

Evaluating (~\ref{V-lead-min1-1}) and (~\ref{V-lead-0-1}), 
we choose an arbitrary unit vector $n\in\R^3$ and differentiate with respect to
$X^\parallel=\langle X,n\rangle_{\R^3}$. The result for $n=-1$ and $n=0$ is  
\eqn
		\partial_{X^\parallel}^2\Big|_{\spvar\rightarrow\unull}V_{\underline{0,1,0,1}}^{(L=2;n)} 
		=\frac{8\pi }{3}g^2C_n  +O(g^3)
\eeqn
with
\eqn
		C_{-1}=\int_0^\infty dx\,\frac{\knull^2(x)\bar\chi_1^2[x]}{1+\frac{x}{2}\bar\chi_1^2[x]}
\eeqn
and 
\eqn
		C_0=\int_0^1 dx\frac{\bar\chi_\rho^2[x]\chi_1^2[x]
		(1-\frac x2\frac{\bar\chi_1^2[x]}{1+\frac x2\bar\chi_1^2[x]})^2}
		{1+\frac{x}{2}\chi_1^2[x]\bar\chi_\rho^2[x]
		-\frac{x^2}{4}\bar\chi_1^2[x]\chi_1^2[x]\frac{1}{1+\frac{x}{2}\bar\chi_1^2[x]}} \;.
\eeqn
A straightforward calculation (using $\chi_1^2+\bar\chi_1^2=1$) shows that
\eqn
		C_{-1}+C_0=\int_0^\infty dx\,\frac{\knull^2(x)\bar\chi_\rho^2[x]}{1+\frac{x}{2}\bar\chi_\rho^2[x]}\;,
\eeqn
which is similar to $C_{-1}$, but with $\bar\chi_1$ replaced by $\bar\chi_\rho$.
This is a consequence of the composition property of the smooth Feshbach map, see Section 
{~\ref{concatlawssect}}.

Thus clearly,
\eqn
		|C_{-1}+C_0-\tilde c_2|<O(\rho g^2)\;,
\eeqn
where $\rho$ is chosen $g$-dependent and small in the end.

\subsubsection{Higher order errors}

We shall next estimate the higher order corrections to the leading term.

\begin{lemma}
\label{gamma-V-diff-lemma-1}
For $n=-1$,
\eqn
		\Big\|\Delta \gamma_{(-1)}+V_{\underline{0,1,0,1}}^{(L=2;-1)}
		\Big\|_{\Tspace}\;<\;cg^3  \;,
		\label{rig-error-bd-1}
\eeqn
while for $n\geq0$, 
\eqn
		\Big\|\Delta \gamma_{(n)}+V_{\underline{0,1,0,1}}^{(L=2;n)}
		\Big\|_{\Tspace}\;<\;cg^3 \frac{(\ccontr\rho)^{3n}}{\rho^2} \;.
		\label{rig-error-bd-2}
\eeqn 
\end{lemma}

\prf
This follows from the calculation in (~\ref{hat-w-pnull-bd-1}), which gives
\eqn    &&\Big\|\sum_{L=3}^\infty(-1)^{L-1}
        \sum_{p_1,q_1,\dots,p_L,q_L: \atop p_\ell+q_\ell\geq1}
        V_{\underline{0,p,0,q}}^{(L;n)}
        \Big[\h^{(n)}[e_{(n,\infty)}]\,\Big|\,\spvar\Big]\Big|_{p=0}\Big\|_{\Tspace}
        \nonumber\\
        &\leq&20\,C_\Theta^2\rho\sum_{L=3}^\infty    (L+1)^2
        \Big(\frac{C_\Theta}{\rho}\Big)^L 
        \nonumber\\
        &&\hspace{0.5cm}\times\, 
        \sum_{p_1,q_1,\dots,p_L,q_L:\atop
        p_\ell+q_\ell\geq1} 
        \prod_{\ell=1}^L\Big\{\Big(\frac{2}{\sqrt{p_\ell}}\Big)^{p_\ell}
        \Big(\frac{2}{\sqrt{q_\ell}}\Big)^{q_\ell}
        \|w_{ p_\ell, q_\ell}^{(n)}\|_\sigma^\sharp
        \Big\}
        \nonumber\\
        &\leq&20\,C_\Theta^2 \rho \sum_{L=3}^\infty    (L+1)^2 (B^{(n)})^L
        \nonumber\\
        &\leq&  \frac{2000 C_\Theta^5}{\rho^2} \|\h_1^{(n)}\|_{\xi,\sigma}^3 
        \label{lead-err-1}
\eeqn 
for $n\geq1$, where we recall that
\eqn
        B^{(n)}:=\frac{C_\Theta}{\rho(1-2\xi)^2 }\|\h_1\|_{\xi,\sigma}
        \leq\frac{4C_\Theta }{\rho}\|\h_1^{(n)}\|_{\xi,\sigma}
\eeqn
for $\xi<\frac14$, and use $\sum_{L=3}^\infty    (L+1)^2 B^L<25 B^3$
for $B<\frac{1}{10}$. 
Since 
\eqn
		\|\h_1^{(n)}\|_{\xi,\sigma}\leq \e_n\;\leq\;(\ccontr\rho)^n\e_0\;,
\eeqn
we get 
\eqn
		(~\ref{lead-err-1}) \leq \frac{2000 C_\Theta^5}{\rho^2} (\ccontr\rho)^{3n}\e_0^3 \;,
\eeqn
and we recall that $\e_0=O(g)$.

Moreover,
\eqn
		\sum_{p\geq2}\Big\|\,V^{(L=2;n)}_{\underline{0,p,0,p}}\,\Big\|_{\Tspace}
		&\leq&20\,\frac{C_\Theta^3}{\rho}
        \sum_{p\geq2}\|w_{0,p}^{(n)}\|_\sigma^\sharp\|w_{p,0}^{(n)}\|_\sigma^\sharp
        \nonumber\\
        &\leq&20\,\frac{C_\Theta^3}{\rho}\|\h_2^{(n)}\|_{\sigma,\xi}^2
        \nonumber\\
        &\leq&20\,\frac{C_\Theta^3}{\rho}\e_n^4\;.
        \label{lead-err-2}
\eeqn
It is clear that in the case $L=2$, only  $V^{(L=2;n)}_{\underline{0,p,0,q}}$
with $p=q$ are non-zero. 
This proves (~\ref{rig-error-bd-2}). 

The bound (~\ref{rig-error-bd-1}) for $n=-1$ is obtained similarly, but uses
the modifications described in Section  {~\ref{firstdecsubsect}} (replacement of operator norm bounds on the
interaction operators by relative bounds of $\tilde W_{(-1)}$ with respect to the free Hamiltonian
$\tilde T_{(-1)}$).
\endprf

\begin{lemma}
\eqn
		\sum_{n=1}^\infty\rho^{-n}\big|\partial_{X^\parallel}^2\Delta\gamma_{(n)}\big|
		&\leq& c\rho g^2 \;,
\eeqn
for $\rho<\frac{1}{2\ccontr^2}\ll1$.
\end{lemma}

\prf
In the case $n=1$, one can explicitly verify that
\eqn
		\Big|\partial_{X^\parallel}^2\Big|_{\spvar\rightarrow\unull}
		V_{\underline{0,1,0,1}}^{(L=2;1)}\Big|
		\leq c \e_1^2
\eeqn
(no inverse factors of $\rho$).
The calculation is essentially the same as for $C_0$.
Combining (~\ref{lead-err-1}) and (~\ref{lead-err-2}) for $n=1$, we thus obtain
\eqn
		\Big\|\partial_{X^\parallel}^2\Big|_{\spvar\rightarrow\unull}
		\Delta\gamma_{(1)}\Big\|_{\Tspace}\leq O(\e_1^2)+O(\rho^{-2}\e_1^3) \;.
\eeqn
For $n\geq2$, we find
\eqn
		|\partial_{X^\parallel}^2\Delta\gamma_{(n)}|
		&\leq&
		\sum_{L=2}^\infty(-1)^{L-1}
        \sum_{p_1,q_1,\dots,p_L,q_L: \atop p_\ell+q_\ell\geq1}
        \Big|\partial_{X^\parallel}^2 V_{\underline{0,p,0,q}}^{(L;n)}
        \Big[\h^{(n)}[e_{(n,\infty)}]\,\Big|\,\spvar\Big]\Big|_{p=0}\Big|        
		\nonumber\\
        &\leq&20\,C_\Theta^2 \rho \sum_{L=2}^\infty    (L+1)^2 (B^{(n)})^L
        \nonumber\\
        &\leq&  \frac{ 960 C_\Theta^4 }{\rho} \|\h_1^{(n)}\|_{\xi,\sigma}^2 
		\nonumber\\
		&\leq& \frac{960 C_\Theta^4 }{\rho}  \e_n^2 \;,
\eeqn
by the same arguments as in the proof of the previous lemma, or 
in Section  {~\ref{codim2contrproofsubsubsect-2}}. 

Thus, recalling that $\e_n=(\ccontr\rho)^n\e_0$,
\eqn
		\sum_{n=1}^\infty\rho^{-n}\big|\partial_{X^\parallel}^2\Delta\gamma_{(n)}\big|
		&\leq& c \sum_{n=1}^\infty\rho^{-n}\e_n^2
		\nonumber\\
		&=&c \sum_{n=1}^\infty(\ccontr^2\rho)^{n}\e_0^2
		\nonumber\\
		&<&c\rho\e_0^2 \;=\; O(\rho g^2)
\eeqn
where $\rho<\frac{1}{2\ccontr^2}\ll1$.
\endprf

\subsubsection{The denominator of (~\ref{part2Eidenty-2})}

Collecting the above results, we find
\eqn
		\Big|\sum_{n\geq-1}\rho^{-n} \partial_{X^\parallel}^2\Big|_{\spvar\rightarrow\unull}
		\Delta\gamma_{(n)}-\tilde c_2\Big|
		&\leq&|C_0+C_1-\tilde c_2|
		\nonumber\\
		&+&\sum_{n=-1,0}\Big\|\Delta \gamma_{(n)}+V_{\underline{0,1,0,1}}^{(L=2;n)}
		\Big\|_{\Tspace} 
		\nonumber\\
		&+&\sum_{n\geq1}\rho^{-n}\big|\,\partial_{X^\parallel}^2\Big|_{\spvar\rightarrow\unull}
		\Delta\gamma_{(n)}\,\big|
		\nonumber\\
		&\leq&O(\rho g^2)+O(\rho^{-2}g^3) 
\eeqn  
for $\rho\leq\frac{1}{(2\ccontr)^{\frac32}}\ll1$.

\subsubsection{The numerator of (~\ref{part2Eidenty-2})}
 
One can straightforwardly verify that
\eqn
		\partial_{X_0}\Big|_{\spvar\rightarrow\unull}V_{\underline{0,1,0,1}}^{(L=2;n)}
		\leq O(g^3)
\eeqn
for $n=-1$ and $n=0$, and 
\eqn
		\big|\partial_{X_0}\Big|_{\spvar\rightarrow\unull}(\Delta\gamma_{(n)}
		-V_{\underline{0,1,0,1}}^{(L=2;n)})\big|&<&
		\Big\| \Delta\gamma_{(n)}
		-V_{\underline{0,1,0,1}}^{(L=2;n)}\Big\|_{\Tspace}
		\nonumber\\
		&<&c\rho^{-2}g^3
\eeqn
from Lemma {~\ref{gamma-V-diff-lemma-1}}.
On the other hand, for $n\geq1$,
\eqn
		\big|\partial_{X_0}\Big|_{\spvar\rightarrow\unull}\Delta\gamma_{(n)}\big|
		&\leq&\Big\|\sum_{L=2}^\infty(-1)^{L-1}
        \sum_{p_1,q_1,\dots,p_L,q_L: \atop p_\ell+q_\ell\geq1}
        V_{\underline{0,p,0,q}}^{(L;n)}
        \Big[\h^{(n)}[e_{(n,\infty)}]\,\Big|\,\spvar\Big]\Big|_{p=0}\Big\|_{\Tspace}
        \nonumber\\
        &\leq&20\,C_\Theta^2 \rho \sum_{L=2}^\infty    (L+1)^2 (B^{(n)})^L
        \nonumber\\
        &\leq&  \frac{960 C_\Theta^4}{\rho} \|\h_1^{(n)}\|_{\xi,\sigma}^2 
        \;,
\eeqn
for $\xi<\frac14$, and using $\sum_{L=2}^\infty    (L+1)^2 B^L<12 B^2$
for $B<\frac{1}{10}$. 
In conclusion,
\eqn
		\sum_{n=0}^\infty \big|\partial_{X_0}\Big|_{\spvar\rightarrow\unull}\Delta\gamma_{(n)}\big|
		&\leq&O(\rho^{-2}g^3)+  \frac c\rho\sum_{n=1}^\infty (\ccontr\rho)^{2n} \e_0^2
		\nonumber\\
		&=&O(\rho^{-2}g^3)+\frac c\rho(\ccontr\rho)^{2}\e_0^2
		\nonumber\\
		&=&O(\rho g^2) +O(\rho^{-2}g^3)\;.
\eeqn
for $\rho\leq\frac{1}{ 2\ccontr }\ll1$.

\subsubsection{Determination of the renormalized mass}

In conclusion, 
\eqn
		m_{ren}(0,0)&=&
		\frac{1+\sum_{n=0}^\infty \partial_{X_0}\Big|_{\spvar\rightarrow\unull}\Delta\gamma_{(n)} }
		{1+\sum_{n=0}^\infty \rho^{-n}
		\partial_{X^\parallel}^2\Big|_{\spvar\rightarrow\unull}\Delta\gamma_{(n)} }
		\nonumber\\
		&=&\frac{1+O(\rho g^2)}{1- \frac{8\pi }{3}g^2\tilde c_2  +O(\rho g^2) +
		O(\rho^{-2}g^3)}
		\nonumber\\
		&=&
		1+ \frac{8\pi }{3}g^2\tilde c_2  +O( g^{\frac73})  \;,
\eeqn
where the bounds have been optimized by choosing $\rho=g^{\frac13}$.
Based on the isospectral renormalization group, 
we have here obtained rigorous error bounds with explicitly computable constants.
\endprf

In the same spirit, 
it is possible, by use of the isospectral renormalization group, to determine 
the renormalized mass to any given level of precision, with
rigorous error bounds and explicitly computable constants.

\subsection*{Acknowledgements}
J.F. thanks V.B. for hospitality at the University of Mainz.
I.M.S. was supported in part by NSF grant DMS-0400526.
T.C. was supported by a Courant Instructorship while being at
the Courant Institute, in part by a
grant of the NYU Research Challenge Fund, and in part by NSF grant DMS-0407644.
T.C. thanks V.B and the Department of Mathematics at the University of Mainz,
J.F. and the Institute for Theoretical Physics at ETH Z\"urich,
and I.M.S. and the Departments of Mathematics at the Universities of Toronto and  
Notre Dame, for their kind hospitality.

\parskip = 0 pt
\parindent = 0 pt

\end{document}